%% file: JME-13-004_temp.tex
\pdfoutput=1

\documentclass[11pt,twoside,a4paper,cmspaper,final,collab]{cms-tdr}
\def\svnVersion{389632}\def\cmsCernNoTag{CERN-PH-EP/2015-305}\def\cmsCernDate{\today}\def\cmsMessage{Published in the Journal of Instrumentation as \href{http://dx.doi.org/10.1088/1748-0221/12/02/P02014}{\doi{10.1088/1748-0221/12/02/P02014.}}}

\begin{document}\cmsNoteHeader{JME-13-004}

\hyphenation{had-ron-i-za-tion}
\hyphenation{cal-or-i-me-ter}
\hyphenation{de-vices}
\RCS$Revision: 389621 $
\RCS$HeadURL: svn+ssh://svn.cern.ch/reps/tdr2/papers/JME-13-004/trunk/JME-13-004.tex $
\RCS$Id: JME-13-004.tex 389621 2017-02-26 11:49:45Z sordini $
\newlength\cmsFigWidth
\ifthenelse{\boolean{cms@external}}{\setlength\cmsFigWidth{0.85\columnwidth}}{\setlength\cmsFigWidth{0.4\textwidth}}
\ifthenelse{\boolean{cms@external}}{\providecommand{\cmsLeft}{top}}{\providecommand{\cmsLeft}{left}}
\ifthenelse{\boolean{cms@external}}{\providecommand{\cmsRight}{bottom}}{\providecommand{\cmsRight}{right}}

\newcommand{\ptoffset}{\ensuremath{\langle p_{\text{T,offset}}\rangle}\xspace}
\newcommand{\zmmjet}{\ensuremath{\PZ (\to \mu \mu)\text{+jet}}\xspace}
\newcommand{\zeejet}{\ensuremath{\PZ (\to \Pe \Pe)\text{+jet}}\xspace}

\cmsNoteHeader{JME-13-004}
\title{Jet energy scale and resolution in the CMS experiment \texorpdfstring{in $\Pp\Pp$ collisions at 8\TeV}{in pp collisions at 8 TeV}}

\date{\today}

\abstract{
Improved jet energy scale corrections, based on a data sample corresponding to an integrated luminosity of 19.7\fbinv collected by the CMS experiment in proton-proton collisions at a center-of-mass energy of 8\TeV, are presented. The corrections as a function of pseudorapidity $\eta$ and transverse momentum \pt are extracted from data and simulated events combining several channels and methods. They account successively for the effects of pileup, uniformity of the detector response, and residual data-simulation jet energy scale differences. Further corrections, depending on the jet flavor and distance parameter (jet size) $R$, are also presented. The jet energy resolution is measured in data and simulated events and is studied as a function of pileup, jet size, and jet flavor. Typical jet energy resolutions at the central rapidities are 15--20\% at 30\GeV, about 10\% at 100\GeV, and 5\% at 1\TeV. The studies exploit events with dijet topology, as well as photon+jet, $\PZ$+jet and multijet events. Several new techniques are used to account for the various sources of jet energy scale corrections, and a full set of uncertainties, and their correlations, are provided.The final uncertainties on the jet energy scale are below 3\% across the phase space considered by most analyses ($\pt>30\GeV$ and $\vert \eta\vert <5.0$). In the barrel region ($\vert \eta\vert <1.3$) an uncertainty below 1\% for $\pt>30\GeV$ is reached, when excluding the jet flavor uncertainties, which are provided separately for different jet flavors. A new benchmark for jet energy scale determination at hadron colliders is achieved with 0.32\% uncertainty for jets with \pt of the order of 165--330\GeV, and $\vert \eta\vert <0.8$.
}

\hypersetup{%
pdfauthor={CMS Collaboration},%
pdftitle={Jet energy scale and resolution in the CMS experiment in pp collisions at 8 TeV},%
pdfsubject={CMS},%
pdfkeywords={CMS, physics, JES, JEC, JER, jet energy scale, jet energy corrections, jet energy resolution}}

\maketitle
\tableofcontents

\section{Introduction}

The state-of-the-art techniques used in the CMS experiment at the CERN LHC for jet energy scale (JES) and jet energy resolution (JER) calibration are presented,
based on a data sample corresponding to an integrated luminosity of 19.7\fbinv collected in proton-proton collisions at a center-of-mass energy of 8\TeV.
Jets are the experimental signatures of energetic quarks and gluons produced in high-energy processes.
Like all experimentally-reconstructed objects, jets need to be calibrated in order to have the correct energy scale: this is the aim of the jet energy corrections (JEC). The detailed understanding of both the energy scale and the transverse momentum resolution of the jets is of crucial importance for many physics analyses, and a leading component of their associated systematic uncertainties. Improvements made in understanding the JES in the recent years have resulted in very precise measurements of, \eg, the inclusive jet cross section~\cite{Chatrchyan:2012bja,CMS:2011ab,Chatrchyan:2011qta,
Chatrchyan:2014gia,Khachatryan:2015xwa},
and the top quark mass~\cite{Chatrchyan:2013xza,Chatrchyan:2013haa,Chatrchyan:2012cz,Khachatryan:2015hba}.
The JES uncertainties presented here propagate to uncertainties of 2--4\% in the jet
cross sections in the central region, and of $\pm 0.35\GeV$ in the top-quark mass determination.

The results in this paper are reported for jets reconstructed with the particle-flow (PF) method \cite{CMS-PAS-PFT-09-001, CMS-PAS-PFT-10-001} using the anti-$\kt$ algorithm \cite{Cacciari:2008gp} with distance parameter $R=0.5$.
The jet energy corrections are calculated using a detailed Monte Carlo (MC) simulation of the detector, and are then adjusted for data using a combination of several channels and data-driven methods. The JEC successively correct for the offset energy coming from multiple proton-proton collisions in the same and adjacent beam crossings (pileup), the detector response to hadrons, and residual differences between data and MC simulation as a function of the jet pseudorapidity $\eta$ and transverse momentum \pt.
The jet \pt is corrected up to the so-called particle-level jets clustered from stable (decay length $c\tau>1\unit{cm}$) and visible (excluding neutrinos) final-state particles.

\begin{figure}[htbp!]
\includegraphics[width=0.32\textwidth]{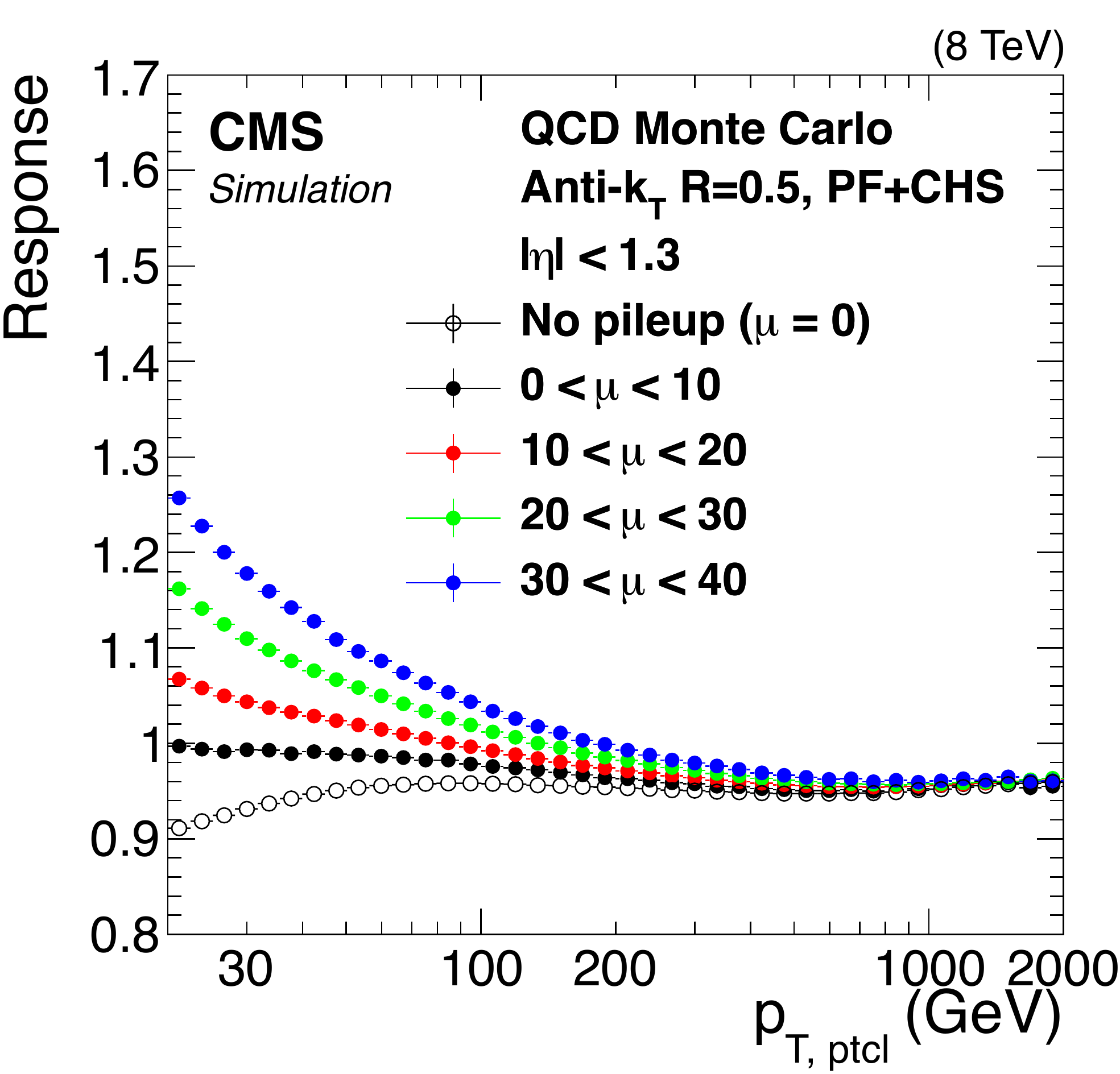}
\includegraphics[width=0.32\textwidth]{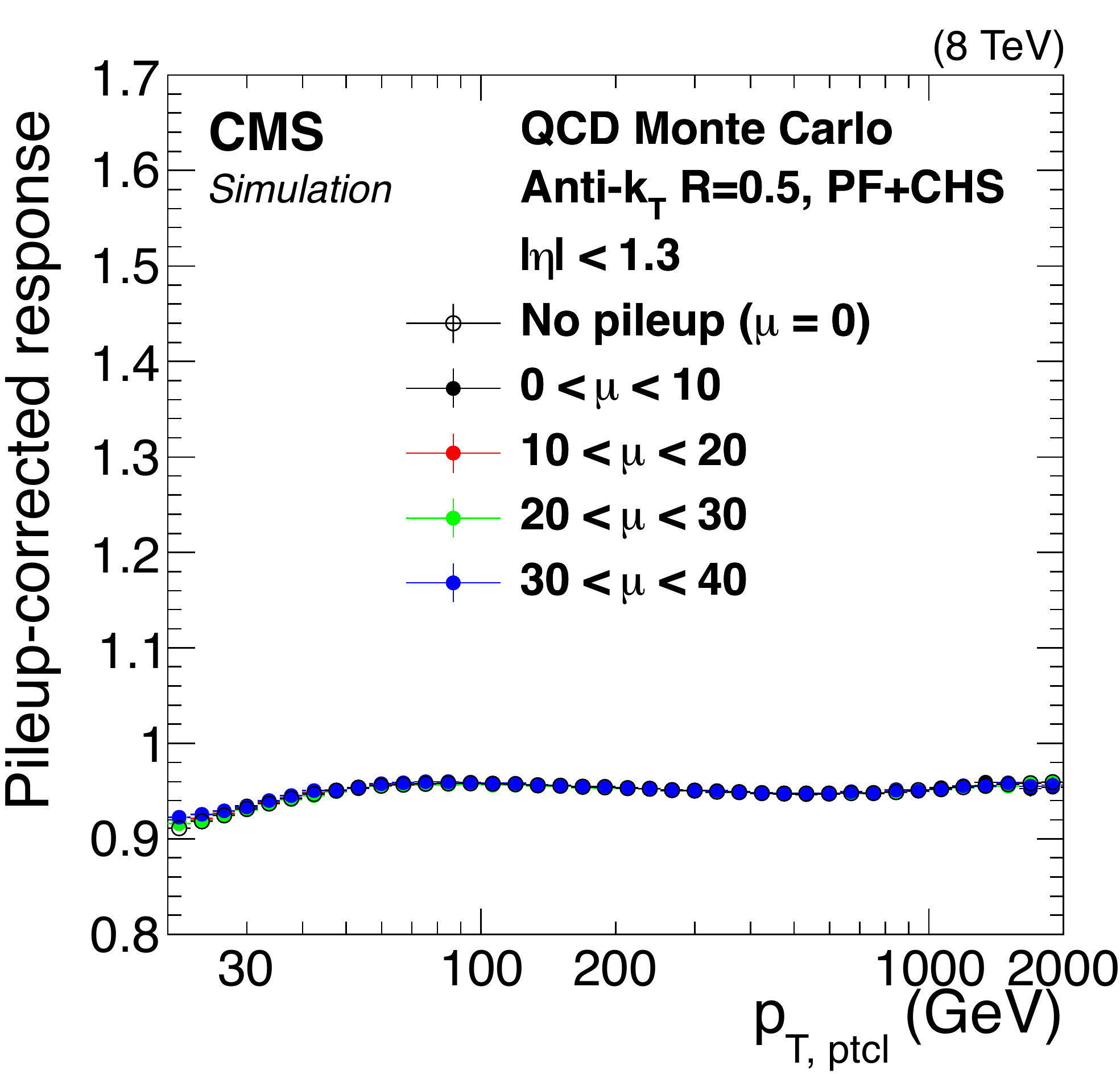}
\includegraphics[width=0.32\textwidth]{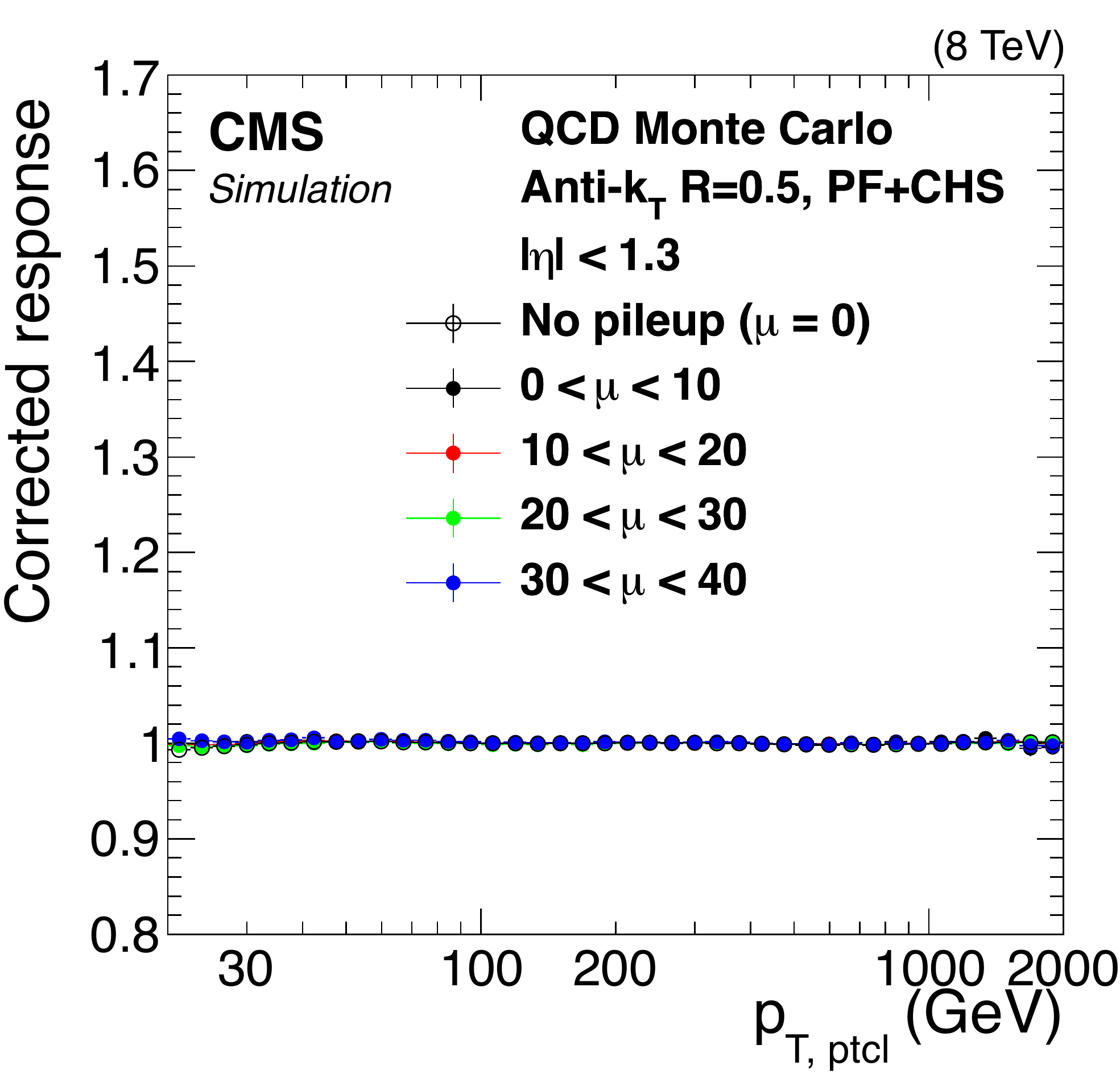}
\caption{Average value of the ratio of measured jet \pt to particle-level jet $p_\text{T, ptcl}$ in QCD MC simulation, in bins of $p_\text{T, ptcl}$, at various stages of JEC:
before any corrections (left),
after pileup offset corrections (middle),
after all JEC (right). Here $\mu$ is the average number of pileup interactions per bunch crossing.
}\label{fig:jecsteps}
\end{figure}

 Corrections depending on jet flavor (for quarks: $\PQu$ and $\PQd$, $\PQs$, $\PQc$ and $\PQb$; and for gluons) and jet
distance parameter $R$ are also presented.  The uncertainties affecting the JES determination are discussed,
and a full set of uncertainties and their correlations are provided.
Figure~\ref{fig:jecsteps} shows the jet response at the various stages of JEC for jets (produced in quantum chromodynamics (QCD) hard-scattering processes) measured at central pseudorapidities ($\vert \eta\vert <1.3$): for each bin in $p_\text{T, ptcl}$, the jet response is defined as the average value of the ratio of measured jet \pt to particle-level jet $p_\text{T, ptcl}$. The response is shown before any correction, after correcting for the effect of pileup, and after all stages of corrections, that will be detailed in the following. Distributions corresponding to different average numbers of pileup interactions per bunch crossing ($\mu$) are shown separately, to display the dependence of the response on the pileup.

The jet \pt resolution, measured after applying JEC, is extracted in data and simulated events. It is studied as a function of pileup, jet size $R$, and jet flavor. The effect of the presence of neutrinos in the jets is also studied.
The typical JER is 15--20\% at 30\GeV, about 10\% at 100\GeV, and 5\% at 1\TeV at central rapidities.

The general principles behind the methods of extraction of the JES,
and the reasons why the JES obtained with the PF algorithm is different from unity,
are discussed. The results and methods are compared to previous CMS studies done for 7\TeV proton-proton collisions ~\cite{JEC_JINST}. Several new techniques are introduced in this paper to account
for \pt-dependent pileup offset, out-of-time (OOT) pileup, initial- and final-state radiation (ISR+FSR), and $\PQb$-quark jet
($\PQb$-jet) flavor response. We also add the information from multijet balancing ~\cite{Aad:2011he} and introduce a new technique that uses it as part of a global \pt-dependent fit which constrains the uncertainties by using their correlations between channels and methods.

Pileup collisions result in unwanted calorimetric energy depositions and extra tracks.
The charged-hadron subtraction (CHS, Section~\ref{subsec:pumitigation}) reduces these effects by removing tracks identified as originating from pileup vertices.
The results in this paper are reported for jets reconstructed with and without CHS.

The JEC are extracted for jets with $\pt>10\GeV$ and $\vert \eta\vert <5.2$, with uncertainties less than or about $3\%$ over the whole phase space. The minimum JES uncertainty of $0.32\%$ for jets with $165 < \pt < 330\GeV$ and $ \vert \eta \vert <0.8$, when excluding sample-dependent uncertainties due to jet-flavor response and time-dependent detector response variations, surpasses the precision of previous JES measurements
at the Tevatron \cite{CDF_JEC,D0_JEC} and the LHC \cite{JEC_JINST,ATLAS_JEC}.

\subsection*{Outline of the paper and overview of the corrections}

The CMS detector and reconstruction algorithms are briefly described in Section~\ref{sec:cms}. The data and MC samples used throughout this document, together with the different selection criteria, are detailed in Section~\ref{sec:datasets}.

The pileup offset corrections, discussed in Section~\ref{Sec:pileup}, are determined from the simulation of a sample of dijet events processed with and without pileup overlay. They are parameterized as a function of offset energy density $\rho$, jet area $A$, jet pseudorapidity $\eta$, and jet transverse momentum \pt. Corrections for residual differences between data and detector simulation as a function of $\eta$ are determined using the random cone (RC, Section~\ref{sec:jetarea}) method in zero-bias events (Section~\ref{sec:event_sel}). The pileup offset corrections are determined both before and after CHS, which removes tracks identified as originating from pileup vertices.

The simulated jet response corrections are determined with a CMS detector simulation based on {\GEANTfour}~\cite{Agostinelli:2002hh} combined with the {\PYTHIA 6.4} \cite{Sjostrand:2006za} tune Z2* \cite{GEN-14-001}, as discussed in Section~\ref{sec:mctruth}.
The corrections are determined for various jet sizes. The default corrections are provided for the QCD dijet flavor mixture as a function of \pt and $\eta$. Uncertainties arising from the modeling of jet fragmentation are evaluated with {\HERWIGpp}~2.3~\cite{Bahr:2008pv} tune~EE3C~\cite{Gieseke:2012ft}, and uncertainties from the detector simulation are evaluated with the CMS fast simulation \cite{Giammanco:2014bza}.

The residual corrections for data are discussed in Section~\ref{Sec:l2l3res}.
The $\eta$-dependent corrections are determined with dijet events, relative to a jet of similar \pt in the barrel reference region $ \vert \eta \vert <1.3$. These corrections include a \pt dependence of the JES relative to the JES of the barrel jet for $\pt>62\GeV$ and up to about 1\TeV, the limit of available dijet data. The absolute scale, together with its \pt dependence within $ \vert \eta \vert <1.3$ for $30<\pt<800\GeV$, is measured combining photon+jet, \zmmjet\ and \zeejet\ events. The \pt dependence at $\pt>800\GeV$ is constrained with multijet events.
Detailed studies are performed to correct for biases in the data-based methods
due to differences with respect to the MC simulation in ISR+FSR as well as in jet \pt resolution.

\begin{figure}[htbp!]
\begin{center}
\includegraphics[width=1.15\textwidth]{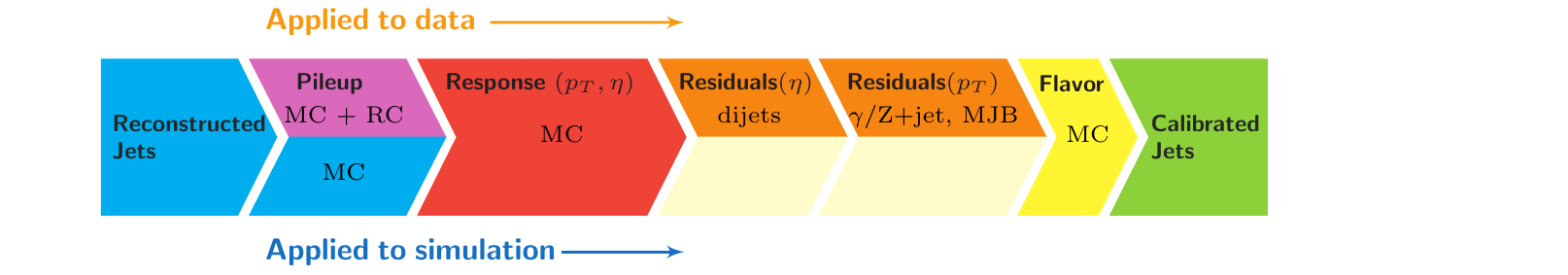}
\caption{Consecutive stages of JEC, for data and MC simulation. All corrections marked with MC are derived
from simulation studies, RC stands for random cone, and MJB refers to the analysis of multijet events.} \label{fig:jecfactorized}
\end{center}
\end{figure}

The optional jet-flavor corrections derived from MC simulation are discussed in Section~\ref{sec:l5} together with the JEC flavor uncertainty estimates based on comparing {\PYTHIA6.4} and {\HERWIGpp}2.3 predictions. These uncertainties are applicable to data vs. simulation comparisons regardless of whether or not the jet-flavor corrections are applied. The flavor corrections and their uncertainties for $\PQb$-quark jets are checked in data with Z+b events.
The consecutive steps of the JEC are illustrated in Fig.~\ref{fig:jecfactorized}.

The jet \pt resolutions are determined with both dijet and photon+jet events, as discussed in Section~\ref{sec:jer}. The reference resolutions obtained from simulation are parameterized as a function of particle-level jet $p_\text{T, ptcl}$ (defined in Section~\ref{sec:cms}) and average number $\mu$ of pileup interactions in bins of jet $\eta$.
Corrections for differences between data and MC simulation are applied as $\eta$-binned scale factors.

The JES uncertainties, discussed in Section~\ref{Sec:uncert}, are provided in the form of a limited set of sources that allow a detailed statistical analysis of uncertainty correlations.
The final uncertainties are below 1\% across much of the phase space covered by these corrections at $\pt>10\GeV$ and $ \vert \eta \vert <5.2$. This sets a new benchmark for jet energy scale at hadron colliders.

In Section~\ref{sec:jetcomp} we describe additional studies made by investigating the particle composition of reconstructed PF jets. These support the overall conclusions drawn from the determination of residual jet energy corrections to be applied on data.

\section{The CMS detector and event reconstruction}
\label{sec:cms}

The central feature of the CMS apparatus is a 3.8~T superconducting
solenoid of 6~m internal diameter.  Within the field volume are the
silicon tracker, the crystal electromagnetic calorimeter (ECAL), and
the brass and scintillator hadron calorimeter (HCAL).  The muon system is
installed outside the solenoid and embedded in the steel flux-return yoke.
CMS uses a right-handed coordinate system, with the origin at the nominal interaction point, the $z$ axis pointing along the direction of the counterclockwise beam,
the $y$ axis pointing up (perpendicular to the plane of the LHC ring), and the $x$ axis chosen to make a right-handed coordinate system.
The polar angle $\theta$ is measured from the positive $z$ axis, and the azimuthal angle $\phi$ is measured in the $x$-$y$ plane in radians.

The CMS tracker consists of 1\,440 silicon pixel and 15\,148 silicon
strip detector modules, with full azimuthal coverage within $ \vert \eta \vert <$ 2.5.
The ECAL consists of 75\,848 lead
tungstate crystals, which provide coverage in pseudorapidity $\vert
\eta \vert< 1.479$ in the central barrel region and $1.479 <\vert \eta \vert < 3.000$ in the two forward endcap regions.
The HCAL is a sampling calorimeter using alternating layers of brass or steel as
absorber and plastic scintillator as active material, it provides a coverage of $\vert\eta\vert <$ 1.3 in the central
region and $1.3< \vert \eta \vert <3.0$ in the endcap regions.  In the forward region ($3.0<  \vert \eta \vert < 5.0$), a different
calorimeter technology is employed in the hadron forward (HF) detector, which uses the Cherenkov light signals collected by short and
long quartz readout fibers to aid the separation of electromagnetic (EM) and hadronic signals. The muon system
includes barrel drift tubes covering the pseudorapidity range $ \vert \eta \vert <
1.2$, endcap cathode strip chambers ($0.9<  \vert \eta \vert < 2.5$), and
resistive-plate chambers ($ \vert \eta \vert < 1.6$).
A detailed description of the CMS detector can be found in Ref.~\cite{CMSdetector}.

Events in CMS are reconstructed using the PF technique~\cite{CMS-PAS-PFT-09-001, CMS-PAS-PFT-10-001}, which reconstructs and identifies
single particles with an optimized combination of all subdetector information.
To suppress noise in the calorimeters, only cells with energies above a given threshold are considered, this procedure is referred to as ``zero suppression''.
The energy of photons is obtained directly from the ECAL measurement, corrected for zero-suppression effects.
The energy of electrons is determined from a combination of the track momentum at the main interaction vertex, the corresponding ECAL cluster energy, and the energy sum of all bremsstrahlung photons associated with the track.
The energy of muons is obtained from the corresponding track momentum.
The energy of charged hadrons is determined from a combination of the track momentum
and the corresponding ECAL and HCAL energies, corrected for zero-suppression effects,
and calibrated for the nonlinear response of the calorimeters.
Finally, the energy of neutral hadrons is obtained from the corresponding calibrated ECAL and HCAL energies.
In the forward region, energy deposits collected by the HF are considered as electromagnetic or hadronic, depending on the respective energy collected by long and short fibers.
The particles reconstructed with the PF algorithm are jointly referred to as PF candidates. Jets are reconstructed by clustering the PF candidates, and the missing transverse momentum \ptvecmiss is the negative vectorial sum of the transverse momenta of all PF candidates reconstructed in an event.

Interaction vertices are reconstructed using track information only, and the primary interaction vertex
is defined as the vertex with the highest sum of the squared transverse momenta of the tracks associated with it.

The first level (L1) of the CMS trigger system, composed of custom hardware processors, uses information
from the calorimeters and muon detectors to select the most interesting events in a fixed time interval
of less than $4\mus$. The high-level trigger (HLT) processor farm further decreases the event rate from
around 100\unit{kHz} to less than 1\unit{kHz} before data storage.

\subsection{Jet reconstruction}

Jets considered in this paper are reconstructed with the anti-$\kt$ clustering algorithm \cite{Cacciari:2008gp}. The nominal results
are obtained for a jet distance parameter, $R=0.5$, which was used in most CMS analyses of 7 and 8\TeV data. Both the JES and JER are also studied for
different values of the $R$ parameter, on simulated events.

The \textbf{simulated particle-level jets} are built by applying the clustering procedure to all stable (lifetime $c\tau>1\unit{cm}$) particles  excluding neutrinos. The lifetime of heavy hadrons (containing c and b quarks) is shorter than $c\tau=1\unit{cm}$, so their decay products are the particles considered for jet clustering.
The exclusion of neutrinos is a convention adopted by CMS, but it is not universally adopted by all experiments in high-energy physics. Indeed, neutrinos are often included at the particle level, but the response is measured from samples
with negligible neutrino content, leading to practically no difference for inclusive JEC.
The CMS convention allows us to define response in a way that is experimentally accessible and significantly reduces response differences
between heavy-flavor ($\PQc$, $\PQb$) and light-quark ($\PQu$, $\PQd$, $\PQs$) or gluon jets, caused by neutrinos produced in semileptonic decays of heavy-flavor hadrons.
It should be noted that the neutrino fraction leads to an additional systematic uncertainty in the heavy hadrons fragmentation relative to the original $\PQb$ and $\PQc$ quarks that is not included in
JEC systematics, but should be considered in, \eg, measurements of the inclusive $\PQb$-jet cross section or of the top quark mass.
The performance of the corrections for $\PQb$ jets is discussed in Section~\ref{sec:Zb}. The variables referring to
particle-level jets are labeled "ptcl" in this document.

We consider two types of reconstructed jets, depending on how the subdetector information is used: calorimeter jets and PF jets.

The \textbf{calorimeter (CALO) jets} are reconstructed from energy deposits in the calorimeter towers alone. A calorimeter tower consists of one or more HCAL
cells and the geometrically corresponding ECAL crystals.
In the barrel region of the calorimeters, the unweighted sum of one single HCAL cell and 5$\times $5 ECAL crystals form a projective calorimeter tower.
The association between HCAL cells and ECAL crystals is more complex in the endcap regions.
A four-momentum is associated with each tower deposit above a certain threshold, assuming zero mass, and taking
 the direction of the tower position as seen from the interaction point.

The \textbf{PF jets} are reconstructed by clustering the four-momentum vectors of PF candidates. The PF jet momentum and spatial resolutions
are greatly improved with respect to calorimeter jets, as the use of the tracking detectors and high granularity of the ECAL
improves the energy resolution through the independent measurements of
charged hadrons and photons inside a jet, which together constitute ${\approx}85$\% of the average jet energy.
In reconstructing the PF candidate four-momentum, photons are assumed massless and charged hadrons are assigned the charged pion mass.

Calorimeter jets result from a relatively simplistic yet robust approach and were widely used in the early CMS publications.
With the improvement of the understanding of the detector and the commissioning of the reconstruction with data,
the performance of the PF reconstruction has proven to be outstanding and reliable.
The event description and reconstruction is more complete and consistent, and for these reasons, we focus here on the PF jets used in the majority of recent CMS analyses.

\section{Event samples and selection criteria}
\label{sec:datasets}

\subsection{Simulated samples}

Simulated samples are generated for QCD dijet and multijet, $\PZ$+jet, and $\gamma$+jet processes.
A sample with single-neutrino production is simulated as well, to reproduce empty events that only contain pileup and detector noise.
The dijet, $\gamma$+jet and single-neutrino samples are generated with {\PYTHIA6.4}~\cite{Sjostrand:2006za}, using the tune Z2*~\cite{GEN-14-001}.
The $\PZ$+jet and multijet samples are generated with the {\MADGRAPH}~4~\cite{madgraph} program matched with parton showers simulated by {\PYTHIA6.4} tune Z2*.
Additional samples for systematic uncertainty studies are available for QCD dijet and $\PZ$+jet processes, both
generated with {\HERWIGpp}~2.3~\cite{Bahr:2008pv}, tune EE3C~\cite{Gieseke:2012ft}.
The single-neutrino sample is compared to zero-bias data (Section~\ref{sec:l1res}).
The dijet sample is used to simulate the jet response (Section~\ref{sec:mctruth}) and also in comparison to data in the dijet balance analysis (Section~\ref{sec:l2res}).
The $\PZ$+jet and $\gamma$+jet simulated samples are used in comparisons of measured response with the corresponding selected samples of data (Section~\ref{Sec:L3residual}).
The multijet sample is used in the multijet balance analysis (Section~\ref{sec:mjb}).

Additional samples are used for the analysis of events with a $\PZ$ boson and a $\PQb$ jet (Section~\ref{sec:Zb}): the {\MADGRAPH}~4 program,
together with \PYTHIA6.4 for the hadronization, is used to simulate top quark pair, W+jets and Drell--Yan+jets (DY+jets) production;
and the \POWHEG~\cite{Frixione:2007vw} program, together with \PYTHIA6.4 for the hadronization, is used for single top quark samples.
A DY+jets sample produced with \HERWIGpp~2.3 is also used for studies of systematic uncertainties.

All generated samples are processed through the CMS detector simulation, based on {\GEANTfour}~\cite{Agostinelli:2002hh}.
Minimum bias events, generated with \PYTHIA6.4 and tune Z2*, are overlayed to all above samples to simulate the pileup.
As will be detailed in Section~\ref{Sec:pileup},
the MC simulation is reweighted to match the distribution of the average number of pileup interactions in data.

\subsection{Data sets and event selection}
\label{sec:event_sel}

The studies presented in this document use the data collected by the CMS experiment in proton-proton collisions at a center-of-mass
energy of 8\TeV, during the year 2012, corresponding to an integrated luminosity of 19.7\fbinv.
In this section we describe the selection criteria used in the different analyses presented in this paper.
Only data collected during stable-conditions collisions with a fully-functioning detector are considered.
Apart from the zero-bias sample, all data samples are required to fulfill some basic event preselection criteria.
The presence of at least one well-reconstructed primary vertex (PV) is required \cite{Chatrchyan:2014fea}, with at least four tracks considered
in the vertex fit, and with $ \vert z(\mathrm{PV}) \vert <24$ cm, where $z(\mathrm{PV})$ represents the position of the PV along the beam axis.
The radial position of the primary vertex, $r_{xy}$(PV), has to satisfy the condition $r_{xy}(\mathrm{PV})<2$ cm.
Finally, the jets used in the analyses are required to satisfy basic identification criteria (``Jet ID'') \cite{CMS-PAS-JME-10-003}, which on simulation are found to retain more than 99\% of genuine jets, while rejecting most of the misreconstructed jets arising from detector noise or cosmic muons.

\subsubsection*{Zero-bias sample}

The zero-bias sample is collected using a random trigger in the presence of a beam crossing with filled bunches, active during the whole data-taking period with stable collisions conditions and a fully-functioning detector. As these events are not triggered by
any specific energy deposit, they generally do not contain any contribution from hard-scattering processes. The main sources of energy
deposits in zero-bias events are detector noise and pileup. The events in the dataset are weighted, according to the luminosity evolution during the running period, in order to be representative of the average pileup conditions of the datasets used in the analyses presented in this paper.

\subsubsection*{Dijet sample}

The dijet sample, composed of events with at least two jets in the final state, is collected using
dedicated HLTs, which accept the events depending on the value of the average \pt
($p_{\mathrm{T}, \text{ave}}=(p_\text{T, 1st jet}+p_\text{T, 2nd jet})/2$) of the two highest-$\pt$ jets in the event, to ensure an unbiased data set.
The HLT uses a PF reconstruction algorithm
with simplified tracking, and
the jet \pt is corrected for nonuniformity of the energy response as a function of the jet $\eta$ and \pt.
Several $p_{\mathrm{T}, \text{ave}}$ thresholds are available, with different prescale factors. Depending on the value of the highest jet \pt in the event,
only the least prescaled fully efficient HLT is used for the decision of keeping or rejecting the event for further analysis.
Events selected with single-jet triggers are also used for the studies of jet composition shown in Section \ref{sec:jetcomp}.

The event selection requires at least one of the two leading jets to have $ \vert \eta \vert <1.3$ and the angular separation
between the two leading jets in the $(x,y)$ plane to be $ \vert \Delta\phi_\text{1st jet, 2nd jet} \vert >2.7$.
Events are rejected if there is any third jet with $p_\text{T, 3rd jet}>5\GeV$
not fulfilling the condition $p_\text{T, 3rdjet}/p_{\mathrm{T}, \text{ave}}=\alpha<0.2$. As will be explained in Section~\ref{sec:l2res}, the
results are studied as a function of the $\alpha$ cut from $\alpha<0.4$ to $\alpha<0.1$ in order to correct for biases from ISR+FSR.

\subsubsection*{The $\PZ$+jet sample}

The \zmmjet~and \zeejet~samples are collected using single-lepton HLTs
with various \pt thresholds.
Events are required to contain either two opposite-sign muons or two opposite-sign electrons, fulfilling standard tight isolation and identification requirements~\cite{cmsmuons,Khachatryan:2015hwa}, with $\abs{\eta}<2.3$ and $\pt>20\GeV$.
The dilepton $(\ell\ell)$ system is required to have $p_{\mathrm{T}, ll}>30\GeV$ and $|m_{ll}-m_\Z|<20\GeV$, where $m_\Z$ is the mass of the $\PZ$ boson. The leading jet in the event
is required to have $\abs{\eta}<1.3$ and $\pt>12\GeV$, and to have a large angular separation in the $(x,y)$
plane with respect to the dilepton system, $|\Delta\phi(\rm Z,1st~jet)|>2.8$.
Events are rejected if there is any second jet with $p_\text{T, 2nd jet}>5\GeV$ not fulfilling the condition $p_\mathrm{T,2nd~jet}/p_\mathrm{T, Z}=\alpha<0.3$. The value of the cut on $|\Delta\phi(\rm Z,1st~jet)|$ is such that it does not bias the distribution of $\alpha$ for $\alpha<0.3$.
As will be explained in Section~\ref{Sec:L3residual}, the requirement on $\alpha$ is tightened from the nominal value of $0.3$ and the results
are studied as a function of its value. In the \zeejet~analysis
an additional requirement is enforced that no electron in the event lie within $\Delta R=\sqrt{\smash[b]{(\Delta\phi)^2+(\Delta\eta)^2}}=0.5$ of a jet.
The $\PZ$+jet selection is also used in Section~\ref{sec:Zb}, with the additional requirement that the jet is tagged as coming from a $\PQb$ quark using the combined secondary vertex tagger \cite{btag}, with a typical tagging efficiency of $70\%$ and a misidentification probability for light-flavor jets of $1\%$.

\subsubsection*{The $\gamma +$jet sample}
The $\gamma +$jet sample is collected with single-photon HLTs with various \pt thresholds and different prescale factors.
Depending on the value of the highest photon \pt in the event, only the least prescaled fully efficient HLT is used for the decision of keeping or rejecting the event for further analysis.
Events are required to contain one, and only one, photon with $\pt>40\GeV$ and $\abs{\eta}<1.3$ that fulfills
the standard tight cut-based photon identification and isolation criteria \cite{Khachatryan:2015iwa}.
The leading jet in the event is required to have $\abs{\eta}<1.3$ and $\pt>12\GeV$ and to have a significant angular separation in the $(x,y)$
plane with respect to the photon, $|\Delta\phi(\gamma,\mathrm{1st~jet})|>2.8$.
Events are rejected if there is any second jet with $p_\text{T, 2nd jet}>5\GeV$ not fulfilling the condition $p_\mathrm{T,2nd~jet}/p_{\mathrm{T}, \gamma}=\alpha<0.3$.
As will be explained in Section~\ref{Sec:L3residual}, the requirement on $\alpha$ is tightened from the nominal value of $0.3$ and the results
are studied as a function of its value.

\subsubsection*{Multijet sample}

The multijet sample is collected with single-jet HLTs with various \pt thresholds and different prescale factors. Depending on the value of the highest jet \pt in the event,  only the least prescaled fully efficient HLT is
used for the decision of keeping or rejecting the event for further analysis.
The event selection is inspired by the analysis described in Ref.~\cite{Aad:2011he}.
Events containing isolated leptons or photons passing standard identification criteria are rejected.
The events are required to have a $\pt>250\GeV$ jet in $\abs{\eta}<1.3$ balanced by a recoil system, composed of two or more low-$\pt$ jets with $25<\pt<750\GeV$, which is within the range
calibrated by the Z$/\gamma$+jet events, and satisfying the condition $p_\text{T, 2nd jet}/p_\mathrm{T, recoil}<0.6$.
The events are also required to have the recoil jets at least
$\Delta\phi(\mathrm{1st~jet, recoil~jet})>1$~radians
away from the leading jet in the transverse plane, and to have the recoil system
back-to-back with the leading jet with $|\Delta\phi(\mathrm{1st~jet, recoil~syst.})-\pi|<0.3$.
As will be explained in Section~\ref{sec:mjb}, all jets with $\abs{\eta}<5$, $\pt>10\GeV$ are considered to be part of the
recoil system; the analysis is also repeated
after changing to $\pt>20$ and 30\GeV the transverse momentum threshold for jets to be considered in the recoil.

\section{Pileup offset corrections}
\label{sec:pileup}\label{Sec:pileup}

The high instantaneous luminosity at the LHC results in multiple proton-proton collisions taking place within a single beam crossing. Such additional pp collisions occurring within the same bunch-crossing as the primary hard interaction produce additional tracks in the tracker and deposit energy in the calorimeters. This contribution is called in-time pileup (IT PU). Due to the finite signal decay time in the calorimeters, the pp collisions occurring in the previous and subsequent beam crossings also contribute to calorimetric energy in the same time window as the primary hard interaction. This contribution is called out-of-time pileup (OOT PU).

The additional contributions to the jet energy and momentum due to pileup are referred to as the "pileup offset", or "offset" in this document. This offset is studied to optimize the subtraction of pileup from the data, with the corrections leading to an improved detector resolution and a more accurate JES.

The observables used for monitoring and correcting pileup are described in Section~\ref{sec:puobs}. The pileup subtraction then proceeds in steps. The OOT PU is mitigated by calorimeter signal processing (Section~\ref{sec:timeslice}), and the IT PU by identifying charged particles originating from pileup vertices and removing them with charged-hadron subtraction (Section~\ref{sec:chs}). The pileup jets are tagged with pileup jet identification (PUJetID) and removed (Section~\ref{sec:pujetid}). The remaining diffuse energy from neutral particles and OOT PU is estimated per event and then subtracted per jet using a calculation of the effective jet area with the extended hybrid jet area method (Section~\ref{sec:jetarea}). The dependence of the particle-level PU offset on jet $\eta$ and \pt for this method is determined from simulation (Section~\ref{sec:pumc}), and the data/simulation offset scale factor is determined from zero-bias data and neutrino gun simulation, with the random cone (RC) method (Section~\ref{sec:rc}). The uncertainties are discussed in Section~\ref{sec:l1uncert} and the results are summarized in Section~\ref{sec:l1summary}.

\subsection{Pileup observables}
\label{sec:puobs}

The amount of pileup present in the event can be estimated by counting the number of good-quality primary vertices $N_\mathrm{PV}$ or by calculating the diffuse offset energy density $\rho$ \cite{fastjet,fastjetB} in the event. It can also be measured using luminosity monitors that estimate the average number of pileup interactions per crossing.

The offset energy density $\rho$ is calculated using the $\kt$ clustering algorithm \cite{KT1,KT2,KT3} with distance parameter $D=0.6$ and $\abs{\eta}<4.7$.
For this calculation, a large number of nonphysical particles (ghosts) with infinitesimal momenta and random direction effectively mapping all the $(\eta,\phi)$ space, is added to the event. When the jet clustering is run on the event, the hard particles in the event are clustered together with such ghosts: a few jets will contain high-momentum particles from the hard-scattering interaction, but most of the jets will be entirely made of ghosts, for which the main real energy contributions come from detector noise and especially pileup.
The offset energy density $\rho$ is defined, in each event, as the median of jet momenta $p_{\mathrm{T},i}$ divided by their area $A_i$, $\rho = \mathrm{ median}(p_{\mathrm{T},i}/A_i)$ ~\cite{Cacciari:2007fd}. For this calculation, no selection on the jet momenta is applied. Using the median instead of the mean makes $\rho$ effectively insensitive to hard jets in the event, and including zero-energy jets composed of only ghost particles reduces bias for low pileup energy densities.
For Run~2, a simpler approach is used to calculate $\rho$, which is evaluated as the median of the energies calculated in a grid of $\eta-\phi$ cells, and does not make use of jet clustering anymore.

The number of good primary vertices $N_\mathrm{PV}$ includes vertices consistent with the luminous region (where the collisions happen) and with a number of degrees of freedom $N_\mathrm{dof}\geq 4$, corresponding to a minimum of four tracks.

The average number of pileup interactions $\mu$ is obtained by multiplying the instantaneous luminosity with the effective minimum bias cross section of $\sigma_\mathrm{ MB}=69.4\unit{mb}$ for 8\TeV (68.0\unit{mb} for 7\TeV) \cite{Chatrchyan:2012nj}. Two detectors are exploited for the luminosity measurement: the hadron forward (HF) calorimeter and the silicon pixel detector. The counting of pixel clusters is used for the offline precision measurement, because of its time stability and very small dependence on pileup. The HF allows for online determination of instantaneous luminosity per bunch crossing. Its results, calibrated offline per luminosity section that corresponds to 23.3~seconds of data, are used for cross-checks \cite{CMS-PAS-LUM-13-001}.

\begin{figure}[htbp!]
\centering
\includegraphics[width=0.49\textwidth]{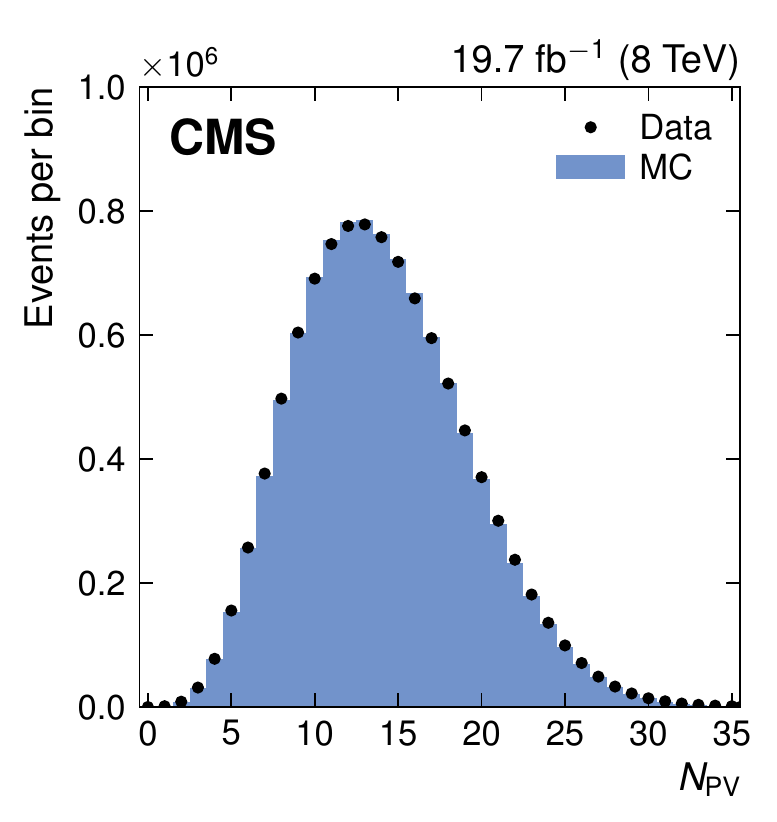}
\includegraphics[width=0.49\textwidth]{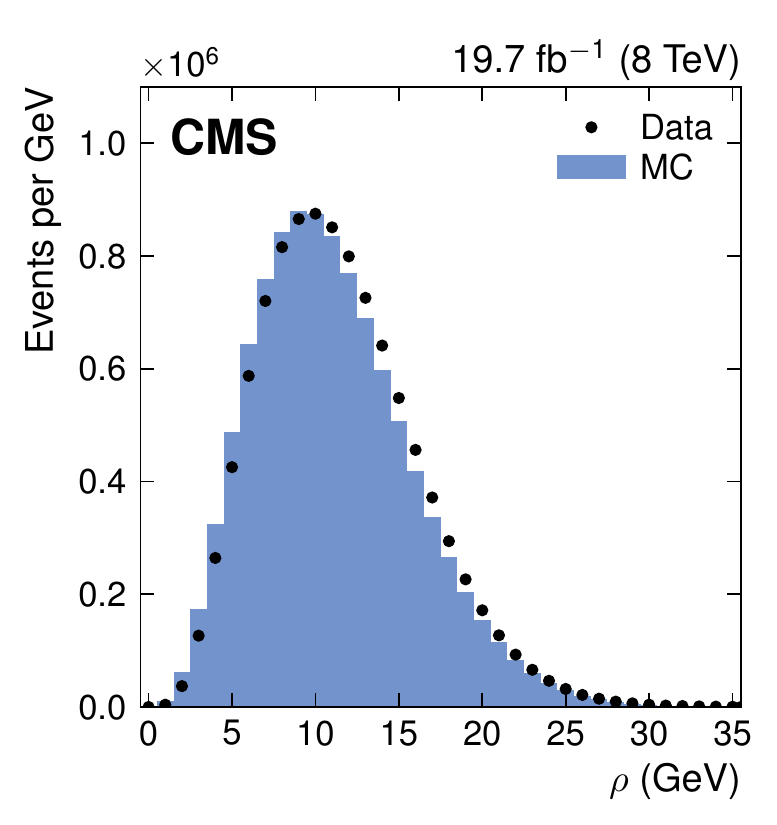}
\caption{\label{fig:rhoandnpv}
Comparison of data (circles) and \PYTHIA6.4 simulation (histograms) for the distributions of the number of reconstructed primary vertices $N_\mathrm{PV}$ (left), and of the offset energy density $\rho$ (right).
}
\end{figure}

The agreement between data and simulation on $N_\mathrm{PV}$ and $\rho$, after reweighting the simulation to match the distribution of the average number of pileup interactions ($\mu$) in data, is shown in Fig.~\ref{fig:rhoandnpv}. The agreement for $N_\mathrm{PV}$ is excellent, while $\rho$ exhibits a small, mostly linear, deviation that is due to different offset densities in data and simulation in the endcap and forward calorimeters.

\begin{figure}[htbp!]
\centering
\includegraphics[width=0.49\textwidth]{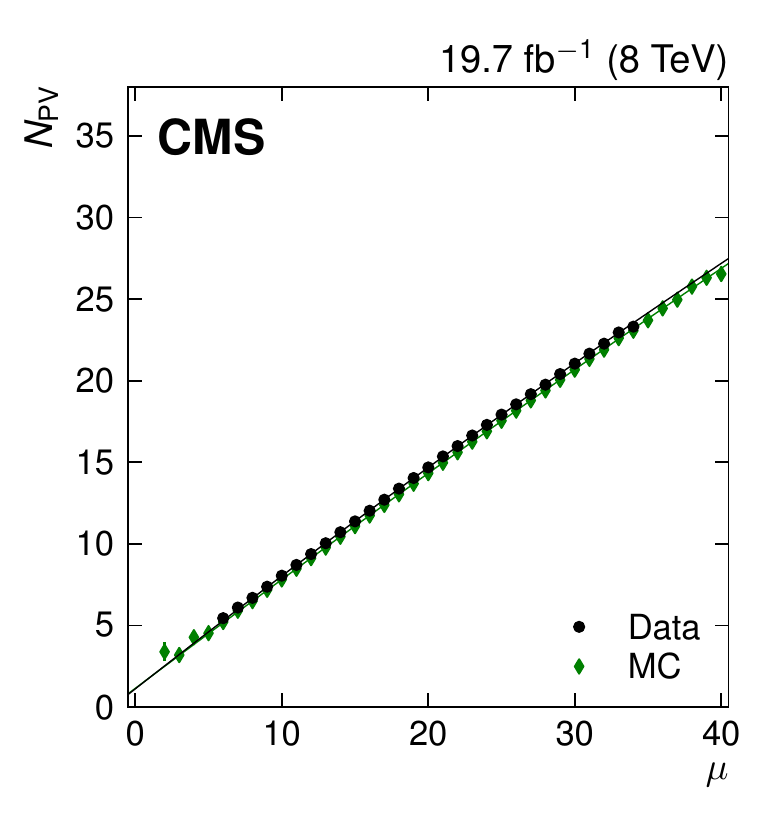}
\includegraphics[width=0.49\textwidth]{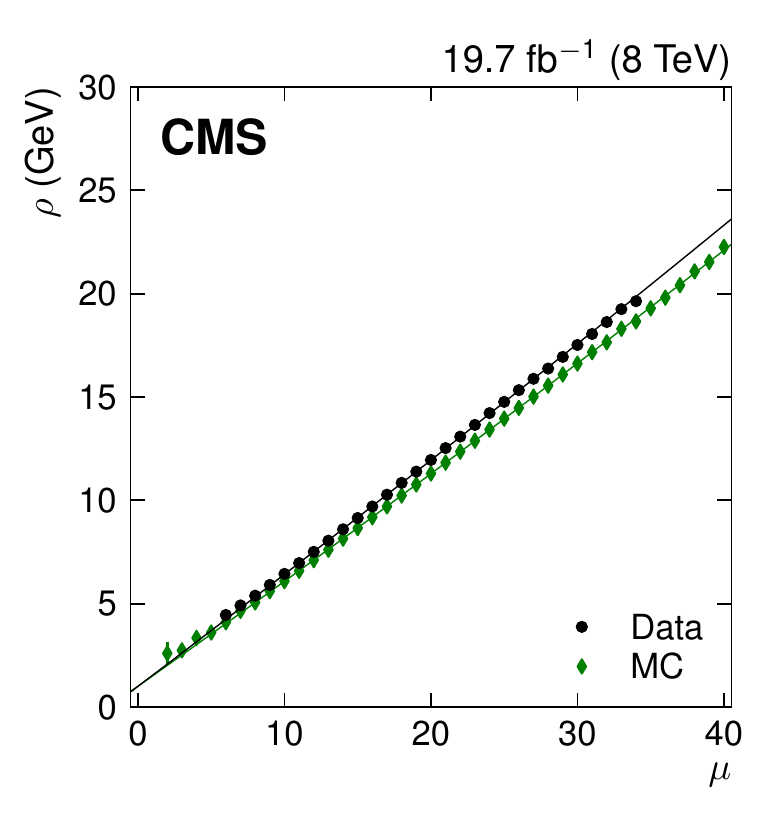}
\caption{\label{fig:rhovsnpv}
Mean of the number of good primary vertices per event, $\langle N_\mathrm{PV}\rangle$ (left), and mean diffuse offset energy density, $\langle\rho\rangle$ (right), versus the average number of pileup interactions per bunch crossing, $\mu$, for data (circles) and \PYTHIA6.4 simulation (diamonds).
}
\end{figure}

Both $N_\mathrm{PV}$ and $\rho$ are very nearly linearly dependent on $\mu$ over the tested range, as shown in Fig.~\ref{fig:rhovsnpv}.
The pileup vertex reconstruction and identification efficiency is about 70\% (while nearly 100\% for hard-scattering events), and IT PU contributes about 0.5\GeV to $\rho$ per interaction.
The vertex $z$ resolution is around 100--300$\mum$ for minimum-bias vertices, improving to tens of microns for hard-scattering events. With a luminous region of root-mean-square (RMS) of about 4\unit{cm} in the $z$ direction, the vertex reconstruction is expected to remain linear up to 100--200 vertices.
The $N_\mathrm{PV}$ versus $\mu$ exhibits a small negative quadratic term due to infrequent merging of pileup vertices, while $\rho$ versus $\mu$ exhibits a similarly small positive quadratic term owing to effects such as effective failed zero-suppression of overlapping calorimeter deposits.
These quadratic terms account for less than 0.5 vertices in $N_\mathrm{PV}$ and 0.5\GeV in $\rho$ at $\mu=20$, respectively.

The correlation between IT PU and OOT PU is modeled by generating the number of interactions for each bunch crossing using a Poisson distribution with the same mean $\mu$. This is a good approximation for 2012 (8\TeV) data, given that the RMS of the bunch-to-bunch variation of $\mu$ within a single luminosity section was only about 8\%.
The value of $N_\mathrm{PV}$ is insensitive to OOT PU, while $\rho$ has a small (${<}5\%$ of the total) OOT PU component with 50\unit{ns} bunch crossings.
The $N_\mathrm{PV}$ variable is highly (${\approx} 94$\%) correlated with the number of IT PU interactions in the event, while $\rho$ is also sensitive to the amount of energy deposited by each interaction, and thus less strongly (${\approx} 85$\%) correlated with the interaction multiplicity.

\subsection{Pileup mitigation}
\label{subsec:pumitigation}
\subsubsection*{Out-of-time pileup}\label{sec:timeslice}

The amount of OOT PU can be reduced by shortening the signal time-integration window and by increasing the separation between bunches.

In HCAL, 68\% of the signal is contained within a 25\unit{ns} time window~\cite{phystdr1}, resulting in about 5\% leakage to a subsequent crossing with 50\unit{ns} bunch spacing and 50\unit{ns} time integration window.
The signal decay time in ECAL is of the order of 100\unit{ns}, but the ECAL reconstruction involves three samples of 25\unit{ns} before the signal and five on the signal, to remove a varying pedestal. This removes OOT PU on average, but with performance depending on the position of the proton bunch within the bunch train, and requiring simulation of up to six preceding bunch crossings ($-300\unit{ns}$). The variation in the offset correction can be up to 10\% in the endcaps when selecting bunches in the front of bunch trains, which represent a small fraction ($<10\%$) of data. The correction is evaluated on the whole dataset and hence this effect averages out.
In HF, the signal is only 10\unit{ns} wide, resulting in negligible OOT PU without any special treatment of the signal.

More advanced techniques are used in Run~2, exploiting the signal timing and pulse shape to fit in-time and out-of-time pulses simultaneously.

\subsubsection*{Charged-hadron subtraction}\label{sec:chs}

The IT PU from charged particles is reduced by identifying which vertex the charged PF candidates originate from, and removing those unambiguously associated with pileup vertices before clustering jets and \ptvecmiss. This method is referred to as charged-hadron subtraction.

The leading primary vertex is chosen based on the largest sum of squares of the tracks transverse momenta ($\sum \vert\pt^\mathrm{track}\vert^2$) associated with the vertex. Subleading PV's, classified as pileup vertices, are required to pass further quality criteria on the compatibility with the luminous region and on their minimum number of degrees of freedom
\begin{equation}
N_\mathrm{dof}=-3+2\sum_{i=1}^\mathrm{nTracks}w_i, \qquad \mathrm{and} \ \ \ w_i\in[0,1],
\end{equation}
where $w_i$ is the weight assigned to the corresponding track by the adaptive vertex fit \cite{Chatrchyan:2014fea}, based on its compatibility with the vertex.
The minimum requirement $N_\mathrm{dof}>4$ corresponds to at least four tracks. Tracks are matched to vertices based on their chi-squared per degree of freedom ($\chi^2/N_\mathrm{dof}$). If $\chi^2/N_\mathrm{dof}<20$ for a vertex, then the track is associated with this and only this vertex. If the track from a charged hadron is associated with a  pileup PV, passing the above quality requirements, it is considered a pileup track, and removed in the CHS procedure. All other tracks, including those not associated with any PV, are kept.

The CHS can remove approximately 50\% of IT PU within the tracker coverage, as illustrated later by the solid red component labeled ``charged hadrons'' in Figs.~\ref{fig:l1pf} and \ref{fig:l1data}. The remaining unassociated charged hadrons are either not pointing to any reconstructed vertex, or are associated with a vertex that did not pass all the quality requirements, or have too large $\chi^2/N_\mathrm{dof}$ for robust vertex association.
The vertex reconstruction and identification inefficiency is about 30\% for pileup vertices, and it is responsible for a large proportion of the unassociated tracks from pileup.

The charged hadrons from PU are typically soft and have an exponentially decreasing \pt distribution, with $\langle \pt \rangle \approx 0.5\GeV$ \cite{dNdeta}.
Many of the unassociated hadrons in contrast have much higher \pt and are often coming from the leading primary vertex, but have too high $\chi^2/N_\mathrm{dof}$ for robust vertex association. This is particularly common for tracks that are of high \pt and therefore very straight and have merged pixel hits within dense jet cores. For jets of several hundred GeV the tracking efficiency within the jet core can fall as low as 60\%, with a correspondingly large increase of the fraction of unassociated tracks.
Future improvements of CHS aimed at removing a higher proportion of pileup tracks, \eg, with more efficient track-vertex association, must therefore maintain a very low misreconstruction rate for tracks from high-$\pt$ jets, or also consider the \pt and local environment of the tracks, as done with the pileup per particle ID ({PUPPI}) method \cite{puppi}.

The PU offset subtraction has been derived with and without CHS, and the later stages of JEC are practically identical after the application of the corresponding offset corrections.
Application of CHS improves the jet \pt resolution, however, as discussed in Section~\ref{sec:jer}.

\subsubsection*{Pileup jet identification}\label{sec:pujetid}

In addition to diffuse energy, PU interactions often generate soft jets with \pt of a few GeV. Overlaying multiple PU interactions in a single beam-crossing leads to nonnegligible probability of two or more of these soft jets overlapping, resulting in hard jets of tens of GeV in \pt, far above the average PU \pt density. These overlaps are referred to as pileup jets, which are particularly problematic for physics analyses as they can pass typical jet \pt requirements, \textit{e.g.}, $\pt>30\GeV$. The pileup jets lack the relatively hard core typically found in hard-scattering jets, and can be identified by using a multivariate analysis (MVA) of the jet shape variables and the fraction of charged particles contributed by pileup vertices. This MVA tool is called PUJetID, can be run on jets with or without CHS, and it is documented in Ref.~\cite{CMS-PAS-JME-13-005}.  For jets in the region $\abs{\eta}<2.5$ and $p_{T}>30$ GeV, the PUJetID efficiency for hard-scattering jets is around $99\%$, at a pileup-rejection of $90-95\%$.

Removing pileup jets can improve the performance of physics analyses, but applying PUJetID has no direct impact on the JEC. PUJetID is currently not used in the JEC measurements to avoid biases arising from the occasional removal of soft jets from the hard-scattering vertex, which affects the ISR+FSR correction. Instead, CHS is used, which indirectly removes most of the jets tagged by PUJetID by significantly lowering their \pt.

\subsection{Hybrid jet area method}\label{sec:jetarea}

The jet area method uses the effective area of the jets multiplied by the average energy density in the event to calculate the offset energy to be subtracted from the jets. This method was introduced in Ref.~\cite{fastjet} and was first used on data in Ref.~\cite{JEC_JINST} with slight modifications to account for the oversubtraction of the underlying event (UE) and for the $\eta$-dependence of the offset. This slightly modified version is referred to as the hybrid jet area method, where the ``hybrid'' in the name derives from the fact that this method combines an $\eta$-dependent average offset $\mathcal{O}(\eta)$ correction versus $N_\mathrm{PV}$, as already used at the Tevatron \cite{D0_JEC}, with the original $\eta$-independent jet area method using only offset \pt density $\rho$ and jet area $A_j$. This is effectively done by replacing $(N_\mathrm{ PV}-1)\mathcal{O}(\eta)$ in the Tevatron method with $(\rho-\rho_\mathrm{UE})(\beta(\eta) A_j)$, where each of the terms $N_\mathrm{PV}$ and $\rho$, $-1$ and $-\rho_\mathrm{ UE}$, and $\mathcal{O}(\eta)$ and $\beta(\eta)\langle A_j\rangle$ have the same basic meaning, which will be detailed in the following.

In this paper we further extend the hybrid method by adding a logarithmic jet \pt dependence. The previous separate UE correction is absorbed in the new $\eta$-dependent constant term.
The full correction formula used as a multiplicative factor for the uncorrected jet transverse momentum $p_\mathrm{T, uncorr}$ at CMS is
\begin{equation}\label{eq:chybrid}
C_\text{hybrid}(p_\text{T, uncorr},\eta,A_j,\rho) = 1 - \frac{\left[\rho_{0}(\eta) + \rho\beta(\eta)\left(1 + \gamma(\eta)\log(p_\text{T, uncorr})\right)\right] A_{j}}{p_\text{T, uncorr}}.
\end{equation}

The input parameters are $p_\text{T, uncorr}$,
jet pseudorapidity $\eta$, jet area $A_j$, and the per-event \pt offset density $\rho$.
In this formula
the parameters $\rho_0(\eta)$, $\beta(\eta)$, and $\gamma(\eta)$ introduce the required shaping of the offset versus $\eta$.
There is no explicit correction for the UE density $\rho_\mathrm{UE}$ as in Ref.~\cite{JEC_JINST}, but that term is effectively absorbed into $\rho_{0}(\eta)$.
Because $\rho\to\rho_\mathrm{UE}$ and $C_\text{hybrid}\to 1$ when $\mu\to 0$, we have $\rho_0(\eta)=-\rho_\mathrm{UE}\beta(\eta)$ at $p_\mathrm{T, uncorr}\to 1\GeV$ in the ideal situation.
The multiplicative factor, $\beta(\eta)$, corrects for the nonuniformity of the IT and OOT PU offsets versus $\eta$, and the residual correction factor, $\gamma(\eta)$, adds their logarithmic jet \pt dependence.

The parameters $\rho_0(\eta)$, $\beta(\eta)$, and $\gamma(\eta)$ are determined from the simulated particle-level offset, and the offset scale factor for the $\rho_0(\eta)$ and $\beta(\eta)$ in data is determined using the random cone method in zero-bias data, as discussed in the following.
The $\rho_0(\eta)$ parameter effectively contains the $\rho_\mathrm{UE}$ for the QCD multijet sample, while $\beta(\eta)$ and $\gamma(\eta)$ parameterize the \PYTHIA6.4 MinBias overlay, which matches data well.
The RC method consists of reconstructing many jets in each event, clustering particles in randomly placed cones, effectively mapping all the $(\eta,\phi)$ space. The average \pt of these jets is a measurement, in each event, of the average energy density that gets clustered in a jet. When the method is applied in events with no contribution from hard scattering, as it is the case for zero-bias events, the main contributions to the jet energies come from noise and pileup. Assuming the noise energy contribution to be negligible with respect to the pileup one, the average \pt of the jets as measured from the RC method indicates the average energy offset due to pileup, for the considered jet algorithm and jet distance parameter.

\subsubsection*{Simulated particle-level offset}\label{sec:pumc}

In simulation, the most direct way to calculate the particle-level offset in jet \pt caused by pileup is to reconstruct the same events with and without pileup overlay and match the reconstructed jets between these samples. This is done on a QCD multijet sample generated with \PYTHIA6.4, tune Z2*.
Some care needs to be taken to reproduce the same signal fluctuations as before the overlaying pileup, to avoid random smearing of jet \pt between these two samples.
All measurements are binned in $\mu$ to decouple pileup reweighting from offset measurement, and to effectively incorporate the correct average amount of OOT PU in the offset correction (OOT and IT PU are correlated through the shared Poisson mean $\mu$). The $\mu$ bins are then mapped to the average measured value of $\langle\rho\rangle$ for parameterizing the correction. Similarly, the \pt dependence of the offset is measured in bins of particle jet \pt ($p_\text{T, ptcl}$) to decouple the offset from the \pt spectrum and JER, before mapping to the average uncorrected measured $\langle p_\mathrm{T, uncorr}\rangle$ for parameterization.

We define the particle-level offset $\langle p_\mathrm{T, offset~ptcl}\rangle$ as the average difference in \pt between matched jets in simulated samples with and without pileup overlay:
\begin{equation}
\label{eq:offsetptcl}
\langle p_\mathrm{T, offset~ptcl}\rangle
(\langle\rho\rangle, \left[ \eta\right], \langle p_\mathrm{T, uncorr}\rangle)
= \langle p_{\mathrm{T},\rm with~PU} - p_{\mathrm{T},\rm without~PU}\rangle
\left[\mu_\mathrm{PU}, \eta, p_\text{T, ptcl} \right].
\end{equation}

The square brackets $[$ $]$ denote the binning variables, while the angle brackets $\langle$ $\rangle$ denote the averages within those bins for the variables that are used to parameterize the corrections. This subtle distinction is made explicit here due to its importance for various observational biases, and due to the fact that the binning and parameterization variables are not the same.
To have an unambiguous particle-level reference, both reconstructed jets are required
to match the same particle jet within a distance less than $\Delta R < R/2$, where $R$ is the jet distance parameter.
The matching efficiency for jets in the without-PU sample to jets in the with-PU sample for $\langle\mu\rangle=20$ is better than 80\% (98\%) for jets of $\pt>10$~(30)\GeV.
In the with-PU sample there is also a large fraction of unmatched jets with $\pt<60\GeV$ that are due to pileup.

\begin{figure}[htbp!]
\includegraphics[width=0.48\textwidth]{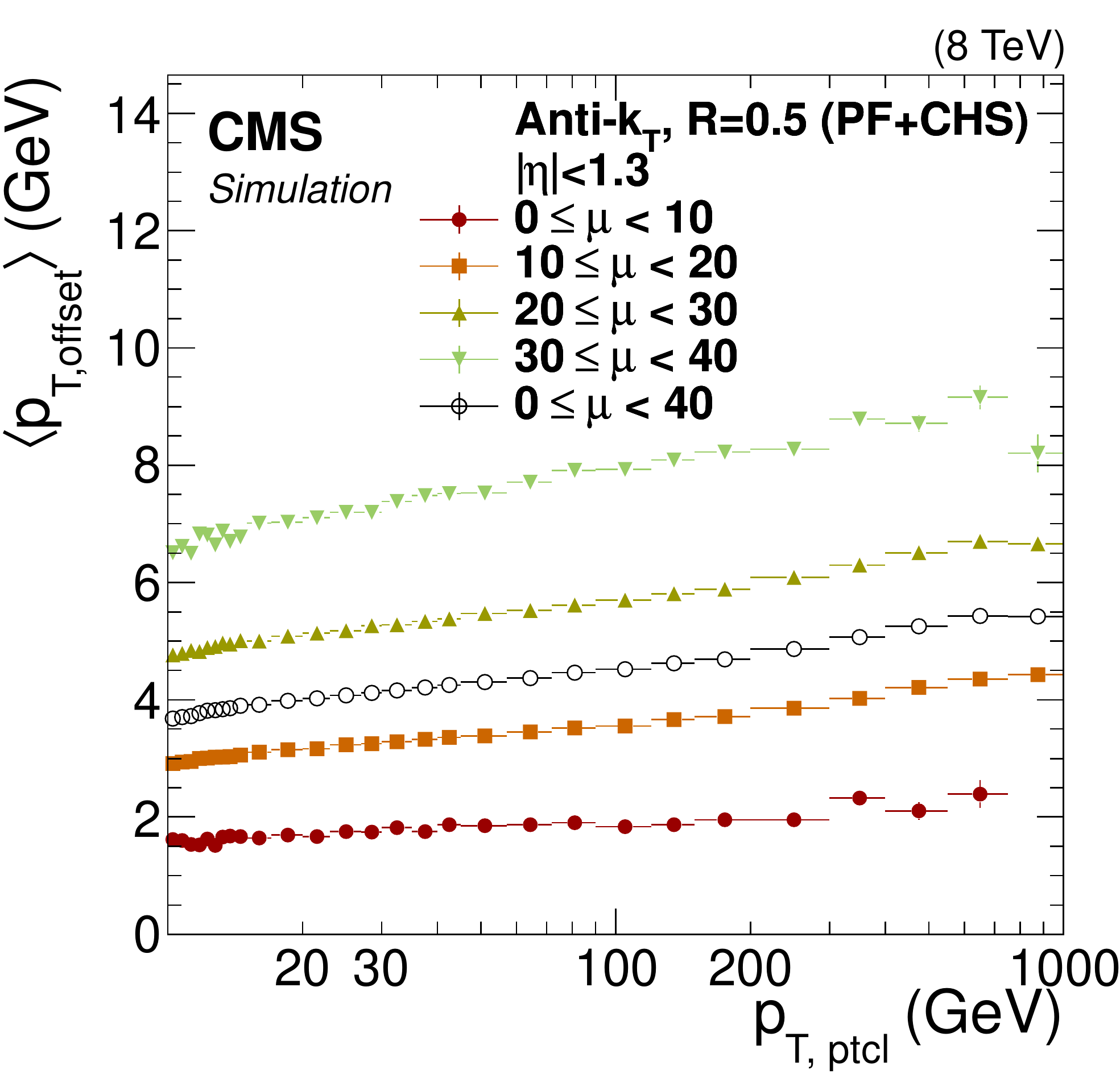}
\includegraphics[width=0.48\textwidth]{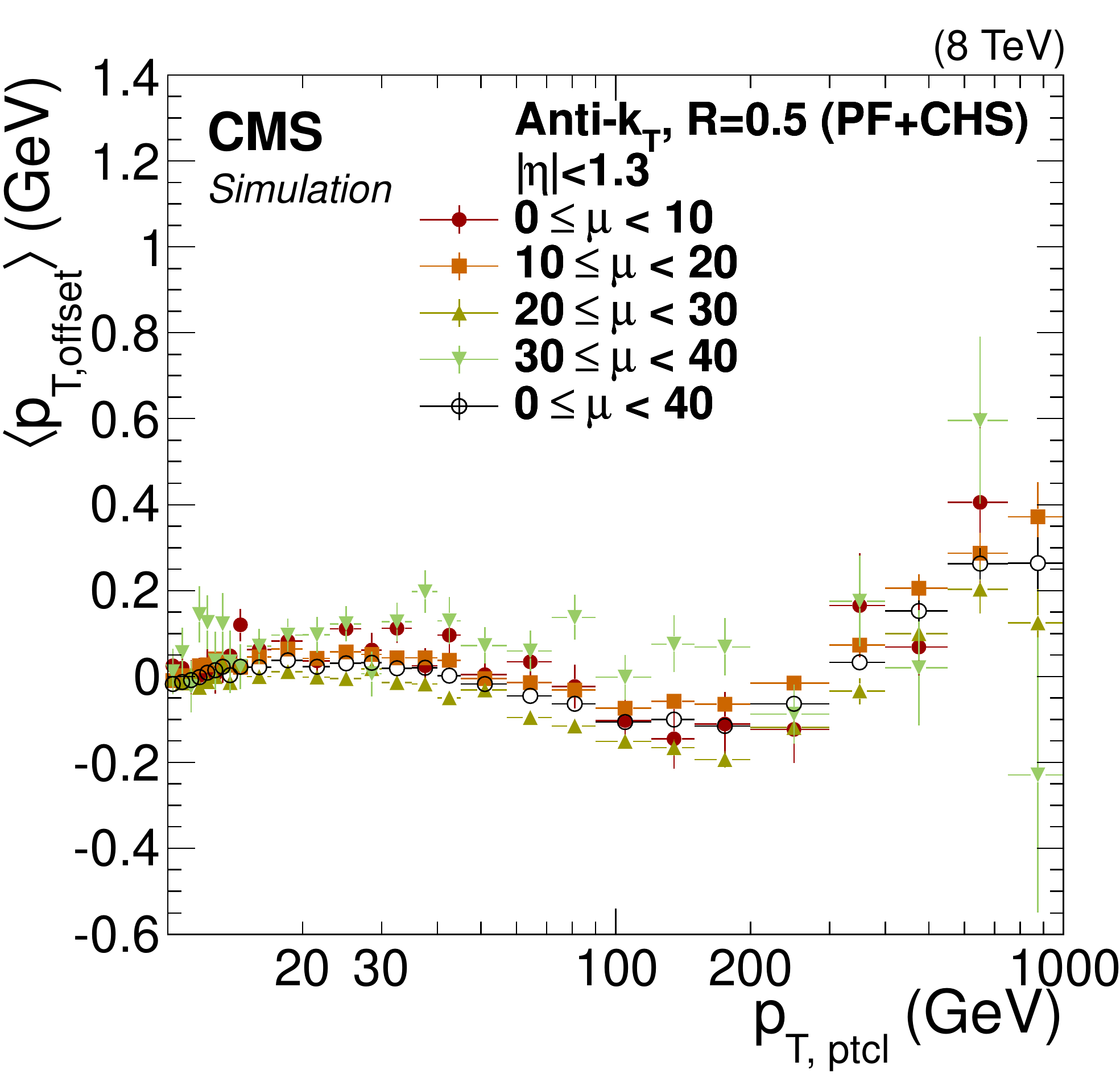}
\caption{\label{fig:offset}
Simulated particle-level offset $\langle p_\mathrm{T, offset~ptcl}\rangle$ defined in Eq.~(\ref{eq:offsetptcl}) (left), and residual offset after correcting for pileup with Eq.~(\ref{eq:chybrid}) (right) for $\abs{\eta}<1.3$, versus particle jet \pt, for different values of average number of pileup interactions per bunch crossing $\langle\mu\rangle$.
}
\end{figure}

The simulated particle-level offset $\langle p_\mathrm{T, offset~ptcl}\rangle$ is parameterized as a function of offset density $\rho$ and jet $\eta$, $p_\mathrm{T, uncorr}$ and area $A_j$
to obtain the $\rho_0(\eta)$, $\beta(\eta)$ and $\gamma(\eta)$ used in Eq.~(\ref{eq:chybrid}), where $C_\text{hybrid} = 1 - \langle p_\mathrm{T, offset~ptcl}\rangle/p_\mathrm{ T, uncorr.}$.
The particle-level simulated offset versus particle jet \pt is shown in Fig.~\ref{fig:offset} (left) for $\vert\eta\vert<1.3$.
The relative slope in offset is parameterized by a logarithmic \pt dependence and is reasonably independent of the level of pileup in the event, while the offset versus $\rho$ is assumed linear.
The resultant level of pileup after applying the corrections is presented in Fig.~\ref{fig:offset} (right), showing the effect of the subtraction. The results are consistent with the absence of additional pileup energy within about 0.2\GeV for the full sample.
For $\mu>30$, small residual offset is visible due to a small unparameterized quadratic dependence of offset on $\rho$.

\begin{figure}[htbp!]
\centering
\includegraphics[width=0.48\textwidth]{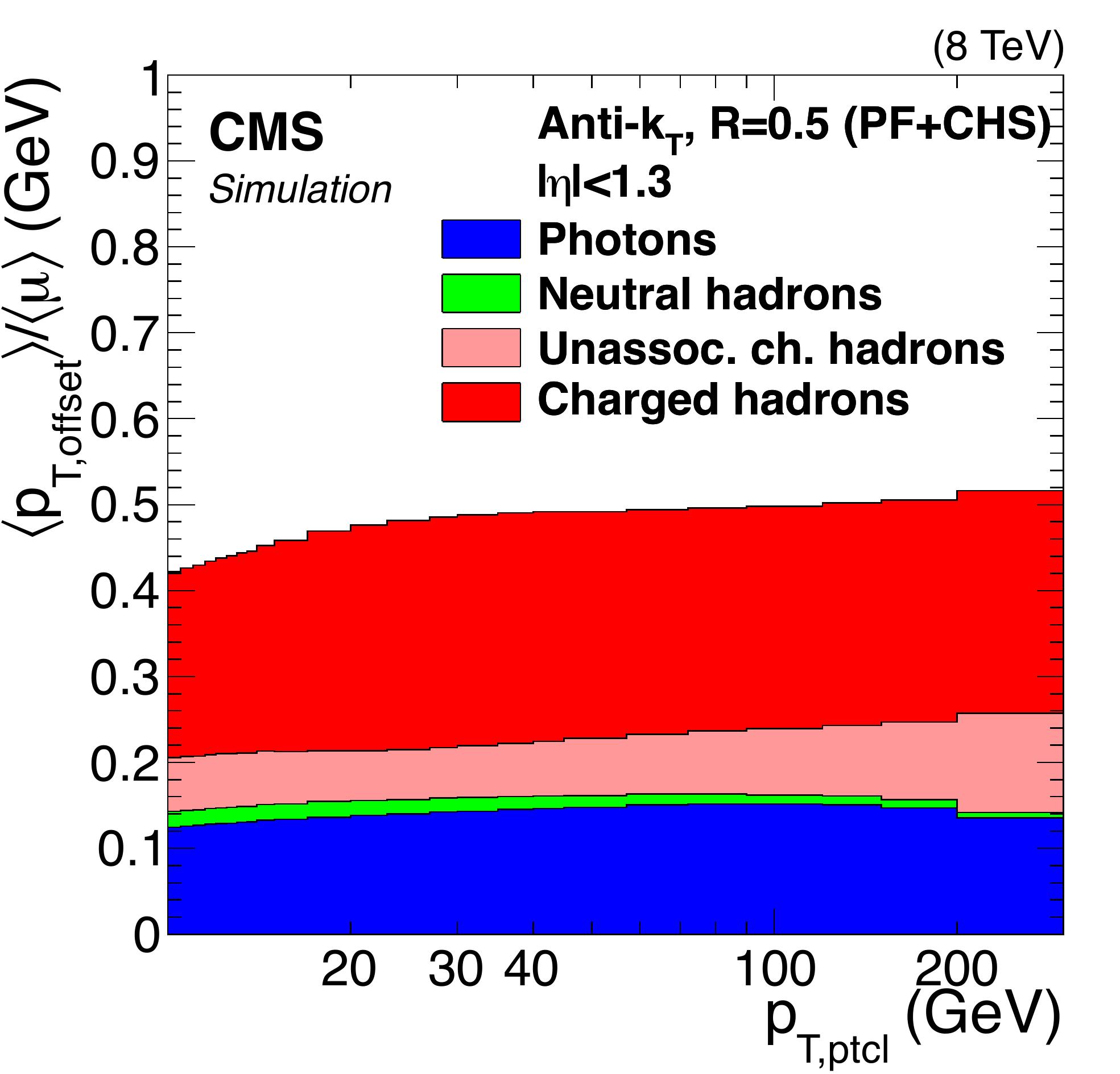}
\includegraphics[width=0.48\textwidth]{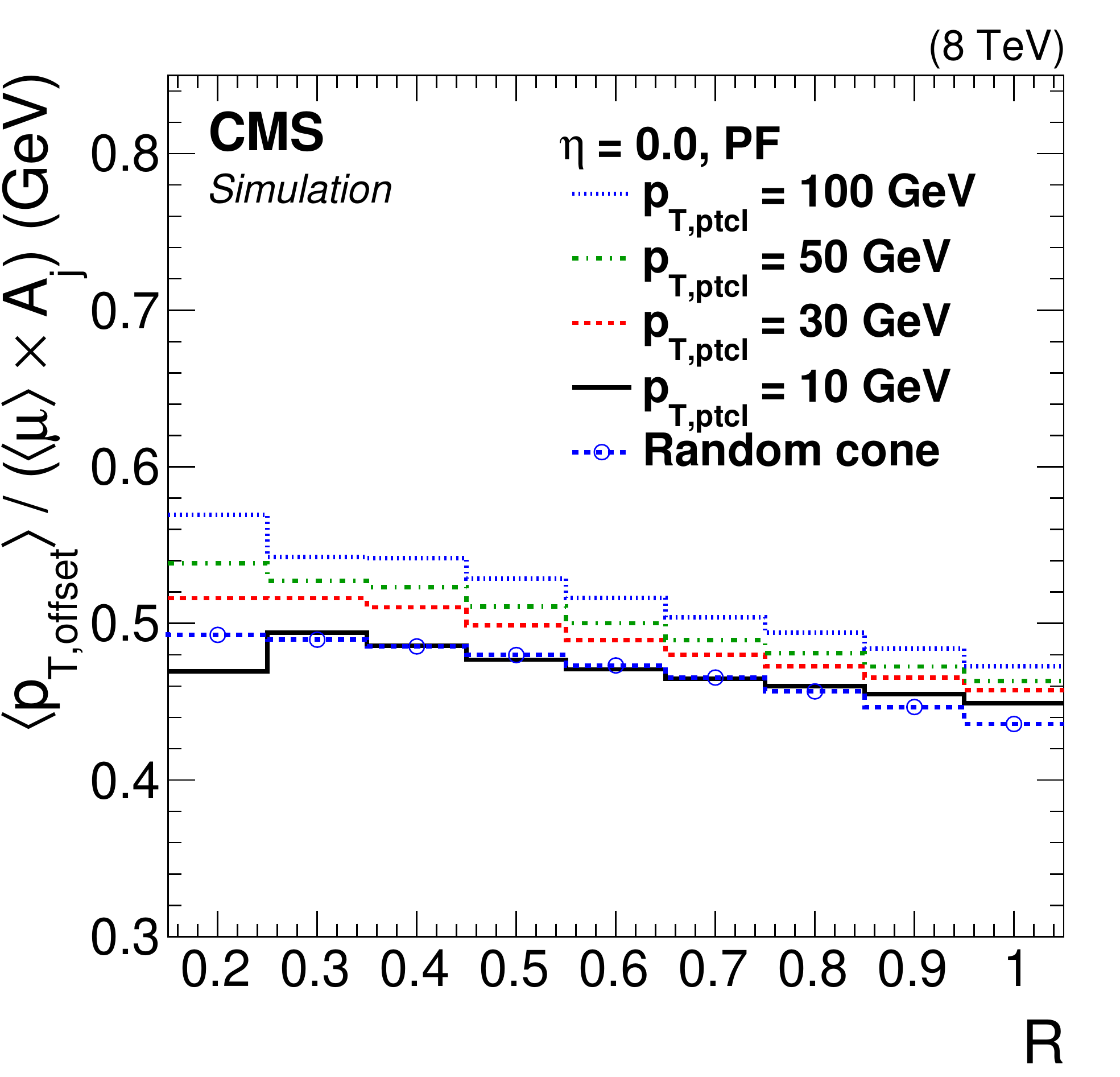}
\caption{\label{fig:l1pf}
Simulated particle-level offset versus \pt separately for each type of PF candidate (left).
Average \pt offset density versus jet distance parameter $R$ for various $p_{\mathrm{T},\rm ptcl}$ compared to a random-cone offset density versus cone radius (right). The jet or cone area $A_j$ corresponds to $\pi R^2$.}
\end{figure}

Figure~\ref{fig:l1pf} (left) shows the \pt dependence of the offset for each PF candidate type. The $\langle p_\mathrm{T, offset}\rangle$ is divided
by the average number of pileup interactions, hence showing the average offset per additional interaction. While the reconstruction thresholds for charged hadrons and photons are of the order of a few hundred MeV, the effective detector reconstruction thresholds for neutral hadrons (mostly $\PKzL$, $\PKzS$, and neutrons) are of the order of 3\GeV. This is far above the typical $\langle p_\mathrm{T, offset}\rangle\approx 5$\GeV for a pileup particle, making the neutral hadron contribution barely visible in Fig.~\ref{fig:l1pf} (left).
The observed \pt dependence comes from an interplay of several effects for overlapping particles, such as failed zero-suppression in calorimeter energy, nonlinearity of PF hadron corrections, fake tracks arising from hit combinations, and misreconstructed tracks arising from pixel hit merging and tracker dynamic inefficiency at high $\mu$.
The rate of overlaps is highest in the jet core, which results in the simulated offset correction depending on the jet size. Figure~\ref{fig:l1pf} (right) shows the average offset density within the jet versus jet distance parameter $R$ and jet \pt. The simulated particle-level offset converges to an RC offset measurement at low \pt, as well as for large jet size parameters. The shallow slope in RC offset versus distance parameter is
due to vector summation of PF candidate momenta, which reduces the offset \pt relative to the offset energy by $\cos(\Delta R)$ at the cone edges.

\subsubsection*{Offset scale factor}
\label{sec:rc}\label{sec:l1res}

The offset data/simulation scale factor is estimated from zero-bias data and simulation using the RC method \cite{JEC_JINST}. Because zero-bias data contain no energy deposition from hard interactions, and the noise contribution is small, the average transverse momentum $\langle p_{\mathrm{T},\rm cone}\rangle(\eta)$ of PF candidates in a randomly placed cone centered at $(\eta,\phi)$ can be identified with the average offset due to pileup, $\ptoffset^\mathrm{RC}(\eta)$:
\begin{equation}
\ptoffset^\mathrm{RC}(\eta, \langle\rho\rangle) = \langle p_{\mathrm{T},\rm cone}\rangle[\eta, \mu].
\end{equation}

As in the case of the simulated particle-level offset, the parameterization variables $(\eta,\langle\rho\rangle)$ and the binning variables $[\eta, \mu]$ are explicitly marked in order to signal their impact on the observational biases.

For deriving the offset scale factor, the RC measurement is fitted with a quadratic function of $\rho$, $\ptoffset^\mathrm{RC} = p_0 + p_1\rho + p_2\rho^2$.
The constant and quadratic terms are small, but are required for a good $\chi^{2}/N_\mathrm{dof}$ of the fit. The constant term has usually a small positive value, because the mean $\langle p_{\mathrm{T},\rm cone}\rangle$ can still have a small nonzero value when the median $\rho$ is already zero. This low-PU behavior of $\rho$ is discussed in Ref.~\cite{Chatrchyan:2012tt}.
The offset scale factor for parameters $\rho_0(\eta)$ and $\beta(\eta)$ in Eq.~(\ref{eq:chybrid}) is defined as
\begin{equation}
\frac{\ptoffset^\mathrm{RC}_\text{data}(\eta, \langle\rho\rangle_\text{data})}
{\ptoffset^\mathrm{RC}_\mathrm{MC}(\eta, \langle\rho\rangle_\mathrm{MC})} .
\end{equation}

Using different $\langle\rho\rangle$ working points for data and simulation is necessary due to the slight difference of about 4\% in $\langle\rho\rangle$ between data and simulation, seen in Fig.~\ref{fig:rhoandnpv}~(right).

\begin{figure}[htbp!]
\centering
\includegraphics[width=0.48\textwidth]{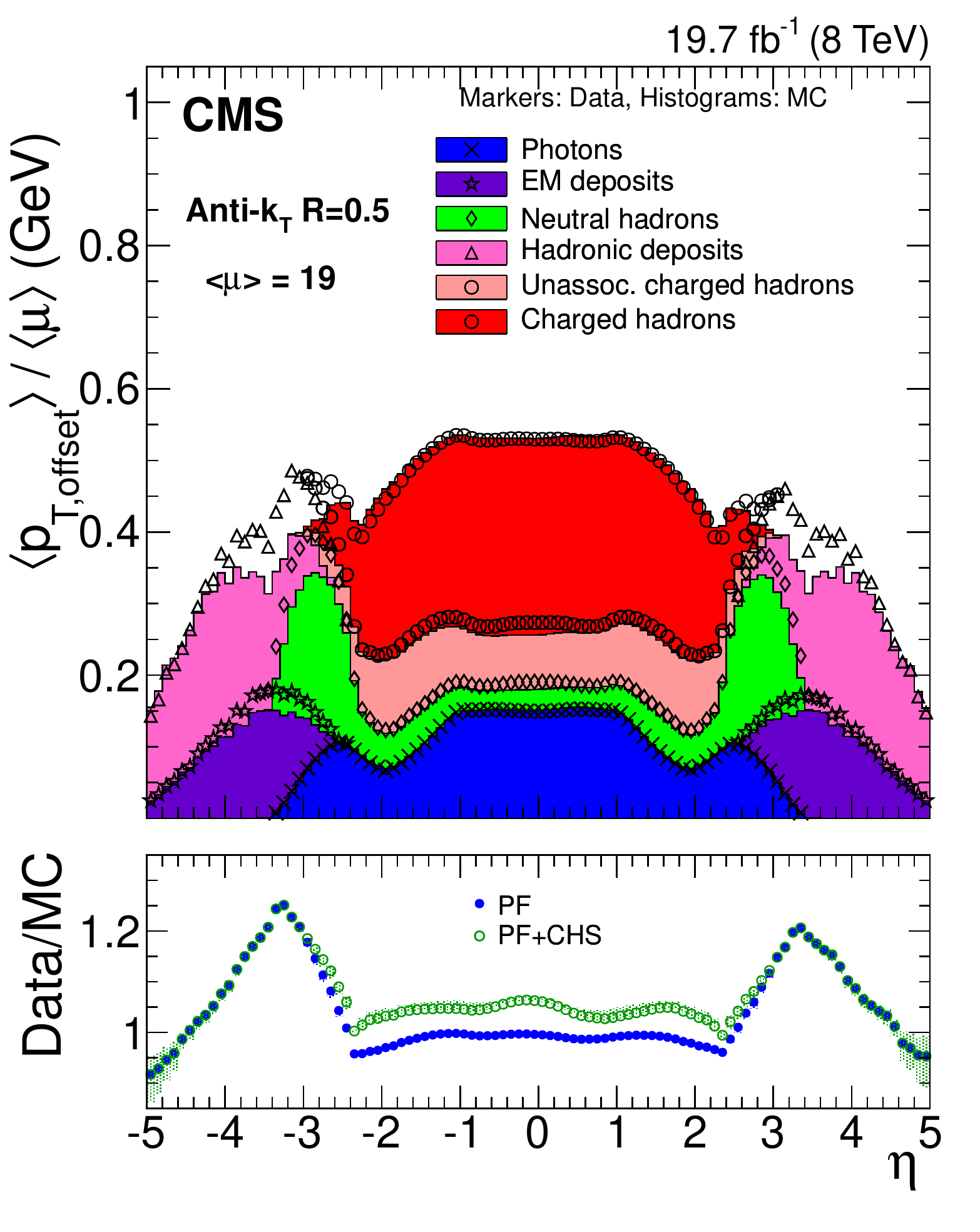}
\caption{\label{fig:l1data}
Random-cone offset measured in data (markers) and MC simulation (histograms) normalized by the average number of pileup interactions $\langle\mu\rangle$, separated by the type of PF candidate. The fraction labeled 'charged hadrons' is removed by CHS. The ratio of data over simulation, representing the scale factor applied for pileup offset in data, is also shown for PF and PF+CHS.
}
\end{figure}

The offsets in data and simulation are shown in Fig.~\ref{fig:l1data} (top), separated by PF candidate type. The offset scale factor for PF and PF+CHS is shown in Fig.~\ref{fig:l1data} (bottom).
The offset scale factor at $\abs{\eta}<2.4$ is less than 5\%, but increases up to 20\% outside of the tracking coverage near the inner edge of HF at $\eta\approx3.2$. The triangular shape is caused by smearing sharp detector effects over a cone area within $\Delta\eta<0.5$. The uncertainty from varying the $\langle\rho\rangle$ working point within the 68\% confidence interval of the $\rho$ distribution is less than 2\% up to $\abs{\eta}<4.7$.

\subsection{Pileup offset correction uncertainties}\label{sec:l1uncert}

The pileup offset correction uncertainties come from two main sources: uncertainty in the offset scale factor used for the $\eta$ dependence in data, and uncertainty in the offset jet \pt dependence that is derived from simulation only. The former uncertainty is evaluated by varying the $\langle\rho\rangle$ working point used for deriving the offset scale factor within one standard deviation of the $\rho$ distribution, while the latter is evaluated using the difference between the simulated particle-level offset and the RC offset. Of these, the jet \pt dependence is the dominant uncertainty across most of the phase space.

Any residual pileup offset is absorbed on average, within the constraints of their respective parameterizations, by the relative $\eta$ and absolute \pt corrections derived from dijet, $\PZ$+jet, $\gamma$+jet and multijet data. Therefore the dominant \pt-dependence uncertainty is propagated through the fit procedure used in the data-based methods to account for this reduction and shaping of pileup offset correction uncertainties. This results in a set of five uncertainty sources:

\begin{itemize}
\item \textit{PileUpEnvelope} is taken as 30\% of the difference between simulated particle-level offset and RC offset. This is the pileup uncertainty we would have if the later calibrations did not reduce the uncertainty. It is not directly included in the JEC uncertainties, but is propagated through the relative $\eta$ and absolute $p_\mathrm{T}$ corrections to give the uncertainties \textit{PileUpPtEta}, \textit{PileUpPtRef} (for $\langle\mu\rangle\approx 20$ data) and \textit{PileUpMuZero} (for $\langle\mu\rangle=0$ data), described below.
\item \textit{PileUpPtEta} (Eta=BB,EC1,EC2,HF) results from the propagation of the \textit{PileUpEnvelope} uncertainty through the $\eta$-dependent correction evaluation from dijet balance. This uncertainty accounts for the residual difference between the \textit{PileUpEnvelope} with shape $(p_0+p_1\log(\pt))/\pt$ and the $\eta$-dependent correction fit in the range of dijet data at $60<\pt<2000/\cosh(\eta)\GeV$ with shape $p_0+p_1\log(\pt)$.
\item \textit{PileUpPtRef} results from the propagation of the \textit{PileUpEnvelope} uncertainty through the evaluation of the absolute-scale \pt dependence from $\PZ$/$\gamma$+jet and multijet data. This uncertainty accounts for the residual difference between the \textit{PileUpEnvelope} and the absolute-scale fit in the range of $\PZ$/$\gamma$+jet and multijet data at $30<\pt<1000/\cosh(\eta)\GeV$.
\item \textit{PileUpDataMC} accounts for uncertainty in the offset scale factor for data, based on variation of the $\langle\rho\rangle$ working point within one standard deviation of the $\rho$ distribution.
\item \textit{PileUpMuZero} is evaluated from the nominal result of the fit for $\eta$- and \pt-dependent data-based corrections, and accounts for the bias that results from deriving them at $\langle\mu\rangle\approx 20$ instead of $\langle\mu\rangle\approx 0$. This uncertainty is to be used for zero-pileup data ($\langle\mu\rangle\approx0$, \eg, in the 2.76\TeV data collected in 2013) and replaces \textit{PileUpPtEta}, \textit{PileUpPtRef} and \textit{PileUpDataMC}.
\end{itemize}

\begin{figure}[htbp!]
\centering
\includegraphics[width=0.49\textwidth]{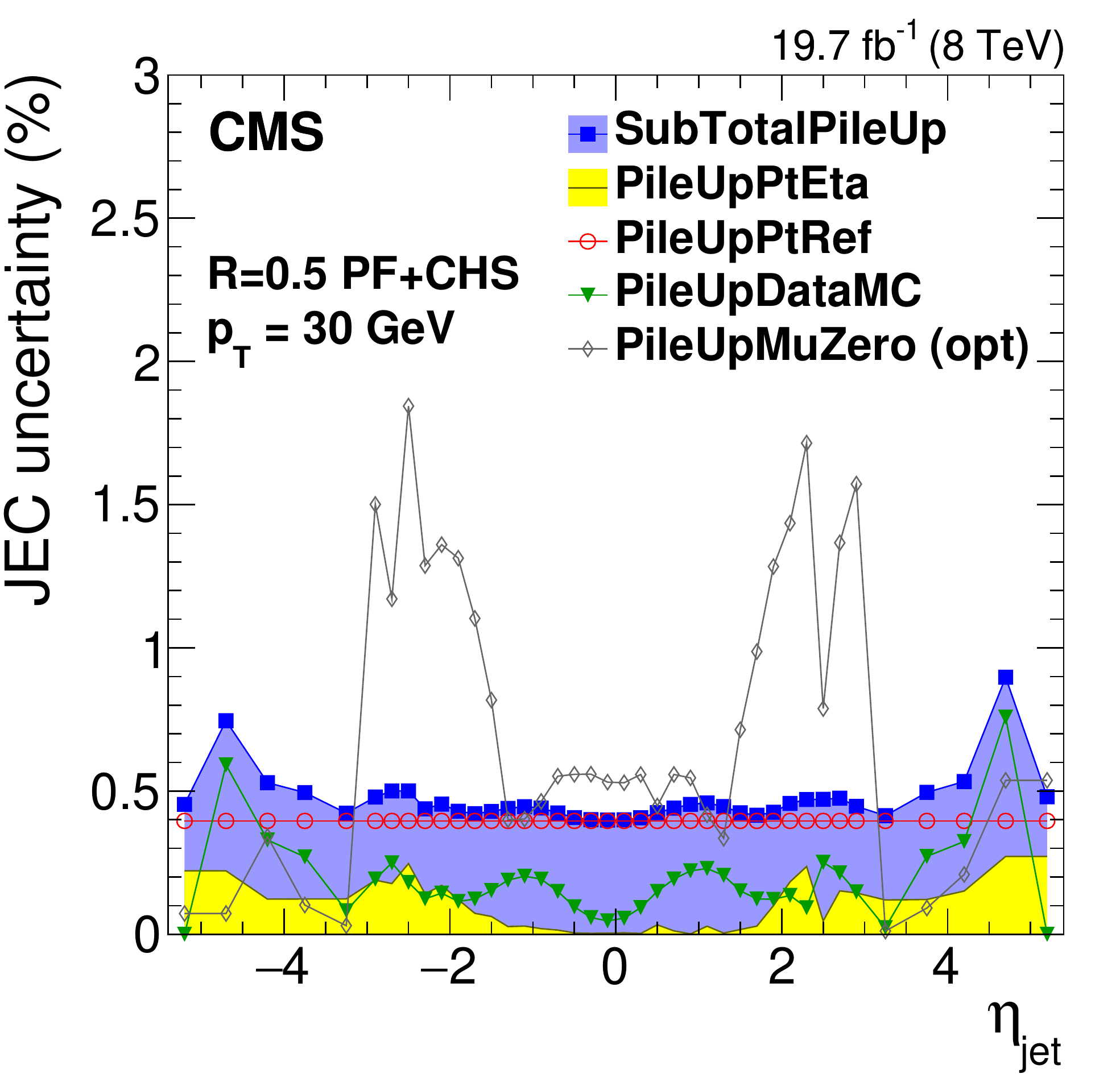}
\includegraphics[width=0.49\textwidth]{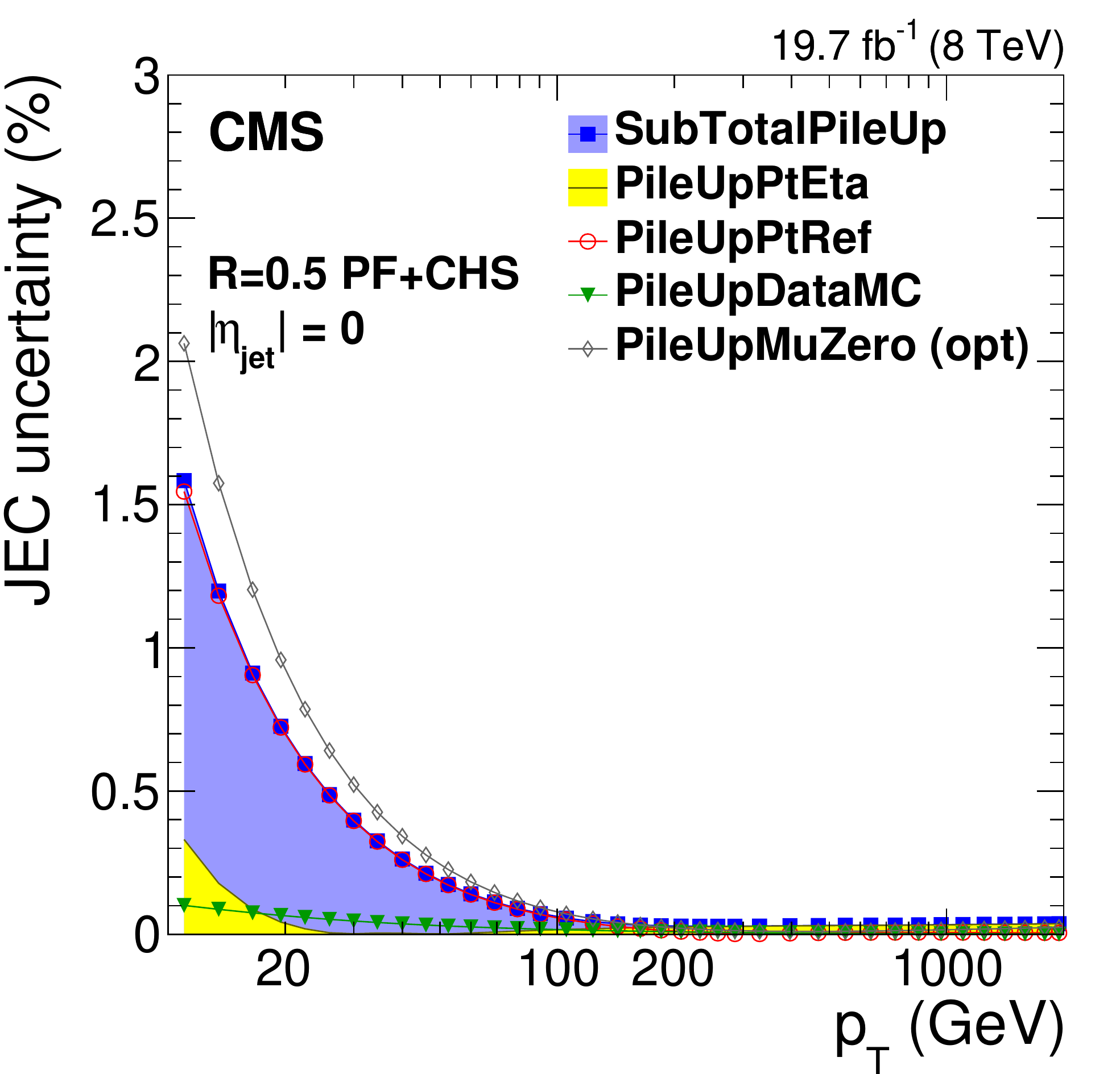}
\includegraphics[width=0.49\textwidth]{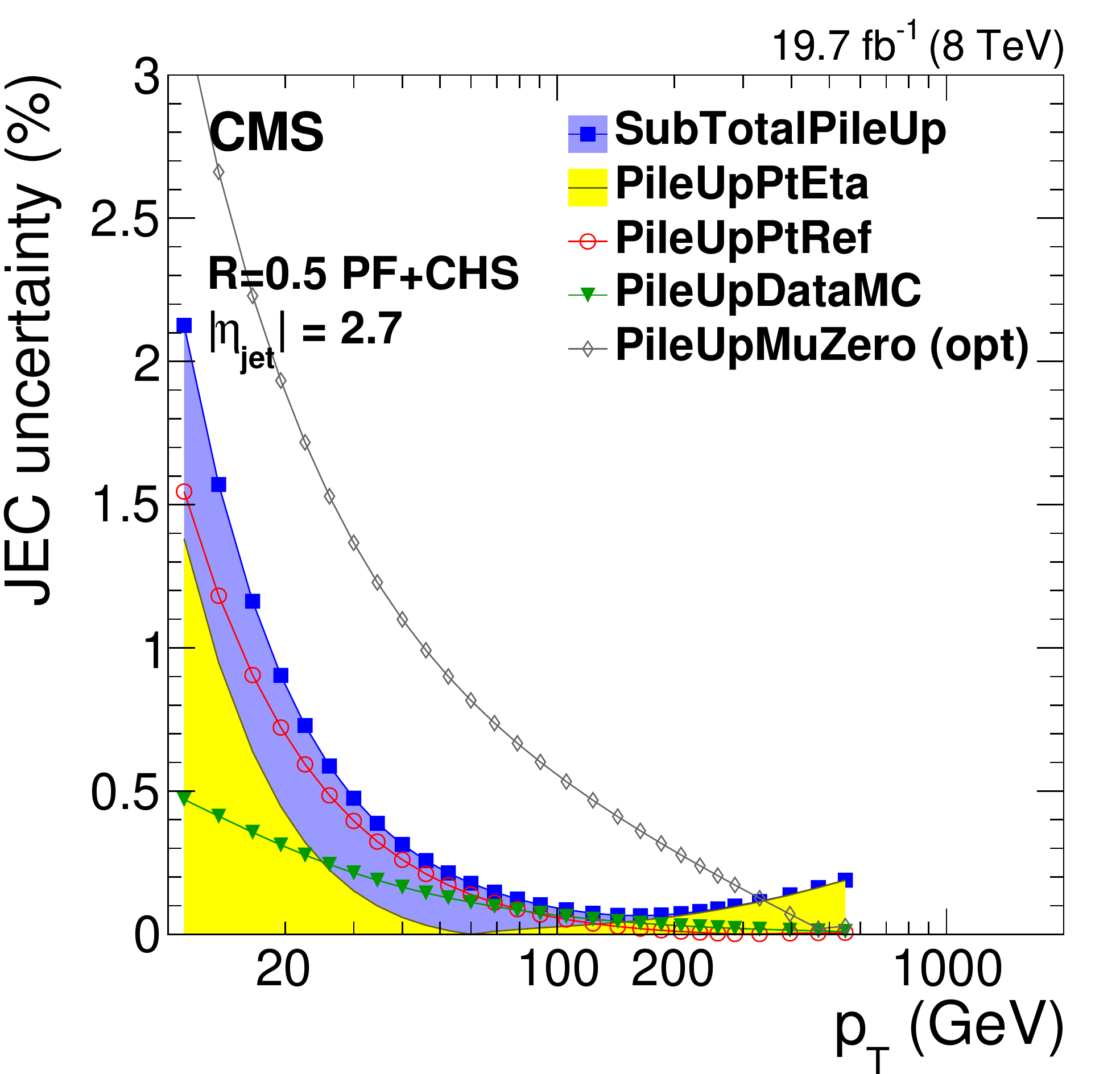}
\includegraphics[width=0.49\textwidth]{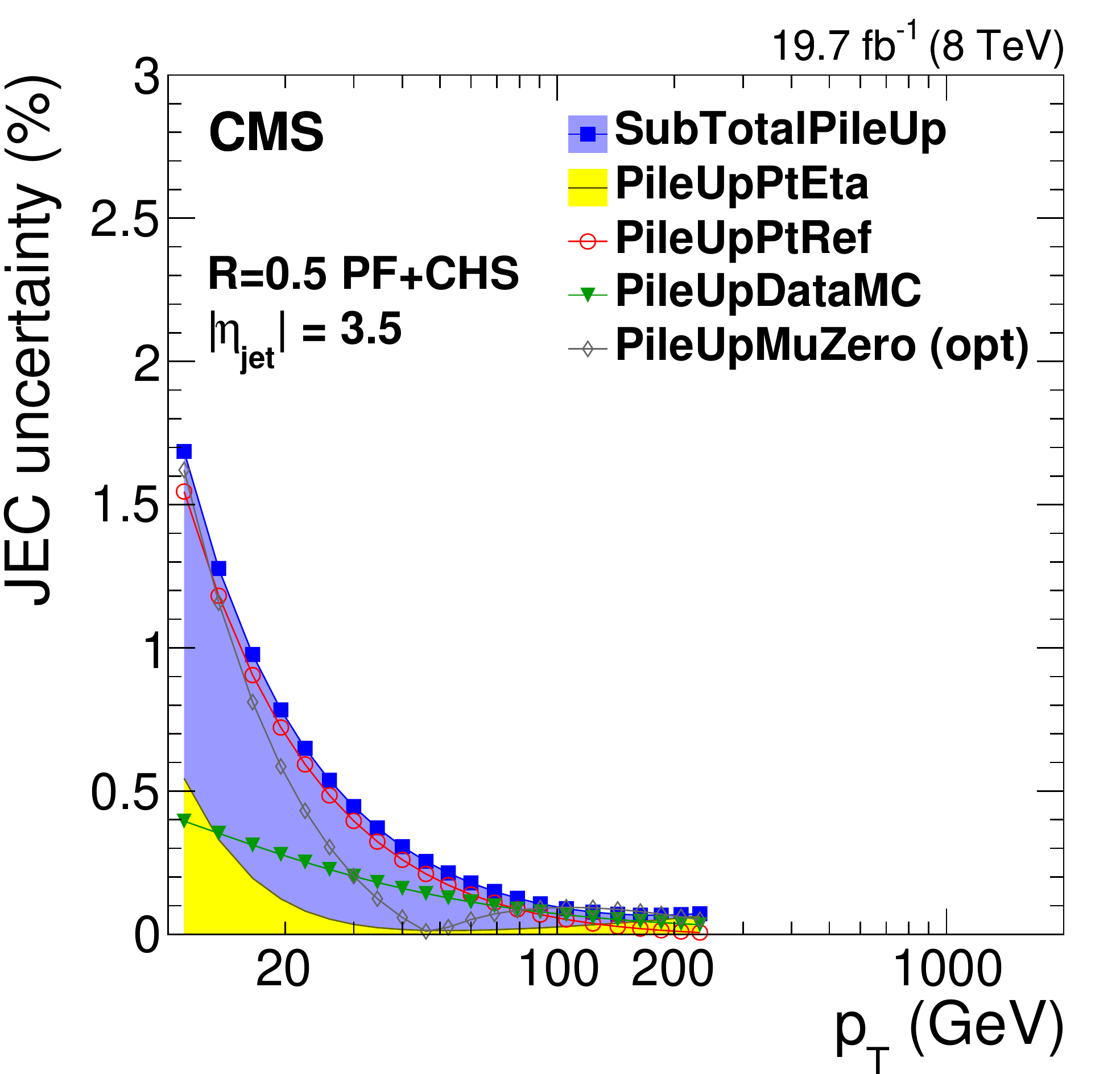}
\caption{\label{fig:l1sys}
Pileup offset correction uncertainties for the average 2012 (8\TeV) conditions for PF jets with CHS and $R=0.5$ as a function of $\eta_\text{jet}$ for fixed $\pt=30\GeV$ (top left) and as a function of jet \pt (top right, and bottom panels). The plots are limited to a jet energy $E=\pt\cosh\eta=4000\GeV$ so as to show only uncertainties for reasonable \pt in the considered data-taking period.
\textit{PileUpMuZero} is an optional alternative uncertainty for zero-pileup ($\langle\mu\rangle\approx 0$) events, and it is therefore not included in the quadratic sum \textit{SubTotalPileUp}. It accounts for the pileup uncertainty absorbed in the residual response corrections at $\langle\mu\rangle\approx 20$, which is particularly prominent at $1.5<\abs{\eta}<3$.
}
\end{figure}

The pileup offset correction uncertainties are summarized in Fig.~\ref{fig:l1sys}. The dominant uncertainty is from the residual jet \pt dependency remaining after the application of the data-based methods. It is at the level of 1\% for $\pt=30\GeV$, and rapidly decreases to the $10^{-3}$ level in the range constrained by the data-based methods. There is a small increase in uncertainty again at high \pt outside the range of data-based methods, where the constrained parameterizations used for data-based residuals result in a small seesaw effect. The uncertainty for $\langle\mu\rangle=0$ data is in many cases similar or even larger than for $\langle\mu\rangle=20$, owing to the absorption of the residual offset into relative $\eta$ and absolute \pt corrections at $\pt>30\GeV$.

\subsection{Summary of pileup offset corrections}\label{sec:l1summary}

\begin{figure}[htbp!]
\centering
\includegraphics[width=0.48\textwidth]{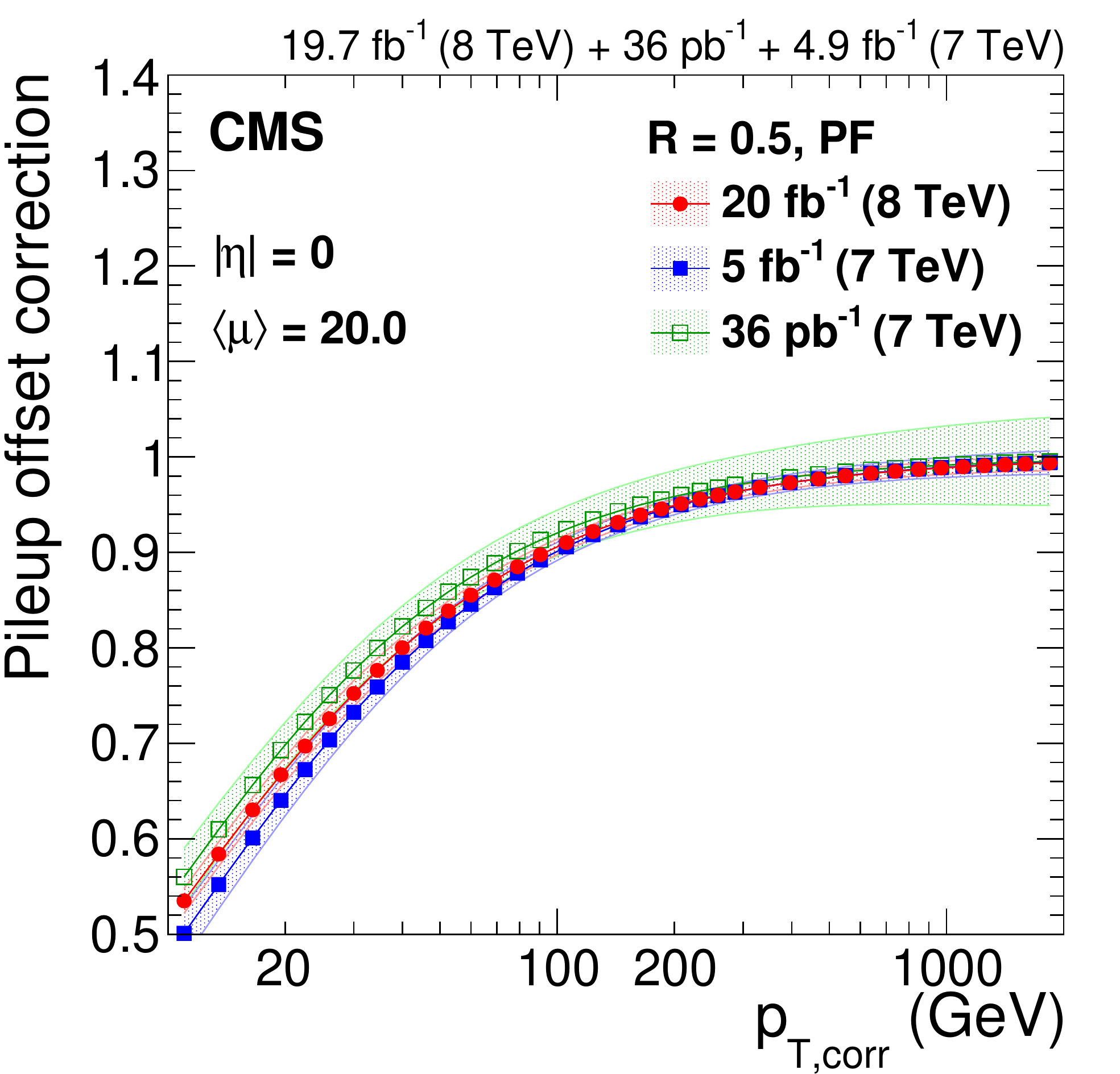}
\includegraphics[width=0.48\textwidth]{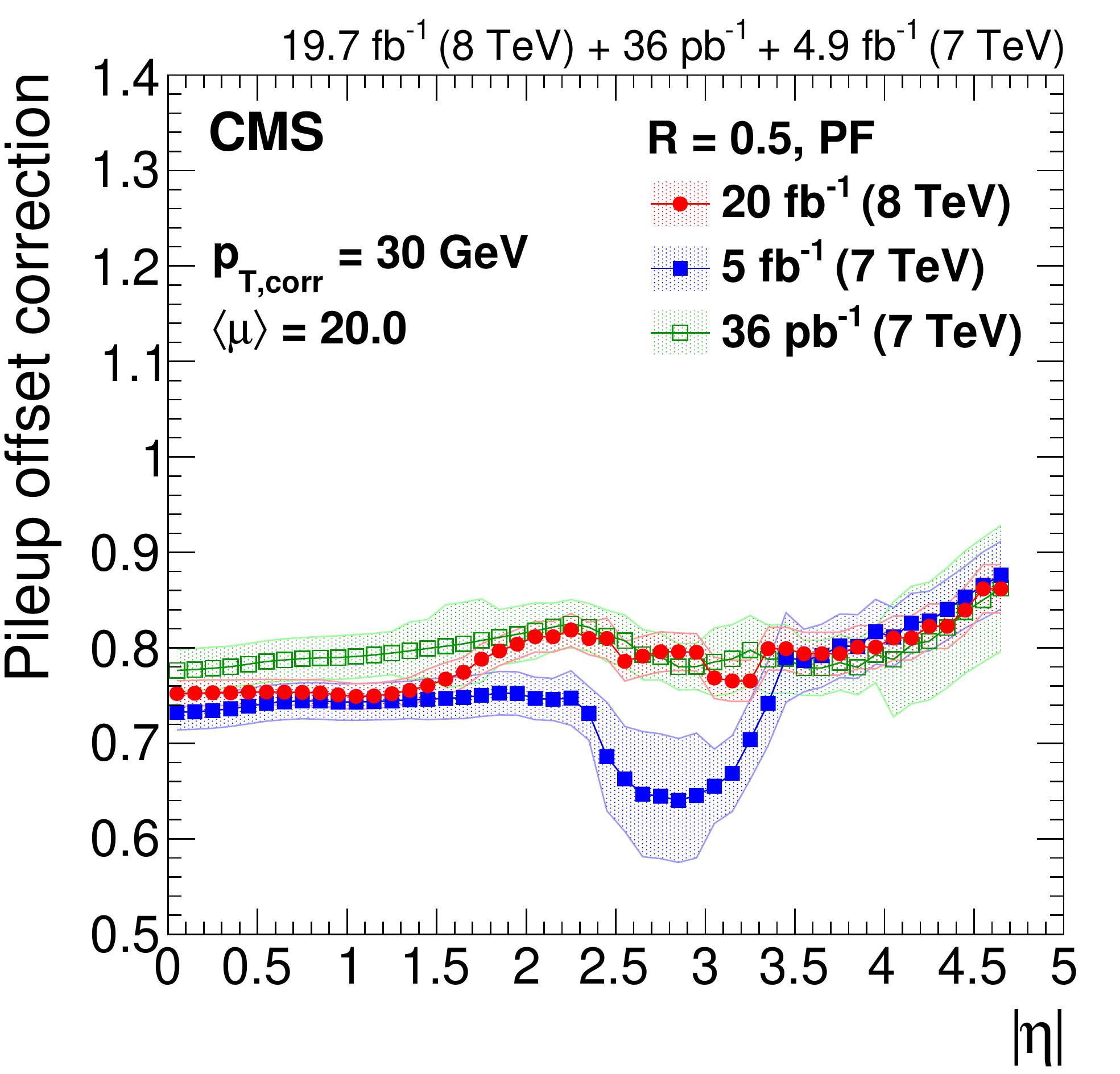}
\includegraphics[width=0.48\textwidth]{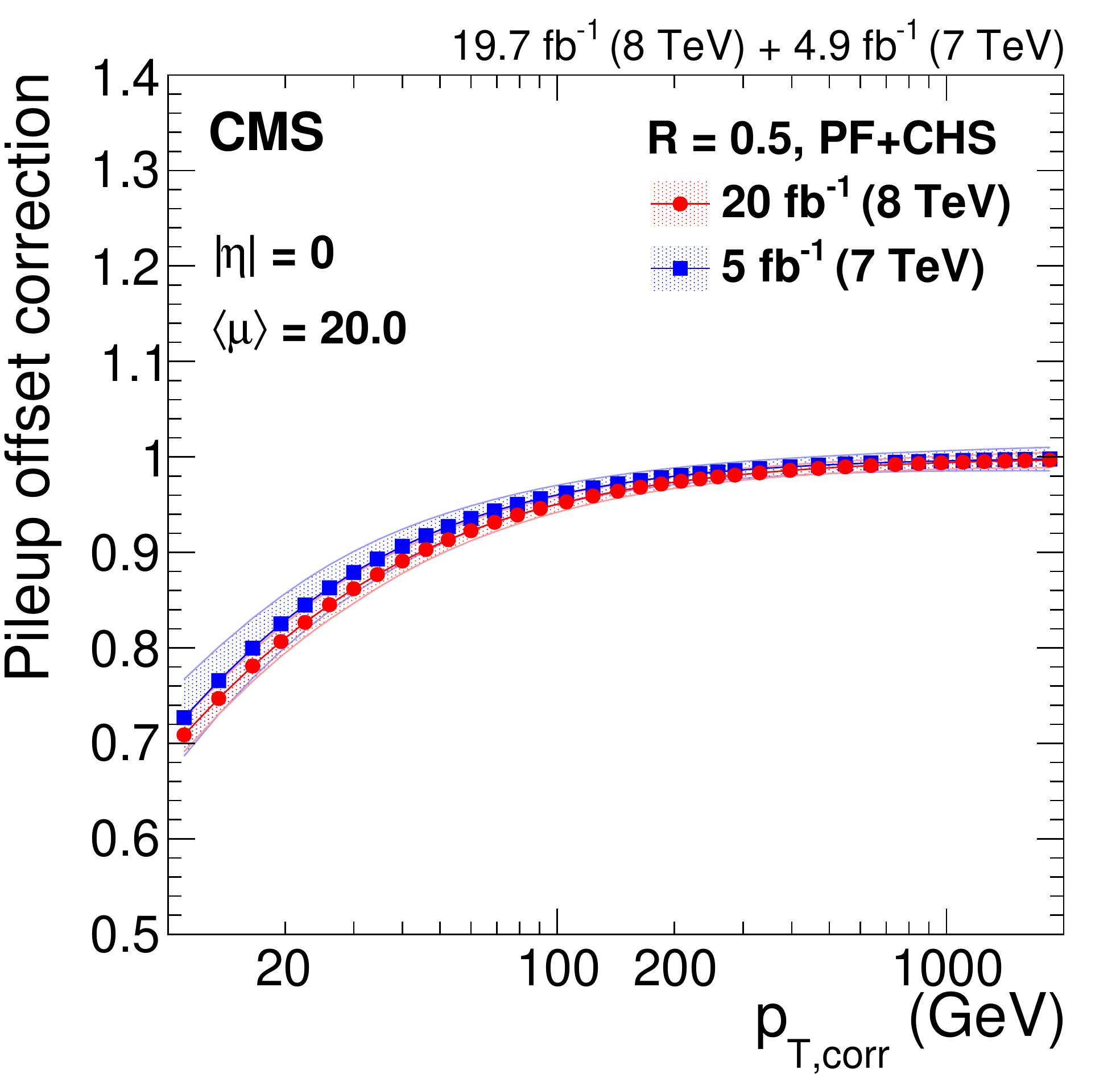}
\includegraphics[width=0.48\textwidth]{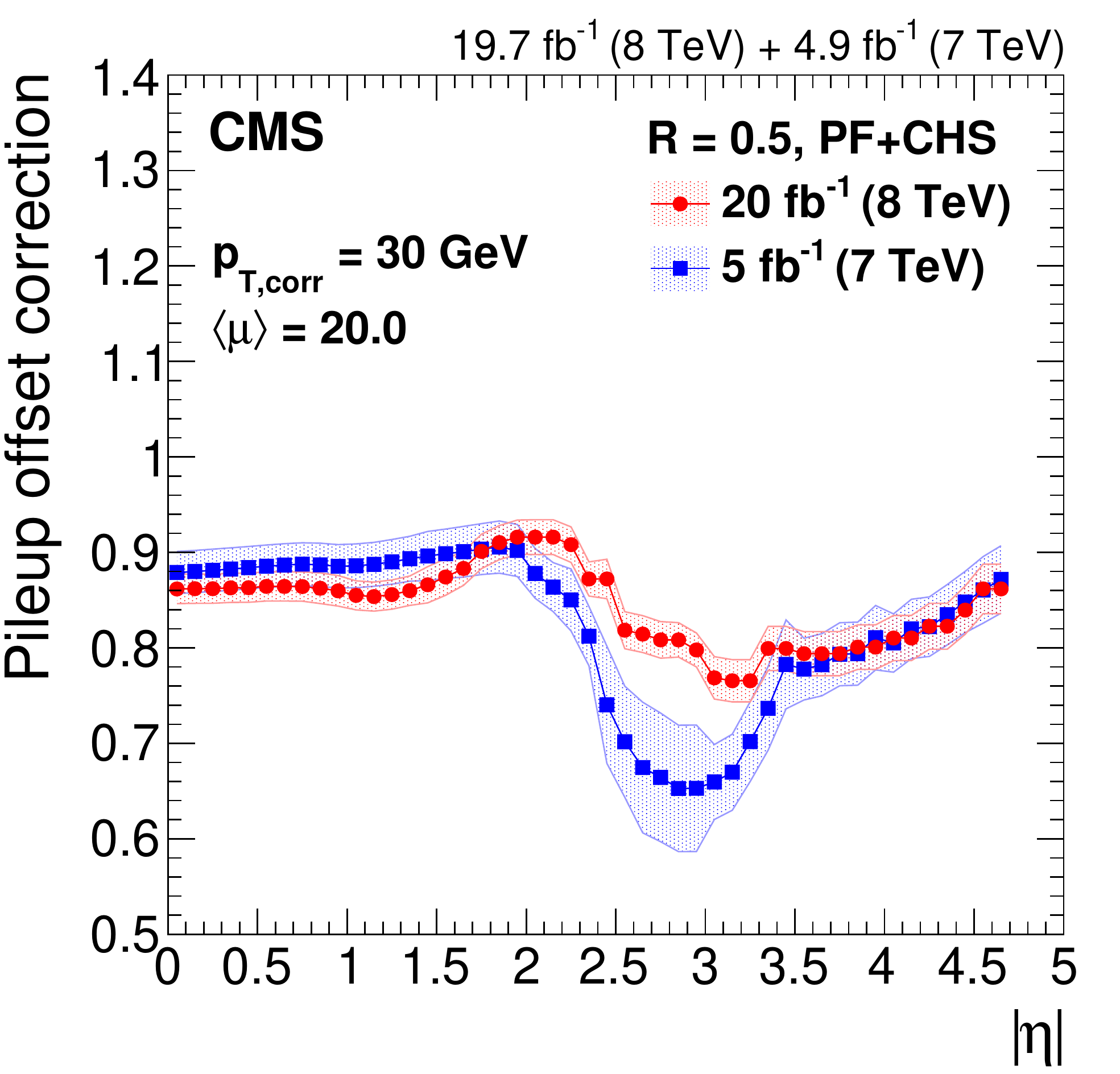}
\caption{\label{fig:l1corrections}
Pileup offset correction $C_\text{hybrid}$ including data/MC scale factors, with systematic uncertainty band, for the average 2012 (8\TeV) conditions of $\langle\mu\rangle=20$ for PF jets without CHS and $R=0.5$ at $\abs{\eta}=0$ versus $p_\mathrm{T, corr}$ (top left), and at $p_{\mathrm{T},\rm corr}=30\GeV$ versus $\abs{\eta}$ (top right),
compared to corrections for 2010 \cite{JEC_JINST} and 2011 \cite{CMS-DP-2012-006} data at 7\TeV after extrapolation to similar pileup conditions.
The same results are also shown for PF jets with CHS and $R=0.5$ at $\abs{\eta}=0$ versus \pt (bottom left), and at $p_{\mathrm{T},\rm corr}=30\GeV$ versus $\abs{\eta}$ (bottom right), compared to corrections for 2011 data at 7\TeV \cite{CMS-DP-2012-006}.
}
\end{figure}

The pileup offset corrections for the anti-$\kt$ algorithm ($R=0.5$) with and without charged-hadron subtraction are summarized in Fig.~\ref{fig:l1corrections} for typical 2012 (8\TeV) conditions of $\langle\mu\rangle\approx20$, compared to corrections for 7\TeV data taken in 2010 and 2011.
The average pileup per interaction for $R=0.5$ is about 0.5\GeV,
adding up to a total of about 10\GeV per jet. This results in a typical offset correction of about 0.75 for a $p_{\mathrm{T},\rm corr}=30\GeV$ ($p_{\mathrm{T},\rm uncorr}=40\GeV$) jet. The CHS removes approximately half of this offset before jet clustering by matching tracks to pileup vertices, reducing the residual offset correction to about 0.85 at $p_{\mathrm{T},\rm corr}=30\GeV$ ($p_{\mathrm{T},\rm uncorr}=35\GeV$). Roughly one third of the remaining pileup is from PF charged hadrons that have not been matched to good pileup vertices, and much of the rest is from PF photons.
The CHS algorithm was only fully commissioned at 7\TeV in 2011, and the 2010 (7\TeV) version of the offset corrections did not yet take into account the remaining unmatched pileup tracks. Therefore only results without CHS are shown for 7\TeV in 2010.

The pileup offset corrections have been relatively stable over time at 7\TeV in 2010 and 2011, when scaled to similar pileup conditions.
This is in part due to the good linearity of the offset corrections for PF and continuous development on the detector side to reduce OOT PU in the calorimeters, and in part due to the adaptability of the jet area method to the prevailing pileup offset.
The largest differences are visible
in the $2.5<\abs{\eta}<3.0$ region, where OOT PU increased at 7\TeV in 2011, but was again brought down in ECAL at 8\TeV in 2012 using more advanced reconstruction algorithms.
The OOT PU is also partially responsible for the 2011--2012 differences in the endcaps within tracker coverage of $1.5<\abs{\eta}<2.5$, and for differences between 2010 and 2011--2012 in the barrel at $\abs{\eta}<1.5$. In addition, the JEC were improved at 7\TeV in 2011 to take into account the difference between the offset outside jets (RC offset) and inside jets (particle-level offset). This increased the offset correction inside the tracker coverage (failed zero-suppression), and lowered it outside (calorimeter response nonlinearity), compared to the 2010 (7\TeV) corrections.

The pileup uncertainties
have been steadily reduced despite rapidly increasing pileup. This can be credited to improvements in the correction methods, more events at high pileup to determine the trends versus pileup, and a reduction of double counting. The 2012 (8\TeV) corrections explicitly take into account the additional constraints
from data-based methods,
which reduce the offset uncertainty in the endcaps by up to 50\% for PF+CHS at $p_{\mathrm{T},\rm corr}=30\GeV$ compared to 7\TeV in 2011. The dominant systematic uncertainty is from the \pt dependence of the pileup offset, which is only indirectly constrained by data.

\section{Simulated response corrections}
\label{sec:mctruth}

The simulated response corrections are derived and applied on jets that have been corrected for pileup offset.
The CMS detector simulation contains a detailed model of the detector geometry, data-based alignment and calibration of the detector elements, and emulation of the readout electronics. It is based upon the {\GEANTfour} package \cite{Agostinelli:2002hh} that simulates the evolution of the electromagnetic (EM) and hadronic showers and their interactions with the detector material. In addition, the \PYTHIA6.4 tune Z2* event generator is used to simulate the fragmentation of the initial quarks and gluons. Together these two components provide an accurate and detailed description of the jet response, which is used for the bulk of the JEC. Data-based methods (Section \ref{Sec:l2l3res}) are needed only for small residual corrections on top of the simulated response and the simulated offset corrections discussed in Section~\ref{sec:pileup}.

The benefit of relying heavily on simulation to derive the jet response is that we are not sensitive to many of the biases inherent in the data-based methods and can cover corners of phase space that are not easily accessible in data.
This includes samples of jets with very low ($\pt<30\GeV$) and very high ($\pt>1$\TeV) momenta,
heavy-flavor jets, and samples with particularly low ($\mu<5$) and high ($\mu>40$) pileup. Describing jet response in terms of variables accessible in simulation also facilitates the understanding of data-based methods, as we can better model the correlation between various samples and corrections.
For the following discussion, jets are assumed to be corrected for the pileup offset as described in Section~\ref{sec:pileup}.

\subsection*{Sample definitions}

We derive the simulated response from a QCD multijet sample of 10~million events generated with \PYTHIA6.4 tune Z2*.
To ensure event generation with efficient coverage of the full kinematic phase space at the LHC with small statistical uncertainty,
the events are generated with a flat \pt spectrum and reweighted by $\hat{p}_\mathrm{T}^{-4.5}$, where $\hat{p}_\mathrm{T}$ is the transverse momentum of the generated $2 \to 2$ hard process, which allows the recovery of the original \pt spectrum
in \PYTHIA6.4 and the production of unbiased results for $p_\text{T, jet}>30\GeV$.
The generated and simulated events are overlaid with pileup generated by \PYTHIA6.4 tune Z2*, with events reweighted such that the pileup distribution matches that found in data.

To estimate pileup offset in jets,
we simulate the same generated events also without additional pileup overlay, as was discussed in Section~\ref{sec:pileup}.
To estimate the jet response dependence on the fragmentation model and jet flavor, a complementary sample is produced with \HERWIGpp~2.3 tune EE3C.
To estimate the jet response dependence on the detector calibration, we also produce additional samples with the CMS fast simulation.

\subsection*{Definition of simulated particle response}

A particle-level jet is matched to the closest reconstructed jet if it is within half of the jet distance parameter $R$. For a distance parameter of $R=0.5$ this corresponds to 0.25. The method ensures a high matching efficiency (reaching 100\% around $p_\mathrm{T}=30$\GeV) and provides a unique match for the anti-$\kt$ jets.
In the present paper, the simulated particle response $R_\mathrm{ptcl}$ is defined as the ratio of arithmetic means of matched reconstructed and particle-level jets transverse momenta,
\begin{equation}
R_\mathrm{ptcl}(\langle \pt\rangle,\eta) = \frac{\langle \pt\rangle}{\langle p_\text{T, ptcl}\rangle}[p_\text{T, ptcl},\eta],
\end{equation}
in bins of particle-level \pt ($p_\text{T, ptcl}$) and reconstructed $\eta$ (where \pt is the transverse momentum of the reconstructed jet).
As in the previous sections, the square brackets $[$ $]$ denote the binning variables, and the angle brackets $\langle$ $\rangle$
indicate the averages within those bins for the variables that are used to parameterize the response.

\subsection{Corrections versus \texorpdfstring{$\eta$ and \pt}{eta and pT}}

\begin{figure}[htbp!]
\centering
\includegraphics[width=0.49\textwidth]{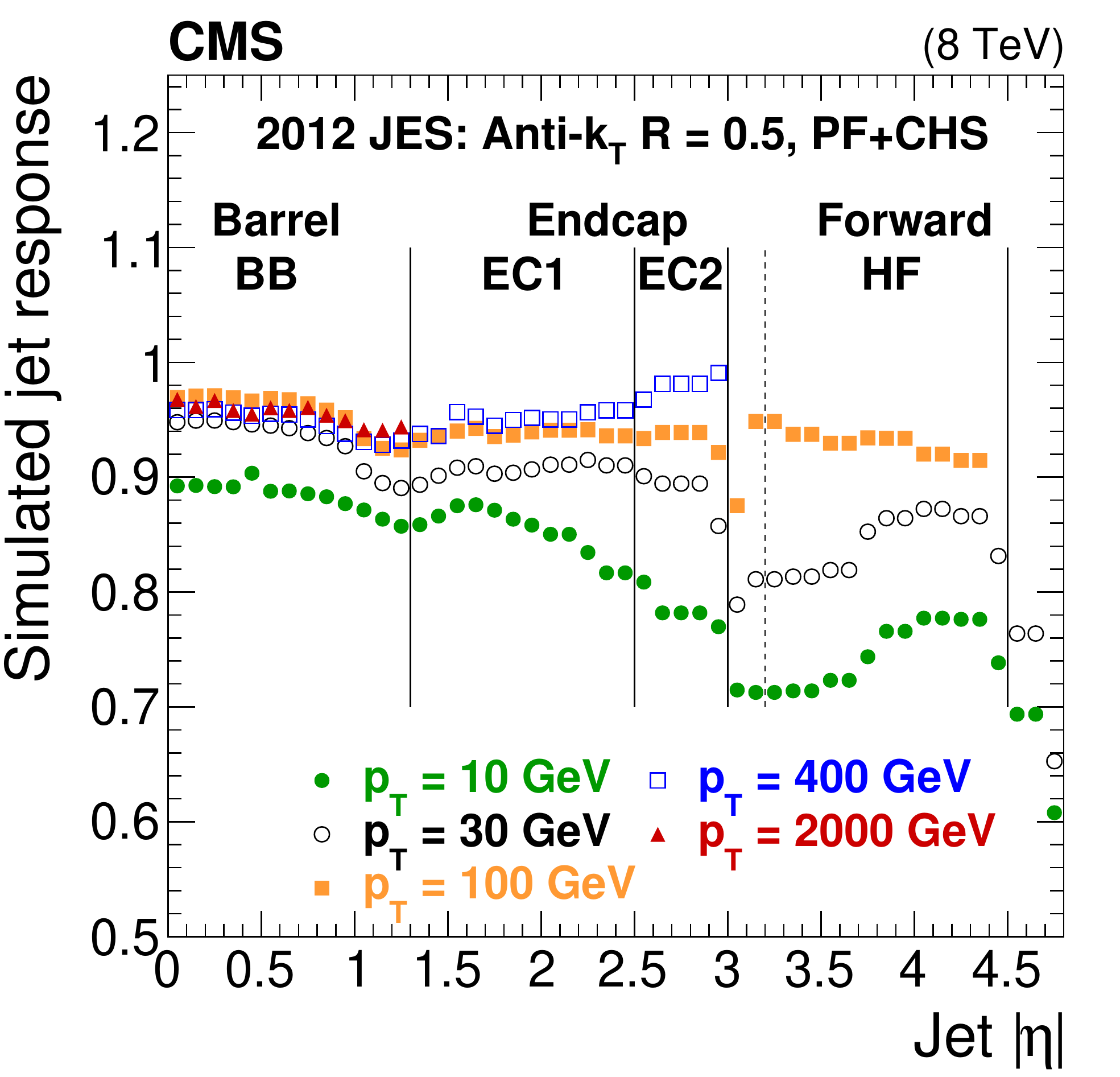}
\includegraphics[width=0.49\textwidth]{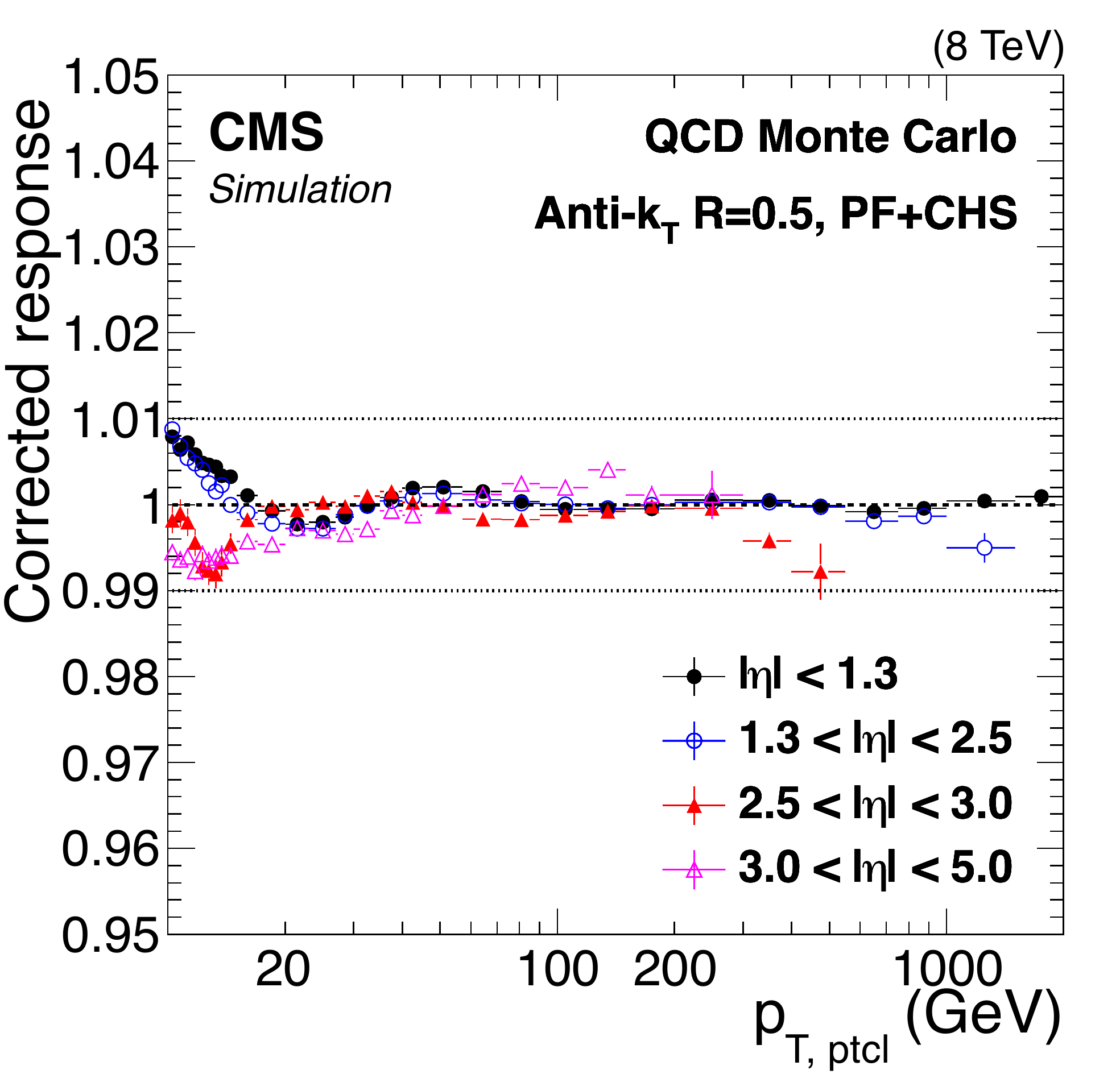}
\caption{\label{fig:response}
Simulated jet response $R_\mathrm{ptcl}$ versus $\abs{\eta}$ for $R=0.5$ (left).
Simulated jet response $R_\mathrm{ptcl}$, after JEC have been applied, versus $p_\text{T, ptcl}$ for $R=0.5$ in various $\eta$ regions, and with statistical uncertainties (right).
}
\end{figure}

Simulated anti-$\kt$ jets, with a distance parameter $R=0.5$, are used to study the detector response as a function of the jet \pt. The simulated particle response is shown in Fig.~\ref{fig:response} (left) as a function of the reconstructed jet $\eta$. The simulated particle response after JEC is shown in Fig.~\ref{fig:response} (right) as a function of the particle-level jet $p_\text{T, ptcl}$
in various $\eta$ regions.
The results show that the response is corrected to within 0.5$\%$ with respect to the particle-level jet, for \pt from about 20\GeV to 2\TeV.

\subsection{Dependence on the jet size}

\begin{figure}[htbp!]
\centering
\includegraphics*[width=0.49\textwidth]{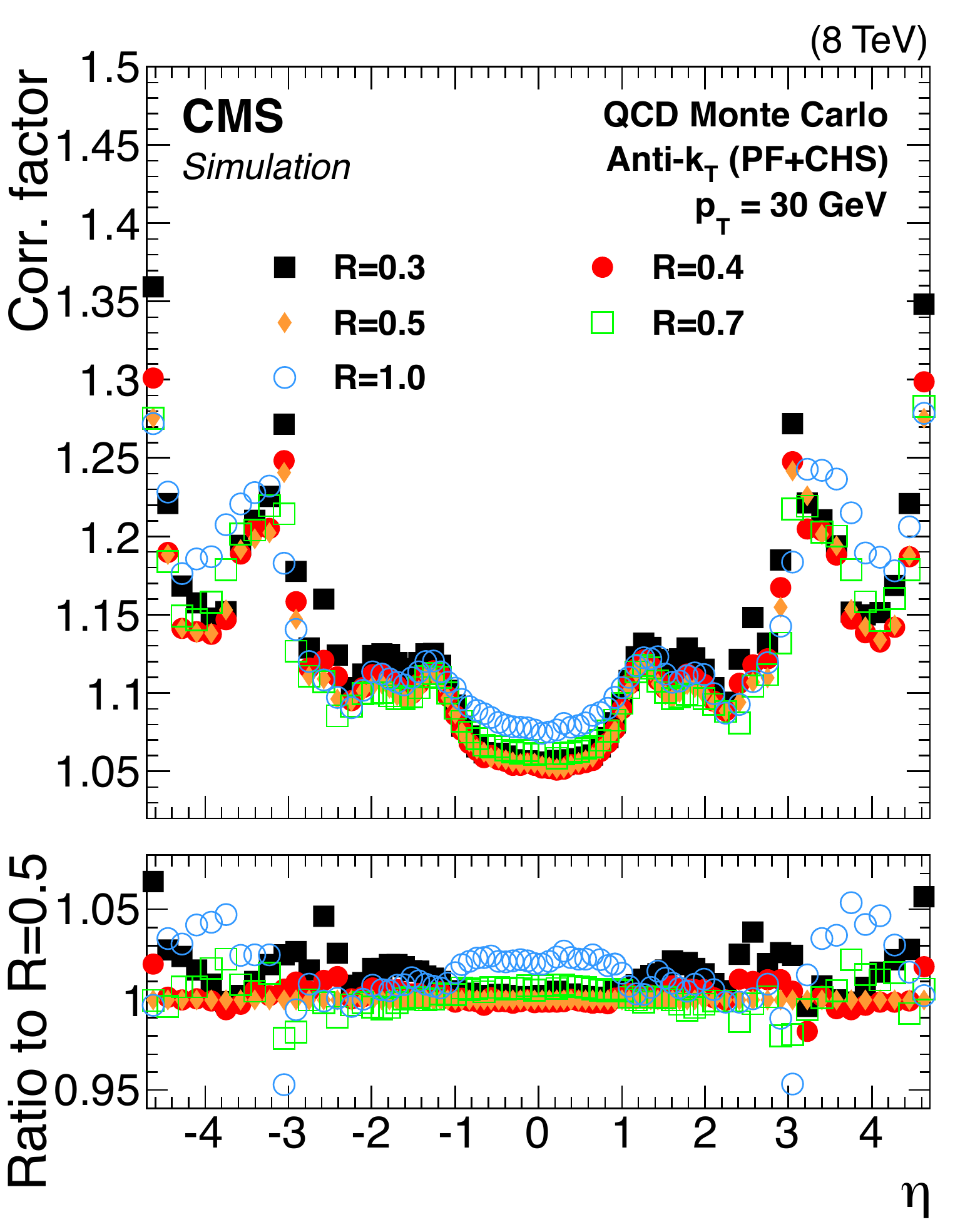}
\includegraphics[width=0.49\textwidth]{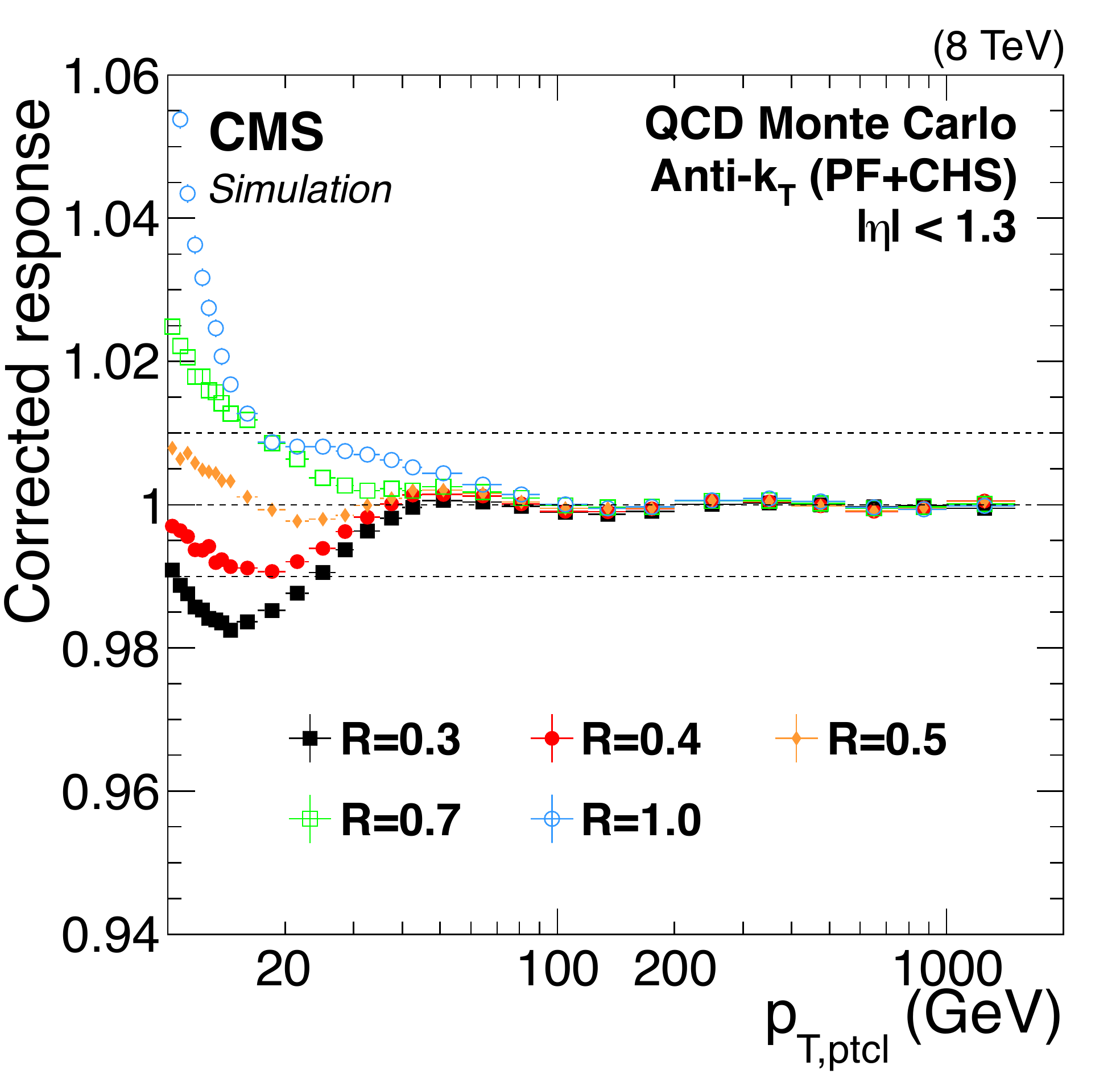}
\caption{\label{fig:rcone}
Jet energy correction factors for a jet with $\pt=30\GeV$, as a function of $\eta$ and for various jet sizes $R$ (left).
Simulated jet energy response $R_\mathrm{ptcl}$ after JEC for $\abs{\eta}<1.3$ as a function of the particle-level jet \pt for various jet sizes $R$ (right).
}
\end{figure}

The dependence of the jet response on the jet distance parameter $R$ has been checked in the range $R=0.3$--1.0.
The response is similar after accounting for the increasing PU offset due to the larger jet area ($A_\text{jet}\approx\pi R^2$). Smaller effects come primarily from two sources:
\begin{itemize}
\item The UE energy within the jet has lower response than the energy from the hard scattering, lowering the response at low \pt for jets with large $R$.
\item A larger distance parameter averages the jet response over a larger area, smearing sharp features in the detector response versus $\eta$.
\end{itemize}
Figure~\ref{fig:rcone} (left) shows the comparison of the JEC factor for various jet sizes at $\pt=30\GeV$. As expected from the larger fraction of UE energy, the corrections rise slightly for larger distance parameters. The very small distance parameter $R=0.3$ is an exception to this rule, because the detector granularity smears some energy out of the cone.
These differences mostly disappear at higher \pt for $R\geq 0.4$, with the smaller jet sizes showing slightly sharper detector features. Simulated jet responses after the application of the JEC are shown in Fig.~\ref{fig:rcone} (right) as a function of jet \pt for a range of distance parameters from 0.3 to 1. The response is consistent with unity within 1$\%$ for $\pt\gtrsim 30\GeV$.
During Run~1 of the LHC, the supported jet size parameters in pp collisions were $R=0.5$ and $R=0.7$. The full jet energy corrections and uncertainties were derived and provided centrally only for these two jet size parameters.

\subsection{Detector simulation uncertainties}

We evaluate several systematic uncertainties using simulation, with the uncertainties further constrained using data-based methods, as discussed later. We discuss here the uncertainties arising from the propagation of detector calibration uncertainties to the jet response. The effects of jet fragmentation and flavor response are discussed in Section~\ref{sec:flavor}.
Because the jet response is later constrained using measurements based on data, these systematics are explicitly set to zero at certain reference points, discussed in Section~\ref{Sec:l2l3res}. They are then used to extrapolate the systematics from these reference points to regions of the phase space not directly calibrated with data.

\subsubsection*{Single-pion response}

\begin{figure}[htbp!]
\begin{center}
\includegraphics[width=0.49\textwidth]{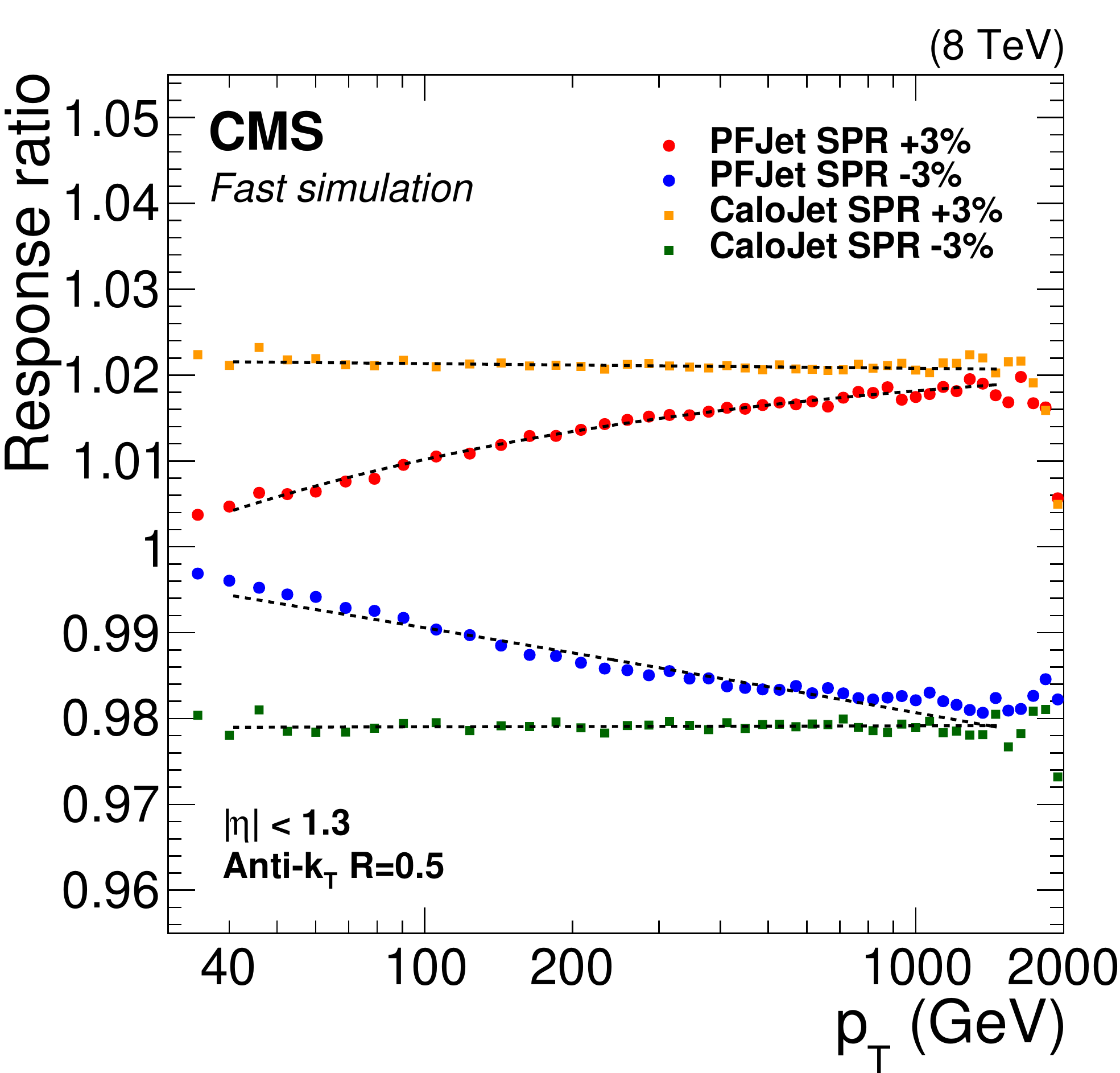}
\end{center}
\caption{\label{fig:spr_simultaneous}
Changes in PF jet and calorimeter jet response resulting from $\pm 3$\% variations of single-pion response in parameterized fast simulation in HCAL+ECAL.
}
\end{figure}

The jet response is sensitive to the underlying detector calibrations. The CMS calorimeters have been calibrated in test beam studies, and the single-pion response (SPR) has subsequently been checked on proton-proton data with charged pions \cite{CMS-PAS-JME-10-008}, confirming good modeling of the barrel response in simulation to within $\pm 3$\%. Because the PF reconstruction relies heavily on tracking for low-$\pt$ jets, the sensitivity to the detector calibration is strongly reduced compared to the calorimeter-only reconstruction. To show this effect, the ratio of the response when varying the SPR with respect to the nominal response is shown in Fig.~\ref{fig:spr_simultaneous}, for jets reconstructed with the PF algorithm and for jets reconstructed with only calorimetric energy deposits, both using the anti-$\kt$ algorithm. For this study, the SPR has been propagated to the JEC using the CMS fast simulation.

\begin{figure}[htbp!]
\includegraphics[width=0.49\textwidth]{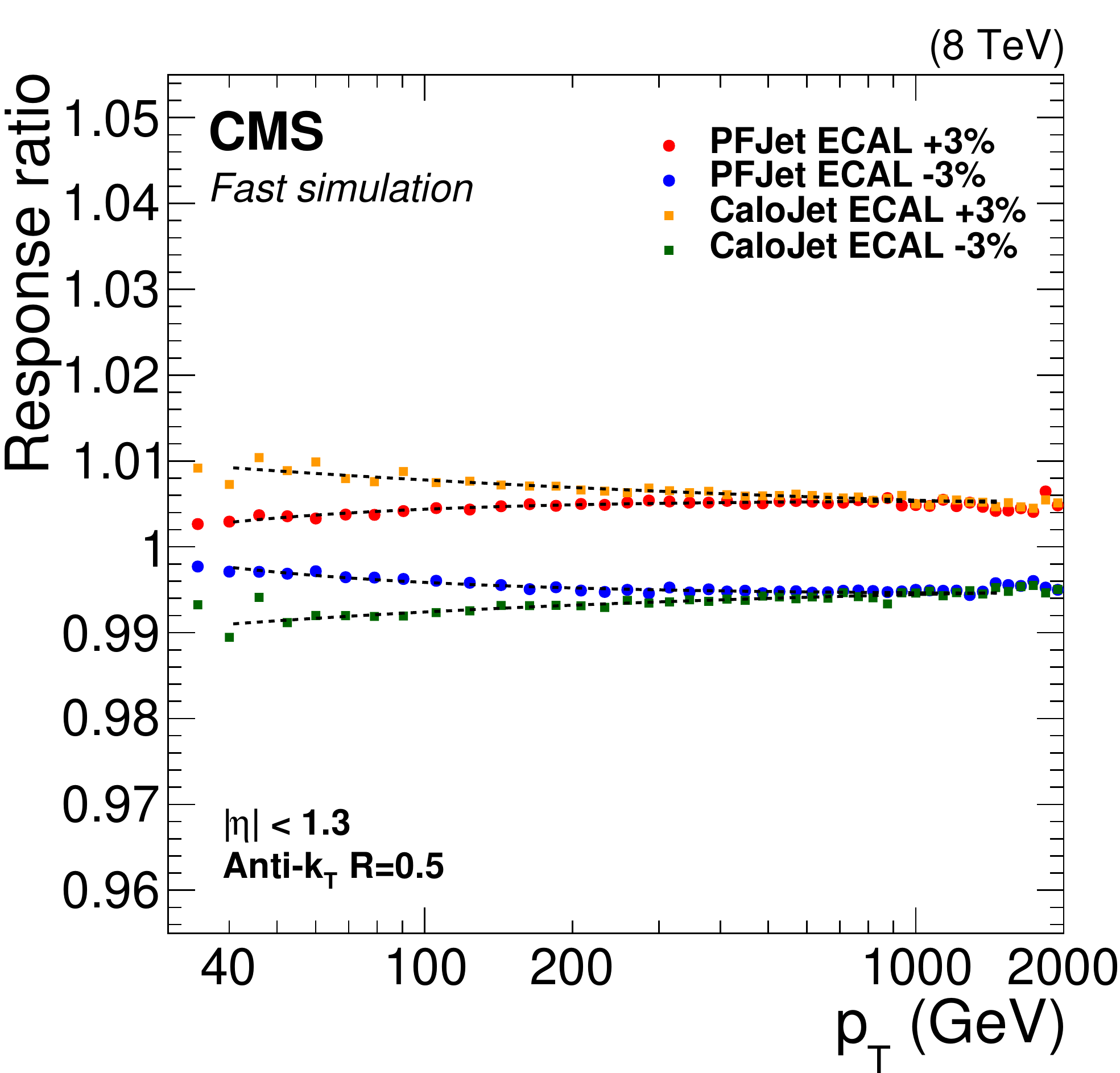}
\includegraphics[width=0.49\textwidth]{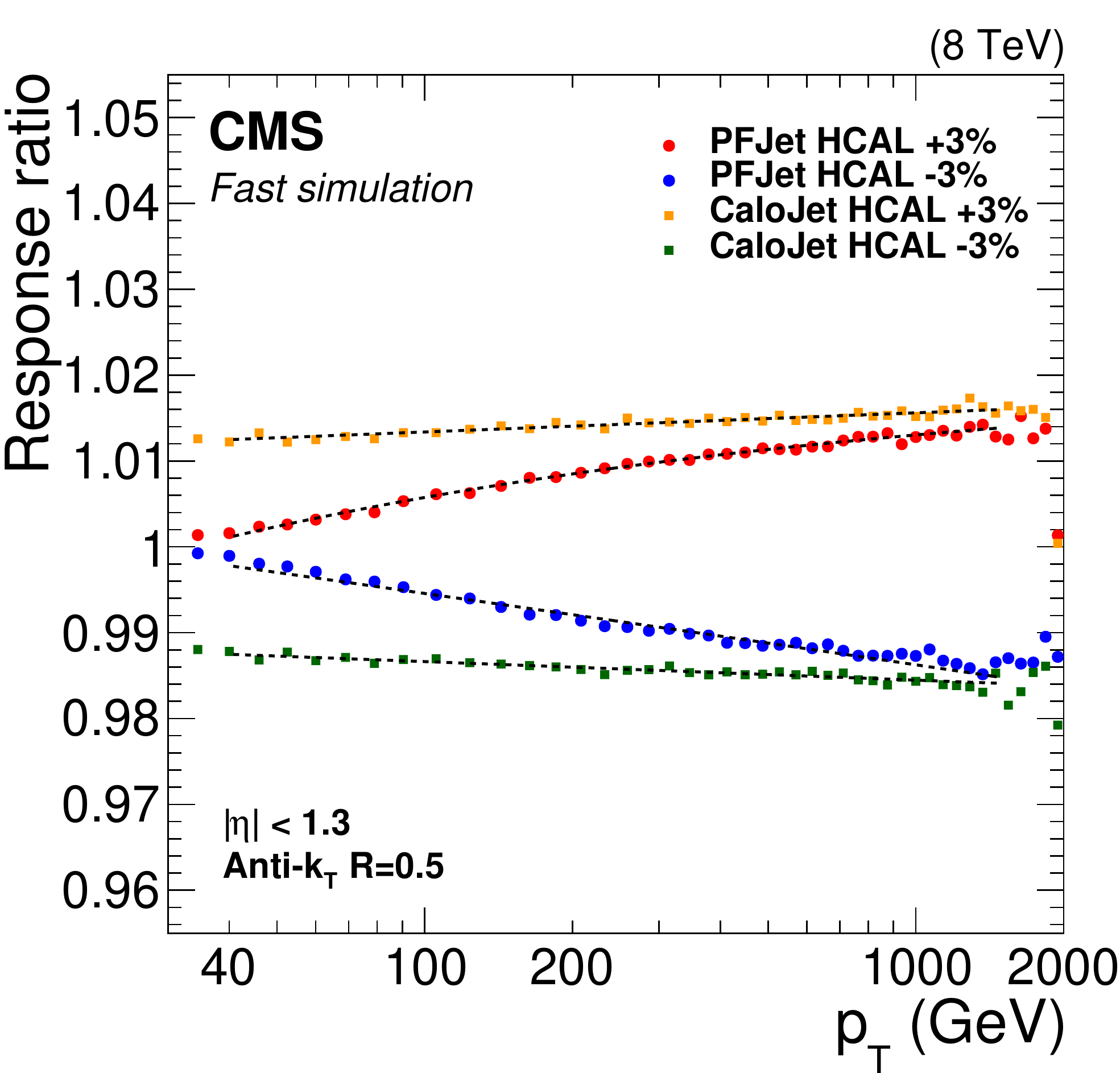}
\caption{\label{fig:spr}
Changes in PF jet and calorimeter jet response resulting from $\pm 3$\% variations of single-pion response in parameterized fast simulation in ECAL (left), and HCAL (right).
}
\end{figure}

At low \pt, PF is directly sensitive to SPR only through  neutral
hadrons, which on average contribute 15\% of the jet energy at particle level, leading
to a sensitivity of about 0.5\% for a simultaneous change of $\pm 3$\% in both ECAL and HCAL SPR. At high \pt the PF performance approaches that of the calorimetric reconstruction, because the tracking efficiency drops in the dense jet core and the leading tracks become too straight for a reliable \pt measurement.
Since 25\% of the jet energy is deposited as photons (Section~\ref{sec:jetcomp}), the JEC sensitivity to a $\pm 3$\% change in SPR is at most 2.3\%. The sensitivity to changes in SPR has been also studied separately for a $3$\% change in the response of the ECAL and HCAL, as shown in Fig.~\ref{fig:spr}. The results are qualitatively similar to an overall change in SPR, but show larger sensitivity to the SPR in HCAL at high \pt. This is because hadronic showers become deeper for high-$\pt$ particles, and deposit a larger fraction of their energy in the HCAL.

\subsection{Jet energy corrections propagation to missing transverse momentum}
\label{sec:typeImet}
The jet energy corrections are propagated to \ptvecmiss by using the so-called type-I correction:
\begin{equation}
\vec{p}^{\, \rm miss}_\mathrm{T, typeI} = \vec{p}^{\, \rm miss}_\mathrm{T, uncorr} + \sum_i \vec{p}_\mathrm{T, uncorr}^{\, i} - \sum_i \vec{p}_\mathrm{T, corr}^{\, i} - \sum_i\vec{\mathcal{O}}_\mathrm{RC}^{i},\label{eq:typeImet}
\end{equation}
where $\vec{p}^{\, \rm miss}_\mathrm{T, uncorr}$ is the uncorrected \ptvecmiss, $\vec{p}_\mathrm{T, uncorr}$ is the uncorrected jet \pt, $\vec{p}_\mathrm{T, corr}$ is the fully corrected jet \pt, and $\vec{\mathcal{O}}_\mathrm{RC}^i$ is the average offset due to pileup, as obtained with the RC method (see Section~\ref{sec:l1res}).
The sum runs over all jets with $p_\mathrm{T, corr}>10\GeV$ in the event.
Including the average RC offset underneath jets in the missing transverse momentum vector sum ensures that the pileup offset remains isotropic and does not bias \ptvecmiss.
The type-I correction is recommended for physics analyses and is used in most CMS results, as well as for deriving residual JEC for data.

\subsection{Summary of simulated response corrections}

\begin{figure}[htbp!]
\centering
\includegraphics[width=0.48\textwidth]{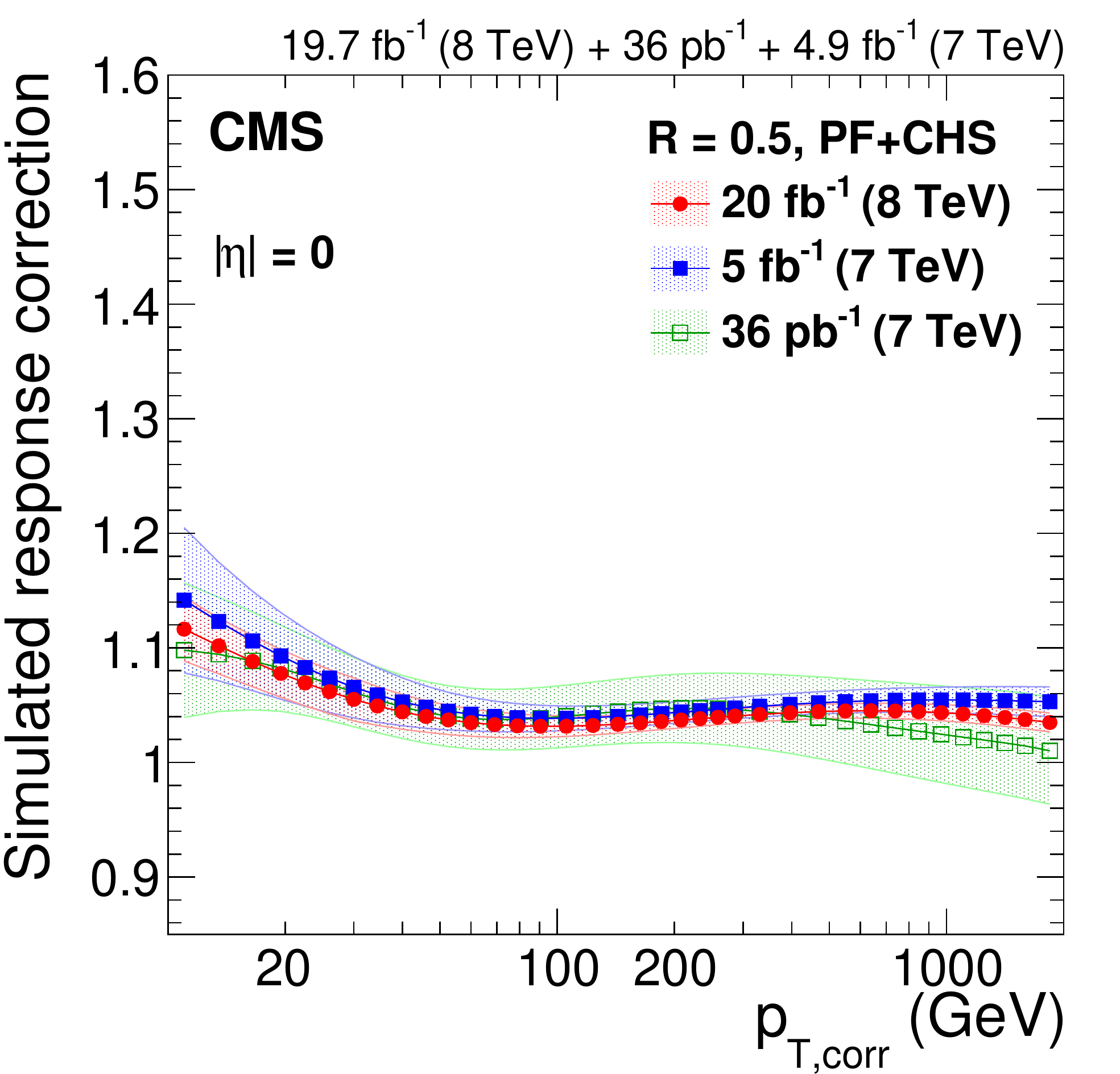}
\includegraphics[width=0.48\textwidth]{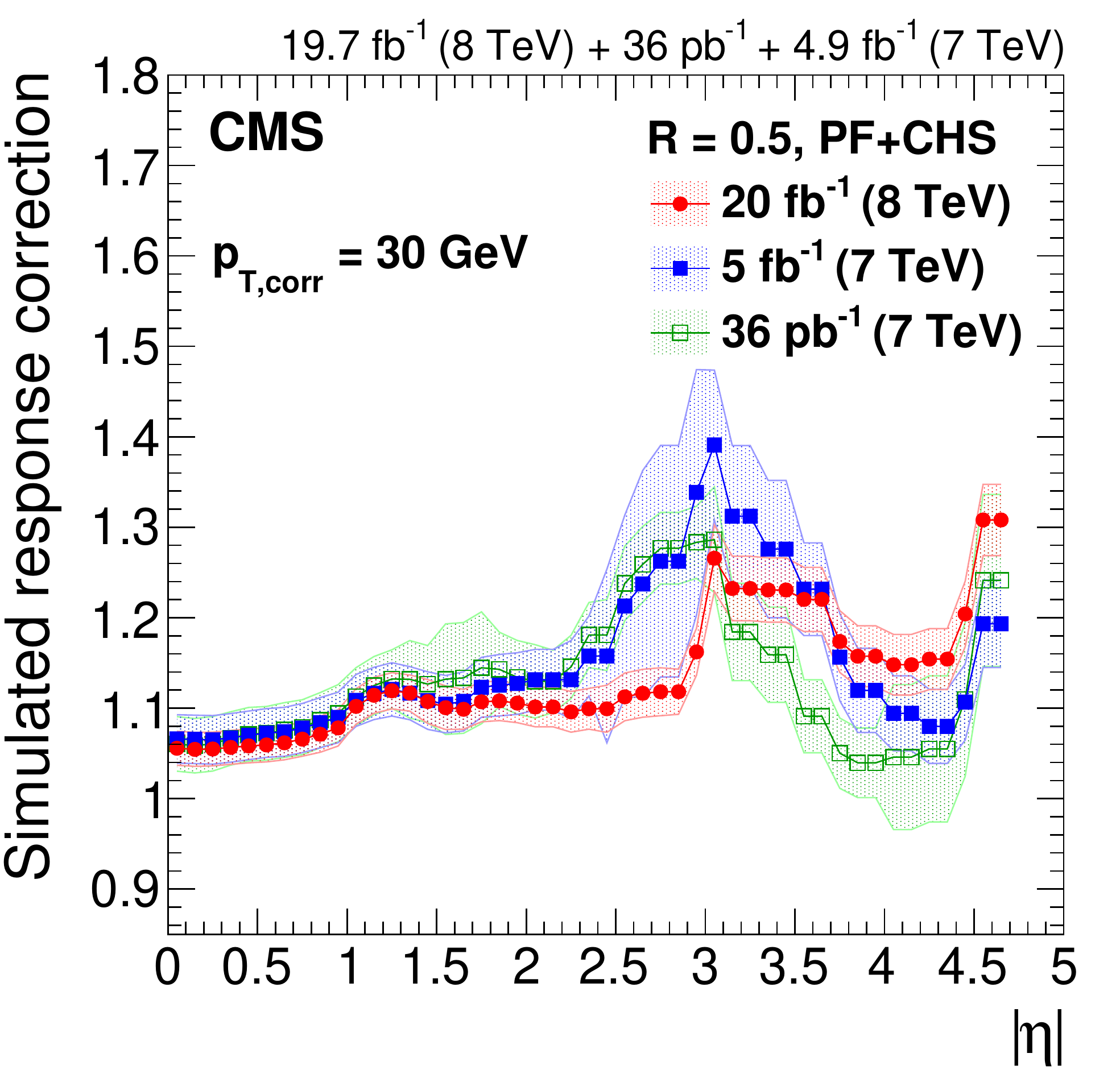}
\includegraphics[width=0.48\textwidth]{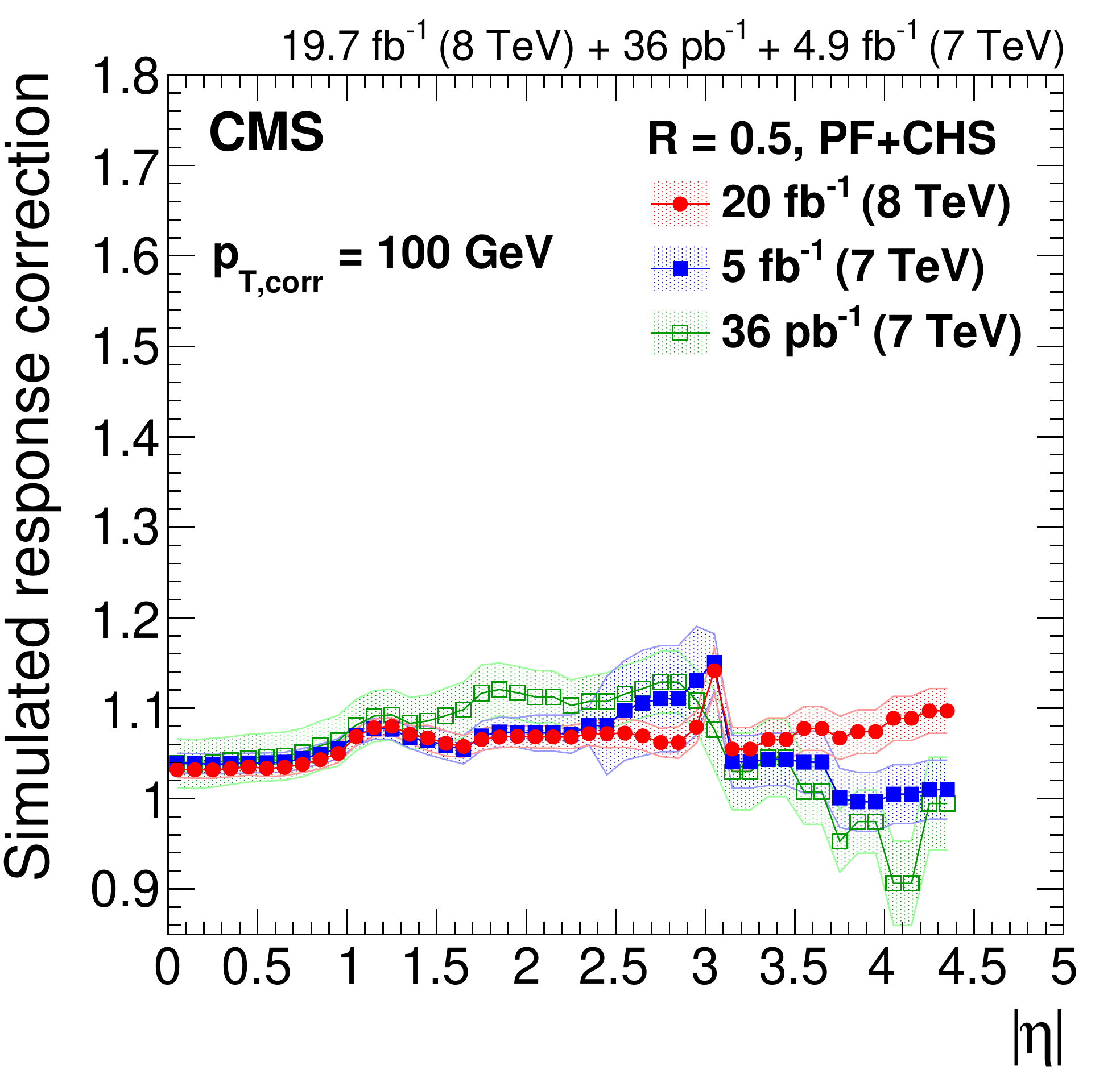}
\includegraphics[width=0.48\textwidth]{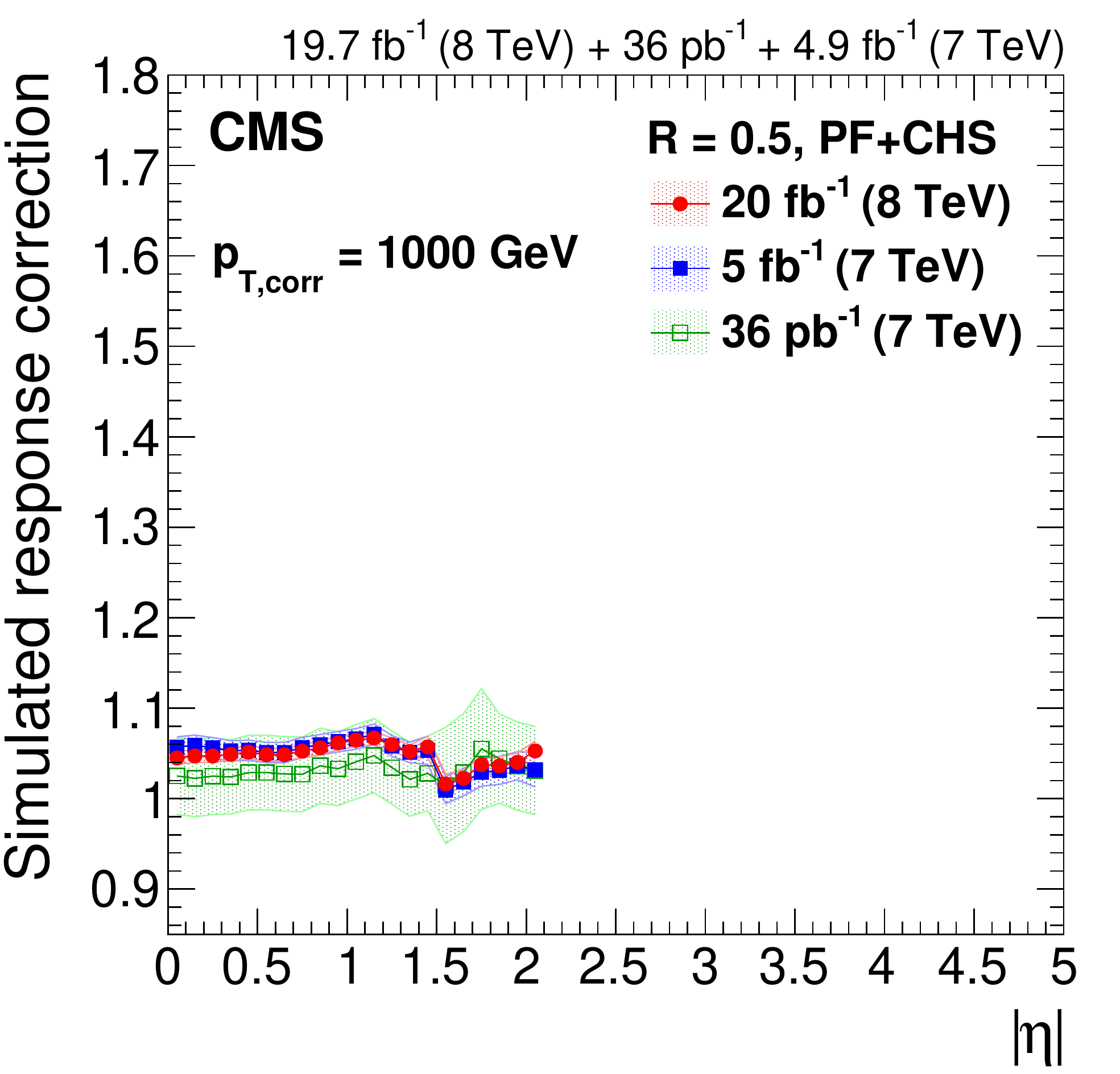}
\caption{\label{fig:l2l3}
Response correction factors with their systematic uncertainty band from simulation for the 2012 data collected at 8\TeV for PF jets with CHS and $R=0.5$, compared to corrections at 7\TeV corresponding to 36~pb$^{-1}$ of data taken in 2010 \cite{JEC_JINST} and 5\fbinv taken in 2011 \cite{CMS-DP-2012-006}. The comparison is shown at $\abs{\eta}=0$ versus $p_\mathrm{T, corr}$ (top left), and as a function of $\abs{\eta}$ at $p_\mathrm{T, corr}=30\GeV$ (top right), $p_\mathrm{T, corr}=100\GeV$ (bottom left) and $p_\mathrm{T, corr}=1000\GeV$ (bottom right). The plots are limited to a jet energy $E=\pt\cosh\eta=3500\GeV$ so as to show only the correction factors for reasonable \pt in the considered data-taking periods.}
\end{figure}

The simulated particle response corrections are summarized in Fig.~\ref{fig:l2l3} for data collected at 8\TeV and compared to corrections for 7\TeV data taken in 2010 and 2011. At low \pt, the JEC rise toward 1.15 due to the 15\% neutral hadron energy that largely falls below calorimeter thresholds. The response is quite flat at $\pt>50\GeV$, where the competing effects of increasing calorimeter response and falling tracking efficiency within the jet core compensate each other. In the barrel and endcap regions, the corrections rise with $\abs{\eta}$, due to the increasing amount of material located in front of the calorimeters, which leads to effects such as an increased rate of nuclear interactions in the tracker. The corrections are higher around $\abs{\eta}=1.3$ and $3.0$ due to the degradation of the response in the transition regions.

Significant improvements in the simulation occurred after the first year of running at 7\TeV in 2010, when in situ collision data became available for tuning the detector simulation.
After that, the simulated particle response corrections have been stable in 2011--2012 despite continuous development of the reconstruction software, and the changes have remained within the steadily-reducing systematic uncertainties. The differences introduced by the change in $\sqrt{s}$ are practically negligible.

\section{Residual corrections for data}
\label{Sec:l2l3res}

The residual data/simulation scale factors for JEC are determined after correcting jets for pileup and simulated particle response.
For consistency, the variations of the jet momenta due to corrections for pileup and simulated response are propagated
to the \ptvecmiss definition \textit{\`a la} Eq.~(\ref{eq:typeImet}).
The residual corrections for data are first determined with a sample of dijet events with low statistical uncertainty, where the response of jets
over a wide range of \pt is corrected relative to the one of jets with $\abs{\eta}<1.3$, and then with a combination of
\zmmjet, \zeejet, $\gamma$+jet, and multijet events for jets with $\abs{\eta}<1.3$ from a
$\pt$ of around 30\GeV to 1\TeV.
The basic idea, in all the considered topologies, is to exploit the transverse momentum balance, at hard-scattering level, between the jet to be calibrated and
a reference object: a jet energy scale different from unity generates imbalance at the reconstructed level.

The jet energy response is studied using the \pt balance and MPF (missing transverse momentum projection fraction) methods \cite{JEC_JINST}. While in the \pt-balance method
the jet response is evaluated by comparing the reconstructed jet momentum ($p_\text{T, jet}$) directly to the momentum of the reference object ($p_\mathrm{T, ref}$),
the MPF method considers the response of the whole hadronic activity in the event, recoiling versus the reference object.
This leads to the following definition of response for the two methods:
\begin{align}
R_{\text{jet}, \pt} &= \frac{p_\text{T, jet}}{p_\mathrm{T, ref}}, \label{eq:ptbalance_def}\\
R_{\text{jet}, \mathrm{MPF}} &= 1 + \frac{\ptvecmiss \cdot \vec{p}_\mathrm{T, ref}}{(p_\mathrm{T, ref})^2}.\label{eq:MPF_def}
\end{align}

The difference and complementarity of the two response determinations will be studied in the following sections.

Part of the transverse momentum imbalance between the jet to be calibrated and the reference object can also come from the
presence of additional jets in the event; this effect depends on the studied topology and is not correlated with the jet energy response.
For this reason, all the corrections are studied as a function of the additional jet activity in the event, quantified by the variable~$\alpha$. This is
defined as the ratio of the most energetic jet that does not originate from the event topology under study, divided by the typical momentum scale
of the event. In other words $\alpha=p_\text{T, 3rd jet}/p_\mathrm{T, ave}$ for dijet events and $\alpha=p_\text{T, 2nd jet}/p_{\mathrm{T}, \gamma/\Z}$ for $\PZ$+jet and $\gamma+$jet events.
The corrections are then extrapolated to the value they would have for $\alpha=0$ in order to address only genuine
jet energy response effects.

\subsection{Relative \texorpdfstring{$\eta$-dependent}{eta-dependent} corrections}\label{sec:relative}
\label{sec:l2res}

Residual $\eta$-dependent corrections to the jet response are obtained using dijet events, where the "tag" jet has $\abs{\eta}<1.3$, and the "probe" jet pseudorapidity is unconstrained.
In this way, the response for all jets is corrected relative to the response for central jets ($\abs{\eta}<1.3$). These residual corrections are derived from jets already corrected with the simulation-based corrections and account for any residual difference between data and simulation, as a function of both $\eta$ and \pt.

For dijet events, where the reference object (barrel jet) has poor resolution,
the biases from JER are
minimized by binning in average jet \pt instead of $p_{\mathrm{T},\rm tag}$: $p_{\mathrm{T},\rm ave}=0.5(p_{\mathrm{T},\rm tag} + p_{\mathrm{T},\rm probe})$.

This symmetric \pt binning also cancels out to first order the relative biases from ISR+FSR.
In general, $\langle y / x\rangle \neq \langle y\rangle/\langle x\rangle$, unless $x$ is constant, which is generally the case only for a sufficiently narrow bin in $x$. To avoid biases in the ratio variables, the denominator must therefore also use $p_{\mathrm{T},\rm ave}$.
This leads to the following definitions for \pt balance and MPF in dijet events:
\begin{align}
R_\text{rel}^{\pt} &=
\frac{1+\langle\mathcal{A}\rangle}
     {1-\langle\mathcal{A}\rangle},\quad\text{where}\\
\mathcal{A} &= \frac{p_\text{T, probe} - p_\text{T, tag}}{2 p_\text{T, ave}},
\quad\text{and}\\
R_\text{rel}^\mathrm{MPF} &=
\frac{1+\langle\mathcal{B}\rangle}
     {1-\langle\mathcal{B}\rangle},\quad\text{where}\\
\mathcal{B} &=
\frac{\ptvecmiss\cdot (\vec{p}_\text{T, tag}/p_\text{T, tag})}
     {2 p_\text{T, ave}}.
\end{align}

With sufficiently fine binning in $p_\text{T, ave}$, and by extrapolating the additional jet activity, not coming from the leading jet, to zero with $\alpha=p_\text{T, 3rd jet}/p_\text{T, ave}$, both variables $R_\text{rel}^{\pt}$ and $R_\text{rel}^\mathrm{MPF}$ reduce to $R_\text{rel}=\langle p_\text{T, probe}\rangle / \langle p_\text{T, tag}\rangle$. Under the assumption that $\langle p_\text{T, probe, ptcl}\rangle=\langle p_\text{T, tag, ptcl}\rangle$,
which is true after correcting for the various small second-order biases from JER and ISR+FSR, this is equivalent to the ratio of the jet responses for the tag and probe jets such that $R_\text{rel} =  R_\text{jet, probe} / R_\text{jet, tag}$. The residual $\eta$-dependent corrections are based on results obtained with the MPF method, the \pt balance results are used as a crosscheck.

\begin{figure}[htbp!]
\centering
\includegraphics[width=0.48\textwidth]{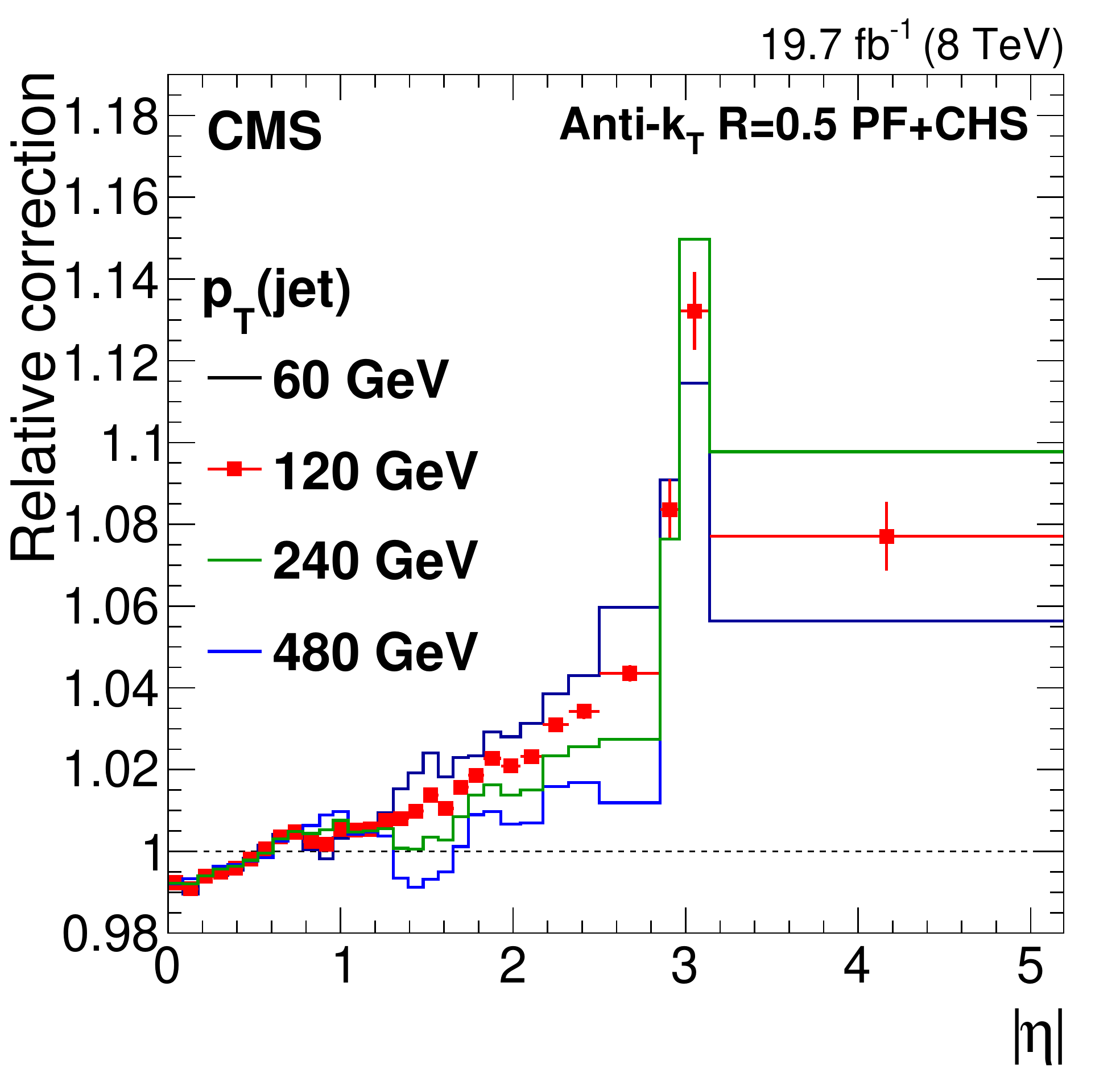}
\caption{\label{fig:L2resPR}
Relative energy scale correction for $\pt=60$, 120, 240 and 480\GeV as a function of $\abs{\eta}$. The residual corrections increase toward high rapidity and low \pt, where effects from nonlinear calorimeter response become more important. The curves are limited to a jet energy $E=\pt\cosh\eta=4000\GeV$ (corresponding to $\eta\approx 2.8$ for a jet with $\pt=480\GeV$) so as to show only the correction factors for reasonable \pt in the considered data-taking period. The statistical uncertainty associated with a constant fit versus \pt is shown for $\pt=120\GeV$ (markers).
}
\end{figure}

As shown in Fig.~\ref{fig:L2resPR}, the relative $\eta$- and \pt-dependent correction $R_\mathrm{rel, MC}/R_\mathrm{rel, data}$ varies between 0.99 and 1.01 in the barrel at $\abs{\eta}<1.3$, between 0.99 and 1.06 at $1.3<\abs{\eta}<2.9$, and increases to $1.15$ in HF. Some \pt dependence is observed in the endcaps relative to the barrel, with the residual corrections approaching unity at high \pt, where nonlinearities in calorimeter response are reduced.
In the following we will review the corrections for ISR+FSR, JER, and jet \pt dependence, as well as the associated uncertainties for the $\eta$-dependent corrections.

\subsubsection*{Initial- and final-state radiation correction}\label{sec:l2fsr}

For central-forward jet pairs there is a higher probability for the ISR to be radiated opposite to the central jet, and the FSR activity may differ slightly for the jets at different $\eta$, which leads to some residual dependence of the measured value of the \pt-balance or MPF response, $R_\text{rel}$, on additional jet activity $\alpha$.
We evaluate this dependence in bins of $\eta$, for the linearly extrapolated $\alpha\to 0$ and $\alpha<0.2$ respectively, and compute the following data/simulation double ratio:
\begin{equation}
\label{eq:kFSRdataOverMC}
 k_\mathrm{FSR}(\alpha=0.2) = {\left(\frac{R_\text{rel}^\text{data}(\alpha\to 0)}{R_\text{rel}^\mathrm{MC}(\alpha\to 0)} \right)} \Bigg/ { \left(\frac{R_\text{rel}^\text{data}(\alpha<0.2)}{R_\text{rel}^\mathrm{MC}(\alpha<0.2)} \right) }.
\end{equation}

\begin{figure}[htbp!]
\includegraphics[width=0.48\textwidth]{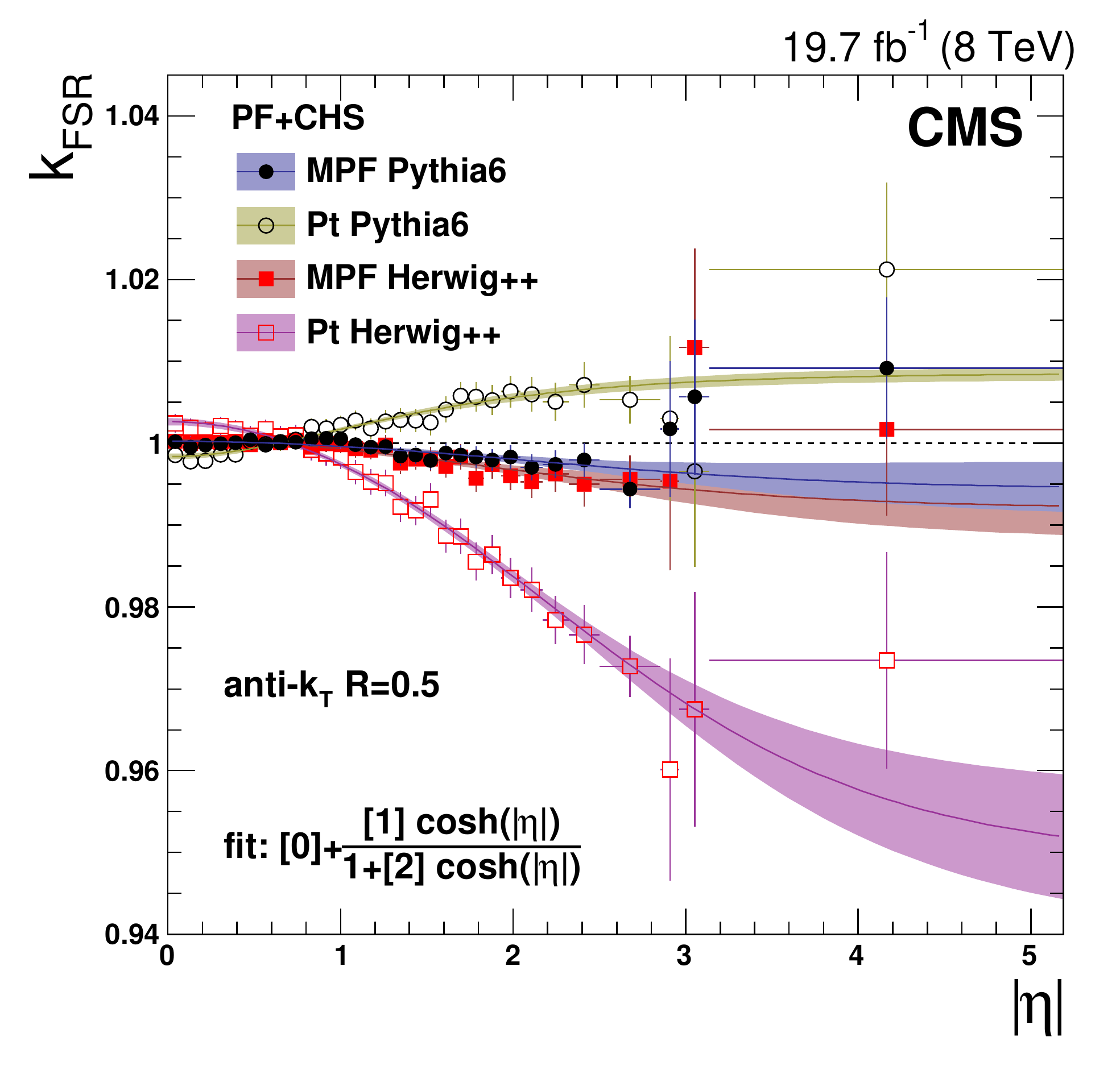}
\includegraphics[width=0.48\textwidth]{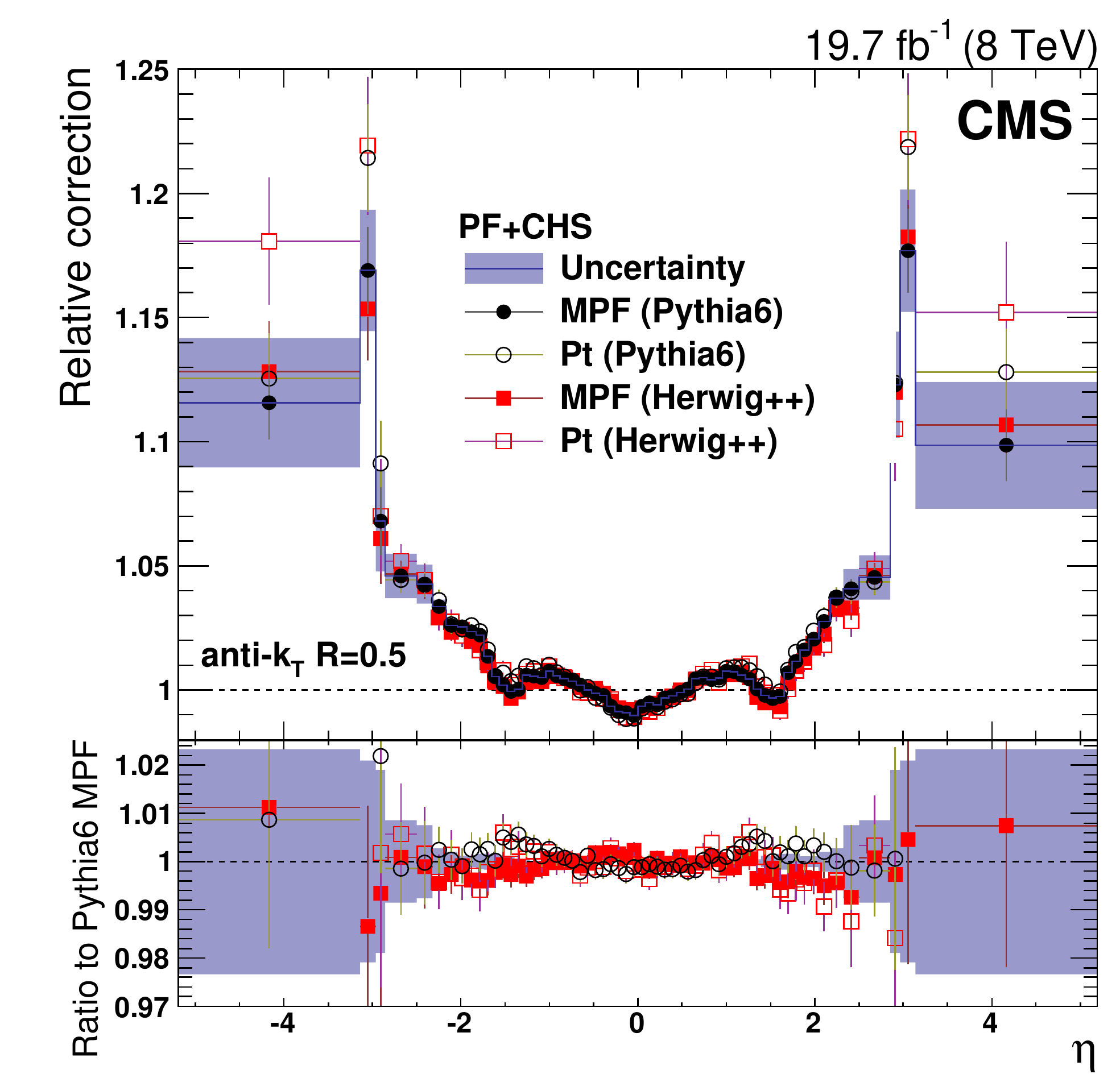}
\caption{\label{fig:kfsr}
The $k_\mathrm{FSR}(\alpha=0.2)$ correction factor (defined in Eq.~(\ref{eq:kFSRdataOverMC})) plotted vs. $\abs{\eta}$ (left). This ratio is used for ISR+FSR corrections that are applied to dijet events with $\alpha<0.2$, for the MPF and \pt-balance methods, and for \PYTHIA6.4 tune Z2* and \HERWIGpp~2.3 tune EE3C. The points are fitted with $f(\eta) = p_0 + {p_1 \cosh(\eta)}/{( 1 + p_2 \cosh(\eta) )}$ as in Ref.~\cite{JEC_JINST}.  Relative $\eta$ corrections obtained with the MPF and balance methods and the \PYTHIA6.4 tune Z2* and \HERWIGpp~2.3 tune EE3C MC generators (right). The results are shown after corrections for ISR+FSR, and compared to the central values, obtained with the MPF method and \PYTHIA6.4 tune Z2* simulated events. }
\end{figure}

The correction factor $k_\mathrm{FSR}$ (we use the subscript FSR instead of ISR+FSR for brevity) is determined separately for the MPF and \pt-balance methods and for \PYTHIA6.4 and \HERWIGpp~2.3, as shown in Fig.~\ref{fig:kfsr}, and is then parameterized versus $\abs{\eta}$ with the same functional form as in Ref.~\cite{JEC_JINST}.
The differences between \PYTHIA6.4 and \HERWIGpp~2.3 for the \pt-balance method are up to 6\% at $\abs{\eta}<5.2$ prior to the application of ISR+FSR corrections, as seen in Fig.~\ref{fig:kfsr} (left). Both agree well after the ISR+FSR correction, as shown in Fig.~\ref{fig:kfsr} (right), but
the MPF method is much less sensitive to ISR and FSR biases than the \pt-balance method, because the entire hadronic recoil is used for the MPF balance.

\subsubsection*{Resolution correction}\label{sec:l2jer}

The MPF and \pt-balance methods are both sensitive to the relative differences in JER between the jets. This bias is expected to cancel out for the data/MC ratio of $R_\mathrm{ rel}$ when the jets in the simulation are smeared to match the measured resolution in data using the relation:
\begin{equation}\label{eq:l2jer}
p_{\mathrm{T},\rm smeared} = \pt \mathrm{Gaussian}(\mu=1, \sigma=\sqrt{k^2- 1}\,\sigma_\mathrm{MC}),
\end{equation}
where $k$ is the data/MC scale factor for JER determined in Section~\ref{sec:jer} and $\sigma_\mathrm{MC}$ is the JER in the MC simulation. The factor $k$ varies between 1.05 and 1.40 depending on $\eta$.
The jet \pt is multiplied by a random number drawn from a Gaussian distribution with mean $\mu=1$ and width $\sigma$, such that the smeared jet has the same resolution $k\sigma_\mathrm{MC}$ as the jets in data.  The smearing is applied on a jet-by-jet basis to all jets in the event, such that the resolution correction is propagated to the $p_\mathrm{T}$-balance and MPF methods in a consistent way.

\subsubsection*{Relative correction: \pt dependence}\label{sec:l2pt}

\begin{figure}[htbp!]
\centering
\includegraphics[width=0.47\textwidth]{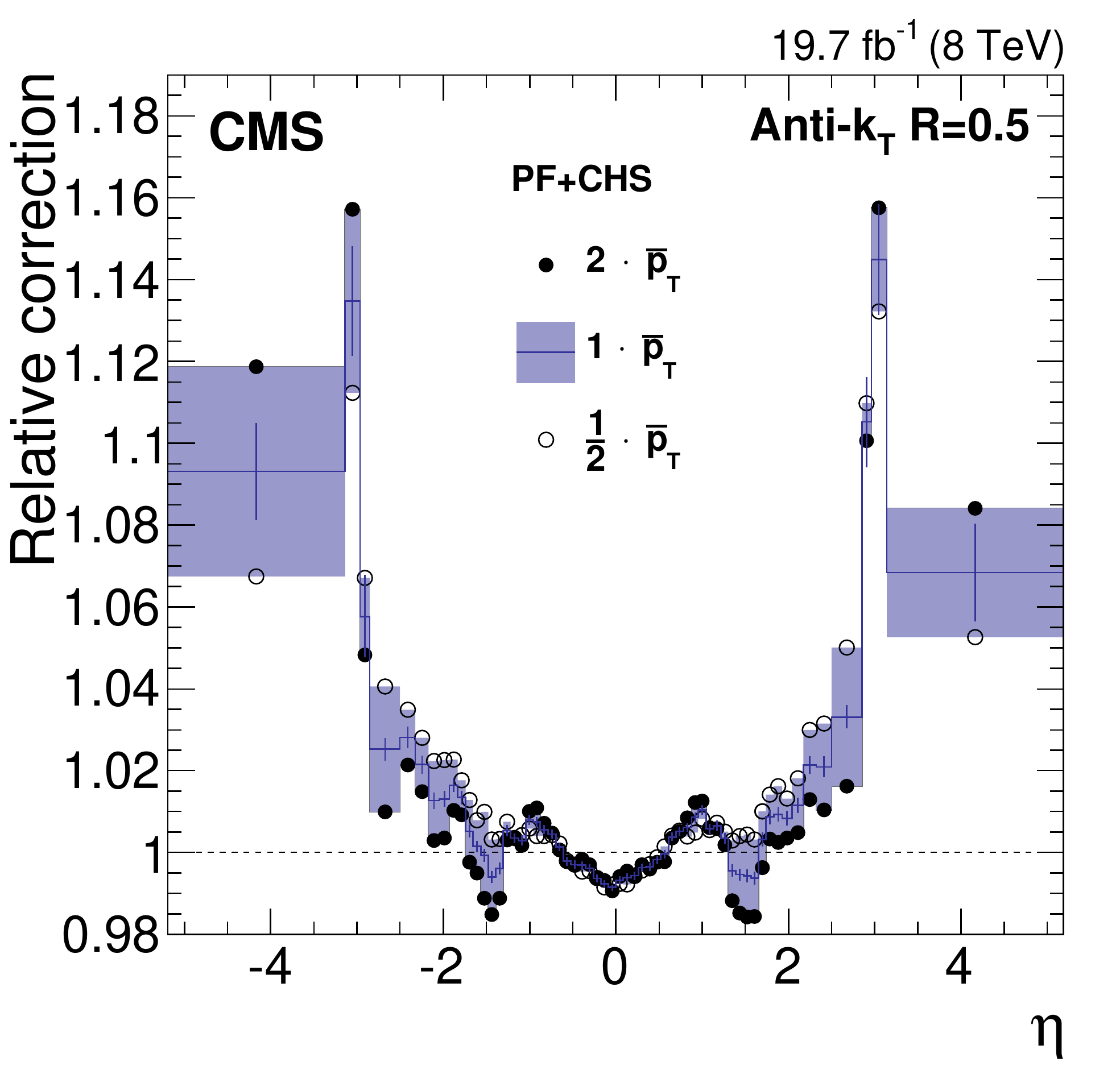}
\includegraphics[width=0.47\textwidth]{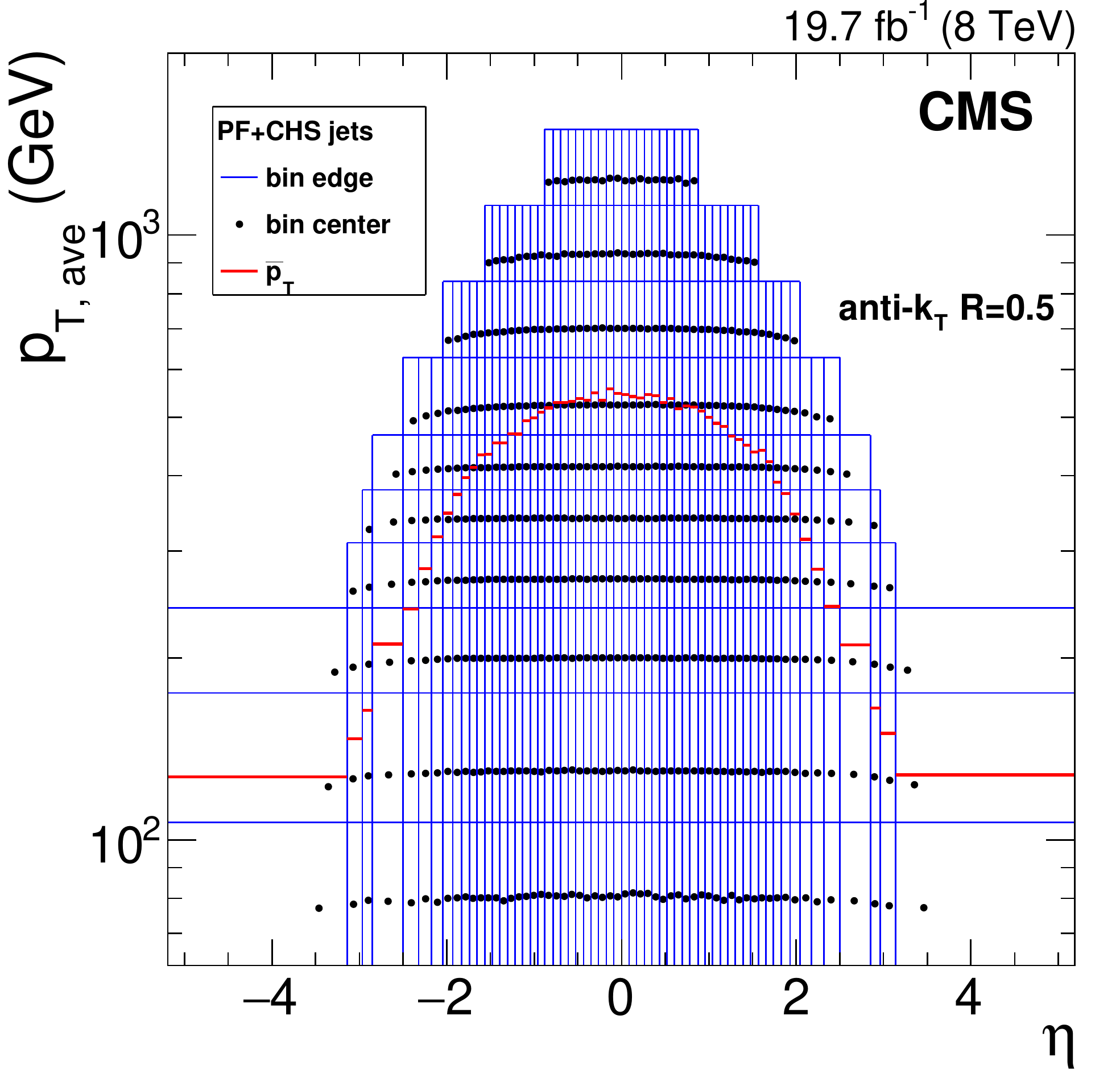}
\caption{\label{fig:l2res}
Relative $\eta$ correction factor at the crossover $\bar{p}_\mathrm{T}$ (defined as the value of \pt where the log-linear and constant fits versus $p_\mathrm{T, ave}$ agree) value, and at half and twice the $\bar{p}_\mathrm{T}$ values (left).  The statistical uncertainty in the constant fit at each value of $\bar{p}_\mathrm{T}$ is also shown.
Distribution of the \pt and $\eta$ bins used in the dijet balance measurement, with a point at the average \pt and $\eta$ for each bin (right). The horizontal red lines indicate the crossover $\bar{p}_\mathrm{T}$ value for each bin.
}
\end{figure}

The $\eta$-dependent corrections are studied in bins of average jet \pt, where a slight \pt dependence is observed. For this reason, the $\eta$-dependent corrections are parameterized with a log-linear \pt-dependence, according to the formula $p_0+p_1\log(\pt)$.
The correction factor as a function of $\eta$, as obtained from the \pt-dependent fit is shown in Fig.~\ref{fig:l2res}~(left), compared to the result from a constant fit. Here, the central value is obtained from evaluating the \pt-dependent correction at the \pt value for which the constant fit and the logarithmic fit agree, $\bar{p}_\mathrm{T}$. The blue band is obtained by varying the \pt at which the logarithmic fit is evaluated between 0.5 times and 2 times $\bar{p}_\mathrm{T}$.
The $\bar{p}_\mathrm{T}$ is typically close to the mean \pt of the dijet samples, and is shown in Fig.~\ref{fig:l2res} (right).
The \pt-dependent fit is used as the central result over the whole $\eta$ range, with the exception of the HF ($\abs{\eta}>$3). For this region, to mitigate the effect of statistical fluctuations (visible \eg in Fig.~\ref{fig:l2res}~(left)), the correction is taken from the constant fit and symmetrized over positive and negative $\eta$ values.

\subsection{Relative correction uncertainties}\label{sec:time}

The largest uncertainties in the relative corrections arise from the following sources:
\begin{itemize}
\item \textbf{ISR+FSR}, ${\leq} 0.2\%$.
The \textit{RelativeFSR} uncertainty in $k_\mathrm{FSR}$ is estimated by using \HERWIGpp~2.3 as "data" and comparing how well the different methods reproduce the ratio of particle-level simulated responses. This uncertainty increases smoothly with increasing $\abs{\eta}$, up to 0.2\% in HF.
\item \textbf{Jet \pt resolution}, ${\leq} 1.4\%$.
The \textit{RelativeJER} systematic uncertainty on the JER correction is estimated by varying the data/MC scale factor $k$ in Eq.~(\ref{eq:l2jer}) within the uncertainties determined in Section~\ref{sec:jer}, which are between 2\% and 20\%, depending on $\eta$. This uncertainty mainly affects the $\eta$ bins in the HF, where JER is poorly constrained from data.
\item \textbf{Relative correction \pt dependence}, ${\leq} 1.4\%$.
Half of the difference between the log-linear and constant fits observed in Fig.~\ref{fig:l2res} is taken as a \textit{RelativePt} systematic uncertainty to account for uncertainties coming from the choice of the log-linear shape for the fit. This is the dominant uncertainty in the barrel and endcaps.
\item \textbf{Statistical uncertainty}, ${\leq} 0.9\%$.
The number of events available in data for the $\eta$-dependent corrections is limited in the endcap and HF regions due to the large prescales applied to the dijet triggers during data taking.
To estimate the impact of this on physics analyses,
the corrections are symmetrized and determined in wide bins of $\abs{\eta}$. The remaining statistical uncertainty of up to 2.5\% is assigned as \textit{RelativeStat} systematic uncertainty.
\item \textbf{Time dependence}, ${\leq} 1.0\%$.
The \textit{TimeEta} systematic uncertainty is estimated as the RMS of the $\eta$-dependent correction factors for a set of about ten data-taking periods, chosen arbitrarily in order to have comparable integrated luminosities.
The variation is assumed to come from residual scale shifts remaining after the radiation damage corrections have been applied to the ECAL and HCAL, and increases toward high rapidities, which suffer larger radiation damage.
\end{itemize}

\subsubsection*{Uncertainty correlations versus $\eta$}

The \textit{RelativeJER}, \textit{RelativePt}, and \textit{RelativeStat} systematic uncertainties are assumed to be correlated versus $\eta$ within the barrel (\textit{BB:} $\abs{\eta}<1.3$), the region of the endcap that is within tracker coverage (\textit{EC1:} $1.3<\abs{\eta}<2.5$), the region of the endcap which is outside the tracker coverage (\textit{EC2:} $2.5<\abs{\eta}<3$), and within the hadron forward calorimeter (\textit{HF:} $3<\abs{\eta}<5.2$), but not between these regions. The \textit{RelativeStat} uncertainty is significant only in the more forward regions, and is only provided for the two latter regions (EC2 and HF, which are considered uncorrelated). All other systematic uncertainties relevant for $\eta$-dependent corrections (\textit{RelativeFSR, TimeEta}) are considered to be fully correlated versus $\eta$. Each correlated region is treated with a separate systematic source, and these are provided separately to the users.

\begin{figure}[htbp!]
\centering
\includegraphics[width=0.49\textwidth]{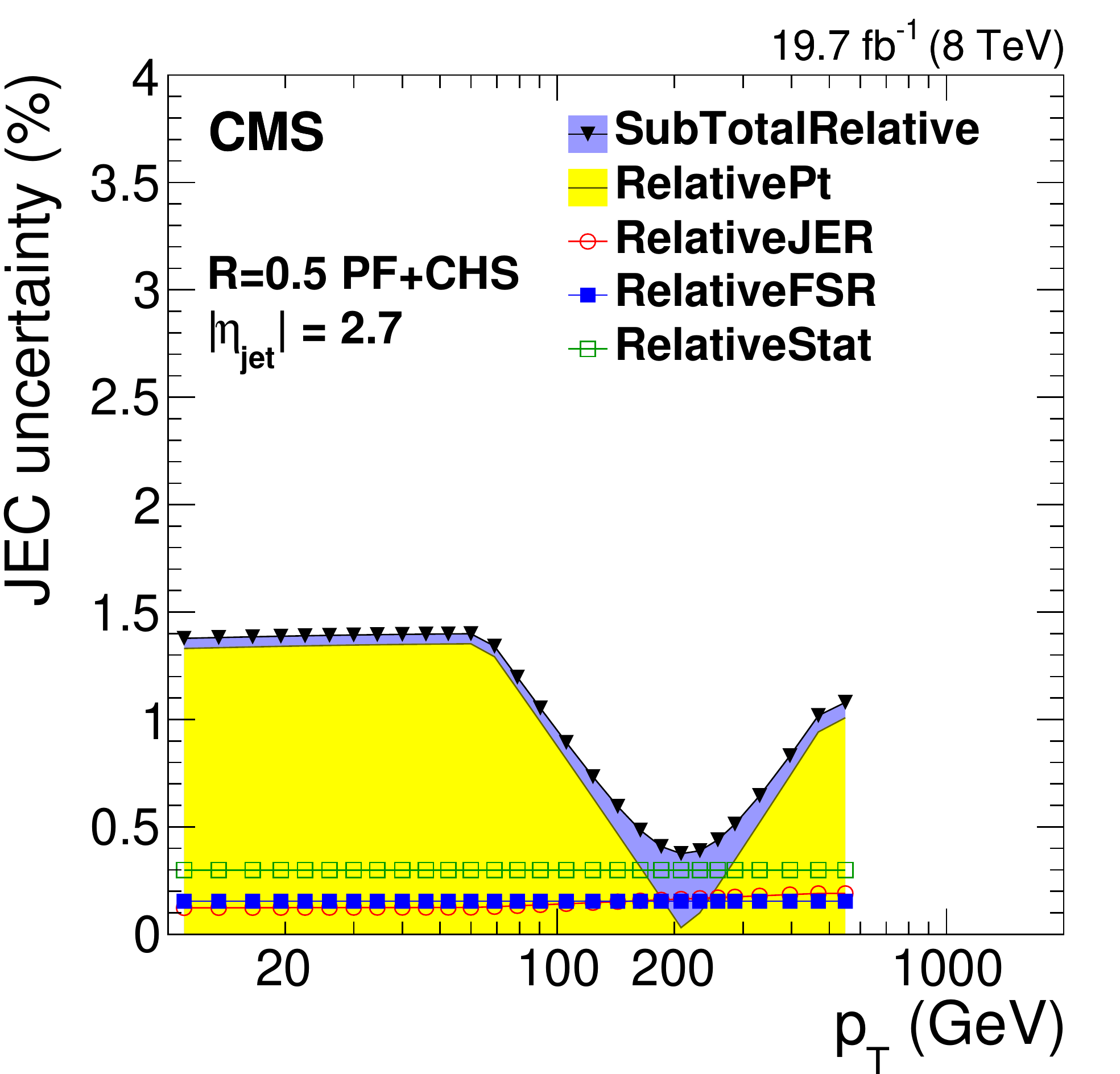}
\includegraphics[width=0.49\textwidth]{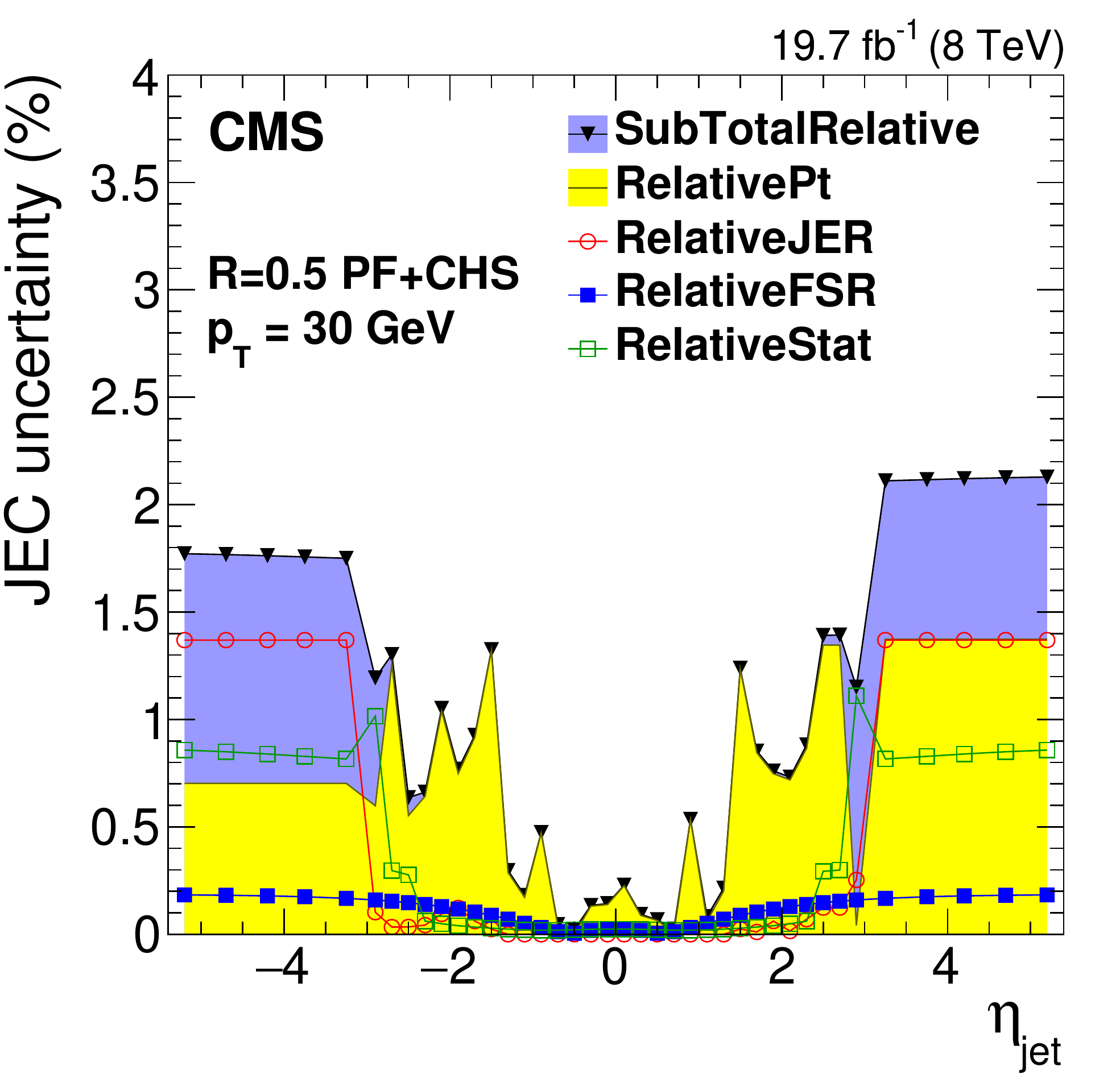}
\includegraphics[width=0.49\textwidth]{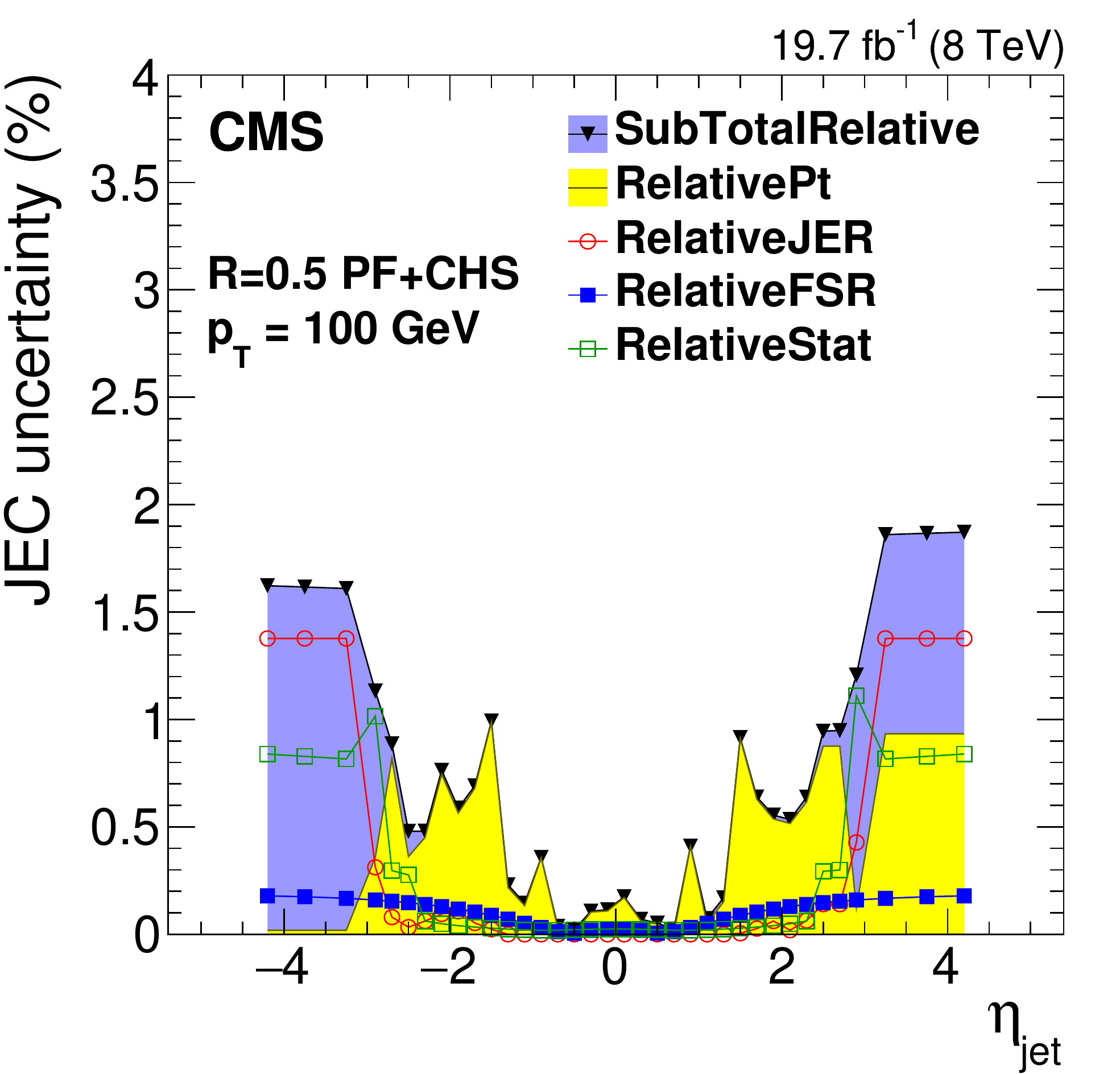}
\includegraphics[width=0.49\textwidth]{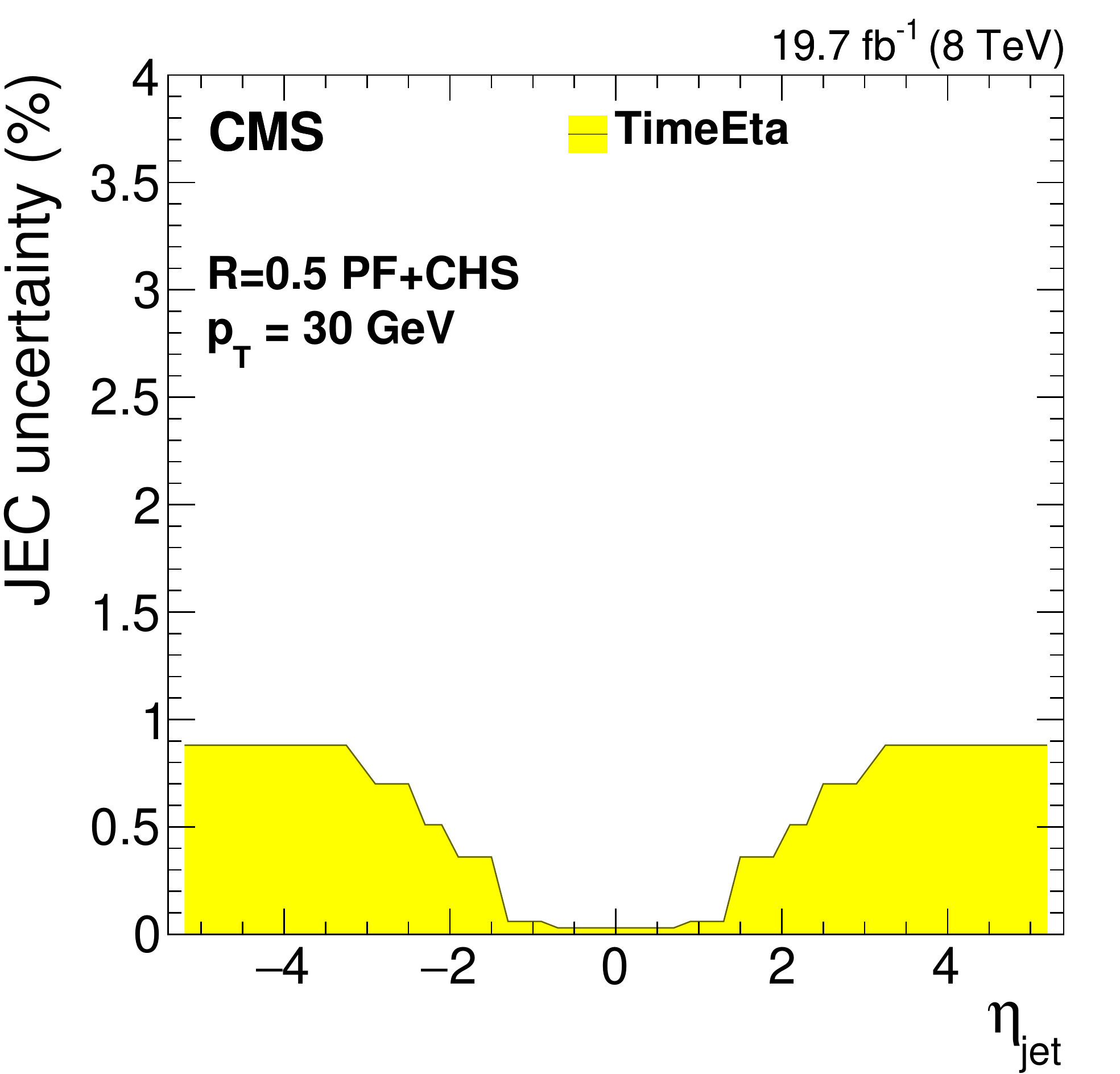}
\caption{\label{fig:etasys}
Systematic uncertainties for the relative $\eta$-dependent corrections as a function of jet \pt (top left) and as a function of jet
$\eta$ for jets with $\pt=30\GeV$
(top right) and for jets with $\pt=100\GeV$ (bottom left). Time-dependent uncertainties
as a function of jet $\eta$ for jets with $\pt=30\GeV$ (bottom right). The plots are limited to a jet energy $E=\pt\cosh\eta=4000$\,GeV so as to show only uncertainties for reasonable \pt in the considered data-taking period.
\textit{SubTotalRelative} is the quadratic sum of \textit{RelativePt}, \textit{RelativeJER}, \textit{RelativeFSR} and \textit{RelativeStat}.
}
\end{figure}

The systematic uncertainties in the relative $\eta$-dependent corrections are summarized in Fig.~\ref{fig:etasys}, for low (30\GeV) and medium (100\GeV) \pt versus $\eta$, and for the outer endcap ($\eta=2.7$) versus \pt.
The time-dependent uncertainties are optional for analyses that are performed on the full 2012 data and are shown separately versus $\eta$ at $\pt=30\GeV$. Among the time-dependent systematic uncertainties only the \textit{TimeEta} is relevant.
The uncertainties are small at high \pt and for central rapidities within the tracker coverage. They increase to 2.1\% at high rapidity mainly due to the limited number of events available in the data for deriving the JEC, JER and ISR+FSR corrections.  The dominant \textit{PileUpPt} uncertainty is inherently asymmetric and has the largest visible differences in the HF region, where the asymmetric log-linear fit is compared to a symmetrized constant fit used for central value in HF. The uncertainty versus $p_\mathrm{T}$ changes sign around 100\GeV for the negative $\eta$ side while it remains same-sign for the positive $\eta$.

\subsection{Absolute corrections}

The absolute JES at $\abs{\eta}<1.3$ is determined with \zmmjet, \zeejet\ and $\gamma$+jet events for jet \pt between 30 and 800\GeV by
comparing the reconstructed \pt of the jet
to that of a precisely measured object (the $\PZ$ boson, or the photon). The response for jets with $\pt>800\GeV$ is constrained using multijet events, where a high-$\pt$ jet in the barrel region is balanced by a recoil system, composed of two or more lower-$\pt$ jets.

For all these analyses, the corrections are derived by comparing the jet energy response (with different methods) in data and simulation, using events in the central region, where jets are already corrected with the simulation-based corrections and $\eta$-dependent residual corrections.
As detailed below,  the response is observed to be slightly lower in data than in simulation. In addition, the ratio of data over the MC prediction of the response shows a \pt dependence. The two effects are factorized and addressed in successive steps. First, a rough estimate of the \pt independent correction is derived from the analysis of \zmmjet\ events. Second, the response and its \pt dependence are determined precisely from a global fit (described in Section~\ref{sec:l3res_globalfit}), with the individual response values obtained from the different channels (\zmmjet, \zeejet,  $\gamma$+jet, multijet) as input.

\subsubsection*{Methods}

\begin{figure}[htbp!]
\begin{center}
\includegraphics[width=0.5\textwidth]{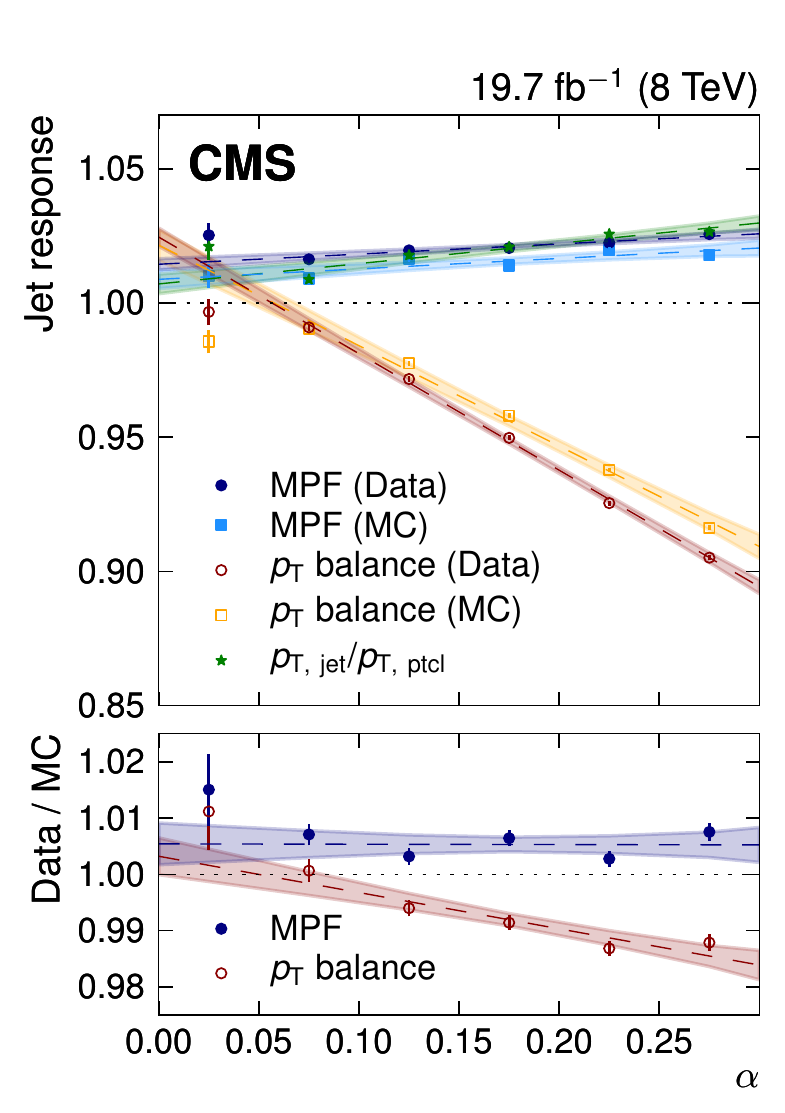}
\end{center}
\caption{\label{fig:RespVSalpha} Jet response obtained with the \pt-balance and MPF methods in $\PZ$+jet events (points), for both data and simulation ({\MADGRAPH}~4+\PYTHIA6.4 tune Z2*), plotted as a function of $\alpha=p_\mathrm{T,2nd~jet}/p_\mathrm{T,Z}$ (top). The response in data is scaled by a factor of 1.02, constant as a function of \pt. A fit to a first-order polynomial (dashed lines) is shown, together with the statistical uncertainty from the fit (shaded bands). Only events with $p_\mathrm{T, Z}>30\GeV$ and $|\eta_\text{jet}|<1.3$ are considered. The ratio of the jet response from the \pt-balance and MPF methods in data and simulation shown in the bottom panel.
The simulated jet response $p_\mathrm{T,jet}/p_\text{T, ptcl}$ is higher than unity because the jets are corrected with JEC from QCD dijet events with lower jet response than $\PZ$+jet events due to higher gluon fraction and larger underlying event.
}
\end{figure}

The absolute jet response is measured relative to a photon or $\PZ$ boson momentum scale, using the \pt-balance ($R_\mathrm{jet, \pt}$) and MPF ($R_\mathrm{jet, MPF}$) methods \cite{JEC_JINST}, as defined
in Eqs.~(\ref{eq:ptbalance_def}) and (\ref{eq:MPF_def}), with $p_\text{T, ref}=p_{\mathrm{T}, \gamma/\Z}$.
The measurements are affected by biases from ISR+FSR, underlying event (UE) and out-of-cone (OOC) showering.
To correct for the FSR+ISR bias, we define a $k_\mathrm{FSR}$ correction as follows:
\begin{equation}\label{eq:krad}
k_\mathrm{FSR}(\alpha) = \frac{R_\text{jet}(\alpha\to 0)}{R_\text{jet}(\alpha)},
\quad \alpha=\frac{p_{\mathrm{T}, \mathrm{2nd~jet}}}{p_{\mathrm{T}, \gamma/\Z}},
\end{equation}
where the jet response $R_\text{jet}$ is measured with the MPF or the \pt-balance method, with separate corrections for each.
As shown in Fig.~\ref{fig:RespVSalpha}, the value of $k_\mathrm{FSR}(\alpha)$ is linearly dependent on $\alpha$ for $0.05<\alpha<0.3$.
Because the average \pt of the $\PZ$ boson decreases with $\alpha$, the particle-level jet response obtained from simulation also shows a dependence on $\alpha$.
Figure~\ref{fig:RespVSalpha} demonstrates that the MPF method is significantly less sensitive to ISR+FSR and the modeling of these processes, than the \pt-balance method: the slope $\Delta R_\text{jet} / \Delta\alpha$ is about $-0.3$ for the \pt-balance method and $+0.06$ for the MPF method.
For the data/MC ratio these slopes are further reduced by an order of magnitude, confirming a good modeling of the OOC and UE effects.
It can be shown that the ratio of MPF and \pt-balance slopes versus $\alpha$ is
\begin{equation}
\frac{ \rd R_{\text{jet}, \mathrm{MPF}}/ \rd \alpha}{ \rd R_{\text{jet}, \pt}/ \rd \alpha} =
1 - \frac{R_\mathrm{FSR+ISR~jets}}{R_\text{jet}}.
\end{equation}

The difference in jet response between the leading jet and the ISR+FSR jets is typically less than 20\%,
as seen in Fig.~\ref{fig:l2l3}, but the sign can be either positive or negative.
The slope of $k_\mathrm{FSR}$ has some dependence on the jet flavor (gluons radiate more than quarks) and it depends, \eg, on the parton shower model used in the MC simulation. As shown in Section~\ref{sec:l3res_globalfit}, determining $k_\mathrm{FSR}$ in narrow bins of $p_\mathrm{T, ref}$ is needed in order to study the \pt dependence of the JES.

The remaining effects of UE and OOC affect MPF and \pt balance slightly differently.
It can be shown that, having corrected for ISR+FSR, the balancing and MPF responses can be written as
\begin{eqnarray}
R_{\text{jet}, \pt}
&=& R_\text{jet}\left(1 - \frac{p_\text{T, ptcl}^\mathrm{OOC}}{p_\text{T, ptcl}} + \frac{p_\text{T, ptcl}^\mathrm{UE}}{p_\text{T, ptcl}}\right),\label{eq:pTbal_oocue}\\
R_\mathrm{jet, MPF}
&=& R_\text{jet}\left[1 - \left(1-\frac{R_\mathrm{OOC}}{R_\text{jet}}\right)\frac{p_\text{T, ptcl}^\mathrm{OOC}}{p_\text{T, ptcl}} + \left(1-\frac{R_\mathrm{ UE}}{R_\text{jet}}\right)\frac{p_\text{T, ptcl}^\mathrm{UE}}{p_\text{T, ptcl}}\right] \label{eq:mpf_oocue},
\end{eqnarray}
where $p_\text{T, ptcl}^\mathrm{OOC}$, $p_\text{T, ptcl}^\mathrm{UE}$ are OOC and UE transverse momenta projected to the reference object axis, and $R_\mathrm{OOC}$, $R_\mathrm{UE}$ are their effective responses. Compared to the \pt balance, the residual biases for MPF are multiplied by a factor that is typically about 10\% or less, and can be safely ignored. The corrections for OOC and UE compensate each other, but for jet radii $R\geq 0.5$ the OOC effect is smaller than the one coming from the UE at low \pt. We can therefore estimate an upper limit on these biases by assuming an UE energy density of about 1\GeV per unit of jet area, which gives a correction of at most ${\approx} 2.6$\% for $p_\text{T, ptcl}=30\GeV$ and jet distance parameter $R=0.5$. This is compatible with the magnitude and sign of the observed difference of less than 2\% between MPF and \pt balance at the $\alpha\to 0$ limit in Fig.~\ref{fig:RespVSalpha}.

\begin{figure}[htbp!]
\includegraphics[width=0.5\textwidth]{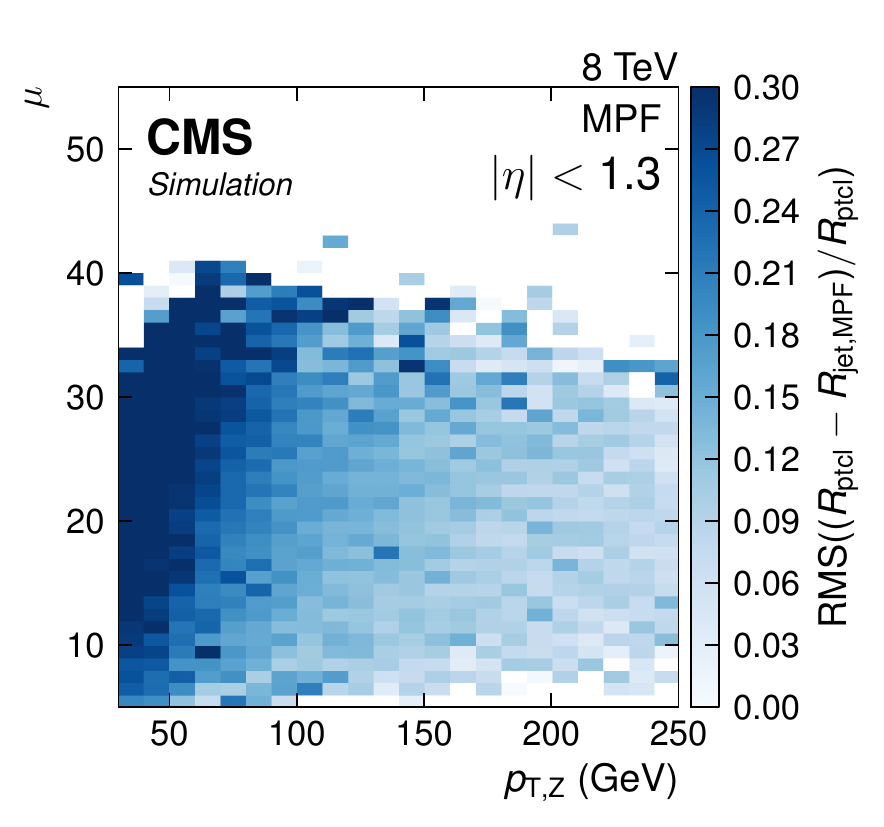}
\includegraphics[width=0.5\textwidth]{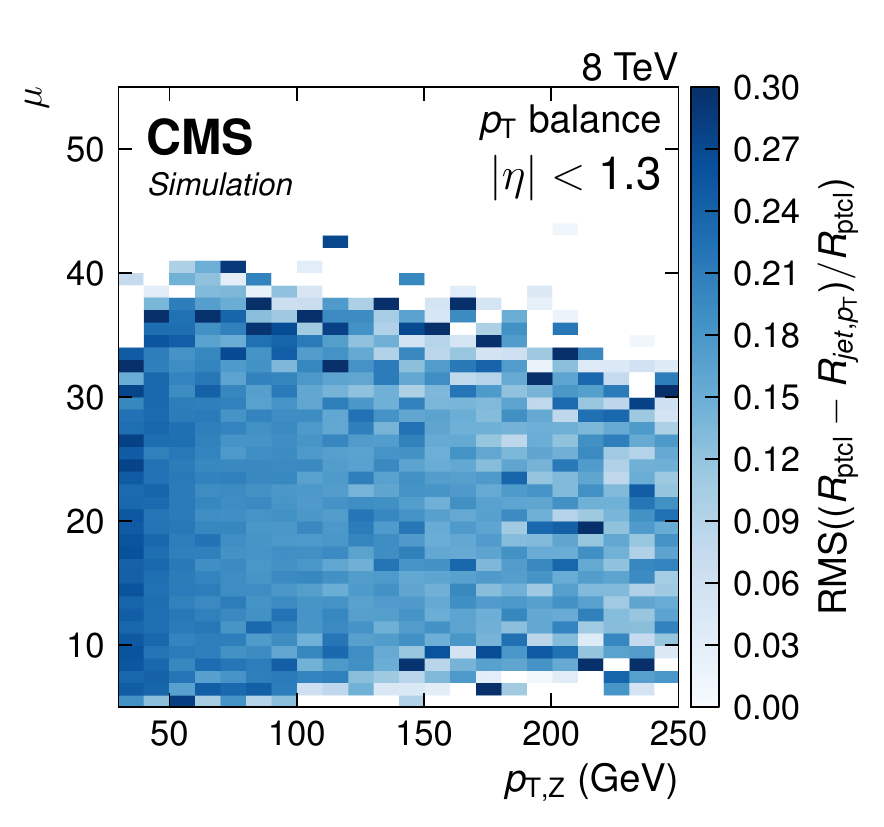}
\caption{\label{fig:ptvsmpf}
Relative resolution (blue scale) in the plane of mean number of pileup events ($\mu$) and $\PZ$ boson transverse momentum ($p_\mathrm{T, Z}$) for the MPF balance (left) and \pt-balance methods (right).
}
\end{figure}

Although the MPF and \pt-balance methods are biased in different ways, both can be corrected for ISR+FSR and are complementary to each other. The remaining biases from OOC and UE (both magnitude and response, see Eqs.~(\ref{eq:pTbal_oocue}) and (\ref{eq:mpf_oocue})) affect the \pt balance and MPF methods differently, and therefore fitting both simultaneously reduces the overall systematic uncertainty in the global fit.  The relative statistical power (quantified by the relative resolution of the measured response, compared to the particle-level response) of \pt balance and MPF depends on the jet \pt and the level of PU, as seen in Fig.~\ref{fig:ptvsmpf}. The MPF method is sensitive to smearing in \ptvecmiss caused by PU, while the \pt balance is sensitive to the smearing in the momentum balance caused by ISR+FSR. The former effect dominates at low \pt, while the latter dominates at high \pt, such that both methods have similar sensitivity at $\pt\approx 100\GeV$ for $\langle\mu\rangle=20$.

\label{Sec:L3residual}
\subsubsection*{$\PZ$+jet and $\gamma$+jet balance}

The event selection is described in Section~\ref{sec:event_sel}. The JES is determined relative to precisely measured muons, electrons, and photons, with a tracker scale uncertainty of 0.2\% for muons at $\abs{\eta}<2.4$ \cite{cmsmuons},
an ECAL scale uncertainty of 0.5\% for electrons at $\abs{\eta}<2.4$~\cite{Chatrchyan:2013dga}, and 0.2\% for photons at $\abs{\eta}<1.3$~\cite{Khachatryan:2015iwa}.

The peak of the invariant-mass distribution of $Z\to\mu\mu ({\Pe\Pe})$ events is used to validate the muon (electron) energy scale between data and simulation. These are found to agree within $0.2\%\,(0.5\%)$.
Additional checks ensure that the \ptvecmiss used in the MPF method is not biased by minimum-ionizing particle deposits of muons in the calorimeters, or by residual leakage of electron and photon energy into ECAL or HCAL not clustered in the reconstructed electron or photon. As the photon energy scale includes corrections for these unclustered contributions, special care is taken in order to avoid double counting of the leakage energy from fully calibrated PF photon superclusters to \ptvecmiss (such double counting will be referred to as electromagnetic footprint effect).

\begin{figure}[htbp!]
\begin{center}
\includegraphics[width=0.49\textwidth]{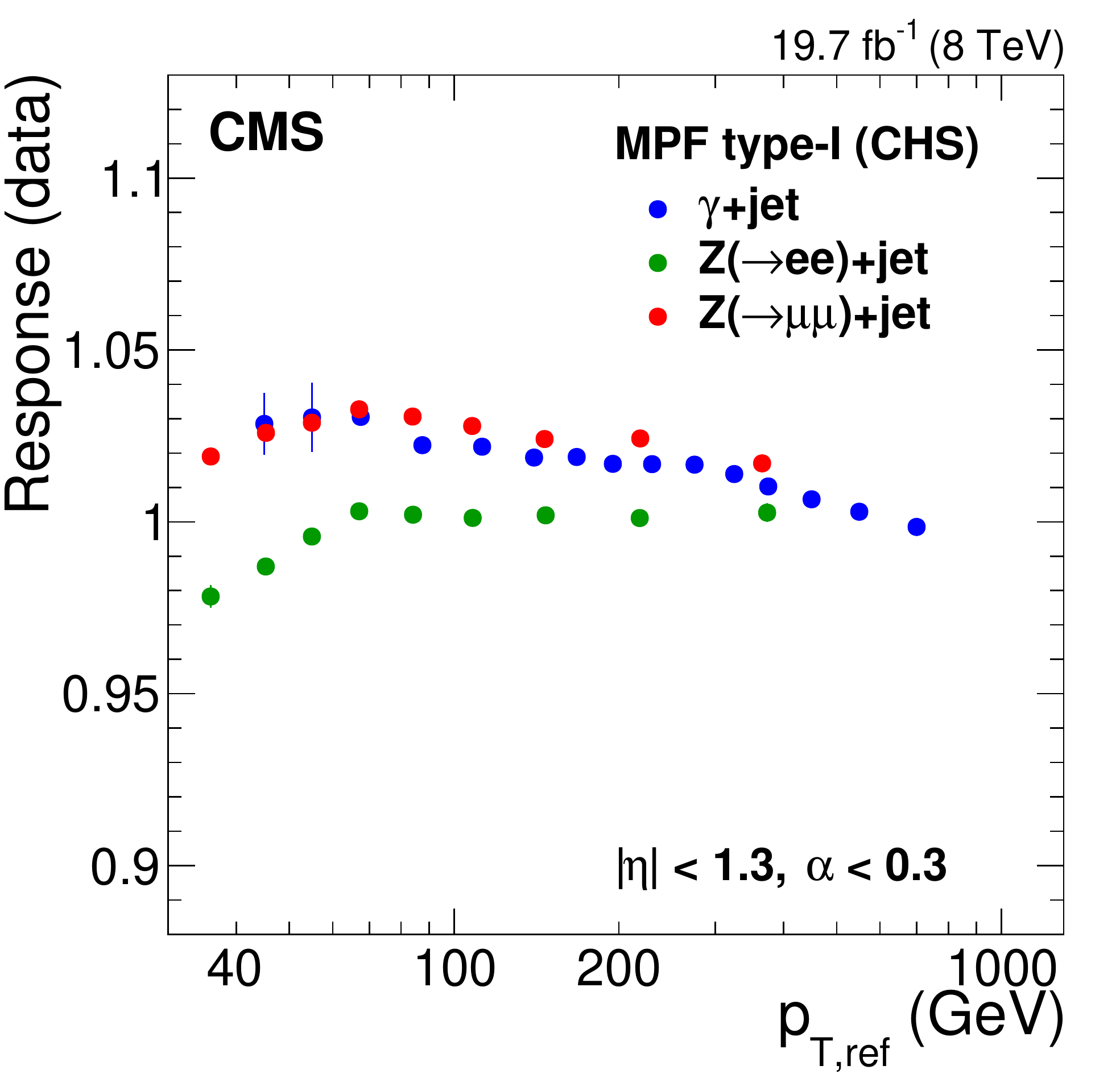}
\includegraphics[width=0.49\textwidth]{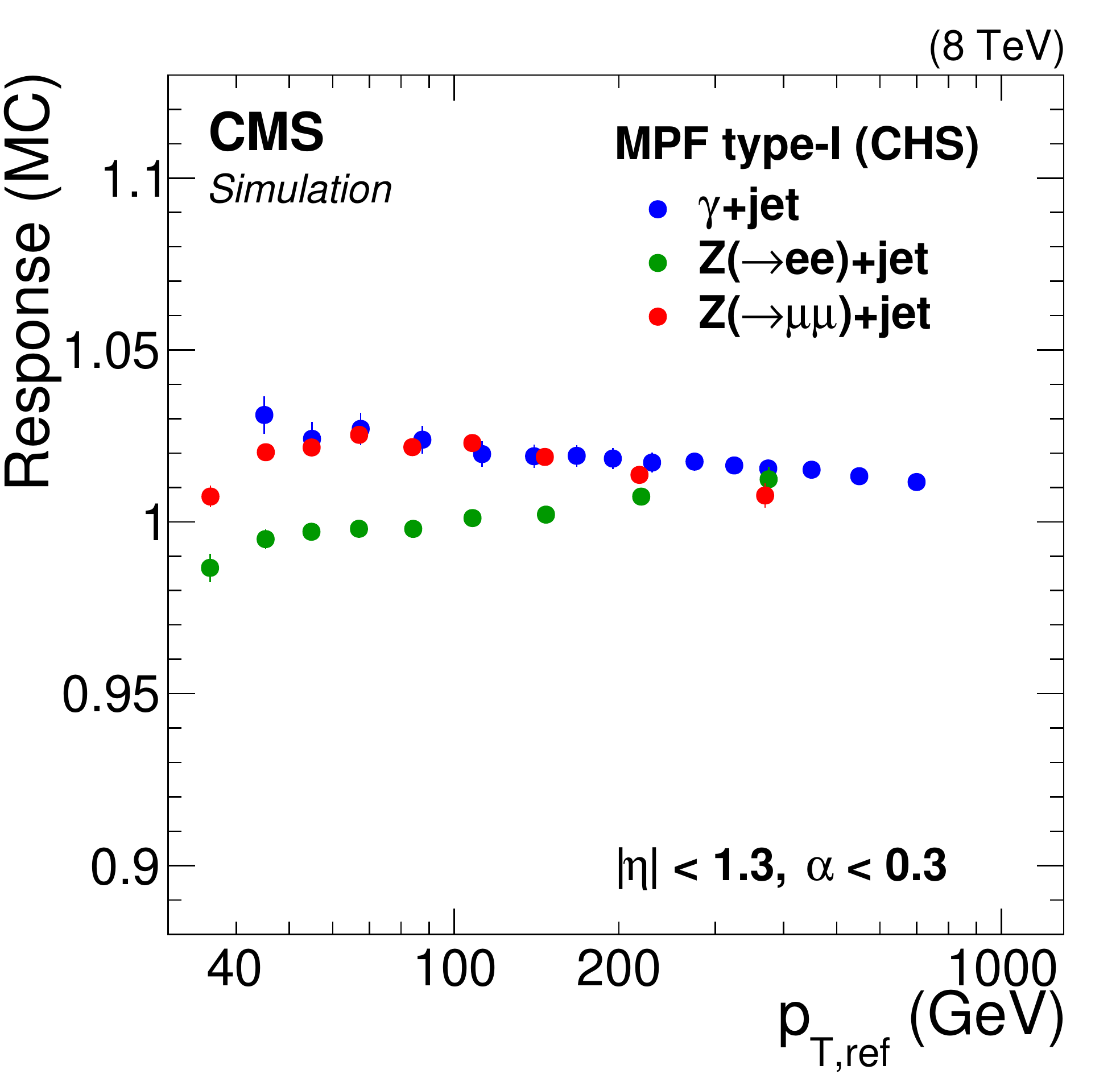}
\includegraphics[width=0.49\textwidth]{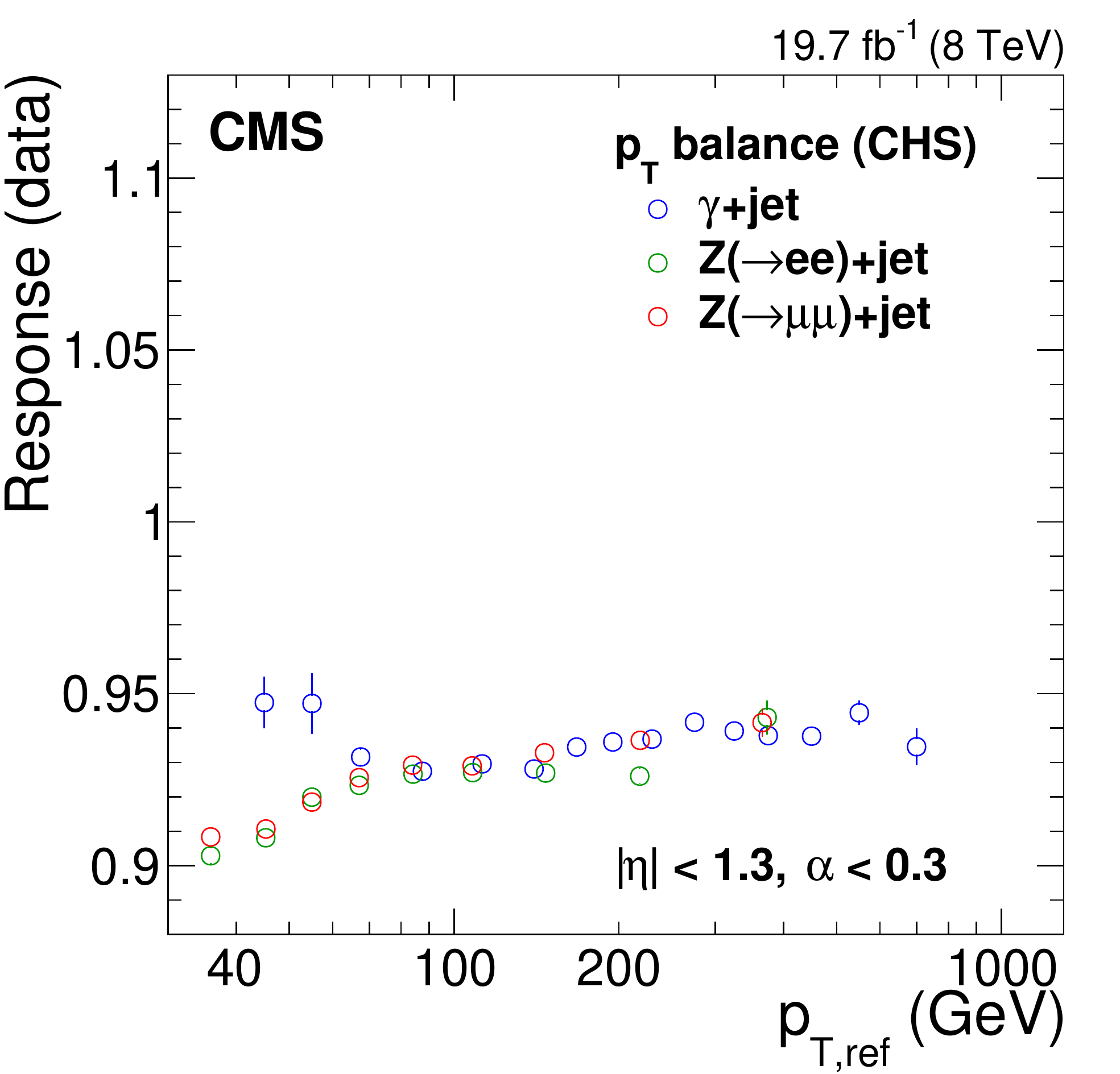}
\includegraphics[width=0.49\textwidth]{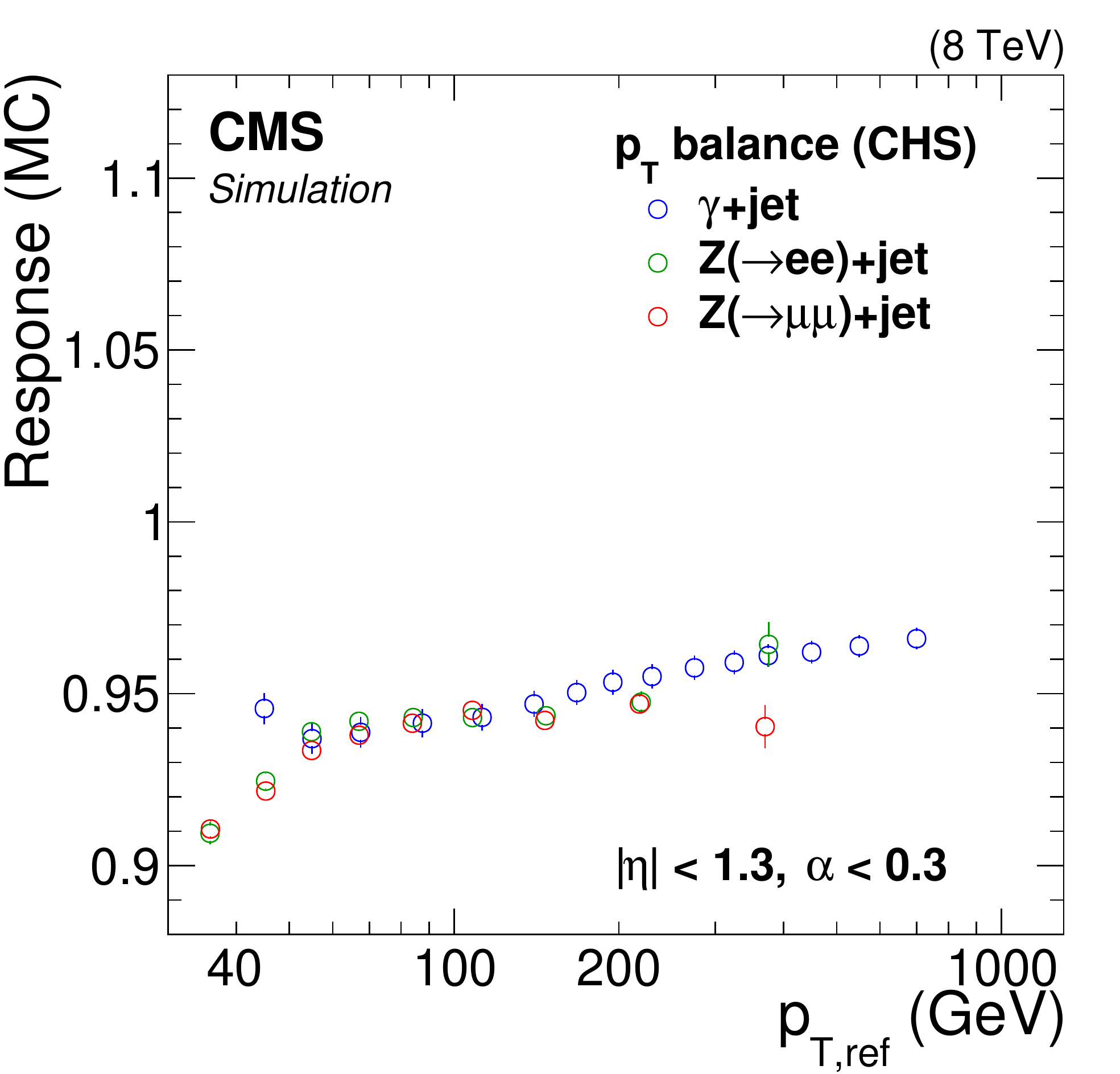}
\end{center}
\caption{\label{fig:absscale}
Comparison of jet response measurements from \zmmjet, \zeejet, and $\gamma$+jet samples as a function of $\PZ$ boson or photon \pt.
The jet response from the MPF method (top) and the \pt-balance method (bottom) is shown as a function of Z and $\gamma$ \pt for data
(left) and simulation (right).
The \zeejet\ sample has not been corrected for the electron EM footprint in \ptvecmiss, explaining the low MPF response in both data and simulation. The footprint effect is absent for muons and corrected for photons.
}
\end{figure}

\begin{figure}[htbp!]
\begin{center}
\includegraphics[width=0.49\textwidth]{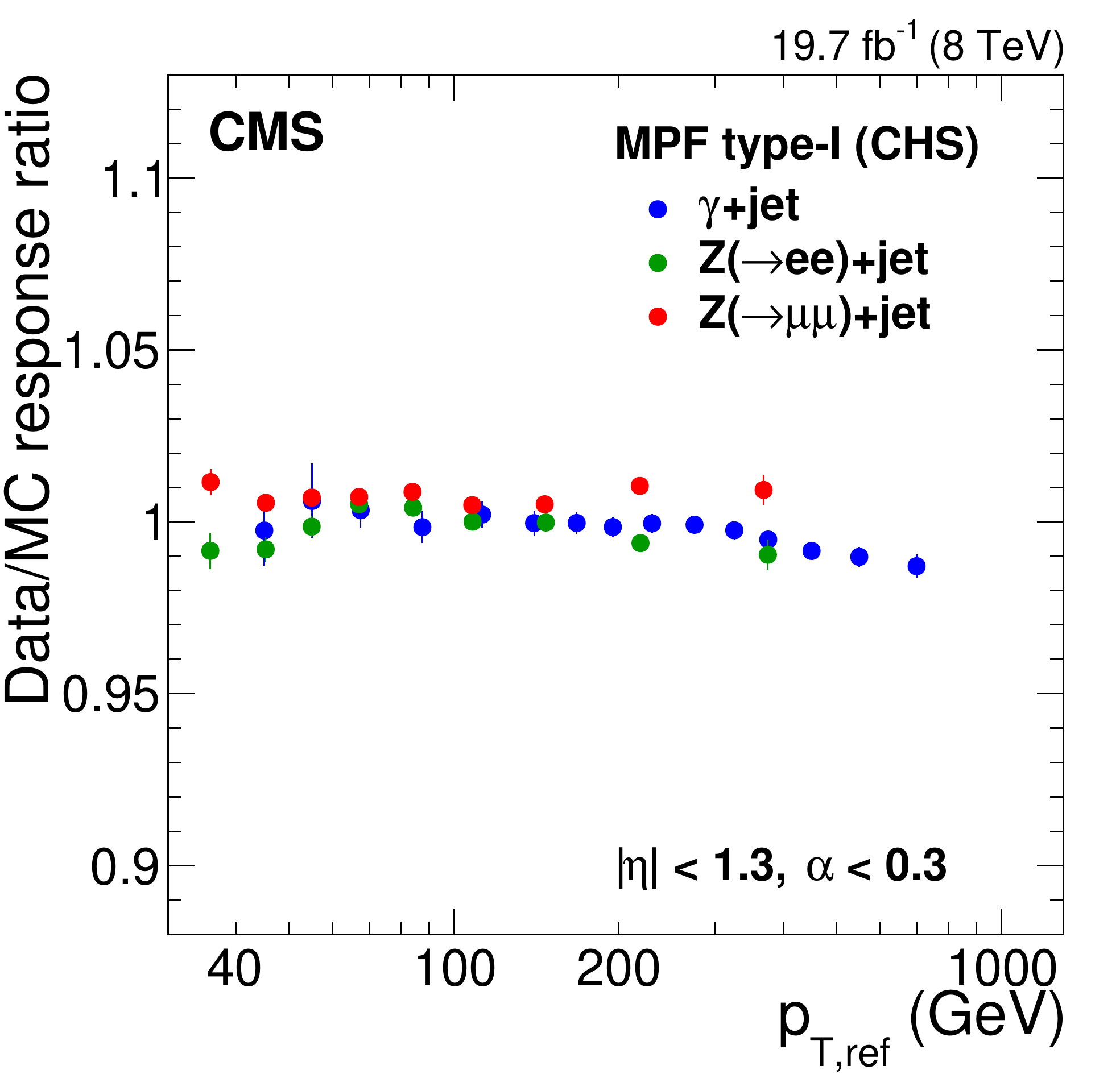}
\includegraphics[width=0.49\textwidth]{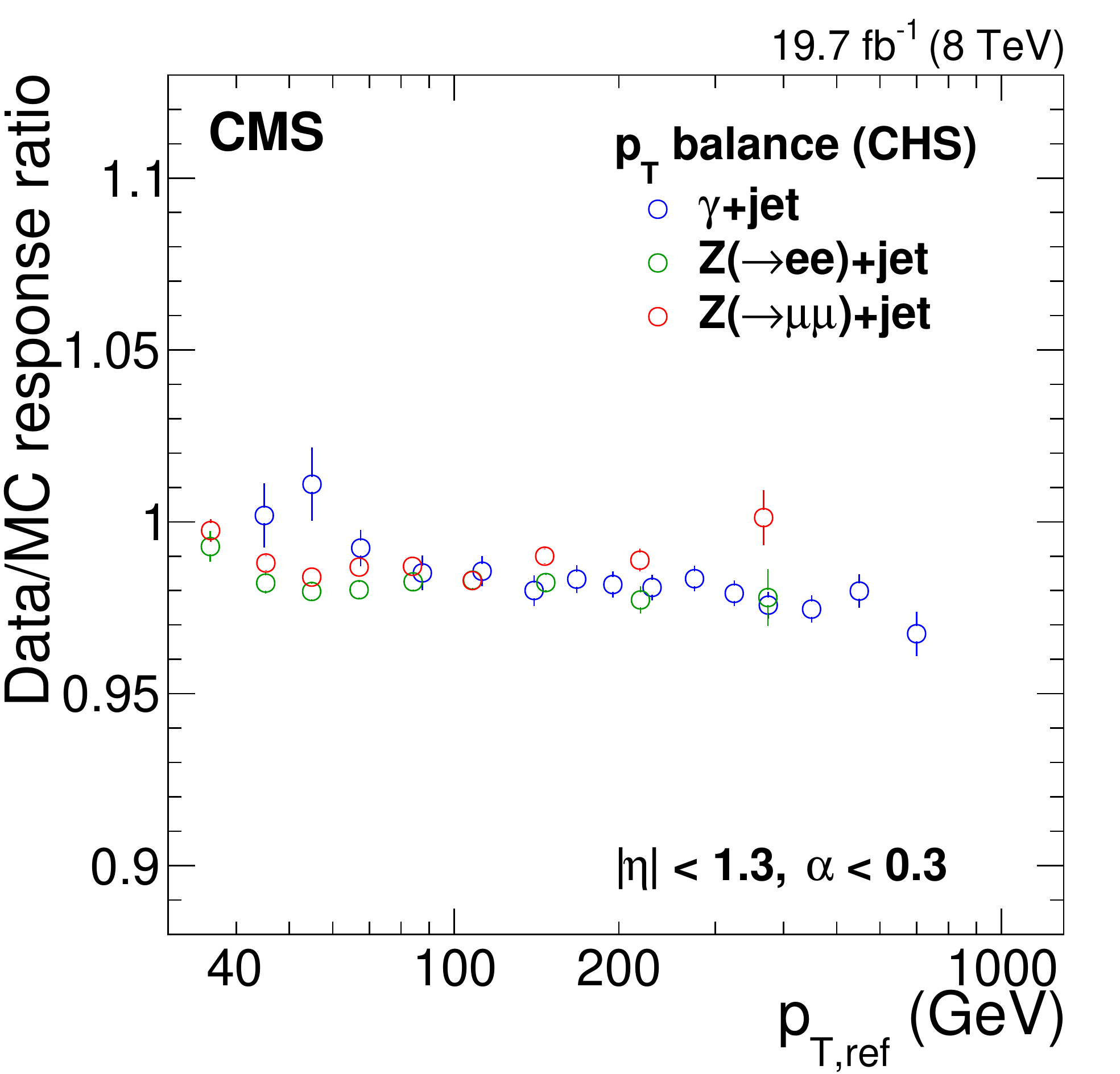}
\end{center}
\caption{\label{fig:absscale_ratio}
Ratio of the jet response measurement obtained from data and simulation with the MPF method (left) and \pt-balance method (right).
Results are shown for the \zmmjet, \zeejet, and $\gamma$+jet samples.
The \zeejet\ sample has not been corrected for the electron EM footprint in \ptvecmiss, but the effect cancels out in the ratio of data over simulation.
}
\end{figure}

Events are binned in photon or $\PZ$ boson \pt to avoid resolution bias from the relatively poorly measured jets.
The asymmetric ISR and FSR is accounted for by extrapolating to zero the additional jet activity (using the $\alpha$ variable, defined in Eq.~(\ref{eq:krad})).
The response for \zmmjet\ events is about 2\% lower in data than in simulation, in agreement with what is observed for the jet response in the 7-TeV data.
 In Figures~\ref{fig:absscale} to ~\ref{fig:multijet}, the jet response in data is first corrected by this factor, independent of the jet \pt.
This correction is reabsorbed in the global fit, that constraints simultaneously the overall normalisation and the $p_\mathrm{T}$-dependence of the ratio of data over simulation responses, to obtain the final correction.

The initial results are obtained after correcting jets and \ptvecmiss for pileup, jet energy response as extracted from simulation studies, and after applying the residual $\eta$-dependent corrections based on the dijet balance and the data/MC scale factor of 1.02 from \zmmjet\ events.
The MPF and \pt-balance methods have different sensitivities to ISR and FSR, which further reduces the uncertainty on the correction.
Jet response measurements obtained from these two methods are shown in Fig.~\ref{fig:absscale} for the nominal working point $\alpha<0.3$. The results are presented for the three event samples (\zmmjet, \zeejet\ and $\gamma$+jet) and are shown for data and simulation. The ratio of data to simulation is shown in Fig.~\ref{fig:absscale_ratio}.
These results are displayed before extrapolating the additional jet radiation to zero and correcting for the \pt dependence of the JES: these effects will be taken into account in the global fit described in Section~\ref{sec:l3res_globalfit}.
The agreement between the measured response in different samples is reasonable for data and MC simulation separately, and most remaining differences cancel out in the data/MC ratio. The differences between \zeejet\ and \zmmjet/$\gamma$+jet events in the MPF method are due to a double counting of leakage from PF electrons and photons to \ptvecmiss (EM footprint effect on \ptvecmiss), which is absent for muons and corrected for photons, but not for electrons. The differences between the MPF and the \pt-balance methods arise from ISR+FSR effects, and largely disappear in the data/MC ratio when $k_\mathrm{FSR}$ corrections are applied.
Residual biases from ISR and FSR, as well as effects from lepton/photon scales and EM footprint in \ptvecmiss, are dealt with in the global fit described in Section~\ref{sec:l3res_globalfit}.

\subsubsection*{Multijet balance}
\label{sec:mjb}

The extrapolation of the JEC to high \pt is constrained using data with the multijet balance (MJB) method introduced in
Ref.~\cite{Aad:2011he}, with events where a high-$\pt$ barrel jet is balanced by a recoil system,
composed of two or more lower-$\pt$ jets. In addition to the traditional balancing variable, $\mathrm{MJB}=p_\mathrm{T, lead}/p_\mathrm{T, recoil}$,
the response is also studied using an MPF method, where $\mathrm{MPF} = 1 + (\ptvecmiss\cdot\vec{p}_\mathrm{T, recoil})/|\vec{p}_\mathrm{T, recoil}|^{2}$.
The jets used in the analysis are corrected for all the previous stages, including residual $\eta$-dependent corrections from dijets, but excluding the final absolute correction versus \pt. The response in data is scaled by a rough factor of 1.02, constant as a function of \pt,  extracted from the study of \zmmjet\ events. The \ptvecmiss is corrected for all jets with $\abs{\eta}<5$ and $\pt>10\GeV$ in the MPF method, but only jets with $\pt>30\GeV$ are used for the event selection and for constructing the  $p_\mathrm{T,recoil}$.
The $\pt>30\GeV$ threshold ensures that the event selection is not biased by pileup jets, and that the recoil is composed of jets directly calibrated with data-based methods.

To interpret the results for JES we define an effective average $p_{\mathrm{T},\rm eff\, ptcl}$ of the jets in the recoil, such that
\begin{equation}
\mathrm{R}_\text{jet}(p_{\mathrm{T},\rm eff \, ptcl})\vec{p}_{\mathrm{T},\rm recoil\, ptcl} = \sum_{i\in\mathrm{recoil\, jets}} \mathrm{R}_\text{jet}(p_{\mathrm{T},i\,\mathrm{ ptcl}})\vec{p}_{\mathrm{T},i\,\mathrm{ptcl}},
\label{eq:ptrecoil_def}
\end{equation}
where $\mathrm{R}_\text{jet}(p_\mathrm{T})$ is the response of a jet of transverse momentum $p_\mathrm{T}$.

This is solved for $p_{\mathrm{T},\rm eff\, ptcl}$ in the log-linear approximation, $\mathrm{R}_\text{jet}(p_\mathrm{T}) = p_0 + p_1\log(p_{\mathrm{T}})$, giving
\begin{eqnarray}
p_{\mathrm{T},\rm eff\, ptcl} &=& p_{\mathrm{T},\rm recoil\, ptcl}\exp\left(\sum_iF_i\log(f_i)\right),\quad\mathrm{where}\\
f_i &=& \frac{p_{\mathrm{T},i\, \mathrm{ptcl}}}{p_{\mathrm{T},\rm recoil\, ptcl}},\quad\mathrm{and}\quad
F_i = f_i\cos(\Delta\phi_i).
\end{eqnarray}

\begin{figure}[htbp!]
\centering
\includegraphics[width=0.48\textwidth]{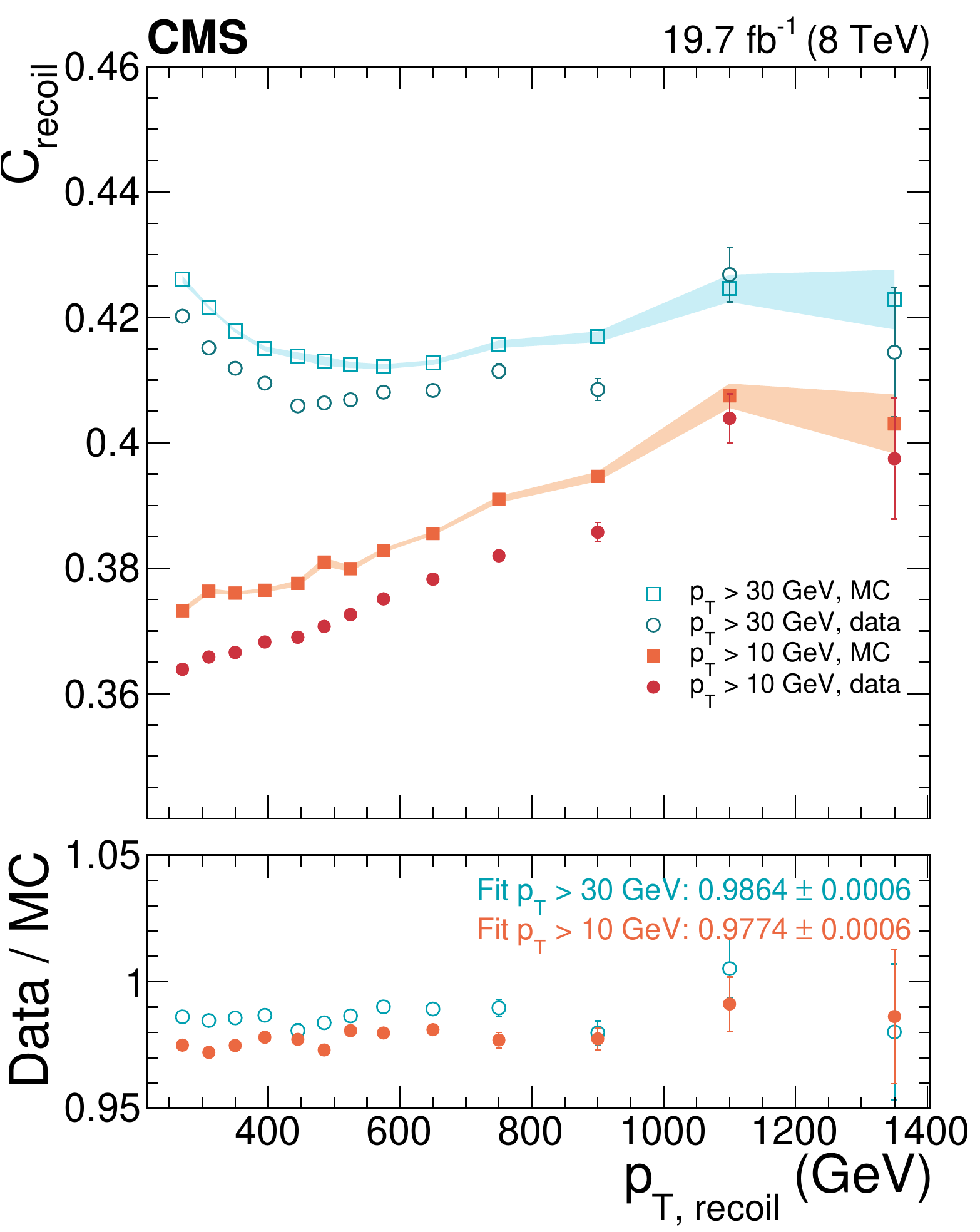}
\includegraphics[width=0.48\textwidth]{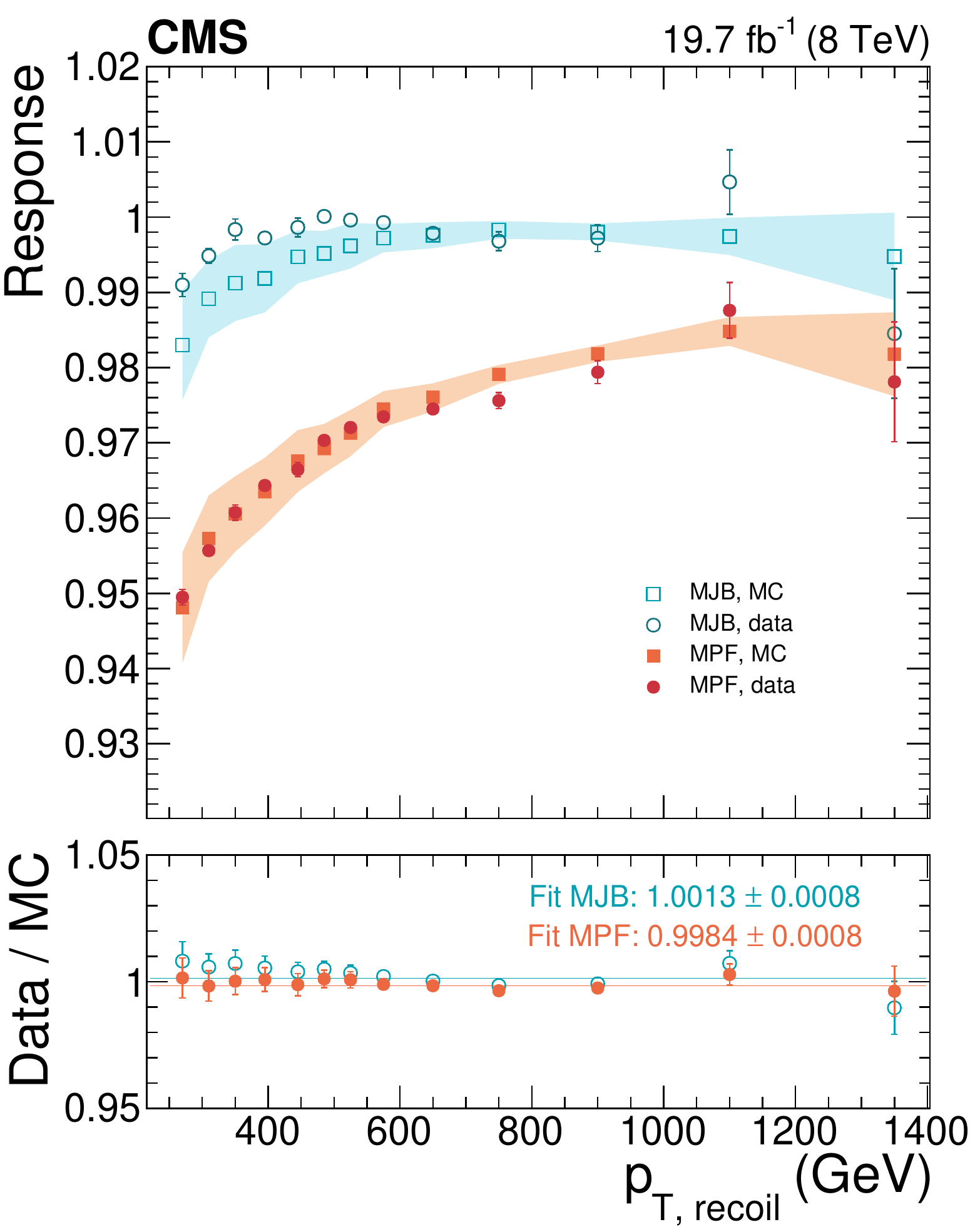}
\caption{\label{fig:multijet}
$C_\mathrm{recoil}$ ratio of the effective jet \pt of jets in the recoil over the total recoil \pt, (Eq.~(\ref{eq:crecoil})), calculated with recoil jets of $\pt>30\GeV$ (for MJB) and $\pt>10\GeV$ (for MPF) in data and MC simulation (left). Multijet balance response calculated with the MJB and MPF methods for data and MC simulation (right). The filled bands show the statistical uncertainty on MC for the left plot and the total (statistical and systematic) uncertainty on MC for the right plot. The error bars show the statistical uncertainty on data.
}
\end{figure}

The $\Delta\phi_i$ is the angle between the jet four-vector and the recoil, and by construction $\sum_i F_i=1$. Labeling further
\begin{equation}\label{eq:crecoil}
C_\mathrm{recoil} = p_{\mathrm{T},\rm eff\, ptcl}/p_{\mathrm{T},\rm recoil\, ptcl},
\end{equation}
we can write
\begin{equation}\label{eq:mjb}
\mathrm{MJB} = \frac{\mathrm{R}_\text{jet}(p_{\mathrm{T},\rm lead\, ptcl})p_{\mathrm{T},\rm lead\, ptcl}}{\mathrm{R}_\text{jet}(p_{\mathrm{T},\rm eff\, ptcl})p_{\mathrm{T},\rm recoil\, ptcl}} \approx \frac{\mathrm{R}_\text{jet}(\pt)}{\mathrm{R}_\text{jet}(C_\mathrm{recoil} \pt)},
\end{equation}
where the last approximation uses $p_{\mathrm{T},\rm lead\, ptcl}\approx p_{\mathrm{T},\rm recoil\, ptcl} \equiv \pt$. The $p_{\mathrm{T},\rm eff}$ represents the effective average jet \pt in the recoil system (see Eq.~(\ref{eq:ptrecoil_def})),
and the high-$\pt$ jets are calibrated relative to $p_{\mathrm{T},\rm eff}$.
In the actual calculation, $C_\mathrm{recoil}$ is evaluated from reconstructed jets $p_{\mathrm{T}}$ (corrected with MC-based corrections and data-driven corrections for pileup effects and $\eta$ dependence of the response), which is equivalent to the particle-level $C_\mathrm{recoil}$ to sufficient precision. $C_\mathrm{recoil}$, shown in Fig.~\ref{fig:multijet} (left), is therefore a measure of
the lever arm of the multijet balance method
in measuring \pt dependence of JES. Because the MPF method indirectly uses jets also below $\pt=30\GeV$, its lever arm $C_\mathrm{recoil}$ is calculated from all jets of $\pt>10\GeV$, while that of MJB is calculated only from jets of $\pt>30\GeV$.
The MPF method is stable with respect to the jet \pt threshold, while the MJB drops below MPF at low \pt when the \pt threshold is lowered to 10--20\GeV.

As the leading jet is calibrated using the recoil system as a reference object, the scale obtained for high-$\pt$ jets using this method is relative to the average scale for the lower \pt jets, which are subject to systematic uncertainties particularly from jet-flavor response, pileup offset and $\eta$-dependent corrections.  Systematic uncertainties arising from JEC, JER uncertainties, and PU modeling are propagated to the multijet analysis.
As shown in Fig.~\ref{fig:multijet} (right), the leading high-$\pt$ jet is well balanced against the recoil within the JEC uncertainties, and the multijet balance method constrains the response of jets with a \pt as high as 1.3\TeV.
The global fit takes as input the measured MJB, MPF and $C_\mathrm{recoil}$, using the formulation of Eq.~(\ref{eq:mjb}).

\subsection{Global fit of absolute corrections}
\label{sec:l3res_globalfit}

\begin{figure}[htbp!]
\includegraphics[width=0.49\textwidth]{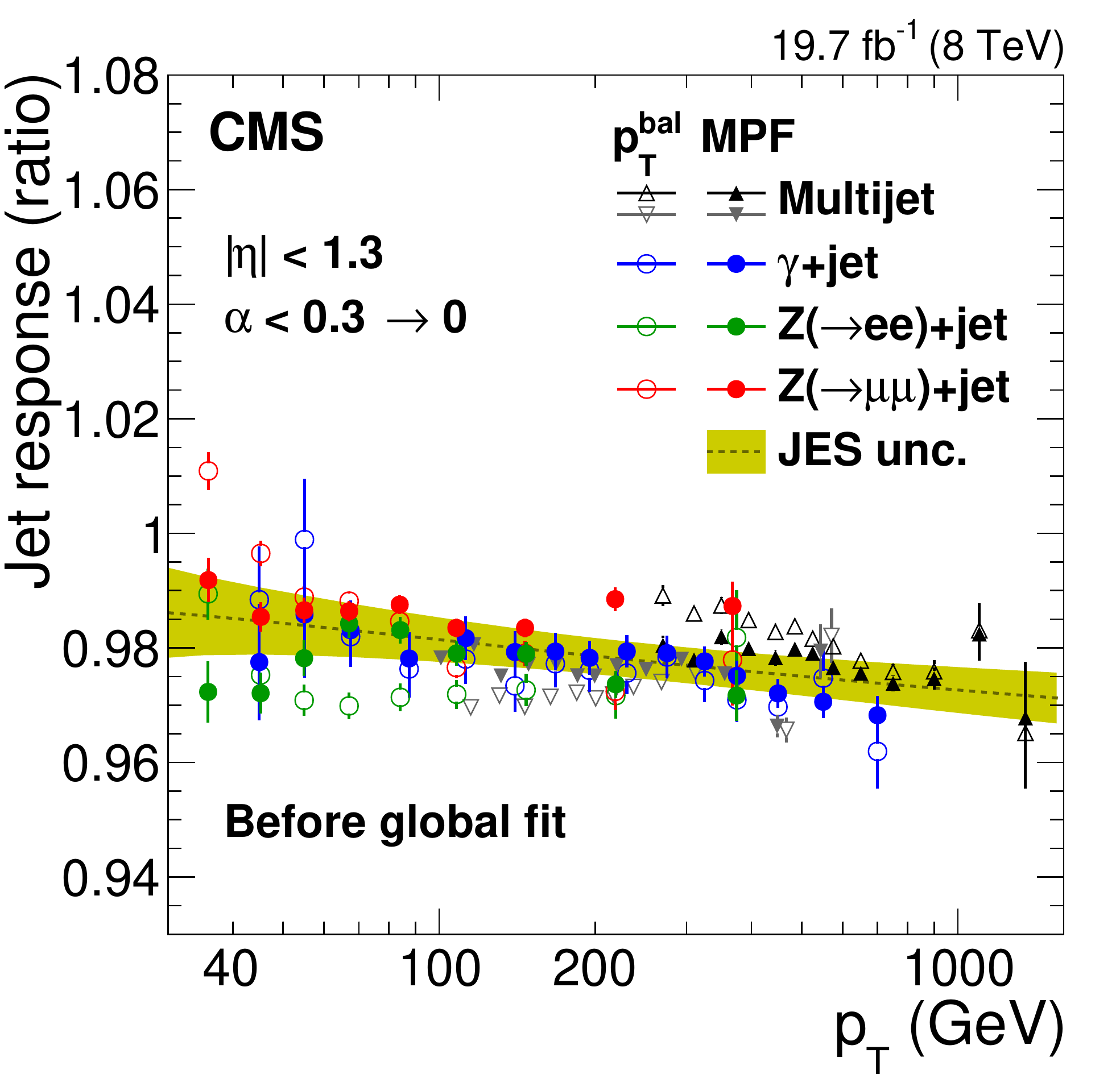}
\includegraphics[width=0.49\textwidth]{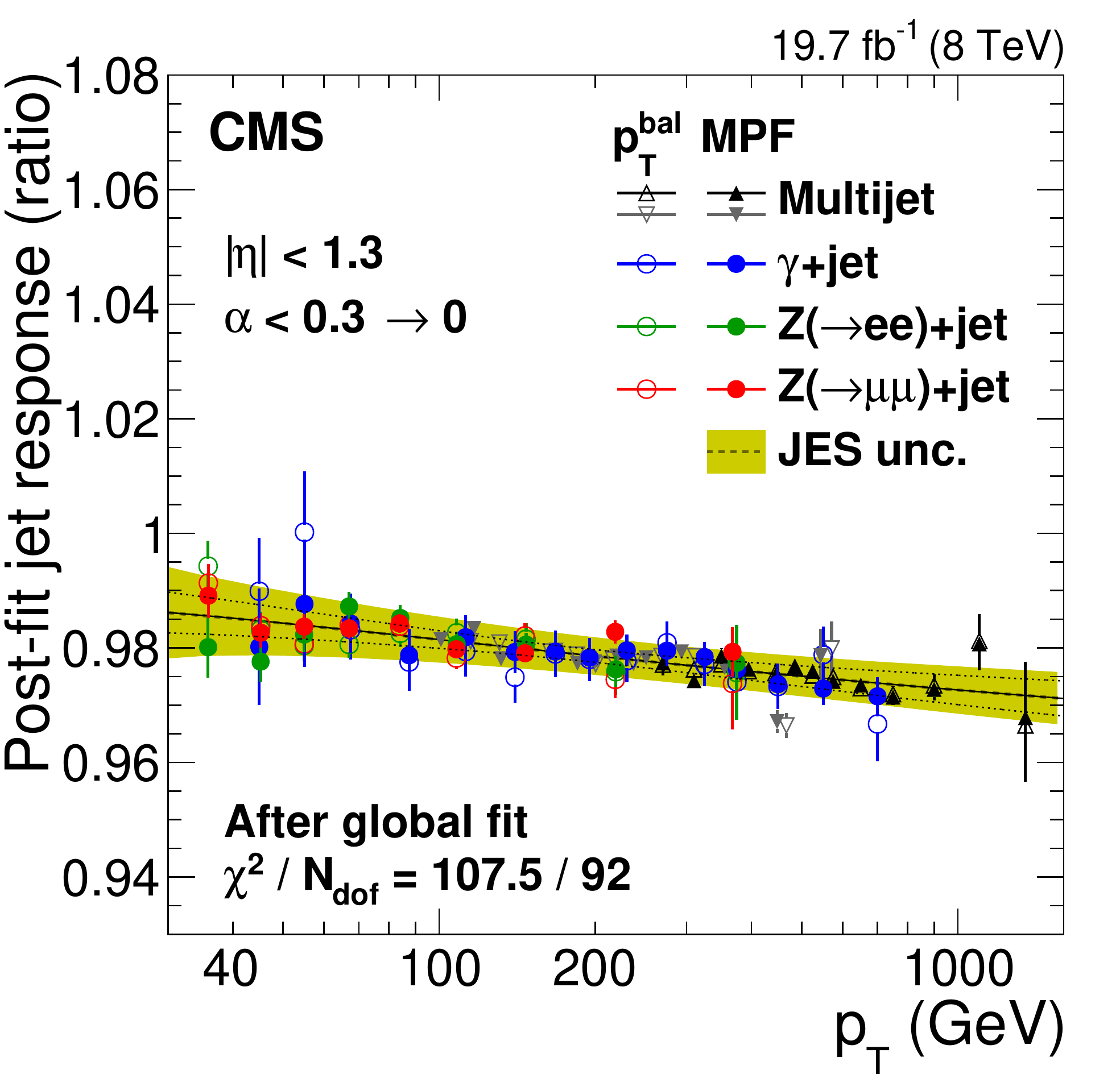}
\caption{\label{fig:abspt}
Comparison of the data-to-simulation ratio of the jet response measurements from \zmmjet, \zeejet, $\gamma$+jet, and multijet samples after applying the corrections for JES and ISR+FSR (left) and after applying, in addition, the nuisance parameter values found by the global fit (right). The uncertainty in the ratio, excluding jet-flavor and time-dependent effects, is shown by the shaded region.  The solid line shows the global fit central value and the dotted curves the statistical uncertainty of the fit. As the multijet analysis connects the energy scale of jets in two different \pt ranges (Eq.~(\ref{eq:mjb})), it can be used to constrain the high-$p_\mathrm{T}$ region given the low-$p_\mathrm{T}$ one (black triangles) or vice versa (grey triangles).
}
\end{figure}

\begin{figure}[htbp!]
\begin{center}
\includegraphics[width=0.49\textwidth]{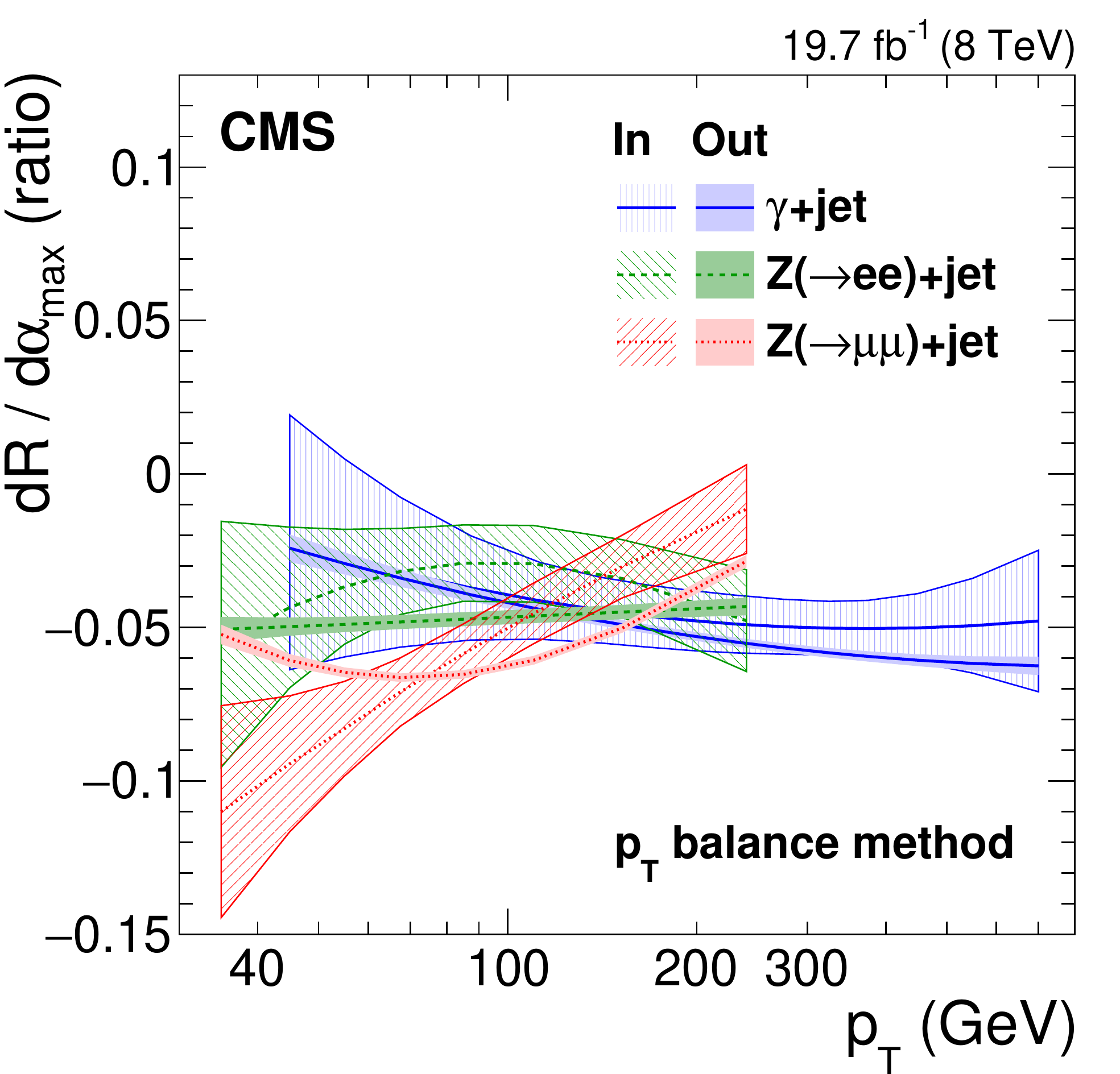}
\includegraphics[width=0.49\textwidth]{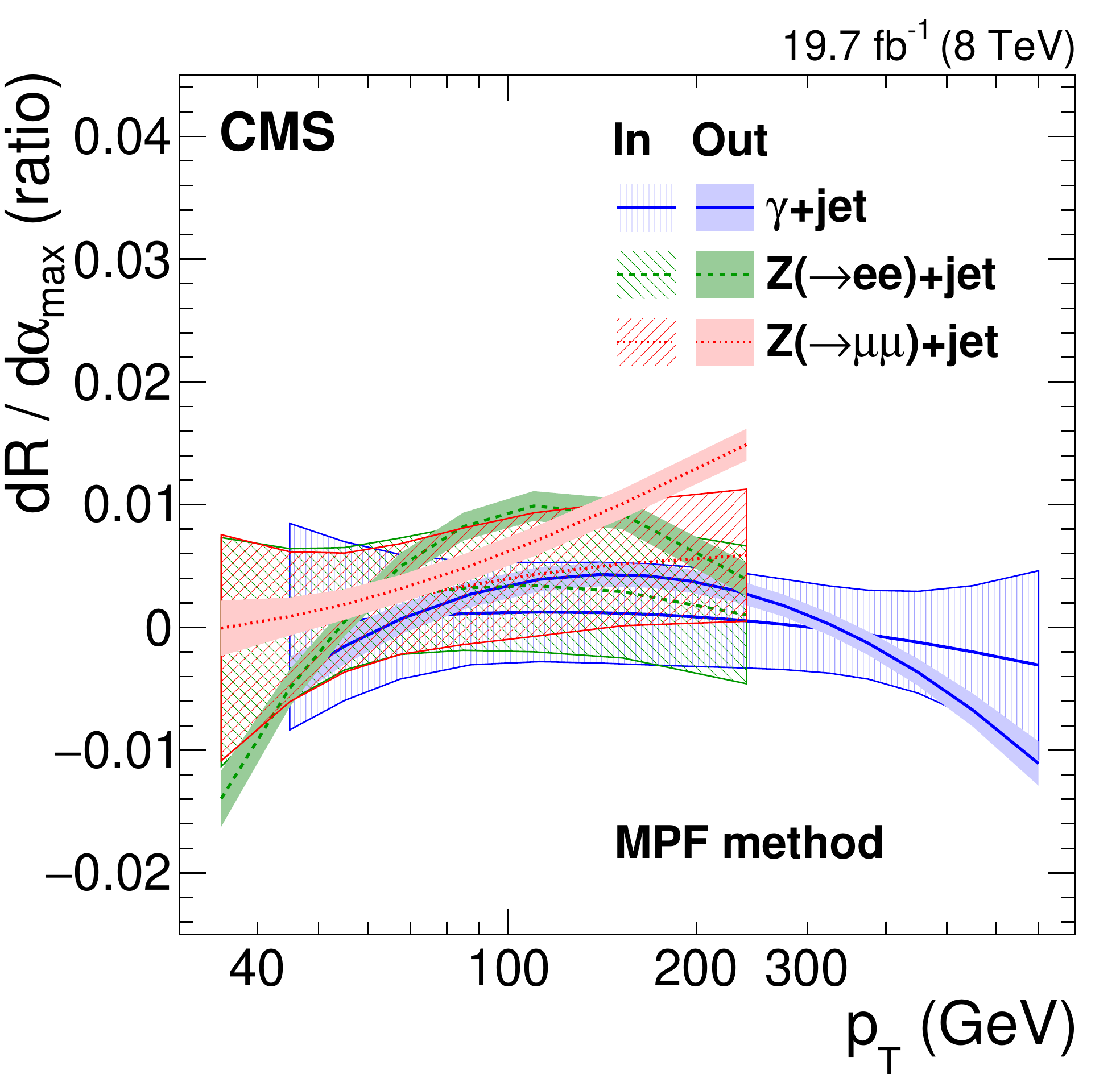}
\caption{\label{fig:kfsr_gfit}Central value of the data-to-simulation ratio of $\rd R/\rd \alpha_\text{max}$, and its 68\% probability region, as a function of jet \pt, for the \pt-balance (left) and MPF (right) methods. The $\rd R/\rd \alpha_\text{max}$ is the derivative of the jet response evaluated in events with $\alpha<\alpha_\text{max}$. The y-axis scale for the MPF method is zoomed by ${\times} 4$ compared to the \pt-balance method, demonstrating the much smaller initial ISR+FSR uncertainty for this method. The shadowed regions show
the input distributions to the global fit, while the full color regions show the post-fit distributions. The uncertainties on $\rd R/\rd \alpha_\text{max}$ before the global fit are labeled 'In', and the uncertainties constrained by the global fit are labeled 'Out'.}
\end{center}
\end{figure}

The absolute jet \pt scale is fitted simultaneously to the muon and electron (for $30<p_{\mathrm{T},Z}<400\GeV$), photon ($40<p_{\mathrm{T},\gamma}<800\GeV$), and multijet ($220<p_{\mathrm{T},\rm recoil}<1300\GeV$) data sets. The muon, photon, and electron scales are allowed to vary within their \textit{a priori} uncertainties of 0.2\%, 0.2\%, and 0.5\%, respectively.
The results used as input to the global fit ($\alpha<0.3$) with initial $k_\mathrm{FSR}$ corrections $\alpha<0.3\to0$ are shown in Fig.~\ref{fig:abspt} (left), overlaid with the final determination of JES and its uncertainties.
The global fit is implemented as a $\chi^2$ minimization.  The fit results depend on some parameters that are known within given uncertainties, these are treated as nuisance parameters in the fit and added quadratically to the $\chi^2$ expression. The nuisance parameters are related to the following effects:
\begin{itemize}
\item \textbf{Lepton/photon scale} uncertainties (0.2\% for $\mu^{\pm}$, 0.2\% for $\gamma$, and 0.5\% for e$^\pm$, assumed to be uncorrelated, and independent of \pt).
\item \textbf{ISR+FSR} uncertainty on the $k_\mathrm{FSR}$ correction fit used in the MPF and \pt-balance methods. The fit is performed in each sample using a three-parameter log-polynomial \pt dependence: its results, with their uncertainties, are used as input to the global fit and are shown as the shaded band labeled 'In' in Fig.~\ref{fig:kfsr_gfit}.
\item \textbf{EM footprint} uncertainty for photons and electrons in the MPF method (0.2\% for photons with footprint correction, and 0.5\% for electrons without footprint correction, independently of each other, and independent of \pt).
\item \textbf{Pileup} uncertainty coming from the difference in the offset calculated inside versus outside of the jet distance parameter.
\item \textbf{Multijet} uncertainties from JES, JER, and pileup, separately for MPF and MJB.
\end{itemize}
The total number of nuisance parameters is three for the lepton/photon scales, $2\times3\times 3=18$ for ISR+FSR, two for EM footprint, one for pileup, and $2\times 3=6$ for multijet balance, for a total of 30 sources.

The global fit has two parameters of interest,
one for fitting the absolute scale and one for fitting the \pt dependence under the assumption that the shape of the response variation is consistent with the one caused by a constant shift in single-pion response in HCAL, shown in Fig.~\ref{fig:spr} (right), and referred to as $f_\mathrm{HCAL}(\pt)$ in the following.
This assumption is supported by the time stability
of charged-pion $E/p$ in HCAL barrel.
The function used to fit the \pt dependence is of the form $a + b(f_\mathrm{HCAL}(p_{\mathrm{T}})-f_\mathrm{HCAL}(p_{\mathrm{T},0}))$, hence introducing a reference momentum $p_{\mathrm{T},0}$ and fitting the shape relative to it. The value of $p_{\mathrm{T},0}=208\GeV$ is chosen to minimize the correlation between $a$ and $b$.

The result of the \pt-dependent fit is shown in Fig.~\ref{fig:abspt} (right). The data points are shifted by the nuisance parameter values found by the global fit in order to demonstrate the good consistency between the data sets. The nuisance parameters are normally distributed, with no outliers beyond the 2$\sigma$ limit. The reduced goodness-of-fit, $\chi^2 / \mathrm{N_\mathrm{dof}}$, is $107.5/92\approx1$ for all degrees of freedom and for data points and nuisance parameters combined, indicating appropriate coverage by the systematic and statistical uncertainties. The statistical uncertainty band of the global fit, which is shown by the dotted curves, can be separated into two independent components, one describing the uncertainty in absolute scale, the other describing the uncertainty in \pt dependence coming from HCAL response. Other JES systematic uncertainties are calculated with respect to the factorization point $\pt=208\GeV$ that also has the smallest fit uncertainty.

The initial ISR+FSR corrections are constrained by the global fit, and their uncertainties before and after the fit are shown in Fig.~\ref{fig:kfsr_gfit} by the solid bands labeled 'In' and 'Out', respectively.
The ISR+FSR correction for $\alpha<0.3$ varies from about 1.5\% for \pt balance ($k_\mathrm{FSR}\approx -5\%$) to less than 0.3\% for MPF ($|k_\mathrm{FSR}|<1\%$). This is consistent with the expectation that MPF is only sensitive to FSR and ISR to second order through differences in the response between the leading jet and the rest of the hadronic recoil, which are expected to be less than 20\%.  The ISR+FSR corrections as constrained by the fit are within the uncertainty of the input values, with the exception of the \pt-balance method for \zmmjet~ events, which stays within twice the input uncertainty and whose initial tension with respect to the \zeejet~ channel is reduced by the fit.

 \subsection{Absolute correction uncertainties}
\label{sec:l3res_globalfit_unc}

\subsubsection*{Scale uncertainty}

The dominant uncertainties for the scale factor arise from the following sources:
\begin{itemize}
\item \textbf{Absolute scale}, 0.11\%. The uncertainty in the fitted absolute scale is driven by the muon scale of 0.2\%, obtained from the position of the $\PZ$ boson mass peak, and by the photon scale uncertainty of 0.2\%, achieved via a precise regression correction~\cite{Khachatryan:2015iwa}.
\item \textbf{HCAL scale}, ${\leq} 0.6$\%. The uncertainty in the fitted HCAL scale is driven by the statistical uncertainty of the $\PZ$+jet sample at low \pt and by the $\gamma$+jet sample at high \pt. This uncertainty is labeled \textit{SinglePionHCAL}.
\item \textbf{Statistical uncertainty}, ${\leq} 0.1$\%.
The statistical uncertainty covers the small residual difference between the global fit uncertainty calculated with the full covariance matrix and the uncertainty calculated from its diagonal elements (absolute scale and HCAL scale, above) only. Only positive contributions are included.
\item \textbf{MPF bias}, 0.28\%. The bias on the MPF method is composed of two subsources:
  \begin{itemize}
    \item \textit{Neutrino production}, 0.2\%. This uncertainty is estimated from the 0.1\% excess of both electron and muon energies in the PF jet compositions (shown in Fig.~\ref{fig:l3pf}), which is assumed to be associated with a neutrino excess from decays of heavy-flavor hadrons, of similar order of magnitude. The estimate is compatible with uncertainties in the fraction of heavy-flavors from gluon splitting.
    \item \textit{ISR outside detector acceptance}, 0.2\%. The extrapolation to zero additional jet activity cannot correct for ISR activity outside detector acceptance, which biases both MPF and \pt balance by the same amount. Phase space constraints limit ISR jet \pt outside the detector to $\pt \leq 30\GeV$, which together with the results on the MPF and \pt-balance response in simulation after the JEC set an upper limit of 0.2\% on this source.
  \end{itemize}
\end{itemize}

\subsubsection*{Single-particle response and fragmentation systematics}

We consider additional shape uncertainties from single-pion response and jet fragmentation using simulation.
These studies were reported in Section~\ref{sec:mctruth}. The JEC at $\pt=208\GeV$ is known to high accuracy from the global fit so the simulation-based shape uncertainties are assumed to be zero at this reference \pt, but increase further away from the reference point. The single-particle response uncertainty of 3\% is implemented using independent variations of $\pm \sqrt{2} 3$\% in responses in the ECAL and HCAL separately. The variation in the HCAL is included in the global fit and constrained to $-3.50\pm1.35$\%.
The fragmentation uncertainty is taken directly as the \PYTHIA6.4/\HERWIGpp~2.3 response difference.

\subsection*{Time dependence}

\begin{figure}[htbp!]
\centering
\includegraphics[width=0.49\textwidth]{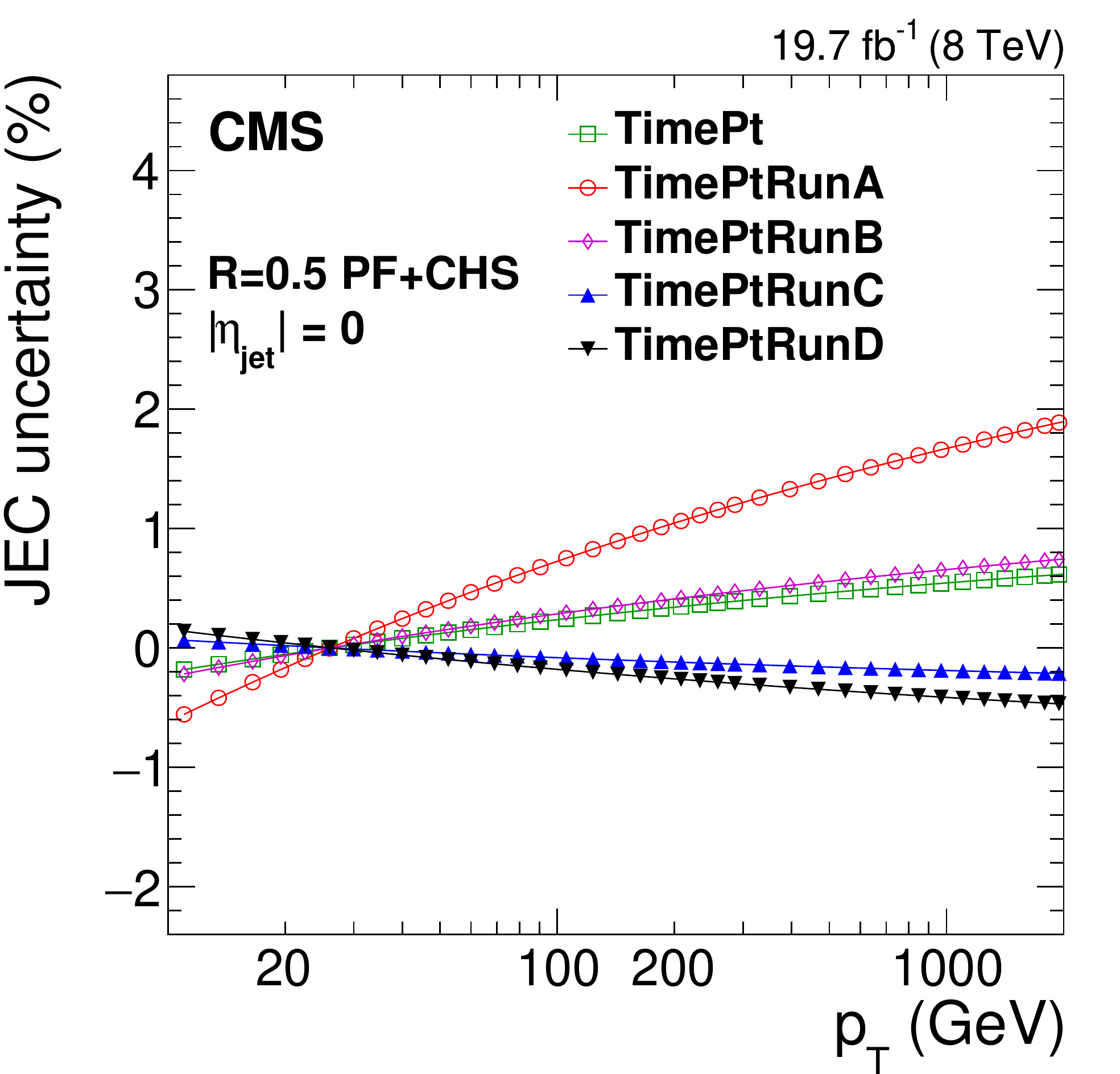}
\includegraphics[width=0.49\textwidth]{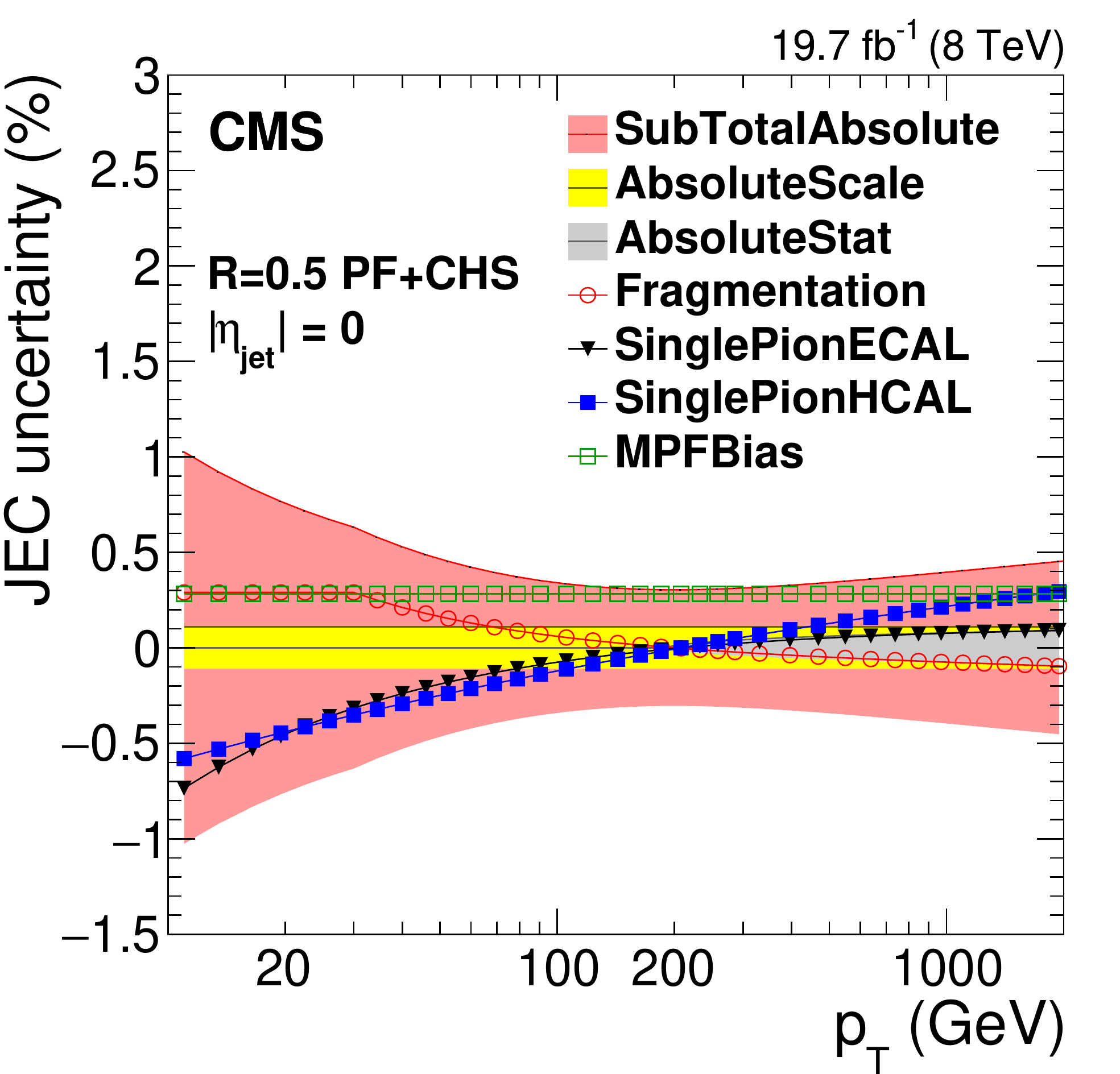}
\caption{\label{fig:l3sys}
Absolute scale time-dependent uncertainty as a function of jet \pt for various data-taking periods (left). Systematic uncertainties for the absolute jet scale as a function of \pt (right).
\textit{SubTotalAbsolute} is the quadratic sum of \textit{AbsoluteScale}, \textit{AbsoluteStat}, \textit{Fragmentation}, \textit{SinglePionECAL}, \textit{SinglePionHCAL} and \textit{MPFBias}.
}
\end{figure}

The JES \pt dependence is believed to originate mostly from a reduction in the HCAL energy scale due to various effects, including uncorrected radiation damage in the front layers of the barrel calorimeter. To estimate the stability of the HCAL scale with time, the ratio of the calorimeter energy to track momentum $E_\mathrm{HCAL}/p_\mathrm{track}$ of isolated barrel hadrons was plotted as a function of time for pions that did not interact in the EM calorimeter.
The 2012 (8\TeV) data sample is divided in four subsequent run periods, with slightly different conditions and trigger requirements: run A (with an integrated luminosity of 0.88\fbinv), B (with 4.41\fbinv), C (with 7.05\fbinv) and D (with 7.37\fbinv). The $E_\mathrm{HCAL}/p_\mathrm{track}$ shows variation of up to 4.5\% between run A and run D for $p=10\GeV$ hadrons.
The HCAL scale variations relative to the full 2012 (8\TeV) data set were taken as uncertainties for each data-taking run (\textit{TimePtRunA}, \textit{TimePtRunB}, \textit{TimePtRunC} and \textit{TimePtRunD} for runs A, B, C and D, respectively), and then propagated to the jet response, as shown in Fig.~\ref{fig:l3sys} (left). The integrated-luminosity-weighted RMS of these run variations is taken as the time-dependent uncertainty in the absolute scale versus \pt, \textit{TimePt}. In a similar fashion, the time-dependent uncertainty in the relative $\eta$-dependent corrections is estimated as the RMS of the correction factors obtained for short data-taking time periods, each weighted by the corresponding integrated luminosity. Neither uncertainty needs to be applied to analyses performed using the complete 2012 (8\TeV) dataset with unprescaled triggers, for which the time-dependent effects average out.

The \textit{SinglePionECAL}, \textit{Fragmentation}, and \textit{MPFBias} uncertainties are summarized in Fig.~\ref{fig:l3sys} (right), together with the \textit{AbsoluteScale}, \textit{SinglePionHCAL}, and \textit{AbsoluteStat} uncertainties, which are effectively the constant, \pt-dependent, and residual components of the statistical uncertainty in the global fit, respectively.

\subsection{Summary of residual corrections}

\begin{figure}[htbp!]
\centering
\includegraphics[width=0.48\textwidth]{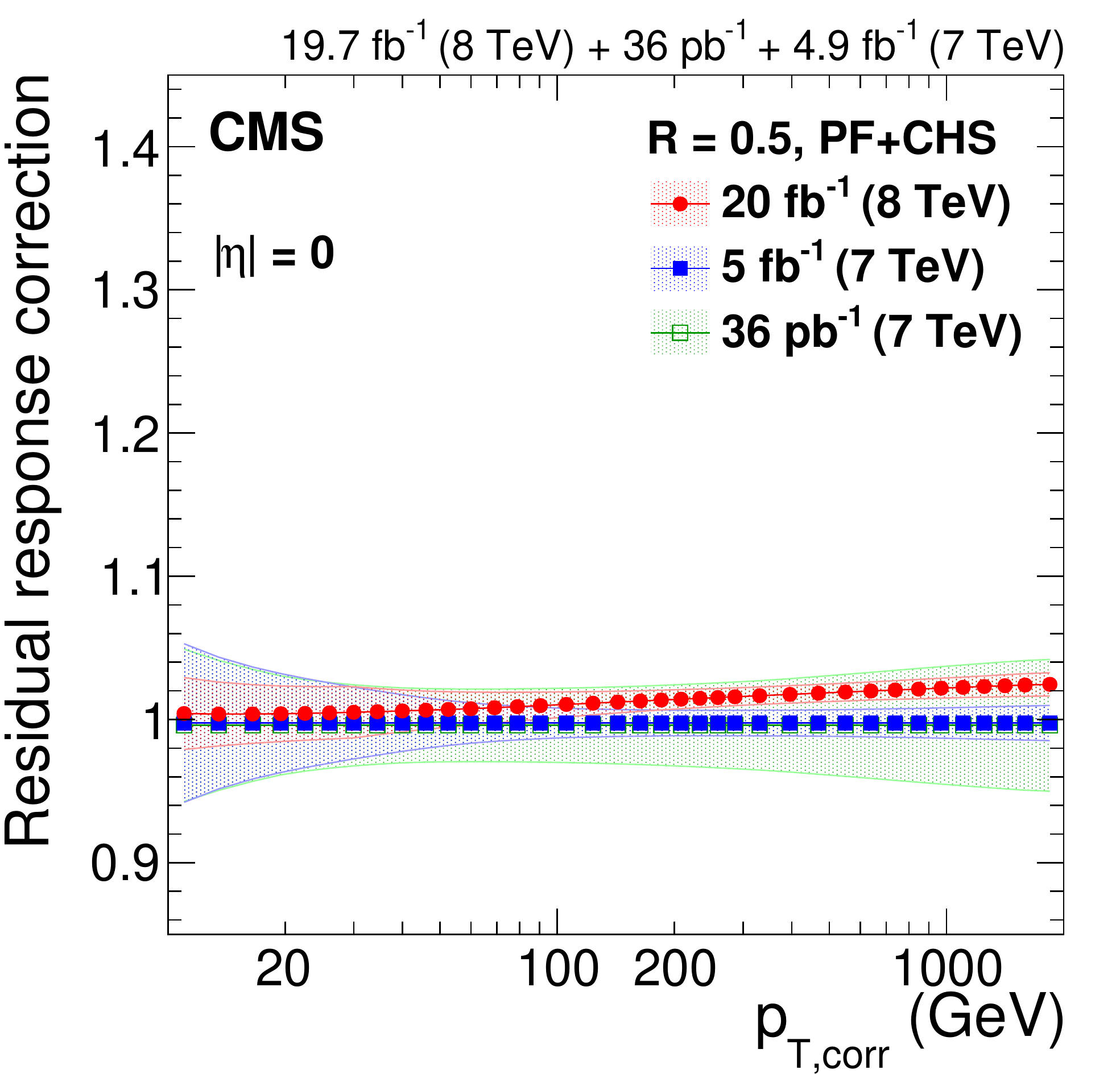}
\includegraphics[width=0.48\textwidth]{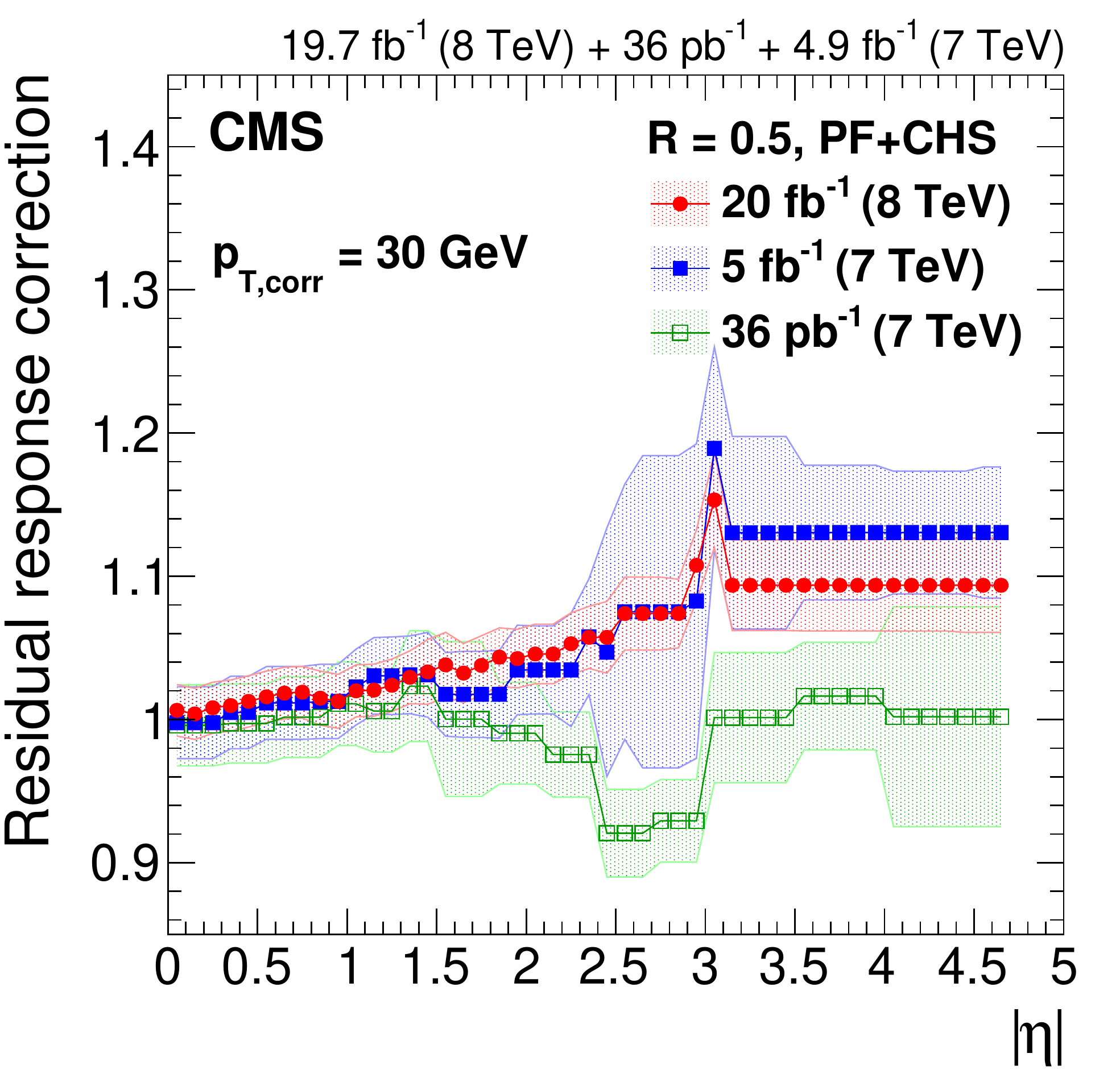}
\includegraphics[width=0.48\textwidth]{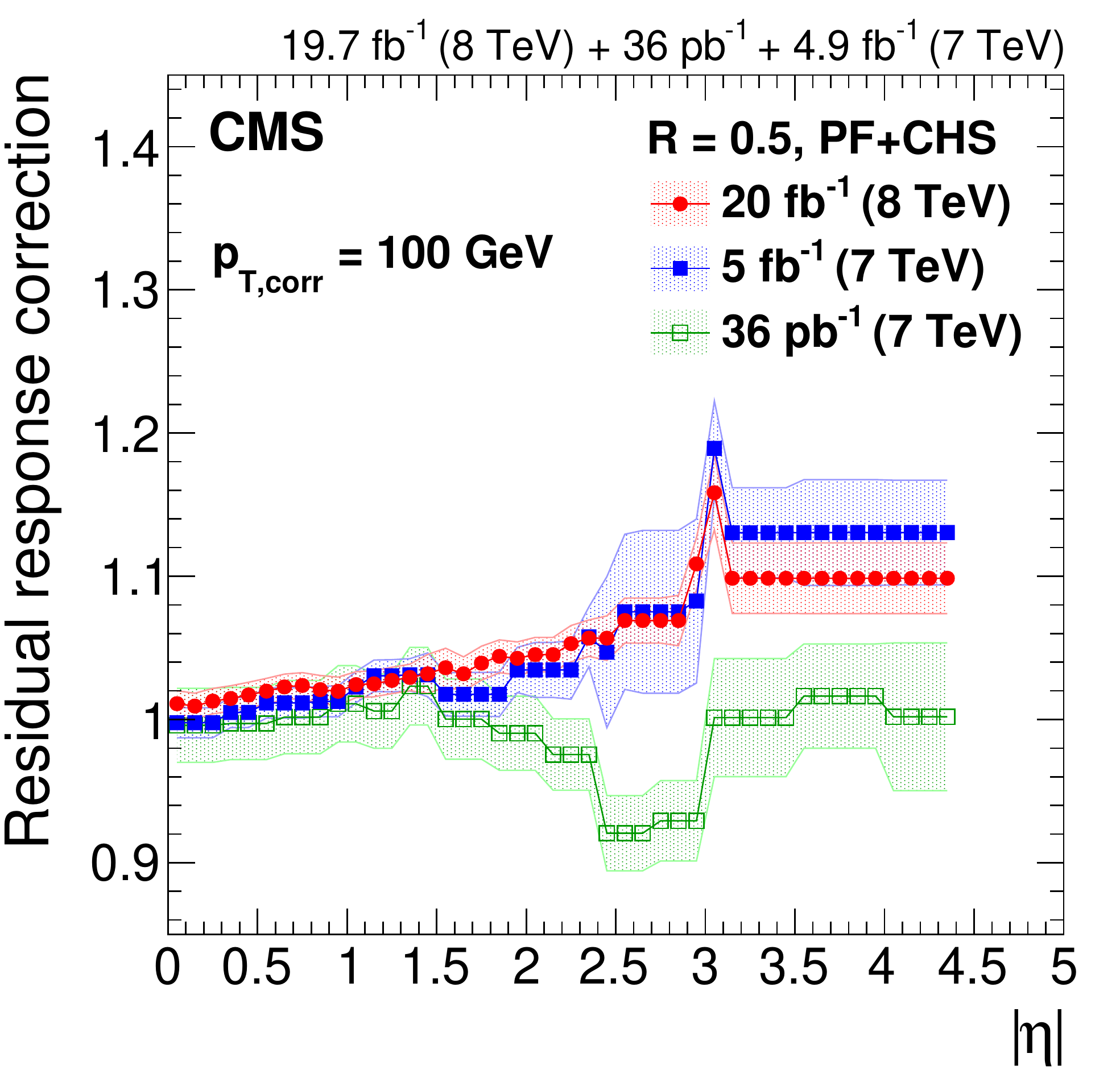}
\includegraphics[width=0.48\textwidth]{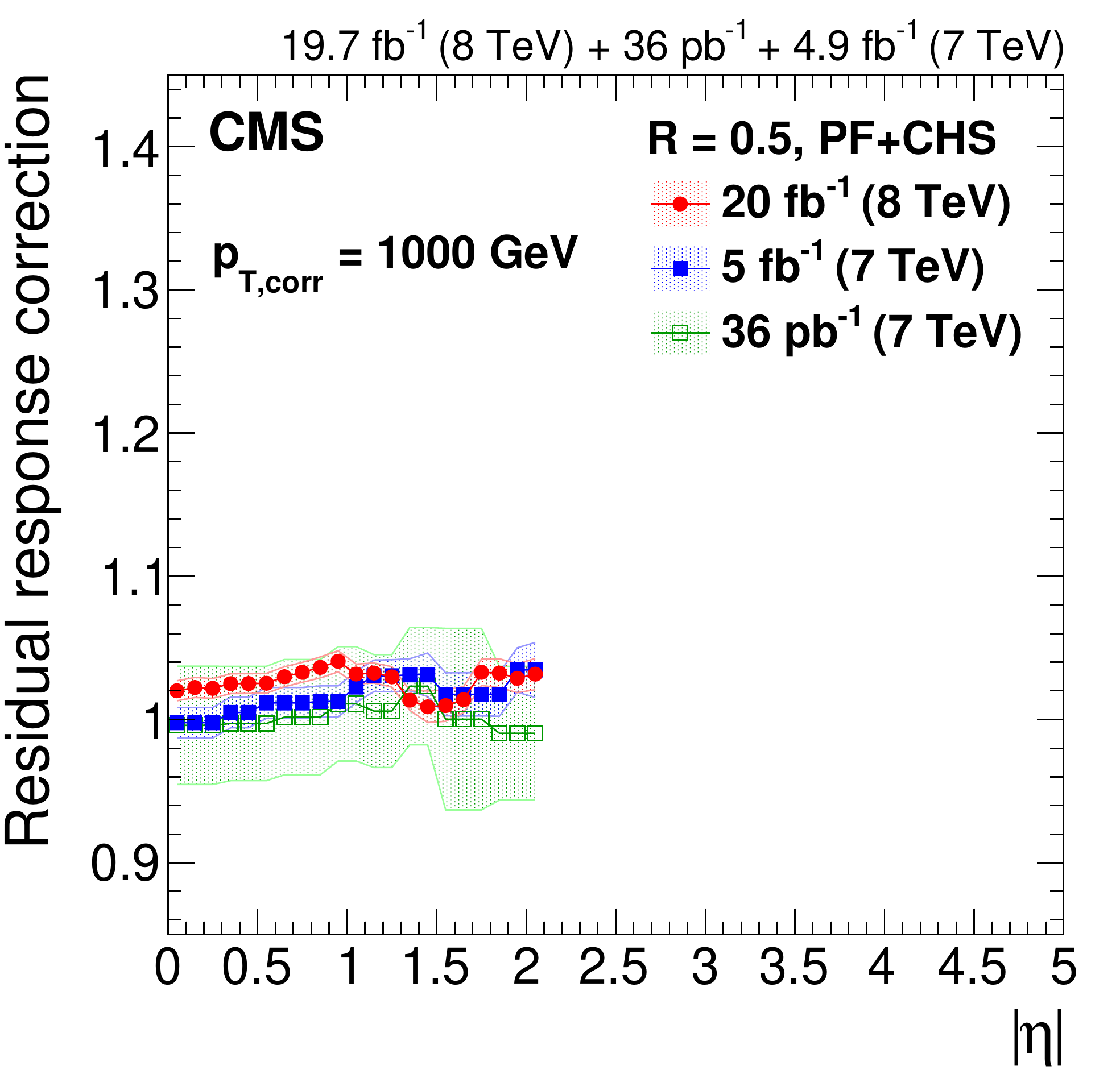}
\caption{\label{fig:l2l3res}
Residual data/simulation response correction factors for the 2012 data collected at 8\TeV for PF jets with CHS and $R=0.5$, compared to corrections at 7\TeV corresponding to 36\pbinv of data taken in 2010 \cite{JEC_JINST} and 5\fbinv taken in 2011 \cite{CMS-DP-2012-006}. The comparison is shown at $\abs{\eta}=0$ versus $p_{\mathrm{T},\rm corr}$ (top left), and as a function of $\abs{\eta}$ for $p_{\mathrm{T},\rm corr}=30\GeV$ (top right), $p_{\mathrm{T},\rm corr}=100\GeV$ (bottom left), and $p_{\mathrm{T}, \rm corr}=1000\GeV$ (bottom right). The plots are limited to a jet energy $E=\pt\cosh\eta=3500\GeV$ so as to show only correction factors for reasonable \pt in the considered data-taking period.}
\end{figure}

The data-based residual corrections for the relative JES versus $\eta$ have been derived with dijet events using the MPF method, and for the absolute JES versus \pt with a global fit combining results from \zmmjet, \zeejet, $\gamma$+jet and multijet analyses with both MPF and \pt-balance methods. The residual corrections are summarized in Fig.~\ref{fig:l2l3res}, and are compared to the results for 7\TeV proton-proton collision data in 2010 and 2011.
The residual response corrections are less than 3\% in the barrel, less than 10\% in the endcaps, and about 10\% in the forward detector, with the exception of the narrow endcap-forward boundary region at $3.0<\abs{\eta}<3.2$.

The uncertainty in JEC \pt dependence, which comes from detector simulation and from fragmentation modeling, is estimated to be less than 1\% for $30<\pt<2000\GeV$. This uncertainty is evaluated after the global fit with $\PZ$+jet and $\gamma$+jet data in the range of about $30<\pt<700\GeV$, and multijet data extending to over 1\TeV.
The uncertainty in JEC $\eta$ dependence derives mainly from ISR+FSR modeling, the effect of JER, and the uncertainty in fitting any additional \pt dependence, with a total uncertainty rising from less than about 0.5\% at $\abs{\eta}<2.5$ up to 2.5\% at $\abs{\eta}>3$.

The residual corrections have been quite stable since 2011, when improvements to simulation and reconstruction algorithms were implemented after the first year of data-taking in 2010, after a hardware intervention in the HF readout helped reducing the rate of anomalous signals in the readout photomultiplier tubes. The most significant changes are seen at high \pt in the barrel, where we attribute the increase in residual corrections to a drop in the HCAL scale in data.
The changes between 7\TeV in 2011 and 8\TeV in 2012 in the endcaps are at least partly attributable to the \pt dependence of the $\eta$-dependent corrections implemented at 8\TeV in 2012 for $\pt>60\GeV$, and the changes in HF are due to statistical fluctuations.

\section{Jet flavor corrections}
\label{sec:l5}

\subsection{Jet flavor definitions}

The QCD dijet sample is enriched in gluon jets, while the $\PZ$+jet and $\gamma$+jet samples are enriched in quark jets, which is important for latest stages of JEC estimation. We define jet flavor in terms of the parton flavor ('physics definition'), where the particle jet is matched to the nearest generator-level parton, considering only partons belonging to the hard scattering  matrix element process, within $\Delta R<0.25$. According to this physics definition, jets resulting from hard gluon radiation without a matching parton have an undefined flavor. This convention is well-defined for both \PYTHIA6.4 and \HERWIGpp~2.3. According to this definition, jets containing heavy-quark hadrons produced through gluon splitting (\eg, $ \Pg \to \bbbar$) are still classified as gluons.

The definition typically used for $\PQb$-tagging purposes uses parton shower flavor ('algorithmic definition') that reclassifies jets with heavy-quark hadrons from gluon splitting as $\PQb$- or $\PQc$-quark jets. However, because this algorithm uses information from the final stage of the parton shower, it currently cannot identify gluon jets in \HERWIGpp~2.3, which forces $ \Pg \to \PQq \PAQq $ splitting for all gluons. The algorithmic definition also has a tendency to reclassify jets with light quarks from gluon splitting as quark jets, and quark jets with early hard gluon radiation as gluon jets, mixing quark and gluon fractions.

\begin{figure}[htbp!]
\centering
\includegraphics[width=0.32\textwidth]{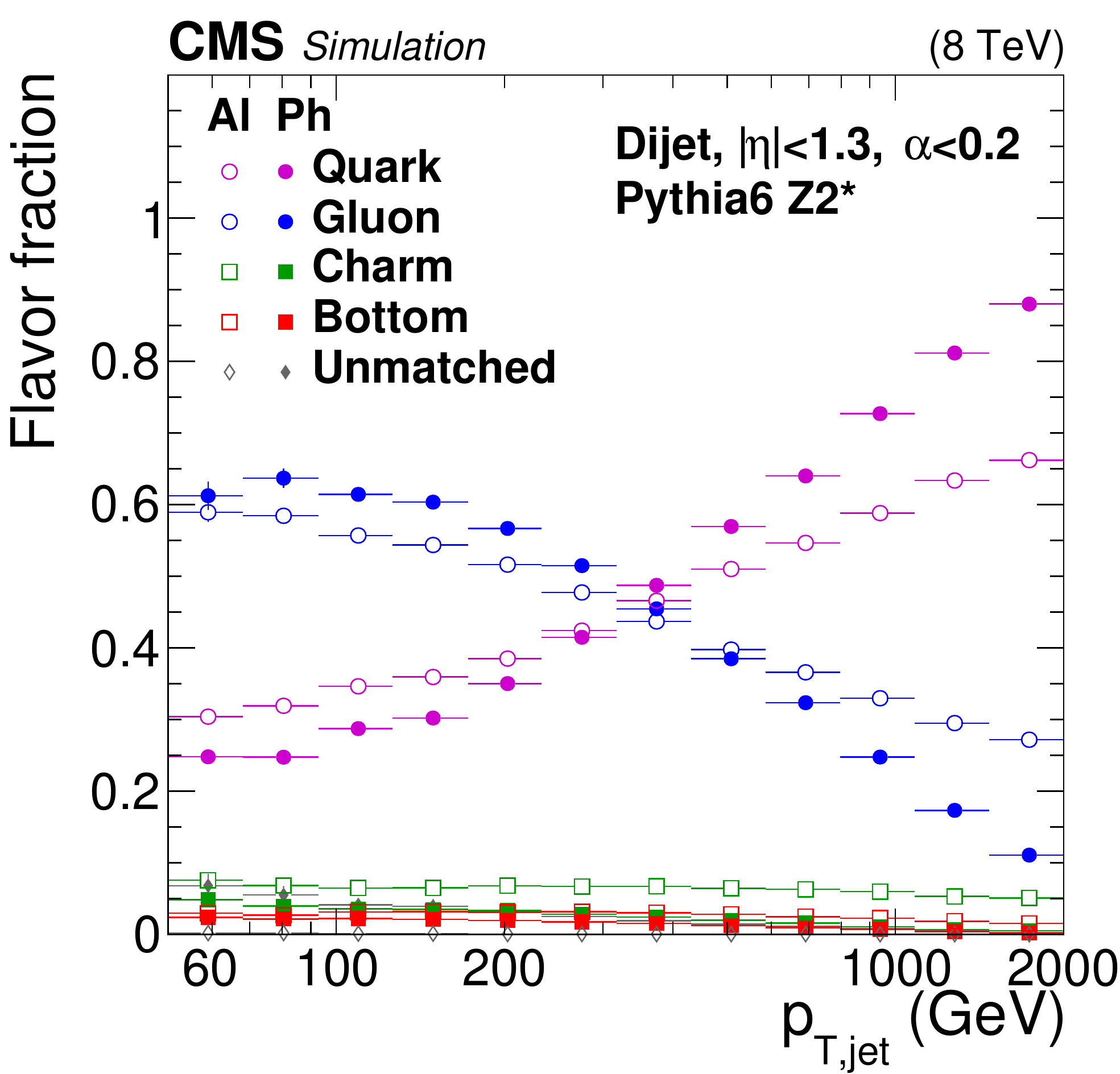}
\includegraphics[width=0.32\textwidth]{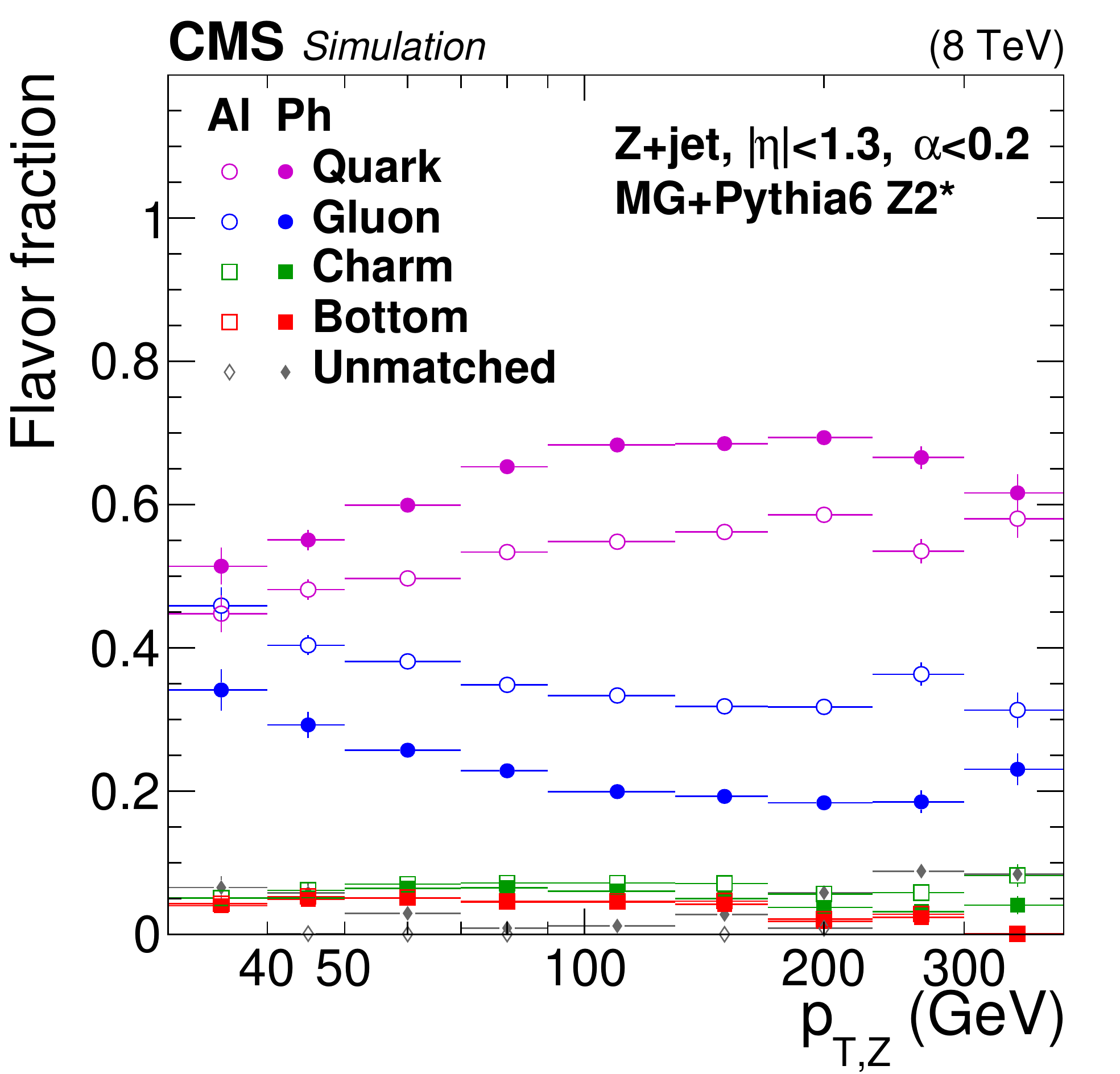}
\includegraphics[width=0.32\textwidth]{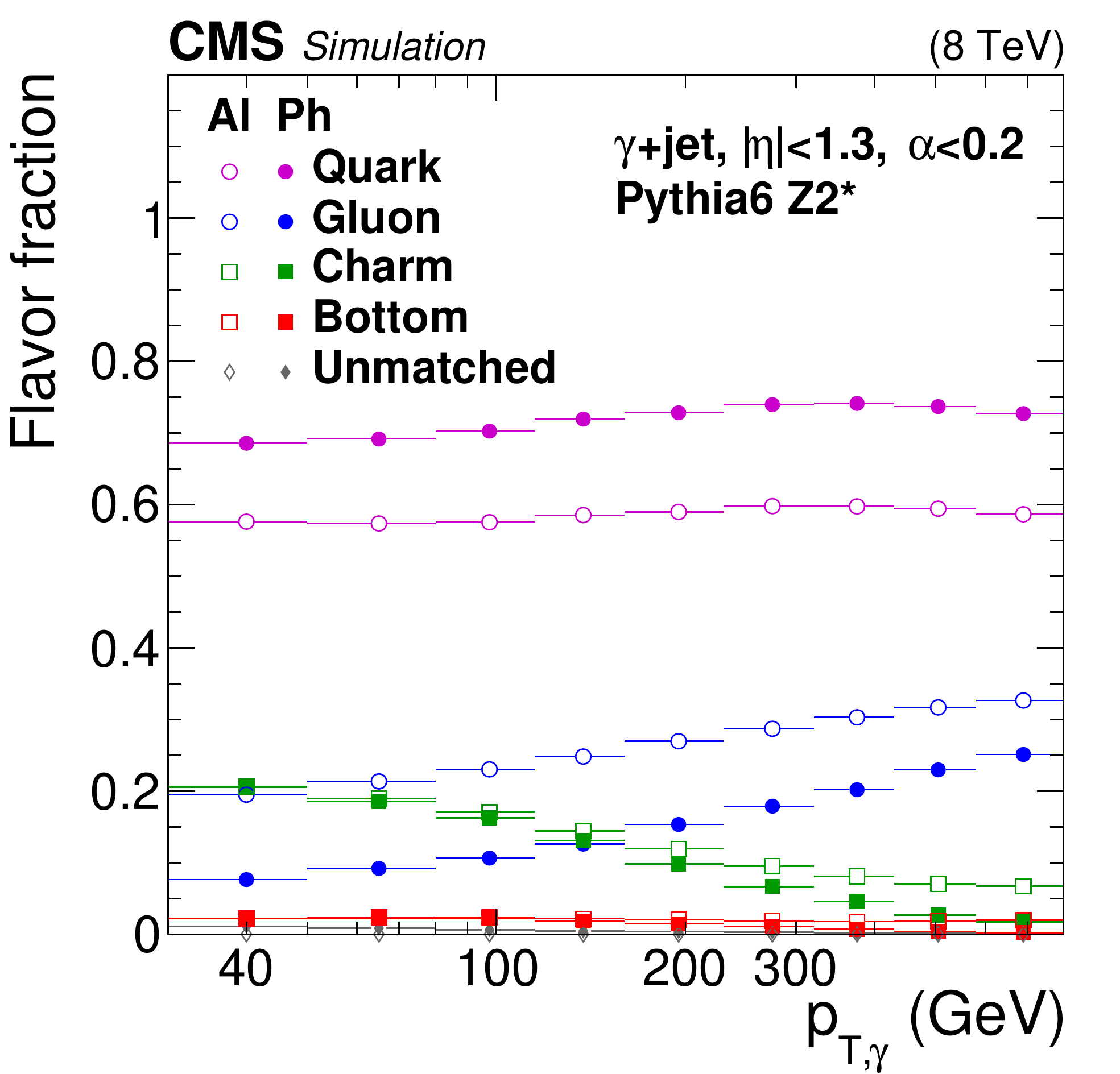}
\caption{\label{fig:flavorfractions}
Jet-flavor fractions in the physics (Ph) and algorithmic (Al) flavor definitions for QCD dijet (left), $\PZ$+jet (middle), and $\gamma$+jet (right) samples. As explained in Section~\ref{Sec:l2l3res}, the variable $\alpha=p_\text{T, 3rd jet}/p_\mathrm{T, ave}$ for dijet events and $\alpha=p_\text{T, 2nd jet}/p_{\mathrm{T}, \gamma/\Z}$ for $\PZ$+jet and $\gamma+$jet events.}
\end{figure}

We show the jet-flavor fractions for both definitions in Fig.~\ref{fig:flavorfractions} for QCD dijet, $\PZ$+jet, and $\gamma$+jet samples. The flavor fractions in the $\gamma$+jet sample are very similar to those of $\PZ$+jet in the $\pt>200\GeV$ range, except for a somewhat larger fraction of directly produced charm jets.

\subsection{Simulated flavor corrections}

The differences in response ($\pt/p_\text{T, ptcl}$) for different jet flavors arise mainly from variations in jet fragmentation energy and variations in particle composition of the jet. Softer jet fragmentation results in more particles outside the detector acceptance. With respect to particle composition, the neutral hadron fraction of the jet, $f_\mathrm{nh}$, is seen to affect the response most.
Jets from $\PQu$ and $\PQd$ quarks have the highest response, while those from gluons have the lowest, as a result of gluons fragmenting into the largest number of soft particles.
The response values for heavy-flavor jets from $\PQc$ and $\PQb$ quarks are in between those for $\PQu$/$\PQd$ and $\Pg$ jets due to additional soft particles from heavy-flavor hadron decays compared to $\PQu$ and $\PQd$ jets. As a reminder, the CMS definition of jet energy response $R$ excludes neutrinos, which would otherwise lower $\PQb$ and $\PQc$ jet response significantly.
The jets from strange quarks also have relatively low response due to the high probability of producing long-lived strange neutral hadrons (mainly $\PKzL$) with significant \pt. This effect is specific to PF jets, which have a larger difference between charged and neutral hadron responses than calorimeter jets.

\begin{figure}[htbp!]
\includegraphics[width=0.48\textwidth]{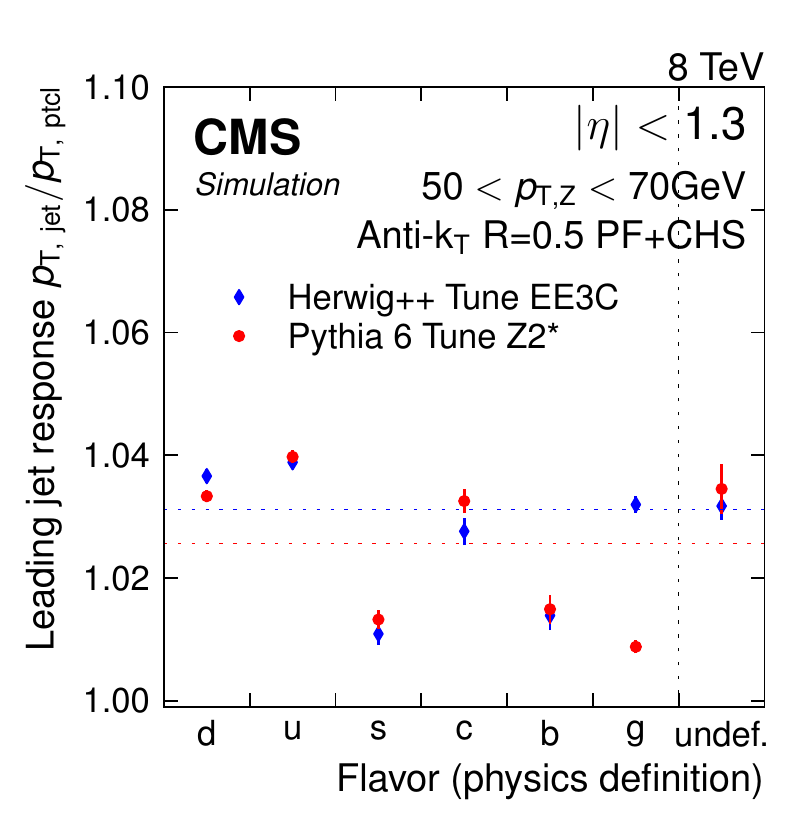}
\includegraphics[width=0.46\textwidth]{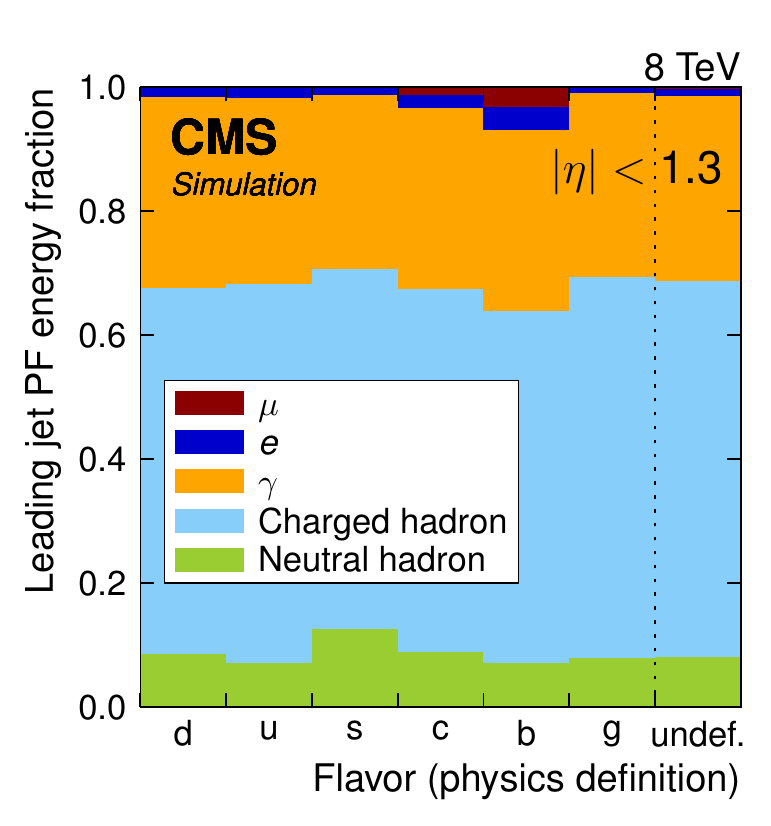}
\caption{\label{fig:pyhw}
Comparison of jet response (left) and simulated composition (right) for different flavors of leading jets in $\PZ$+jet events with $50<\pt^Z<70\GeV$, $|\eta_\text{jet}|<1.3$, and $\alpha=0.3$ (defined in Eq.~(\ref{eq:krad})). The response values are compared for \PYTHIA6.4 and \HERWIGpp~2.3, the composition is from \PYTHIA6.4.}
\end{figure}

\begin{figure}[htbp!]
\centering
\includegraphics[width=0.49\textwidth]{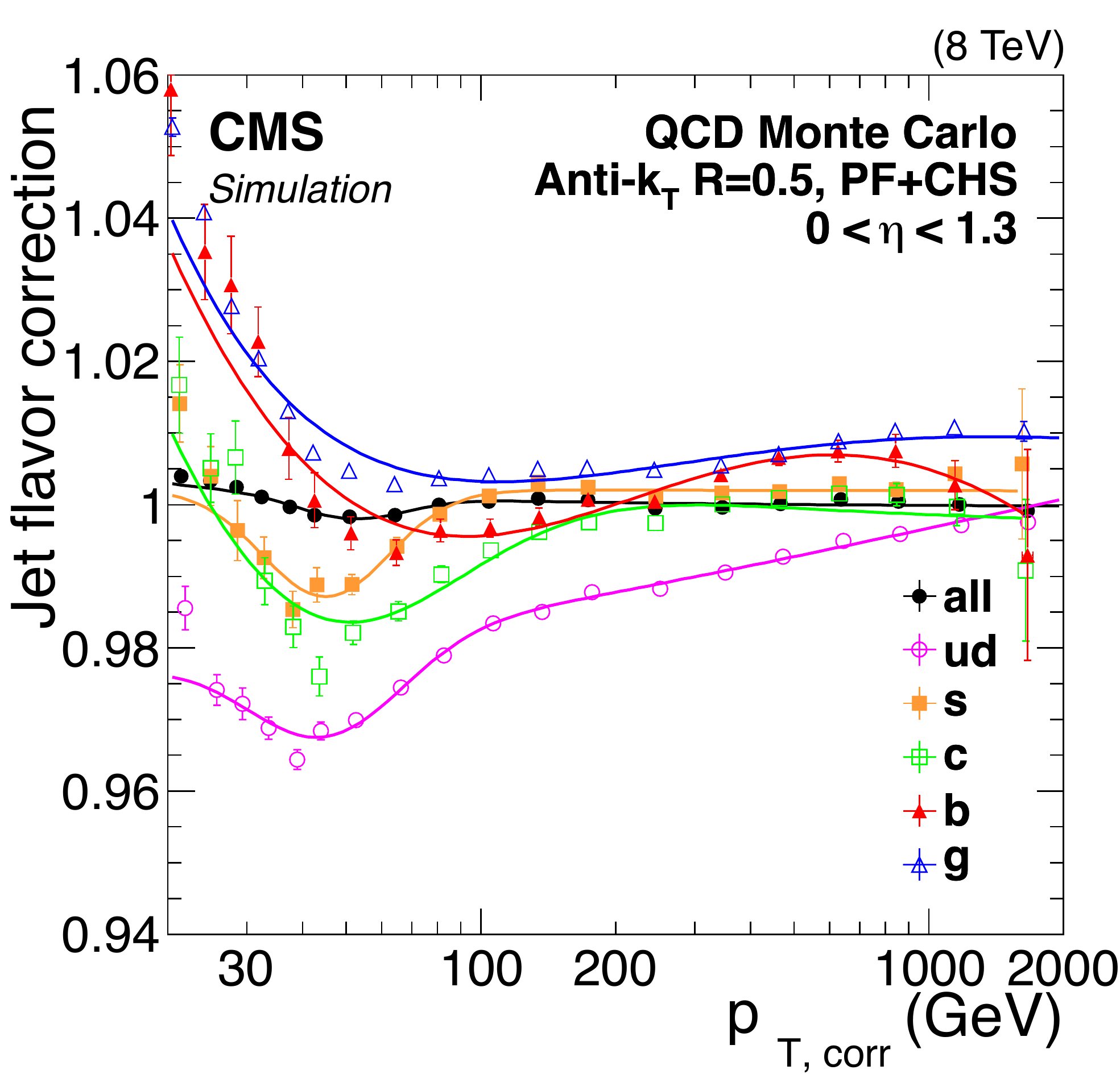}
\includegraphics[width=0.49\textwidth]{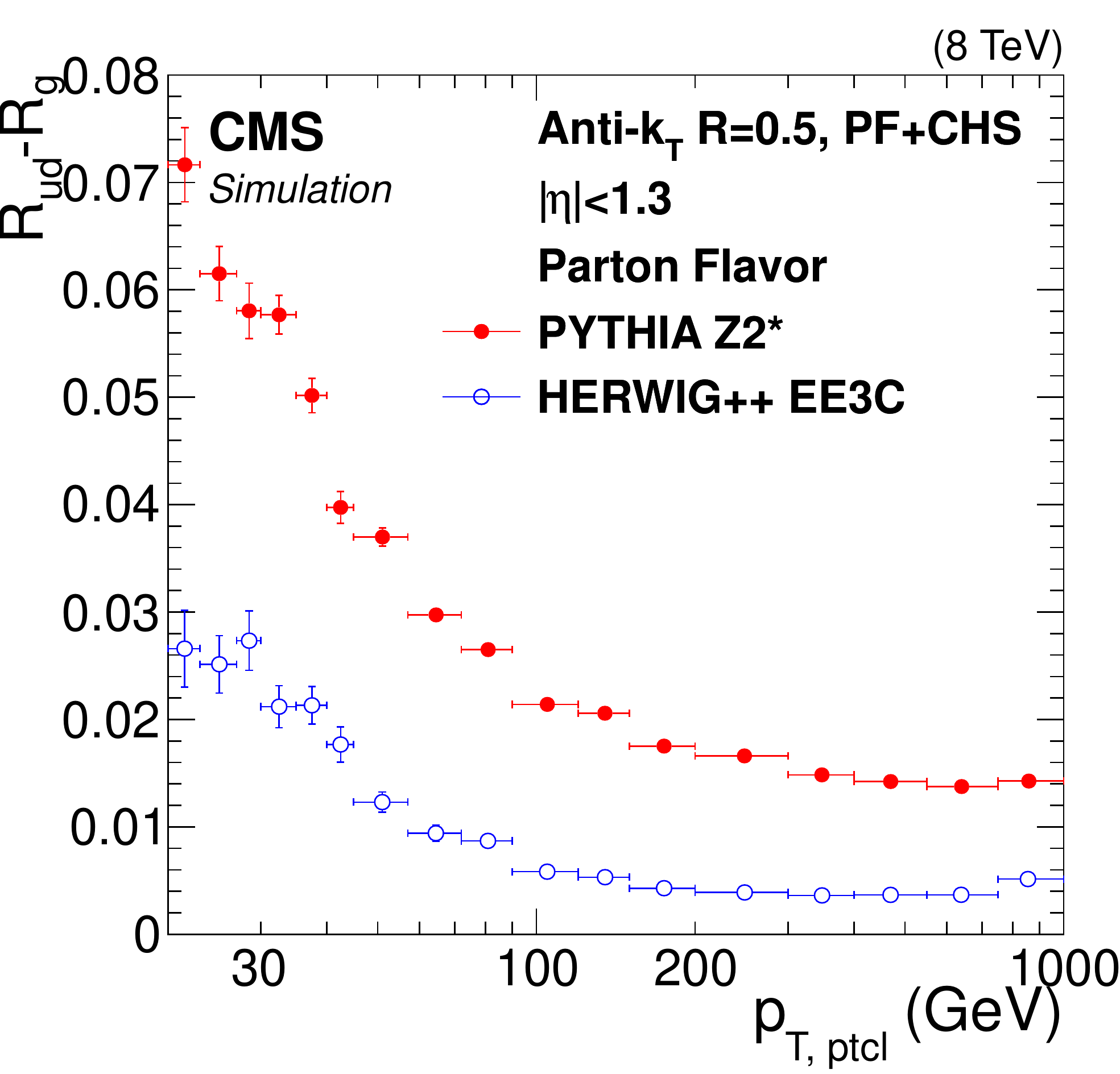}
\caption{\label{fig:flavorresponse}
Residual jet-flavor correction factor as a function of jet $p_\mathrm{T,~corr}$ from \PYTHIA6.4 tune Z2*, derived on top of inclusive JEC and defined relative to the QCD flavor mixture (left).
The neutrinos are excluded from particle jets, which brings $\PQc$- and $\PQb$-jet response in between that of light quarks and gluons. The lines show the parameterizations used for residual jet-flavor corrections.
Difference in light-quark and gluon jet response as a function of jet $p_\mathrm{T,~corr}$, as predicted by \PYTHIA6.4 and \HERWIGpp~2.3 (right).
}
\end{figure}

These effects are quantitatively demonstrated for jet response in Fig.~\ref{fig:pyhw} (left) for $\PZ$+jet events with
$50<p_\mathrm{T, Z}<70\GeV$, $|\eta_\text{jet}|<1.3$, and $p_\mathrm{T, jet2}<0.3\pt^Z$, where the jets are fully calibrated with the corrections relevant for simulated jets, as discussed in the previous sections.
The biggest differences between \PYTHIA6.4 and \HERWIGpp~2.3 are observed for the gluon jets. The response is higher than unity for all flavors, since $\PZ$+jet events have lower contributions from the underlying event
than dijet events, leading to a small bias from the hybrid jet area method, defined in Section~\ref{sec:jetarea}.
Moreover, as shown in Fig.~\ref{fig:flavorfractions}, $\PZ$+jet events have a smaller gluon fraction (25$\%$ for $\pt=60\GeV$) with respect to dijet (65$\%$ for $\pt=60\GeV$) events, and the different response of gluons and light-quark jets, visible in Fig.~\ref{fig:flavorresponse} (right), raises the average response for $\PZ$+jet events.
Figure~\ref{fig:pyhw} (right) demonstrates the simulated jet composition for different jet flavors. The s-quark jets have higher neutral hadron fractions due to $\PKzL$ production, while the neutrino production in $\PQc$ and $\PQb$ jets is evident through the associated muons and electrons. The softer fragmentation of gluon jets has relatively little impact on the particle composition, but results in low effective response for the neutral hadrons $R_\mathrm{nh}$.

Figure~\ref{fig:flavorresponse} (left) shows the inverse of the response for different flavors versus \pt, relative to the one for the QCD flavor mixture.
The relative behavior of the differences has a weak dependence on \pt, but the absolute differences become smaller at high \pt. This can be explained by the asymptotic rise of the neutral hadron response towards unity at high \pt and detector acceptance effects becoming less significant for high-$\pt$ jets.

While \PYTHIA6.4 and \HERWIGpp~2.3 agree well on quark flavor response, there are significant differences in the gluon response modeling. A useful metric for the JES sensitivity to flavor response modeling is the difference in light-quark (uds) and gluon jet response, shown in Fig.~\ref{fig:flavorresponse} (right).
The flavor sensitivity of the CMS PF algorithm is much reduced with respect to the CALO jets reconstruction, as was demonstrated in Ref.~\cite{JEC_JINST}.

\subsection{Flavor uncertainties}
\label{sec:flavor}

\begin{figure}[htbp!]
\centering
\includegraphics[width=0.49\textwidth]{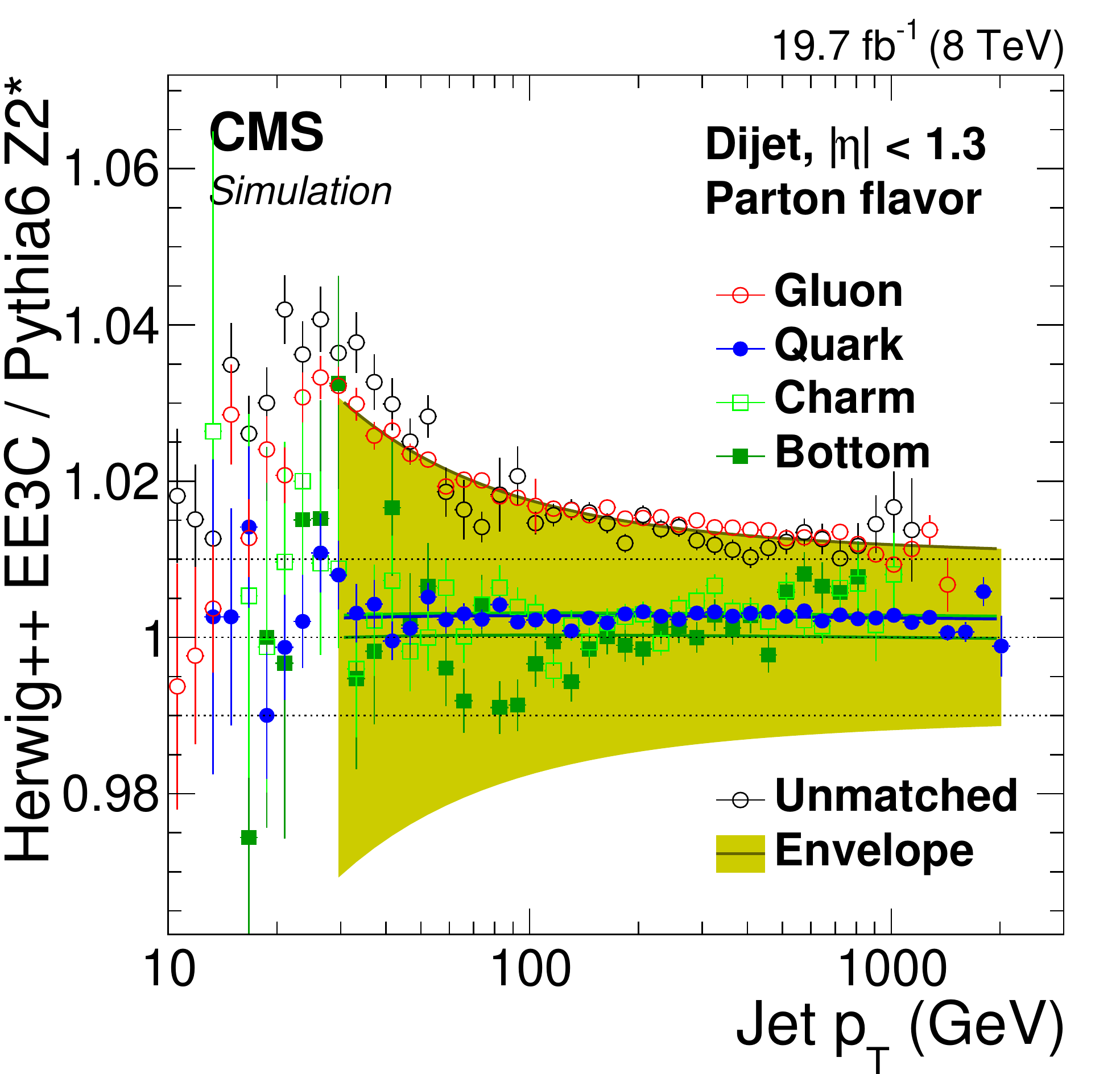}
\includegraphics[width=0.49\textwidth]{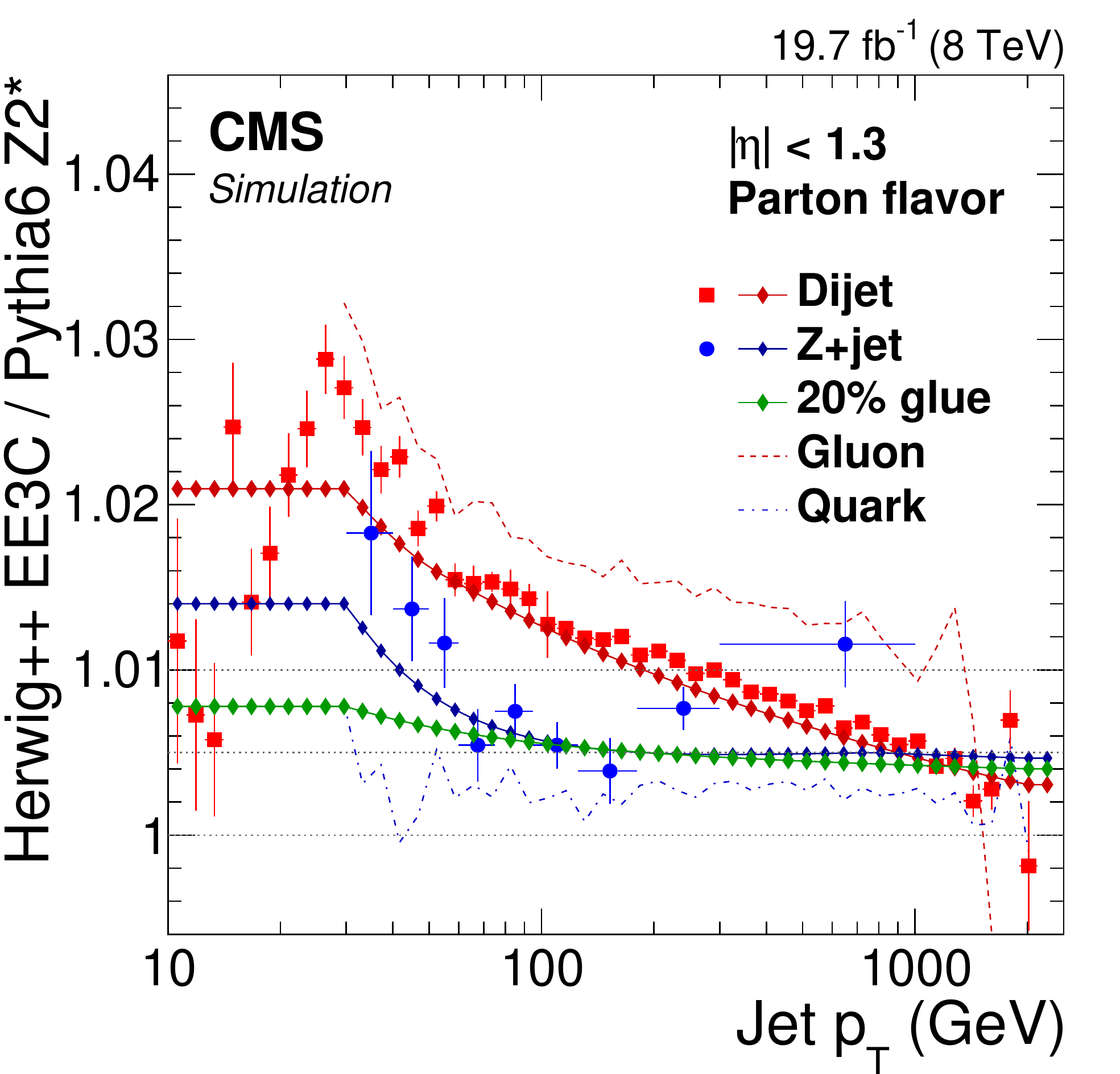}
\caption{\label{fig:flavor}
Left: Ratio of jet responses in \PYTHIA6.4 (tune Z2*) and \HERWIGpp~2.3 (tune EE3C) versus jet \pt, for pure jet flavors selected using the physics definition, where the shaded envelope highlights the largest differences observed for the gluon jets.
Right: \PYTHIA6.4/\HERWIGpp~2.3 response differences as a function of jet \pt for QCD dijet and $\PZ$/$\gamma$+jet flavor mixtures calculated from the parameterized flavor response differences (diamonds) and compared to the full simulation for dijet (squares) and $\PZ$+jet (circles) samples. The ``20\% glue'' corresponds to the effective $\PZ$/$\gamma$+jet flavor mixture at $\pt=200\GeV$, which has 20\% of gluons. The ratio of the responses is also shown for pure gluon (dashed curve) and quark (dot-dashed curve) jets.
}
\end{figure}

We investigate the jet fragmentation and flavor response differences by comparing \PYTHIA6.4 tune Z2* and \HERWIGpp~2.3 tune EE3C in balanced QCD dijet events. These two tunes have been shown to cover differences between data and simulation in many studies of jet structure and fragmentation, in particular for the variables used for quark and gluon tagging \cite{CMS-PAS-JME-13-002}.
The jet flavors are tagged with the matching parton flavor, based on the physics definition.
As shown in Fig.~\ref{fig:flavor}, we observe the largest response differences for the gluon jets, while the light-quark and heavy-flavor jets are in good agreement in both MCs.

The parameterized response differences as a function of $\eta$ and \pt, combined with the flavor fractions in Fig.~\ref{fig:flavorfractions}, are propagated through the fitting procedure used for data-based residual corrections to evaluate the systematic uncertainties from jet flavor.
Jets in the barrel reference region $\abs{\eta}<1.3$ have flavor uncertainty only when the flavor mixture differs from the $\PZ$/$\gamma$+jet flavor mixture used in the data-based methods, which is roughly 20\% gluons at $\pt=200\GeV$.
The $\eta$-dependent corrections influence the jet-flavor systematics through both the tag and the probe jet. The flavor mixture of the central tag jet is different from that in $\PZ$/$\gamma$+jet calibration samples, which leads to a flavor uncertainty for dijets within the reference region in the barrel. This barrel uncertainty for the dijet measurement is then propagated to the more forward regions through the dijet-based $\eta$-dependent corrections, with additional flavor uncertainties relative to the dijet flavor mixture for other samples.

The total flavor uncertainty $\Delta R_\mathrm{flavor}$ for any given flavor mixture is estimated based on pure flavor response $R_f$ (in \PYTHIA6.4 Z2* and \HERWIGpp~2.3) and flavor fractions $F_f$ in \PYTHIA6.4 Z2*:
\begin{eqnarray}
\Delta R_\mathrm{flavor}(\pt, \eta) &=& \hat{R}_\mathrm{mix}(\pt, \eta) - R_\mathrm{ref}(\pt, \eta),
\end{eqnarray}
where
\begin{eqnarray}
\hat{R}_\mathrm{mix}(\pt,\eta) &=&
\sum_f (R_{f,\rm \HERWIGpp}(\pt,\eta) - R_{f,\rm {\PYTHIA6}}(\pt,\eta))
 F_{f,\rm mix}(\pt,\eta), \label{eq:rhat}\\
R_\mathrm{ref}(\pt, \eta) &=& \hat{R}_\mathrm{Z+jet}(200\GeV, 0)
+ \left(\hat{R}_\mathrm{dijet}(\pt, \eta) - \hat{R}_\mathrm{dijet}(\pt, 0)\right).
\end{eqnarray}

\begin{figure}[htbp!]
\centering
\includegraphics[width=0.49\textwidth]{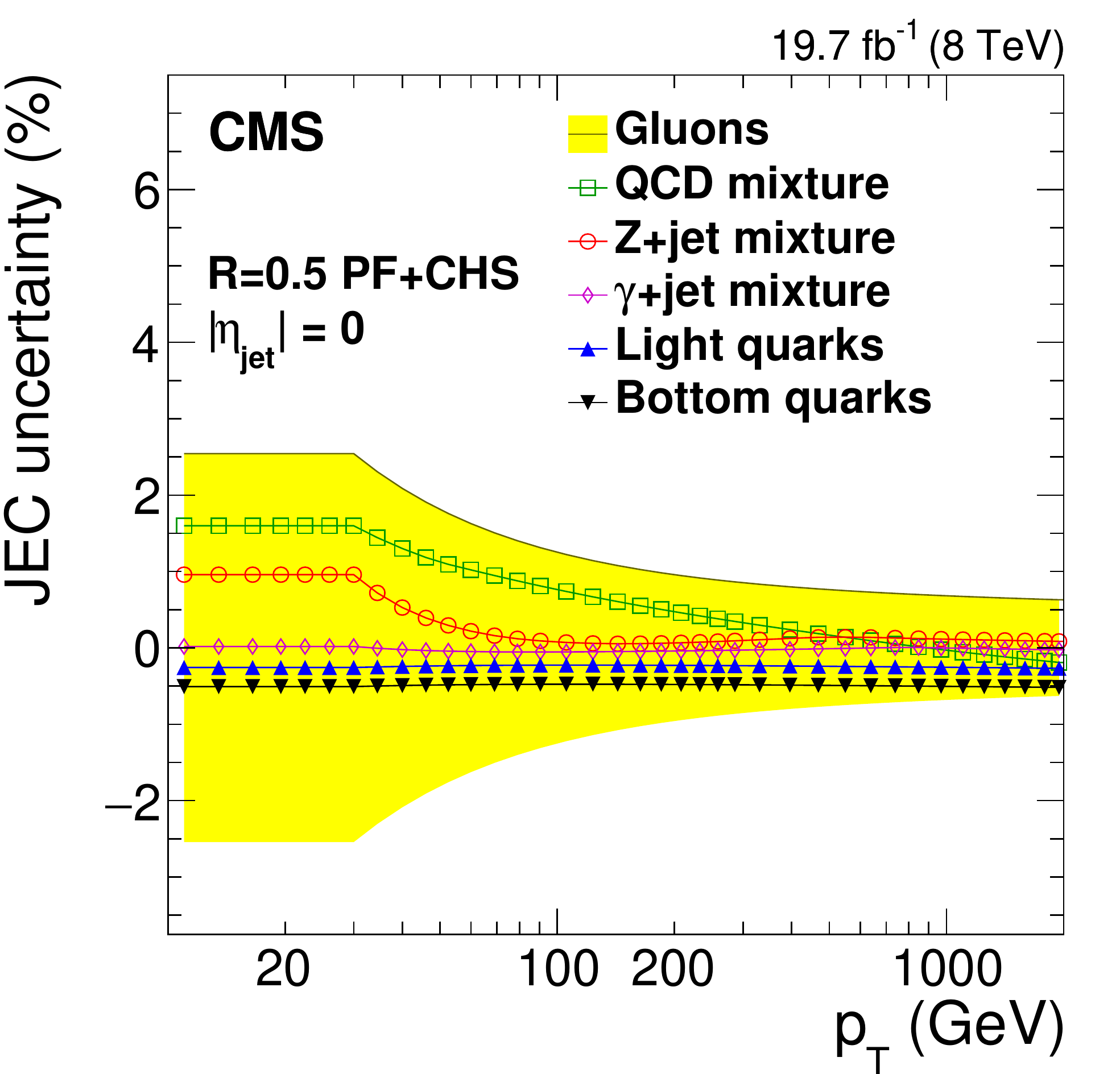}
\includegraphics[width=0.49\textwidth]{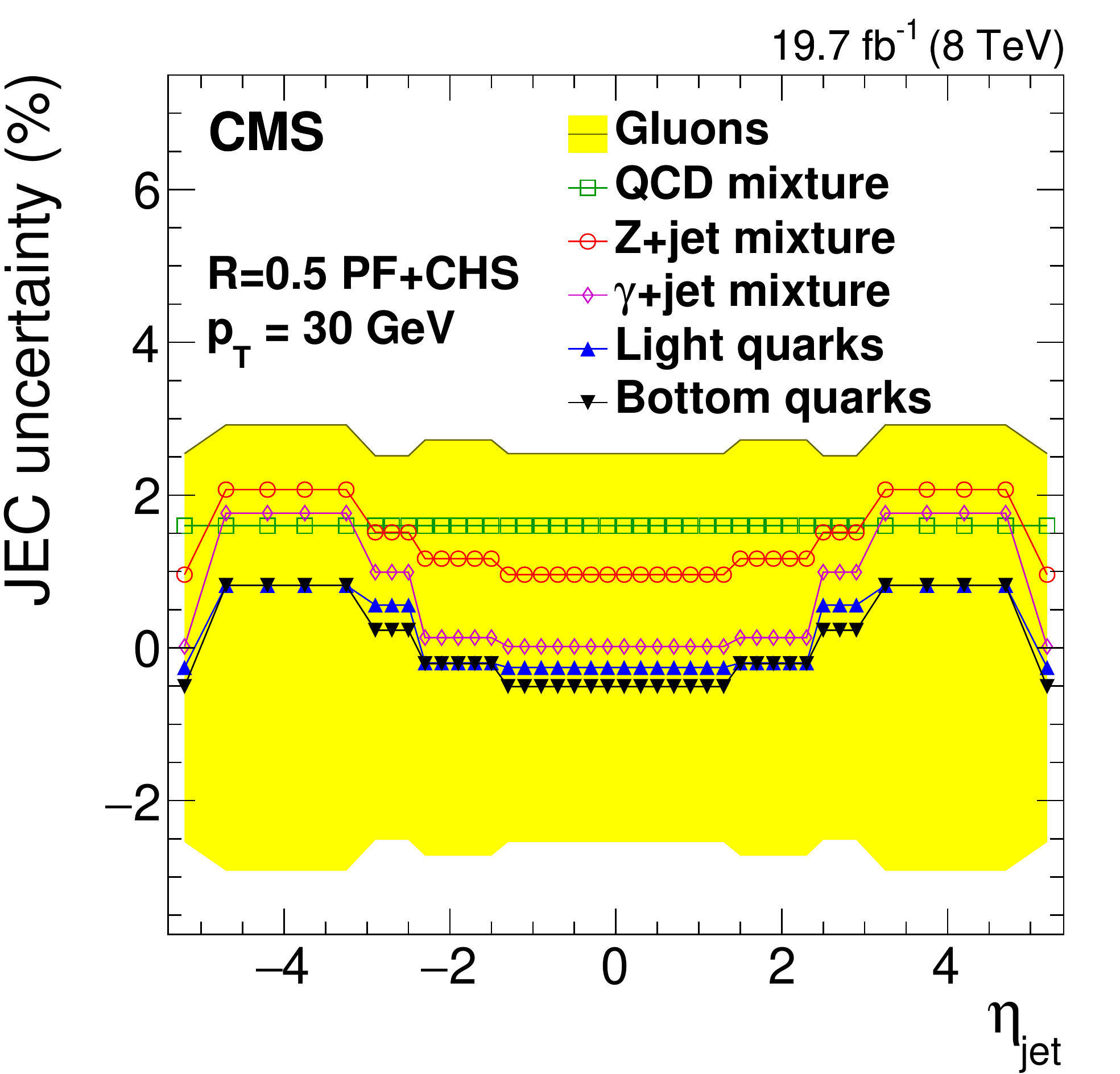}
\includegraphics[width=0.49\textwidth]{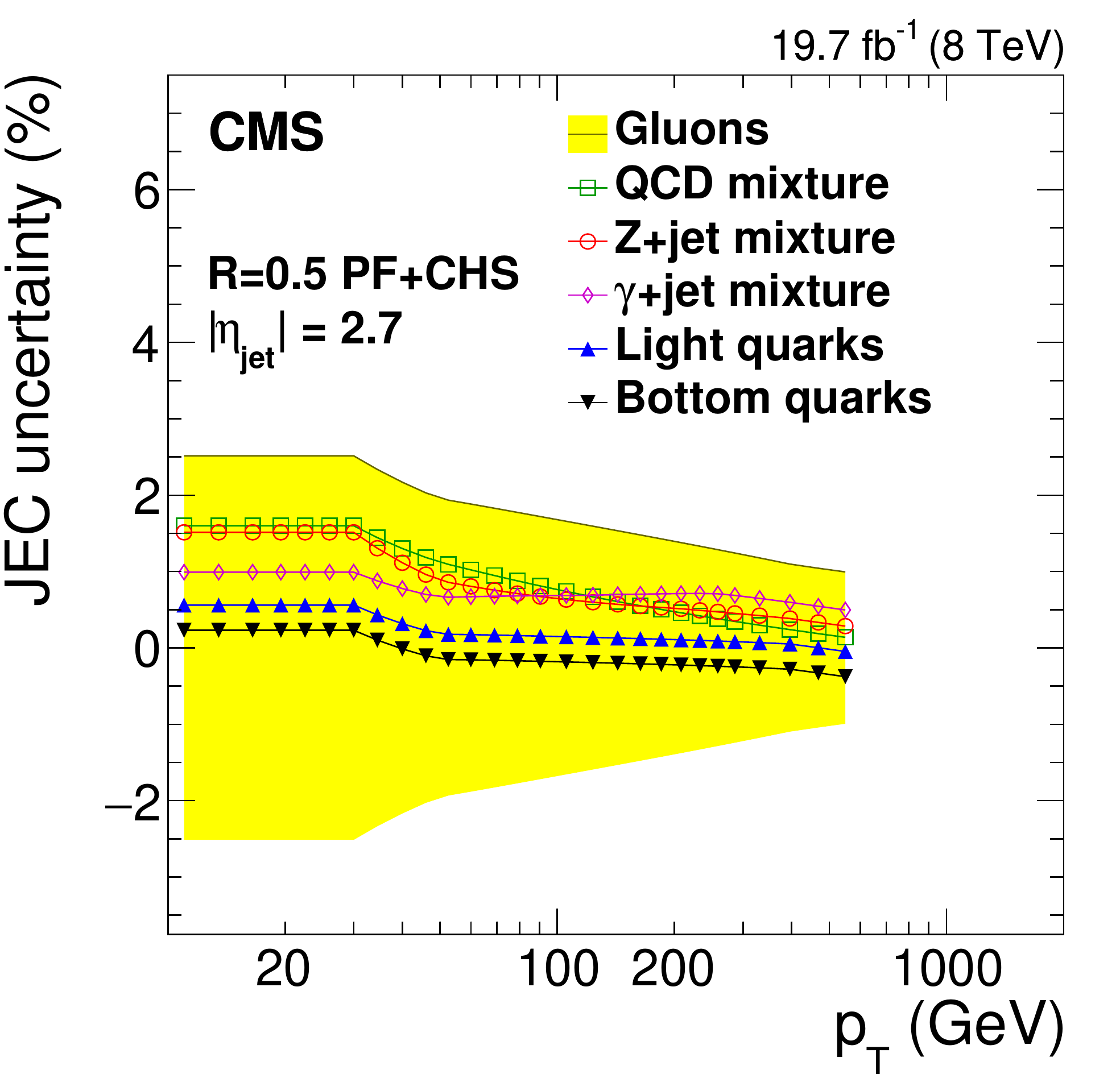}
\includegraphics[width=0.49\textwidth]{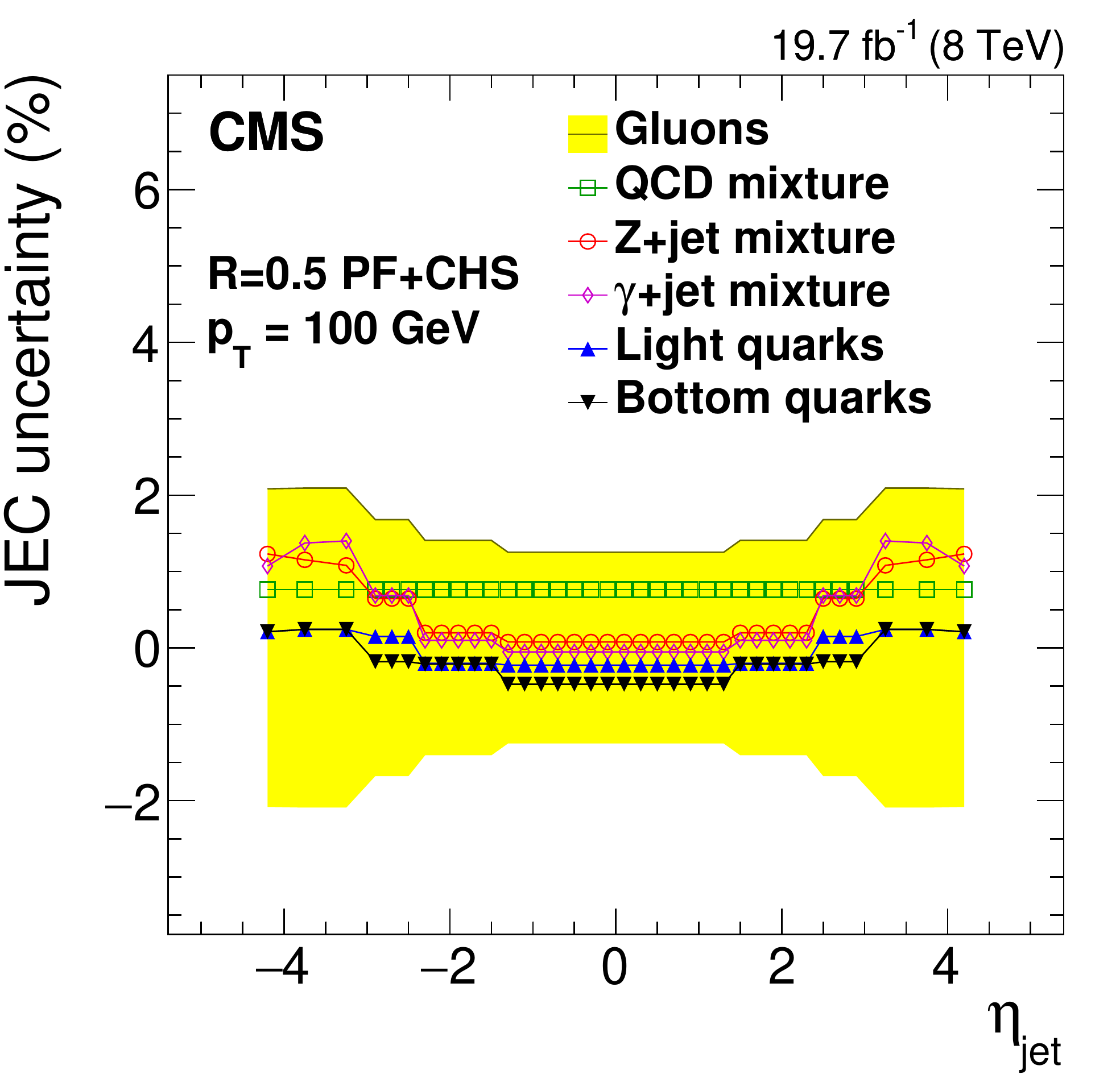}
\caption{\label{fig:flavorsys}
Systematic uncertainties in jet energy corrections for various flavor mixtures (QCD dijets, $\PZ$+jet and $\gamma$+jet) and pure flavors (gluons, light quarks and bottom quarks) as a function of jet \pt (left, for fixed $|\eta_\text{jet}|=0$, top, and $|\eta_\text{jet}|=2.7$, bottom) and $\eta_\text{jet}$ (right, for fixed $\pt=30\GeV$, top, and 100\GeV, bottom).
The sign of the systematic source indicates the sign of the \PYTHIA6.4 tune Z2* and \HERWIGpp~2.3 tune EE3C difference. The shaded band shows gluon flavor response uncertainty symmetrically around zero.
}
\end{figure}

The symbol $\hat{R}$ indicates the linear combination of response differences for a given flavor mixture, and $R_\mathrm{ref}$ is relative to the Z$/\gamma+$jet flavor mixture used in the data-based methods.
The various flavor uncertainties are compared in Fig.~\ref{fig:flavorsys}. This calculation returns by design zero flavor uncertainty for the $\PZ$/$\gamma$+jet flavor mixture at $\pt=200\GeV$ and $\eta=0$ (about 20\% gluons), which is where the constant absolute data/simulation scale factor is effectively determined in the ($\pt$, $\eta$, flavor) space. For the dijet sample the flavor uncertainty is constant versus $\eta$ by design, but increases with \pt away from the effective flavor reference point. For other samples the flavor uncertainty depends both on \pt and $\eta$. The respective flavor uncertainties are provided for the $\PZ$+jet (\textit{FlavorZJet}), $\gamma$+jet (\textit{FlavorPhotonJet}), and dijet (\textit{FlavorQCD}) mixtures as well as for the pure flavors (\textit{FlavorPureGluon, FlavorPureQuark, FlavorPureCharm, FlavorPureBottom}).
The flavor uncertainties are reliably determined only at $\pt>30\GeV$, where the flavor assignment for the leading jets has a small ambiguity.
However, the flavor differences are reduced at $\pt<30\GeV$
in PF so the uncertainty at $\pt=30\GeV$ is used as a conservative upper limit on flavor uncertainties at lower \pt as well.

\subsection{\texorpdfstring{$\PZ$+$\PQb$-jet}{Z+b-jet} balance}
\label{sec:Zb}

\begin{figure}[htbp!]
\centering
\includegraphics[width=0.48\textwidth]{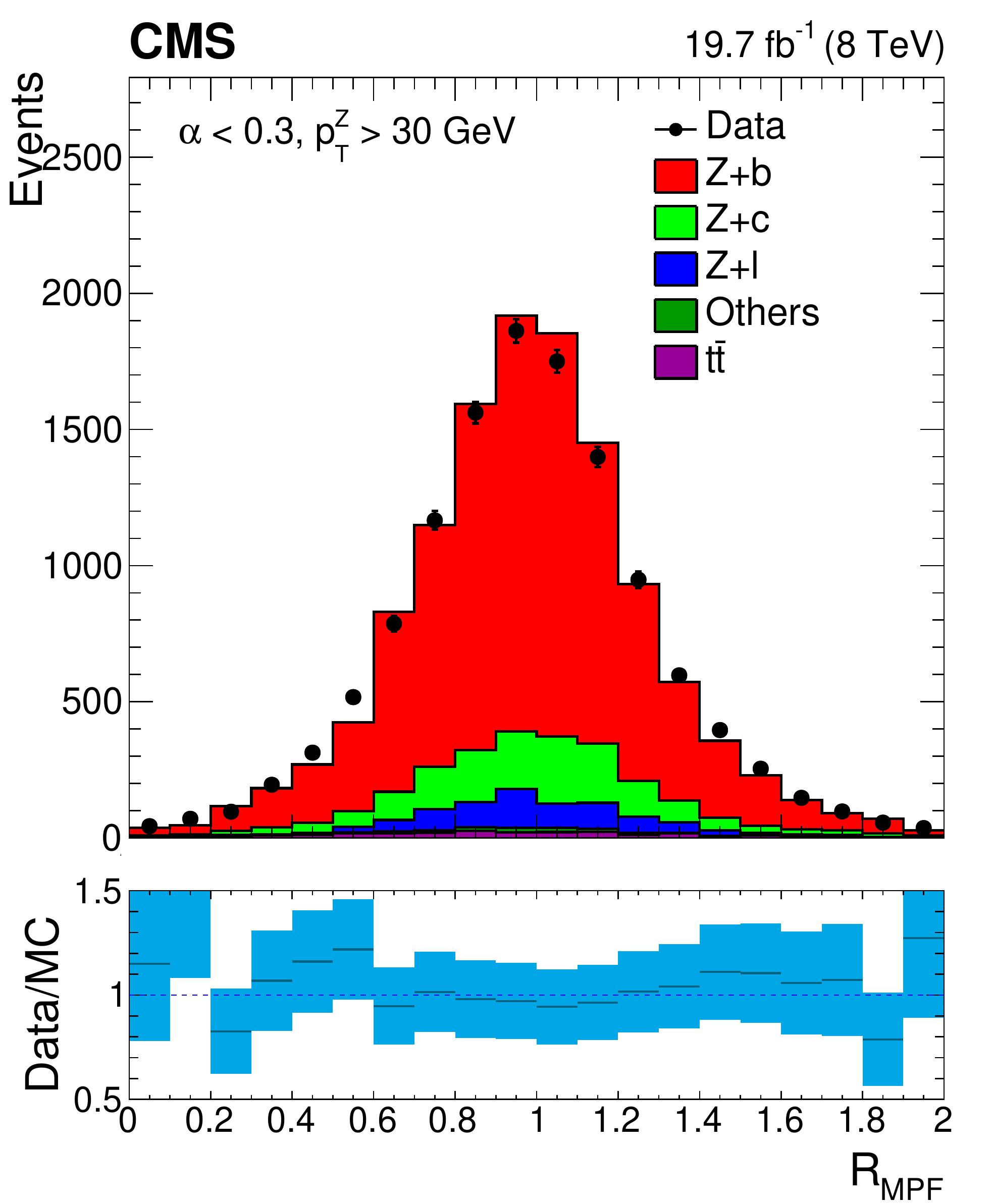}
\includegraphics[width=0.48\textwidth]{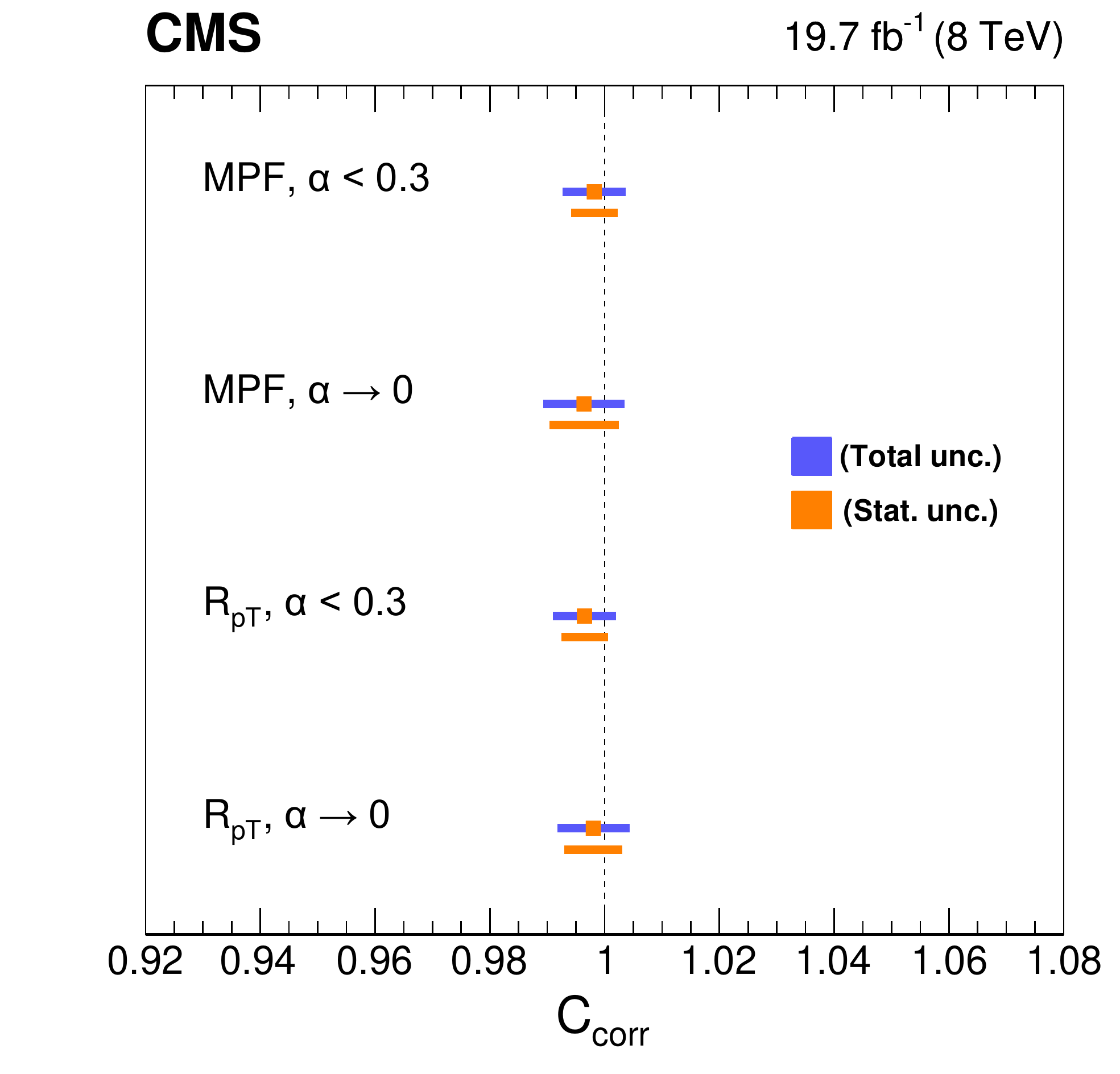}
\caption{\label{fig:zb}
Distribution of the $\PZ$+$\PQb$-jet response using the MPF method with a fixed requirement $\alpha<0.3$ (left).
Data-to-simulation ratio of the $\PZ$+$\PQb$-jet response relative to the inclusive $\PZ$+jet sample with the MPF and the \pt-balance methods (right).
}
\end{figure}

To check the flavor corrections and their uncertainties in data, we use a $\PZ$+$\PQb$-jet \pt-balancing technique \cite{CMS-PAS-JME-13-001} that is a straightforward extension of the $\PZ$+jet balancing methods described in Section~\ref{Sec:L3residual}. In addition to the $\PZ$+jet event selection criteria described in Section~\ref{sec:event_sel}, the leading jet is required to be $\PQb$-tagged with the combined secondary vertex tagger at the medium working point \cite{btag}, and the $\PQb$-jet response is determined with both the MPF and the \pt-balance methods, as in the nominal $\PZ$+jet analysis described in Section~\ref{Sec:L3residual}. The purity of the Z+b sample is about 70--80\% and the MPF response distribution is well modeled by simulation, as shown in Fig.~\ref{fig:zb}. To reduce systematic uncertainties, the main result is reported as a ratio of response in the $\PZ$+$\PQb$-jet sample to that in inclusive $\PZ$+jet sample.
This check is particularly important for the top quark mass measurements, which uses light-quark jets from $\PW$ boson decays to constrain the energy scale for $\PQb$ jets from the top quark decay $\PQt \to \PW \PQb$, assuming the same momentum scale for light-quark jets and $\PQb$ jets with similar kinematics \cite{Chatrchyan:2013xza}.

The most precise result is obtained using the MPF method with a fixed requirement $\alpha<0.3$ ('MPF, $\alpha<0.3$' in Fig.~\ref{fig:zb}). This gives a residual $\PQb$-jet correction of $C_\text{corr} = 0.998 \pm 0.004 \stat \pm 0.004 \syst$ relative to \PYTHIA6.4 tune Z2*,
which is consistent with unity and comparable in precision to the current $\PQb$-jet flavor uncertainty of about 0.5\%, shown in Fig.~\ref{fig:flavorsys}. The results from the response extrapolated to $\alpha\to 0$ (bottom symbols in Fig.~\ref{fig:zb} right) and from the \pt-balance method ($R_{\pt}$) are consistent with the central MPF result.
The systematic uncertainties, which include $\PQb$-tagging efficiency and mistag rate, lepton scales, ISR- and FSR-related effects, and jet fragmentation, are dominated by the uncertainty in the fraction of \pt carried by the neutrinos produced in semileptonic decays of heavy-quark hadrons, which contributes $0.32\%$ compared to $0.21\%$ from all the other sources combined.

The corresponding result, as derived from \HERWIGpp~2.3 tune EE3C, is
\begin{eqnarray}
C_\text{corr} = 1.005 \pm 0.006 \stat \pm 0.004 \syst,
\end{eqnarray}
and has a systematic uncertainty comparable with the one obtained for \PYTHIA6.4. The results of $C_\text{corr}$ for \PYTHIA6.4 and \HERWIGpp~2.3 differ by 0.5\%, which is consistent with the expected flavor differences shown in Fig.~\ref{fig:flavor}.

\begin{figure}[htbp!]
\centering
\includegraphics[width=0.48\textwidth]{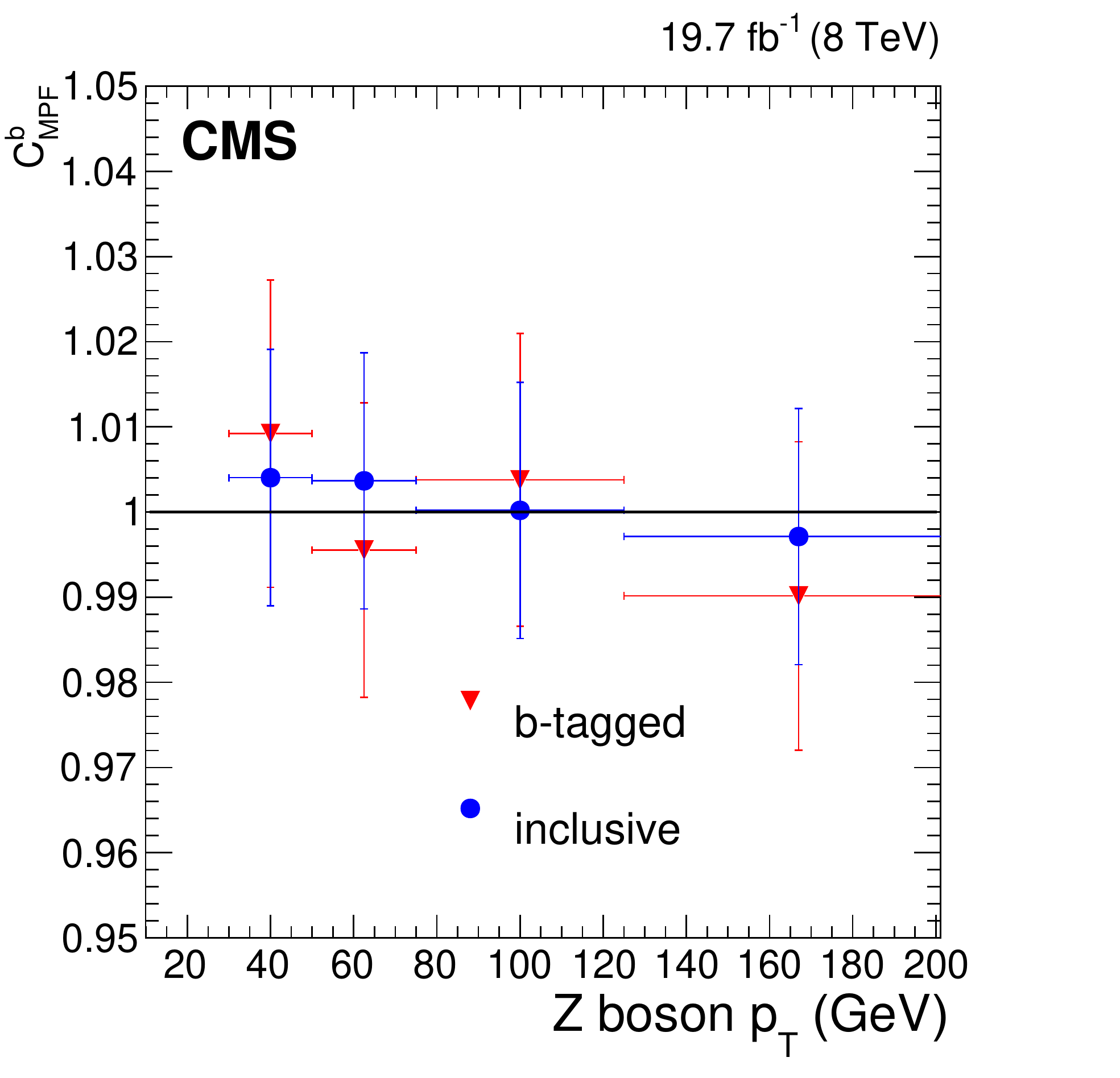}
\includegraphics[width=0.48\textwidth]{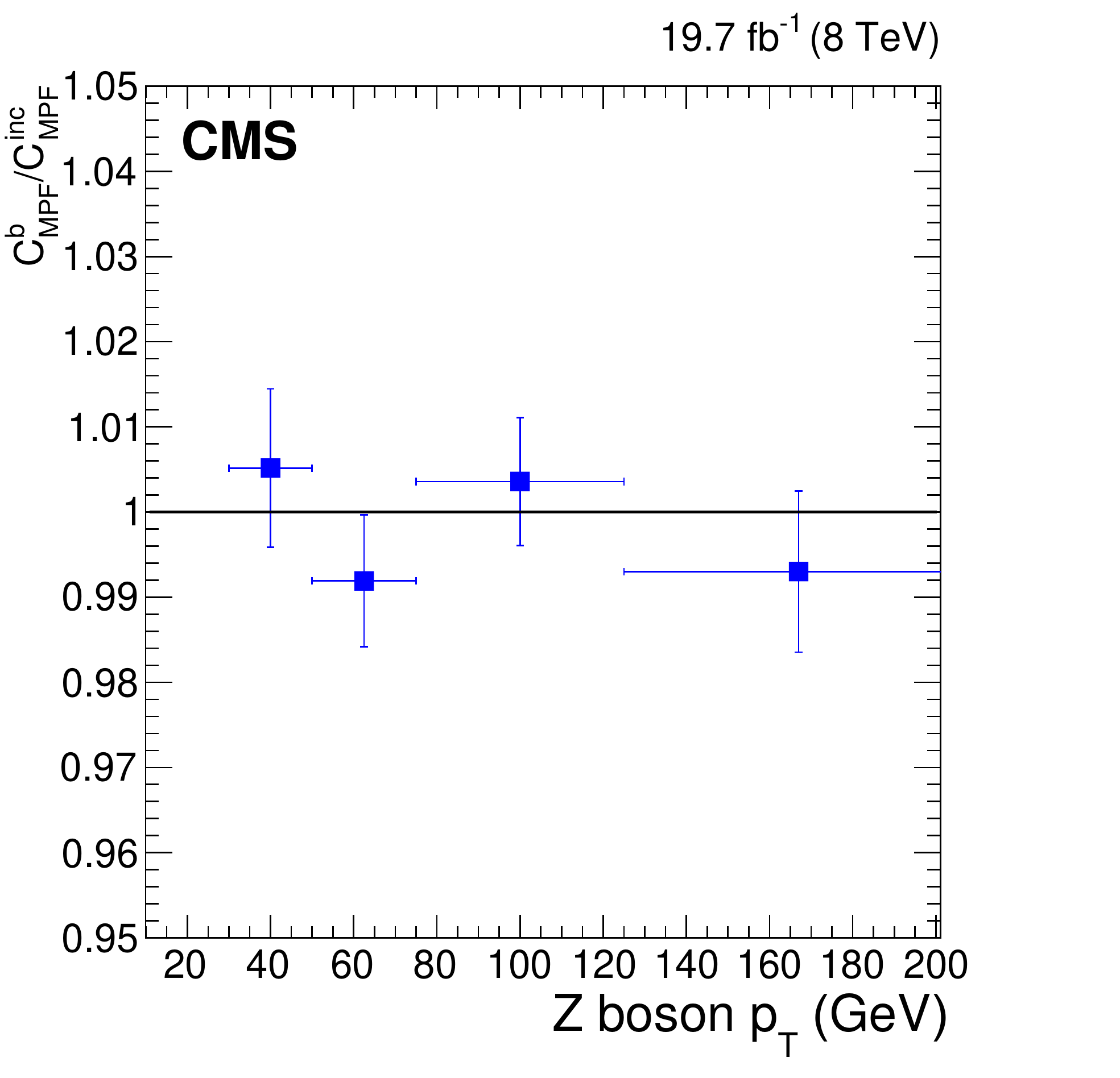}
\caption{\label{fig:zbpt}
Residual correction factors (calculated as the ratio of the MC and data MPF response) as a function of $\PZ$ boson \pt, for $\PZ$+$\PQb$-jet and $\PZ$+jet events with $\alpha<0.3$ (left), and their ratio (right).
}
\end{figure}

To check for a possible \pt-dependence, the residual $\PQb$-jet correction is shown as a function of \pt in Fig.~\ref{fig:zbpt}, where the uncertainties are statistical and systematic. This figure shows the MPF results with the fixed $\alpha<0.3$ requirement for the $\PZ$+$\PQb$-jet and inclusive $\PZ$+jet events separately (left) and for their ratio (right), where most systematic uncertainties cancel out. As a result, while the uncertainties in Fig.~\ref{fig:zbpt} (left) are dominated by the systematic contributions, the uncertainties in Fig.~\ref{fig:zbpt} (right) are mostly statistical. The separate results are not corrected for the \pt-dependence of the absolute JES, which cancels in the ratio. The ratio is compatible with the assumption of no \pt-dependence, although it is limited by statistical uncertainties at the expected level of the systematic flavor uncertainties.

\section{Jet \texorpdfstring{\pt}{pT} resolution}
\label{sec:jer}

The jet \pt resolution is relatively poor compared to the resolution of many other physics objects (electrons, muons, photons), and the biases caused by jet resolution smearing can be important for steeply falling spectra and for resonance decays. In this Section we present a determination of particle-level JER from MC simulation and the results from data-based methods for extracting a data/MC scale factor. The particle-level JER is defined as the width (estimated with a Gaussian fit) of the distribution of $p_\text{T, reco}/p_\text{T, ptcl}$, where $p_\text{T, reco}$ and $p_\text{T, ptcl}$ are the transverse momenta of the reconstructed and corresponding particle-level jets. JEC are applied before deriving JER.

\subsection{Methods}
Our measurement of JER is an extension of the methods used for measuring JES, but instead of looking at the mean of the response distribution, we are interested in its width. In addition, we need to correct for effects that do not produce an overall shift in the mean, but can widen the distribution.

For $\gamma$+jet (and $\PZ$+jet) events we can expand the \pt balance as follows:
\begin{equation}\label{eq:ratioproduct}
\mathcal{B}=\frac{p_\text{T, jet}}{p_{\mathrm{T}, \gamma}} =
\frac{p_\text{T, jet}}{p_\text{T, jet ptcl}}
\frac{p_\text{T, jet ptcl}}{p_{\mathrm{T}, \gamma~\mathrm{ptcl}}}
\frac{p_{\mathrm{T}, \gamma~\mathrm{ptcl}}}{p_{\mathrm{T}, \gamma}},
\end{equation}
where $p_\text{T, jet}$ and $p_{\mathrm{T}, \gamma}$ are the reconstructed jet and photon transverse momenta, and $p_\text{T, jet ptcl}$ and $p_{\mathrm{T}, \gamma~\mathrm{ ptcl}}$ the corresponding transverse momenta at particle level.
For a sum of independent and identically distributed (IID) random variables, the quadratic summation of widths applies. For the product of Eq.~(\ref{eq:ratioproduct}) we can take logarithms and apply $\log(1+X)\approx X$ for $X \ll 1$, which means that the normally distributed variables are also log-normally distributed with the same width $\sigma$. In the following, the symbols $\oplus$ and $\ominus$ indicate quadratic sum and subtraction respectively. For the widths $\sigma$ we then have
\begin{eqnarray}\label{eq:rawjer}
\sigma\left(\frac{p_\text{T, jet}}{p_{\mathrm{T}, \gamma}}\right) &=&
\sigma\left(\frac{p_\text{T, jet}}{p_\text{T, jet ptcl}}\right)\oplus
\sigma\left(\frac{p_\text{T, jet ptcl}}{p_{\mathrm{T}, \gamma~\mathrm{ptcl}}}\right)\oplus
\sigma\left(\frac{p_{\mathrm{T}, \gamma~\mathrm{ptcl}}}{p_{\mathrm{T}, \gamma}}\right),\quad{\ie}\\
\sigma_{\mathcal{B}} &=& \frac{\sigma_{\pt}}{\pt}\oplus
\sigma_\mathrm{UE+OOC+ISR+FSR} \oplus \sigma_{\gamma}. \label{eq:rawjer2}
\end{eqnarray}

In Equation (\ref{eq:rawjer2}), all widths, apart from $\sigma_{\pt}$, are relative quantities.
The first part on the right-hand side is identified with the particle-level JER, $\sigma_\mathrm{JER} = \sigma_{\pt}/\pt$. The second term is the combined effect of UE, OOC, ISR+FSR, and presence of neutrinos.
The final term is
due to the photon \pt resolution and FSR, $\sigma_\gamma$. After extrapolating the secondary jet activity to zero, or equivalently correcting the measured resolution with the effective
$k_\text{rad}=\sigma_{\mathcal{B}}(\alpha\to 0) / \sigma_{\mathcal{B}}$,
the effects of ISR and FSR become negligible. In this case the second term reduces to the effects from UE and OOC alone, and is referred to as the particle level imbalance (PLI), $\sigma_\mathrm{PLI}$ :
\begin{equation}\label{eq:rawjerpli}
\sigma_{\mathcal{B}} k_\text{rad}= \sigma_\mathrm{JER} \oplus
\sigma_\mathrm{PLI}\oplus \sigma_{\gamma}.
\end{equation}

Equation~(\ref{eq:rawjerpli}) can be rearranged as follows:
\begin{equation}
\sigma_\mathrm{JER} = \sigma_{\mathcal{B}} k_\text{rad}\ominus \sigma_\mathrm{PLI} \ominus \sigma_\gamma.
\end{equation}

$k_\text{rad}$ is determined from data by fitting the resolution as a function of secondary jet activity $\alpha=p_\text{T, 2nd jet}/p_{\mathrm{T,}\gamma}$, while $\sigma_\mathrm{ PLI}$ and $\sigma_\gamma$ are taken from MC simulation, after extrapolating to $\alpha\to 0$ for consistency. Since $k_\text{rad} \sigma_\gamma$ is negligible compared to other effects, it is
implicitly absorbed in the $\sigma_\mathrm{PLI}$ factor.

The quadratic addition and subtraction of widths is strictly correct only for RMS, or for resolutions that are approximately Gaussian. In practical cases, the dominant terms ($\sigma_{\pt}$, $\sigma_{\mathcal{B}} k_\text{rad}$) can be determined consistently from the RMS or a fit to the Gaussian core of the distribution, while the smaller contributions ($\sigma_\mathrm{PLI}$, $\sigma_\gamma$) should typically use RMS. This is because the small contributions (with sufficiently steeply falling power-law tails) get folded into the Gaussian core of the distribution, as predicted, \eg, by the Central Limit Theorem (CLT), which states that the sum of multiple IID random variables tends to a Gaussian distribution, provided their power-law tails fall off sufficiently steeply ($x^{-m}$, with $m>3$). For a variable that can be thought of as a multiplicative product of many independent random variables, the CLT applies in the logarithmic domain.

For dijet events the factorization is similar, except that JER now appears once for each jet, and PLI is derived from the asymmetry between two particle jets and includes the UE and OOC contributions twice. The dijet asymmetry \cite{JEC_JINST} is defined as
\begin{eqnarray}
\mathcal{A} &=& \frac{p_\text{T, 1st jet}-p_\text{T, 2nd jet}}{p_\text{T, 1st jet}+p_\text{T, 2nd jet}}\label{eq:dijetasym_def},
\end{eqnarray}
and its resolution $\sigma_{\mathcal{A}}$ is given by
\begin{equation}
  \sigma_{\mathcal{A}} k_\text{rad}=
\frac{\sigma_\mathrm{JER,probe}}{2}
\oplus\frac{\sigma_\mathrm{JER,tag}}{2}
\oplus \sigma_\mathrm{PLI,dijet} , \label{eq:dijet_reso}
\end{equation}
where $\sigma_\mathrm{PLI,dijet}$ is the resolution of the asymmetry variable, built with the momenta of particle-level jets, and extrapolated to zero additional jet activity $\sigma_\mathrm{PLI,dijet}=\sigma_{\mathcal{A}}^{ \rm ptcl} k_\text{rad}^\mathrm{ptcl}$. It is related to the single jets PLI resolutions through $2\sigma_\mathrm{ PLI,dijet}=\sigma_\mathrm{PLI,tag} \oplus \sigma_\mathrm{PLI,probe}$.

For the special case where both jets are in the same region and share the same JER ($\sigma_\mathrm{JER,probe}=\sigma_\mathrm{JER,tag}=\sigma_\mathrm{JER}$), we obtain:
\begin{equation}\label{eq:jersymm}
\sigma_\mathrm{JER} = \sqrt{2} (\sigma_{\mathcal{A}} k_\text{rad}\ominus \sigma_\mathrm{PLI,dijet}) .
\end{equation}

For the case of one central and one forward jet, we can solve for the forward JER by subtracting the central JER determined from Eq.~(\ref{eq:jersymm}):
\begin{equation}
\sigma_\mathrm{JER,forward} = 2\sigma_{\mathcal{A}} k_\text{rad}\ominus 2\sigma_\mathrm{PLI,central-forward} \ominus \sigma_\mathrm{JER,central}.
\label{eq:jer_fwd_ctr}
\end{equation}

In this case the $\sigma_\mathrm{PLI,central-forward}$ is $\sigma_\mathrm{PLI,dijet}$, determined consistently for the same combination of central and forward jets.

\subsection{Simulated particle-level resolution}

\begin{figure}[htbp!]
\centering
\includegraphics[width=0.49\textwidth]{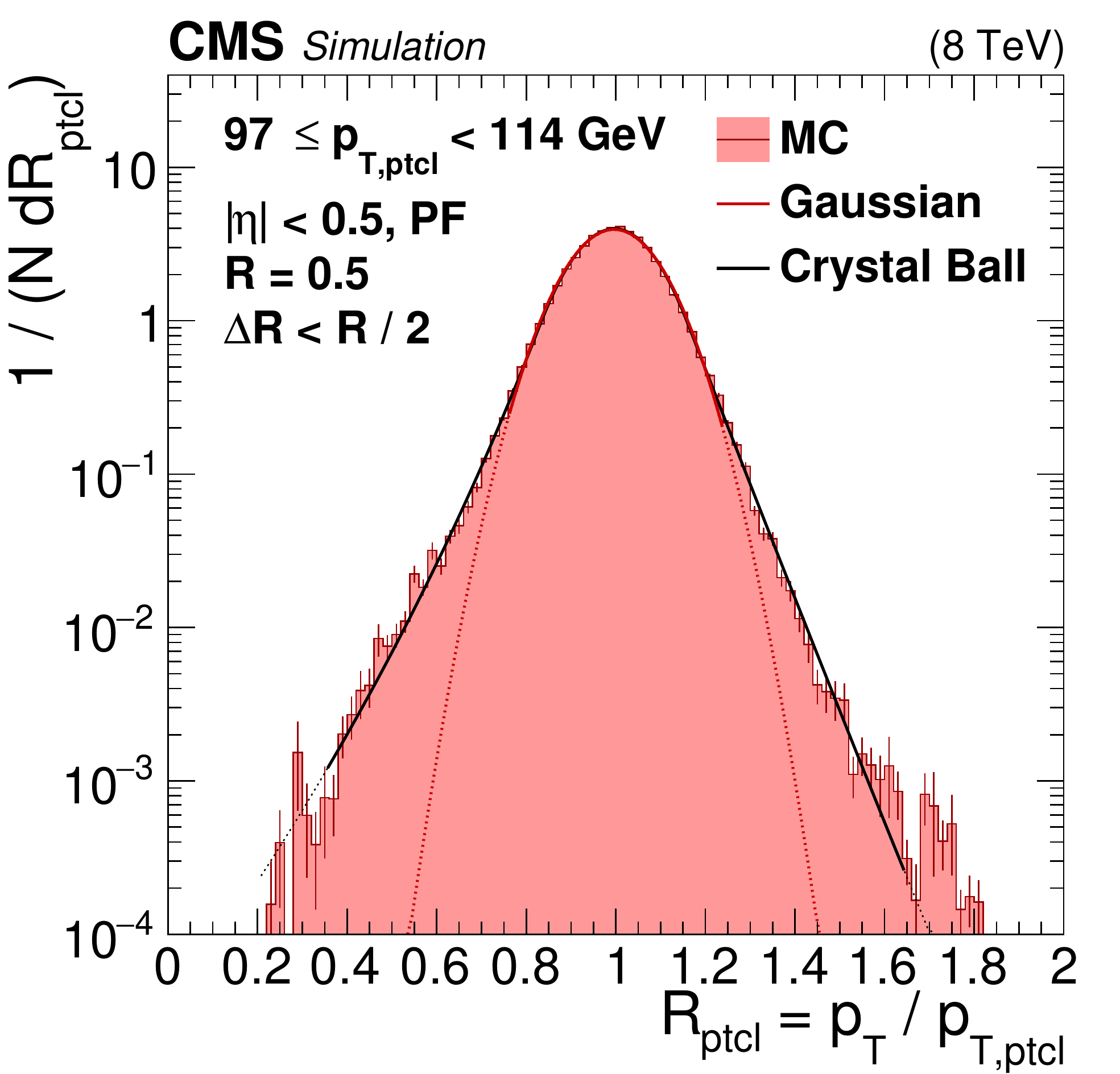}
\includegraphics[width=0.49\textwidth]{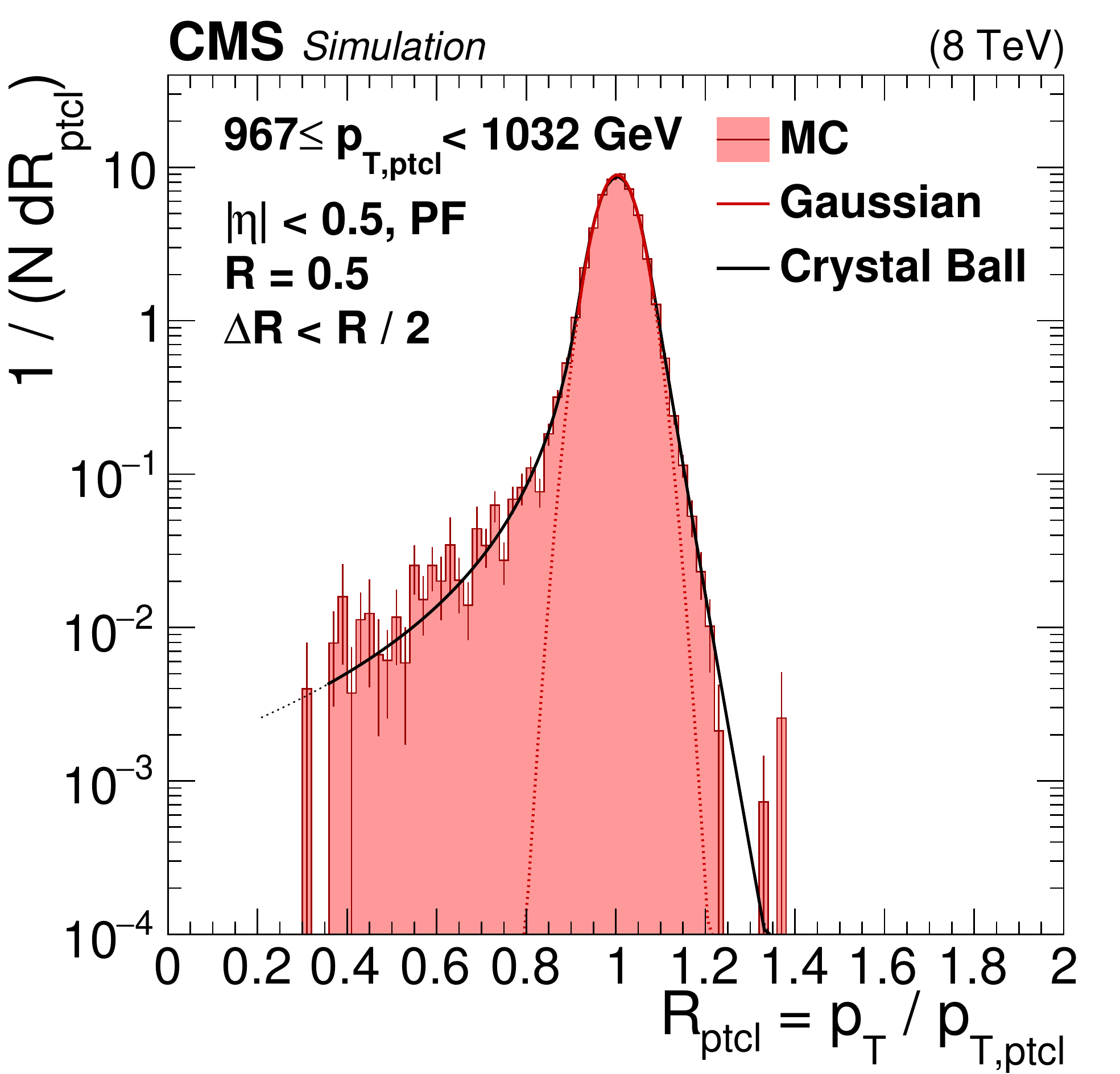}
\caption{\label{fig:jertails}
Jet \pt resolution distributions in the barrel for two bins of jet \pt. $\Delta R$ indicates the distance parameter value used for matching reconstructed jets to the corresponding particle-level jets. The nongaussian tails due to inactive areas of the ECAL and HCAL punchthrough become more visible for narrow high-$\pt$ jets with small core resolution. The Gaussian core resolution is fit to within ${\pm}2\sigma$ (solid line) and its extrapolation is indicated with a dotted line.
The tails are well modeled by a double-sided Crystal Ball function.
}
\end{figure}

The jet \pt resolution is reasonably Gaussian, although some nongaussian low-response tails are present, \eg, due to rare detector effects such as inactive areas of the ECAL and to high-$\pt$ particles punching through the HCAL. At low \pt symmetric tails appear due to combinations where two generator jets produce a single reconstructed jet, or vice versa. Such effects are typically well-modeled by a double-sided Crystal Ball function \cite{crystalball}, as seen in Fig.~\ref{fig:jertails}.
Low tails in response measurements are also commonly produced by neutrinos from semileptonic decays of heavy-flavor hadrons. This does not apply to particle-level resolutions, because CMS particle jets exclude neutrinos, but this does impact the dijet balance method used to measure JER in data.

We define the particle-level JER in simulation as the $\sigma$ of a Gaussian fit to the $p_\text{T, reco}/p_\text{T, ptcl}$ distribution in the range $[m-2\sigma,m+2\sigma]$, where $p_\text{T, reco}$ and $p_\text{T, ptcl}$ are the reconstructed jet \pt and generated particle-level jet \pt, respectively, and $m$ and $\sigma$ are the mean and width of the Gaussian fit, determined with an iterative procedure. To maximize matching efficiency while still ensuring a unique match, the reconstructed and the generated jets are required to be within $\Delta R<R/2$ of each other, with $R$ being the jet distance parameter.

The nongaussian tails increase the RMS of the distribution,
and the differences affect the data-based dijet asymmetry, where two JER distributions are folded together with other (non) Gaussian distributions. The dijet asymmetry is effective in symmetrizing the tails, and according to the CLT the folded distribution will asymptotically approach a Gaussian distribution.
The treatment of these effects will be discussed in more detail in the next section.

\begin{figure}[htbp!]
\centering
\includegraphics[width=0.49\textwidth]{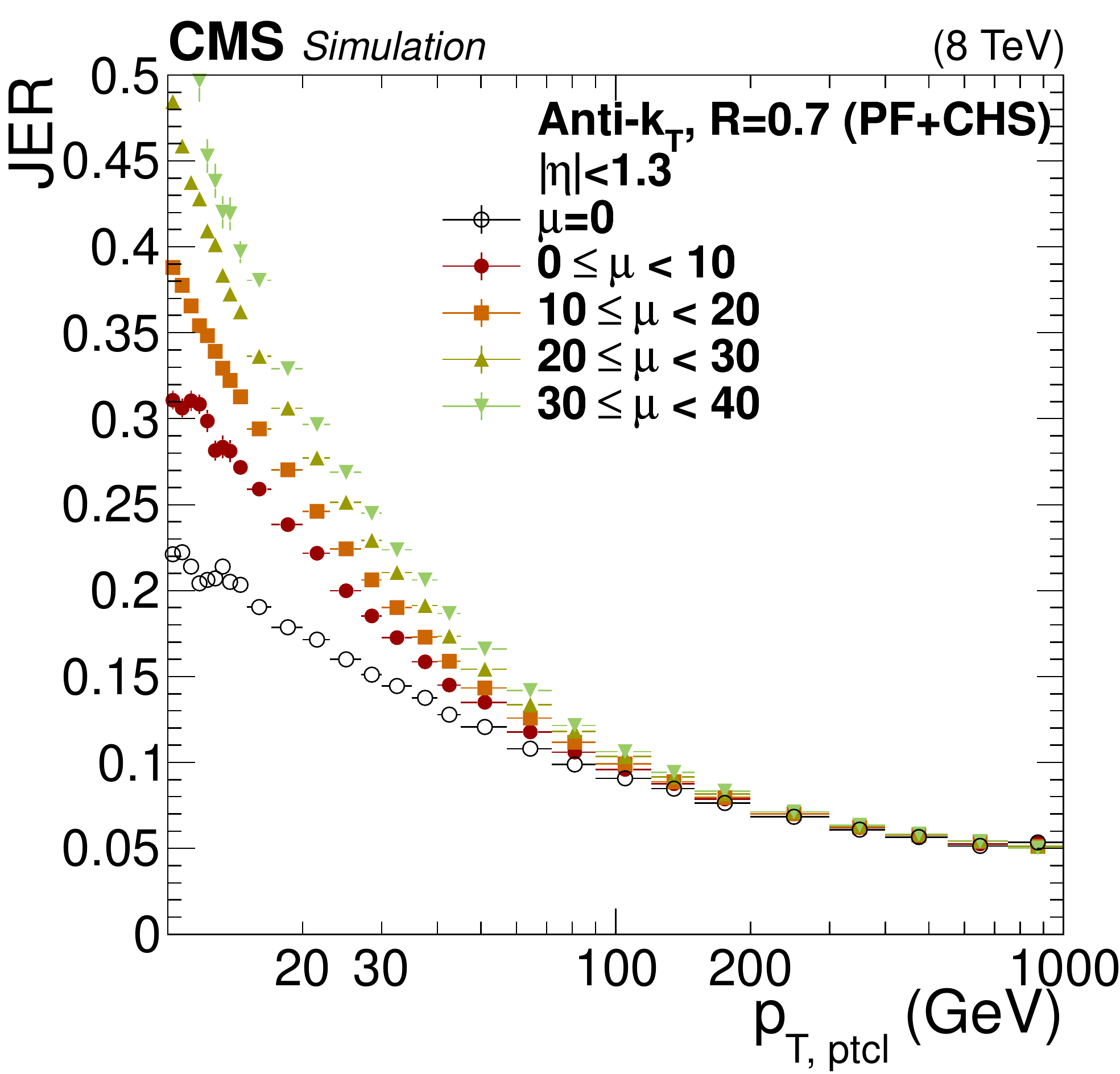}
\includegraphics[width=0.49\textwidth]{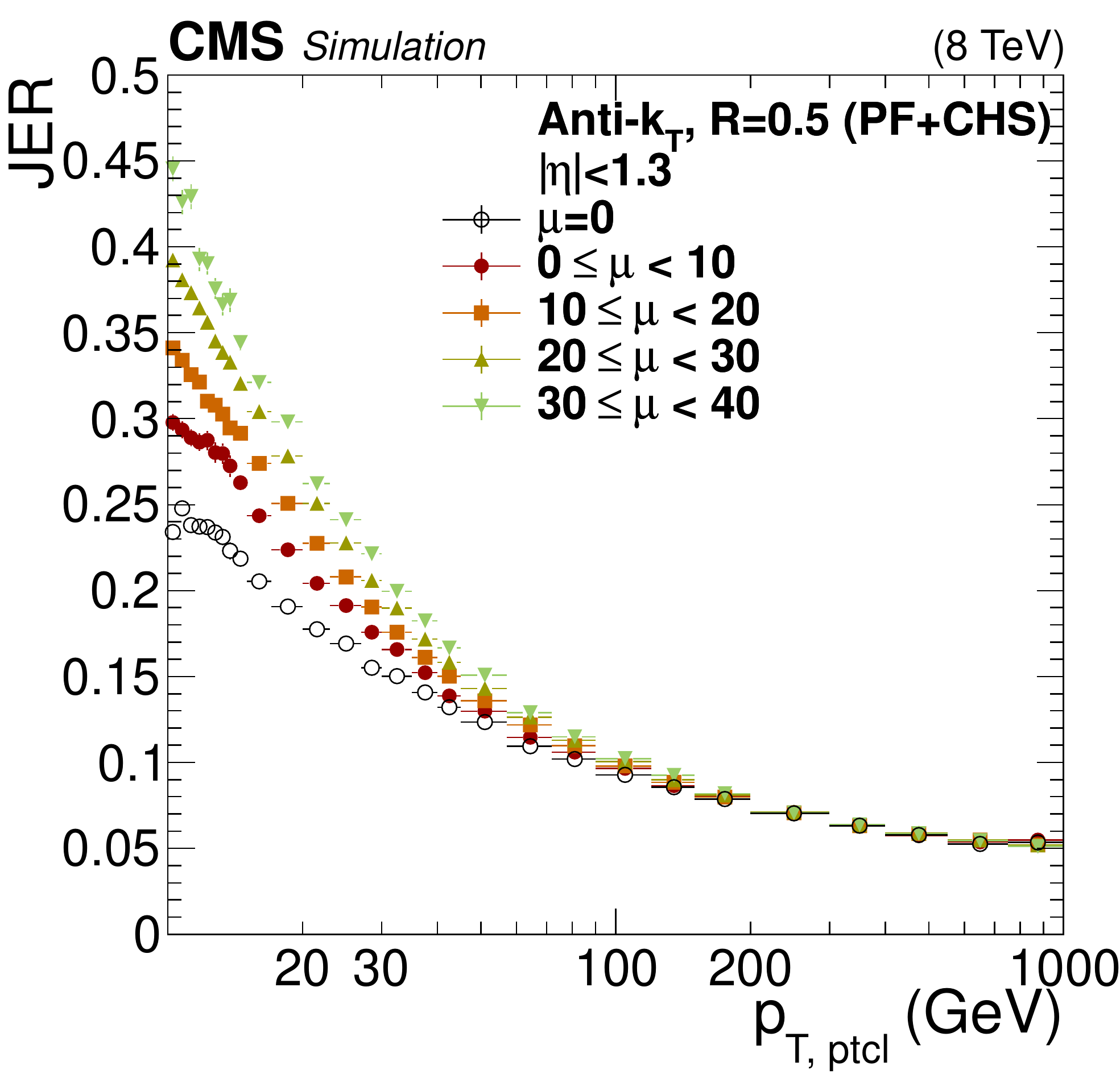}
\caption{\label{fig:mcjer}
JER versus \pt in the barrel for varying levels of pileup $\mu$.
The results are shown separately for PF+CHS jets with size $R=0.7$ (left), and for PF+CHS jets with size $R=0.5$ (right).
}
\end{figure}

The particle-level JER in simulation with a pileup profile matched to 2012 (8\TeV) data is presented in Fig.~\ref{fig:mcjer} in bins of true number of PU interactions $\mu$.
The particle-level JER in simulation is parameterized with the ``$NSC$'' fit for calorimeter resolutions, where $N$ is for noise (and pileup), $S$ is for stochastic fluctuations that scale as $1/\sqrt{E}$, and $C$ is a constant term with no \pt scaling (\eg, intercalibration):
\begin{equation}
\frac{\sigma_{\pt}}{\pt} = \sqrt{\frac{\sgn(N)N^2}{\pt^2} + \frac{S^2}{\pt} + C^2}.
\end{equation}

As shown already in \cite{JEC_JINST}, for PF jets the possibility of having a negative $N^2$ term improves the description of the jet resolutions at low PU.
As visible in Fig.~\ref{fig:mcjer}, in the absence of pileup larger jets have better resolution due to several effects. These include better jet angular resolution (relevant for $p_\mathrm{T}<30$\GeV) and a smaller constant term from intercalibration, due to the jet energy being spread over more particles. These small gains are quickly negated in the presence of pileup, to which larger jets are more sensitive.

\begin{figure}[htbp!]
\centering
\includegraphics[width=0.68\textwidth]{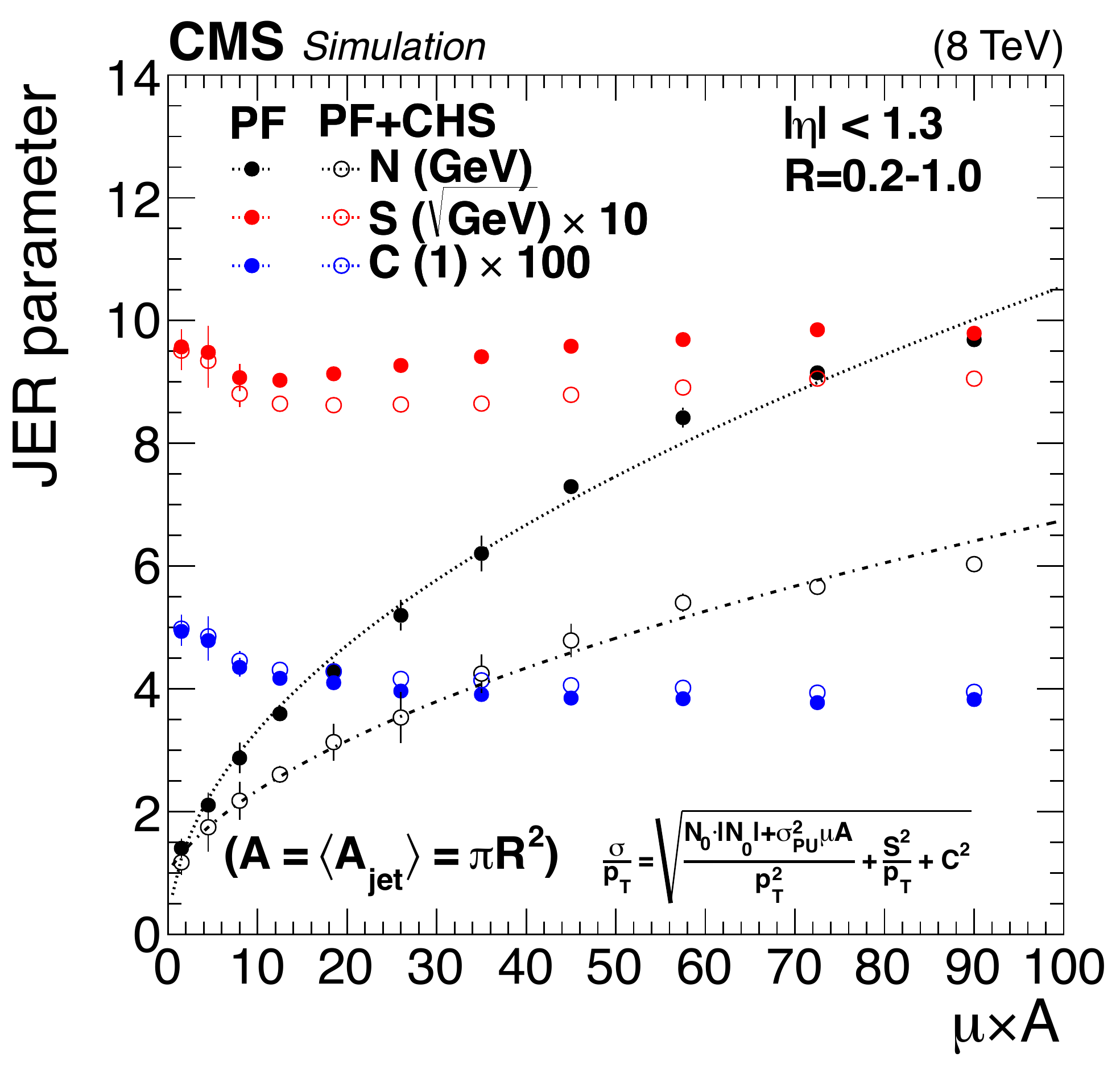}
\caption{\label{fig:jercone}
JER parameters ($N$, $S$, $C$; see text) fitted in bins of $\mu$ for various values of the distance parameter $R$ ($R=0.2$--1), as a function of their average value of pileup times jet area ($\mu  A$). The results are compared between PF (solid symbols) and PF+CHS (open symbols).
The dotted and dash-dotted curves represent the fit for PF and PF+CHS jets, respectively.
}
\end{figure}

The noise term is very sensitive to the PU in jets, with each additional PU interaction contributing about 1\GeV of smearing in quadrature for $R=0.5$ jets. The PU offset increases approximately linearly with
number of collisions and jet area,
$\mu A$, so that the noise from pileup is proportional to $\sqrt{\mu A}$. In contrast, the stochastic and constant terms are stable with respect to pileup, as shown in Fig.~\ref{fig:jercone}.

\begin{figure}[htbp!]
\centering
\includegraphics[width=0.49\textwidth]{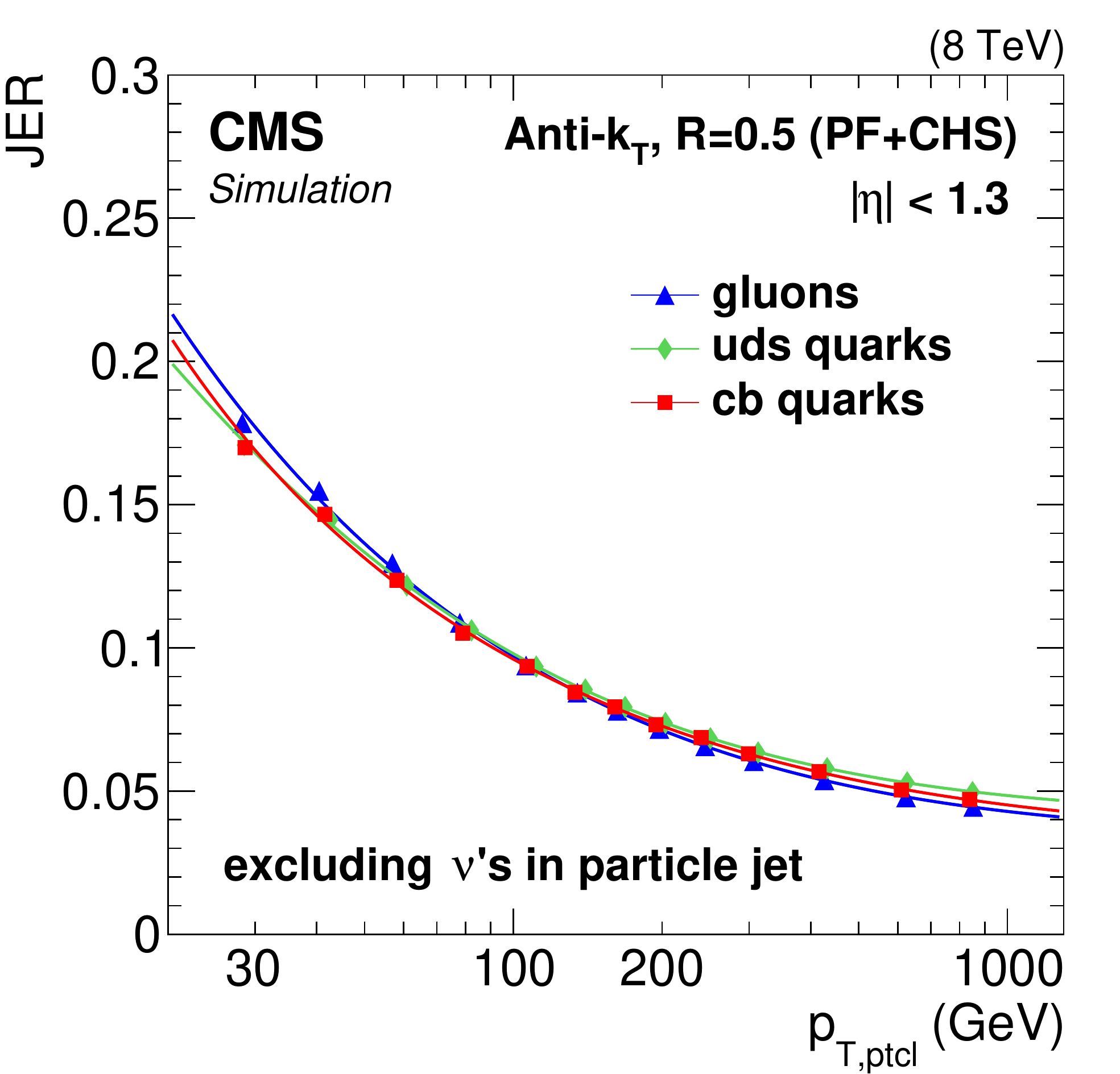}
\includegraphics[width=0.49\textwidth]{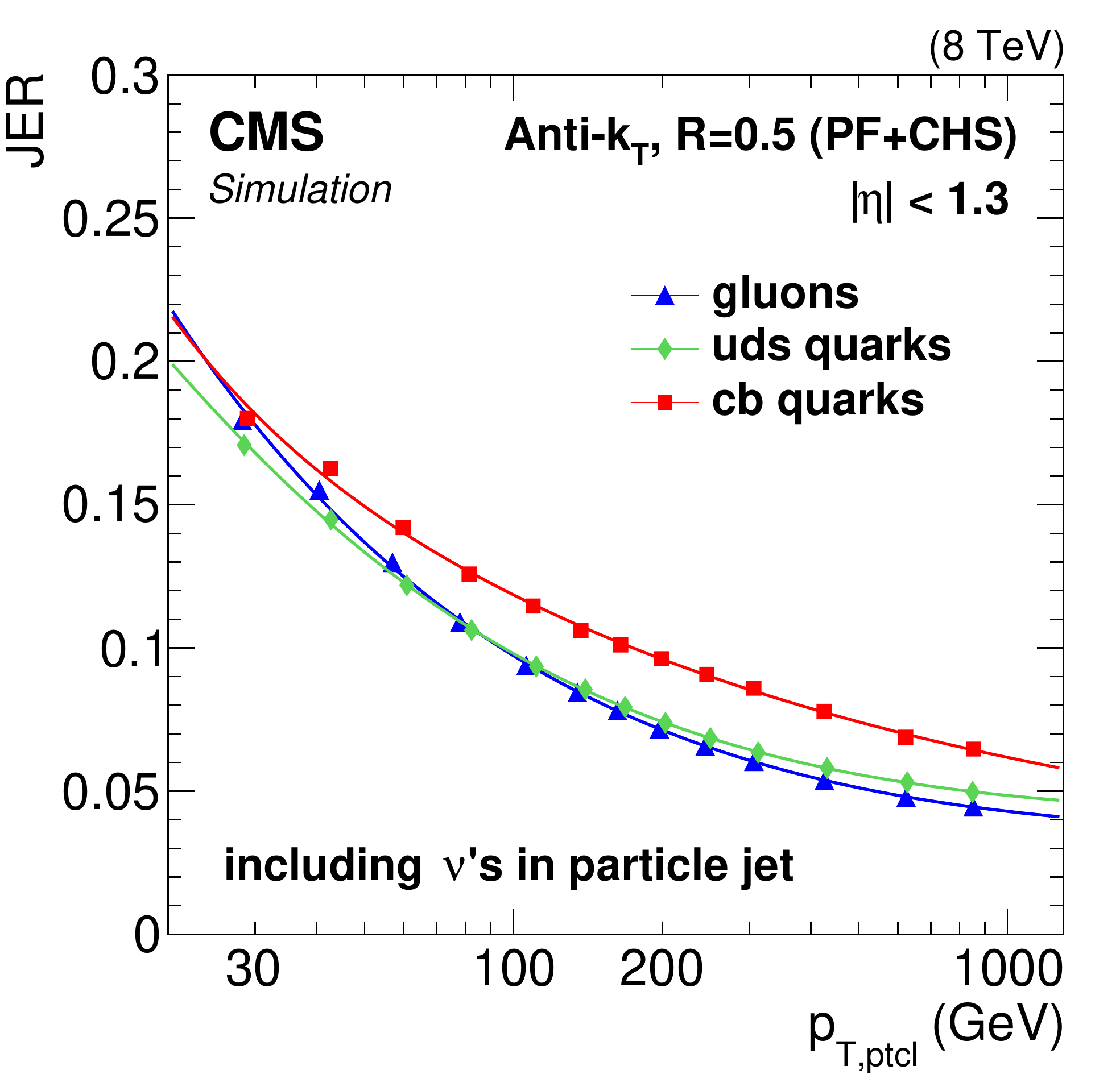}
\caption{\label{fig:jerflavor}
True JER in simulation for different jet flavors in the $\gamma$+jet sample, for jets with $\abs{\eta}<0.5$. The distributions are shown for particle-level jets with no neutrinos (left), and with neutrinos exceptionally included (right) to demonstrate the large fluctuations this induces for $\PQc$ and $\PQb$ jets.
}
\end{figure}

The JER also depends to some extent on the jet flavor. Gluon jets are wider than quark jets, and therefore less sensitive to local $\eta$--$\phi$ intercalibration of the detector. This is manifested as a smaller constant term and better JER at high \pt, as seen in Fig.~\ref{fig:jerflavor} (left), where the constant term is $0.0338\pm 0.0002$ for gluons, and $0.0406\pm 0.0002$ for quarks. The intrinsic JER of $\PQc$ and $\PQb$ jets is similar to uds jets when excluding neutrinos at particle level. However, when neutrinos are included, as in Fig.~\ref{fig:jerflavor} (right), the neutrinos produced in semileptonic decays of heavy-quark hadrons produce long response tails for the data-based measurements and cause substantial smearing of JER.

\subsection{Dijet asymmetry}

The jet \pt resolution in data is measured with the dijet asymmetry method \cite{JEC_JINST} versus $p_\mathrm{T, ave}$:
\begin{equation}
\mathcal{A}= \frac{p_\text{T, 1st jet}-p_\text{T, 2nd jet}}{p_\text{T, 1st jet}+p_\text{T, 2nd jet}} = \frac{p_\text{T, 1st jet}-p_\text{T, 2nd jet}}{2p_\text{T, ave}},
\label{eq:asymm}
\end{equation}
where the width of the asymmetry distribution $\sigma_\mathcal{A}$ is related to the single jet resolutions as explained in Equations~(\ref{eq:dijetasym_def})--(\ref{eq:jer_fwd_ctr}).

\begin{figure}[htbp!]
\centering
\includegraphics[width=0.49\textwidth]{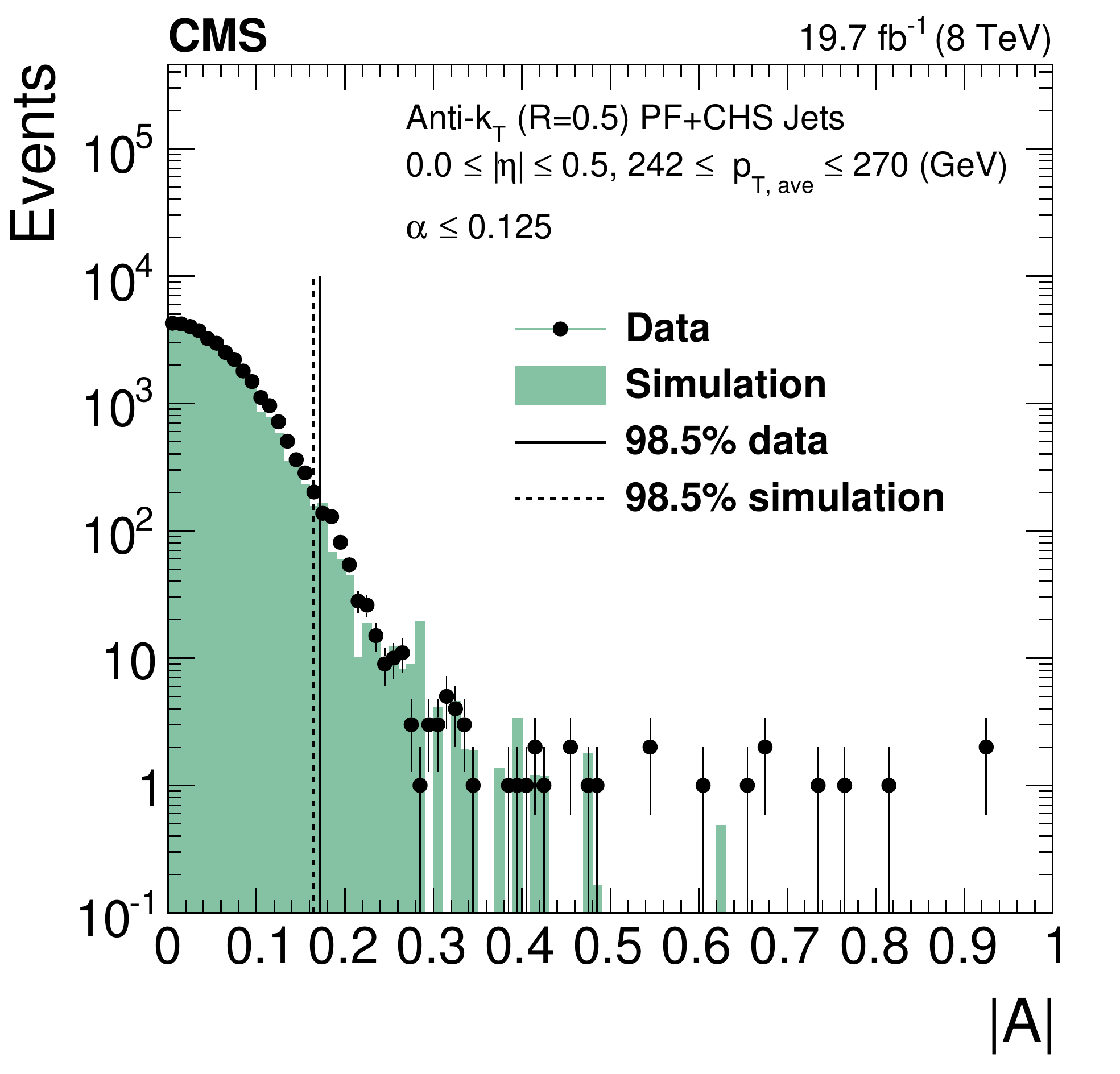}
\includegraphics[width=0.49\textwidth]{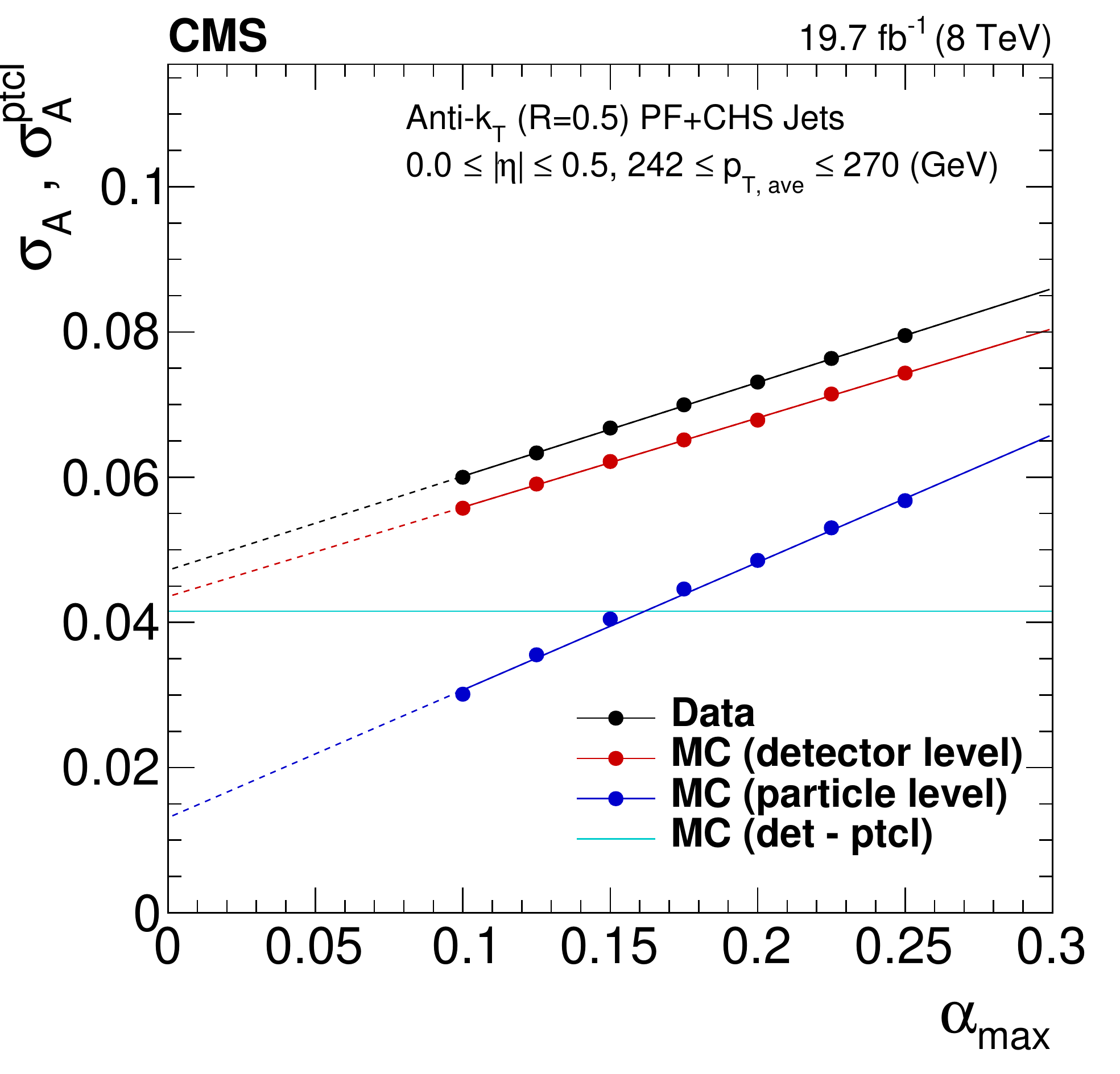}
\caption{\label{fig:asymm}
Left: Asymmetry distribution, Eq.~(\ref{eq:asymm}), for data and simulation for jets with $\pt \approx 250\GeV$ and $\abs{\eta}<0.5$.
Right: Asymmetry measured for various thresholds $\alpha_\text{max}$, extrapolated to zero additional jet activity, for jets with $\pt \approx 250\GeV$ and $\abs{\eta}<0.5$ in data and MC simulation at the detector- and particle-level. The light horizontal line indicates the average particle-level resolution obtained as the difference in quadrature of MC simulation reconstructed asymmetry and particle-level imbalance, extrapolated to zero additional jet activity.
}
\end{figure}

The measured asymmetry distributions are Gaussian-like at low \pt, but some tails are evident at high \pt, as seen in Fig.~\ref{fig:asymm} (left). These tails are reasonably well modeled by simulation. The particle-level JER in simulation is defined as a Gaussian fit to the core of the distribution within 2$\sigma$ so we limit the impact of these tails and of outlier events by using a truncated RMS with 98.5\% of the events in the core of the distribution for data and simulation. The truncation value is chosen to ensure that the core of the distribution is reasonably well described by a Gaussian.
The resolutions are then extrapolated to zero secondary jet activity for data, reconstructed MC simulation, and generated MC simulation, as shown in Fig.~\ref{fig:asymm} (right).

\begin{figure}[htbp!]
\centering
\includegraphics[width=0.55\textwidth]{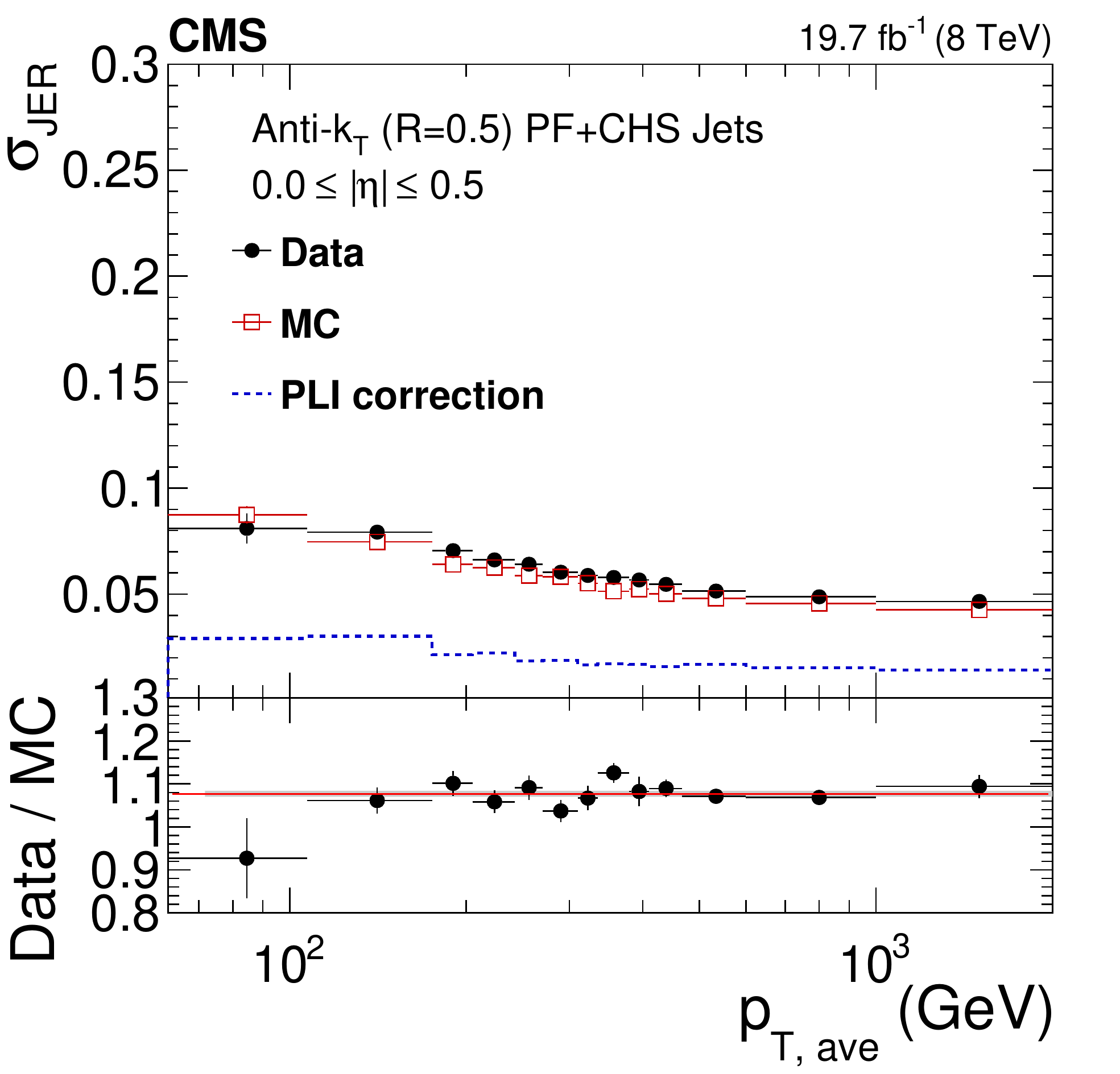}
\caption{\label{fig:datamcjer}
Extrapolated JER as a function of jet \pt obtained with the asymmetry method on dijet events for data (solid circles), reconstructed MC simulation (open squares), and particle-level simulation with PLI (dashed line). The bottom plot shows the ratio of data over MC.
}
\end{figure}

\begin{figure}[htbp!]
\centering
\includegraphics[width=0.55\textwidth]{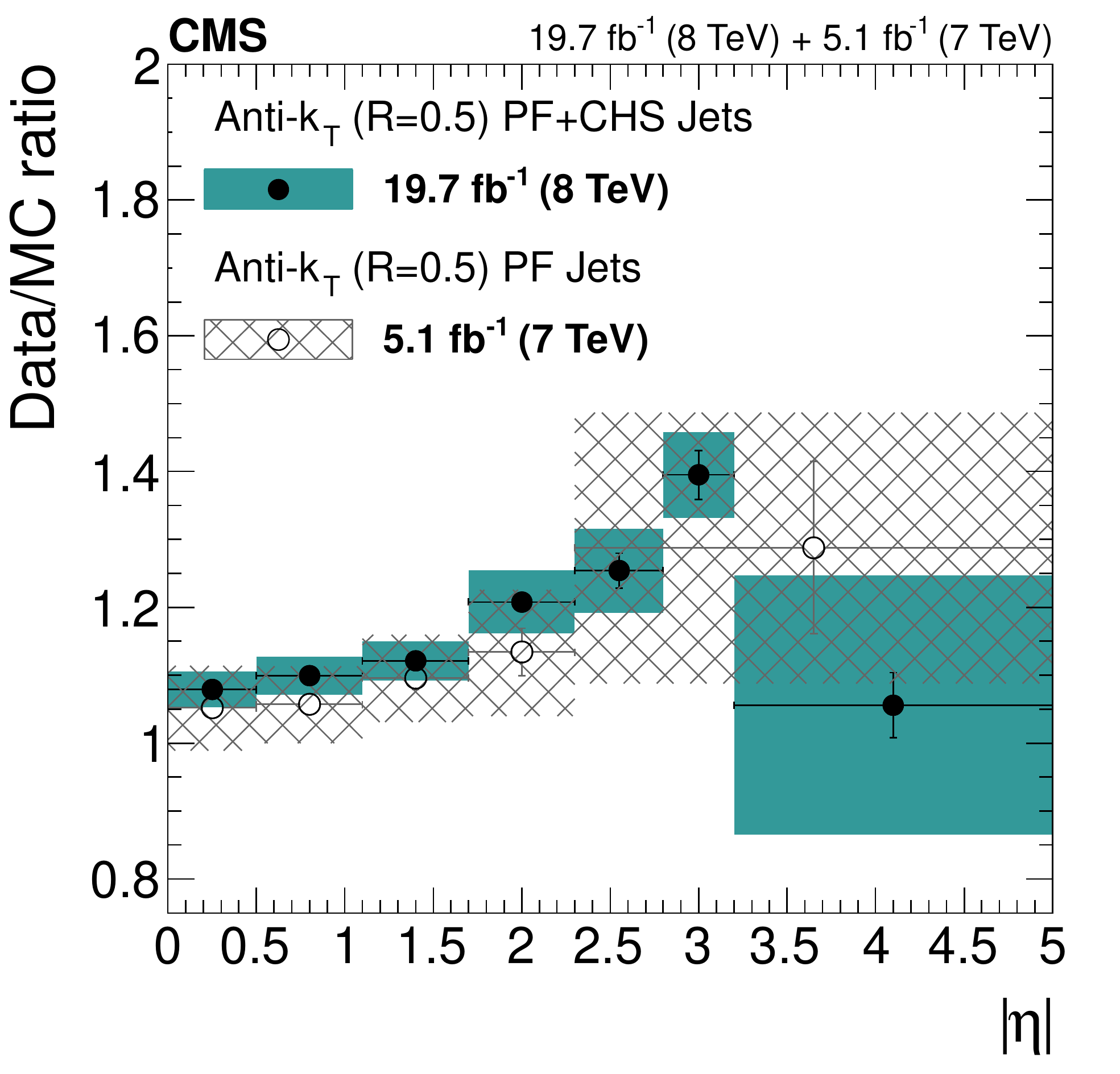}
\caption{\label{fig:kjervseta}
Jet energy resolution data/MC scale factor versus $\abs{\eta}$ for dijet data collected at 8\TeV (closed circles, solid area) compared to results at 7\TeV (open circles, dashed area).
}
\end{figure}

The extrapolated asymmetry is finally corrected for the residual PLI effects, which are significantly smaller than the asymmetry, as shown in Fig.~\ref{fig:datamcjer}. This final stage does not significantly affect the data/MC ratio, as is also shown in Fig.~\ref{fig:datamcjer}, because the same PLI correction is applied to both data and MC simulation. The main effect of the correction is a slight increase in the data/MC ratio.
The data/MC ratio is well-modeled by a constant fit in all the rapidity regions. These constant data/MC scale factors are summarized in Fig.~\ref{fig:kjervseta}. The fit is also repeated assuming different scale factors for the $N$, $S$ ($k_{NS}$) and $C$ ($k_{C}$) terms, which gives reasonable results although the statistical uncertainties are too big to distinguish the two cases. The difference between the measured scale factors $k_{NS}$ and $k_{C}$ and the central value obtained by the constant fit is considered as a systematic uncertainty.

\begin{figure}[htbp!]
\centering
\includegraphics[width=0.49\textwidth]{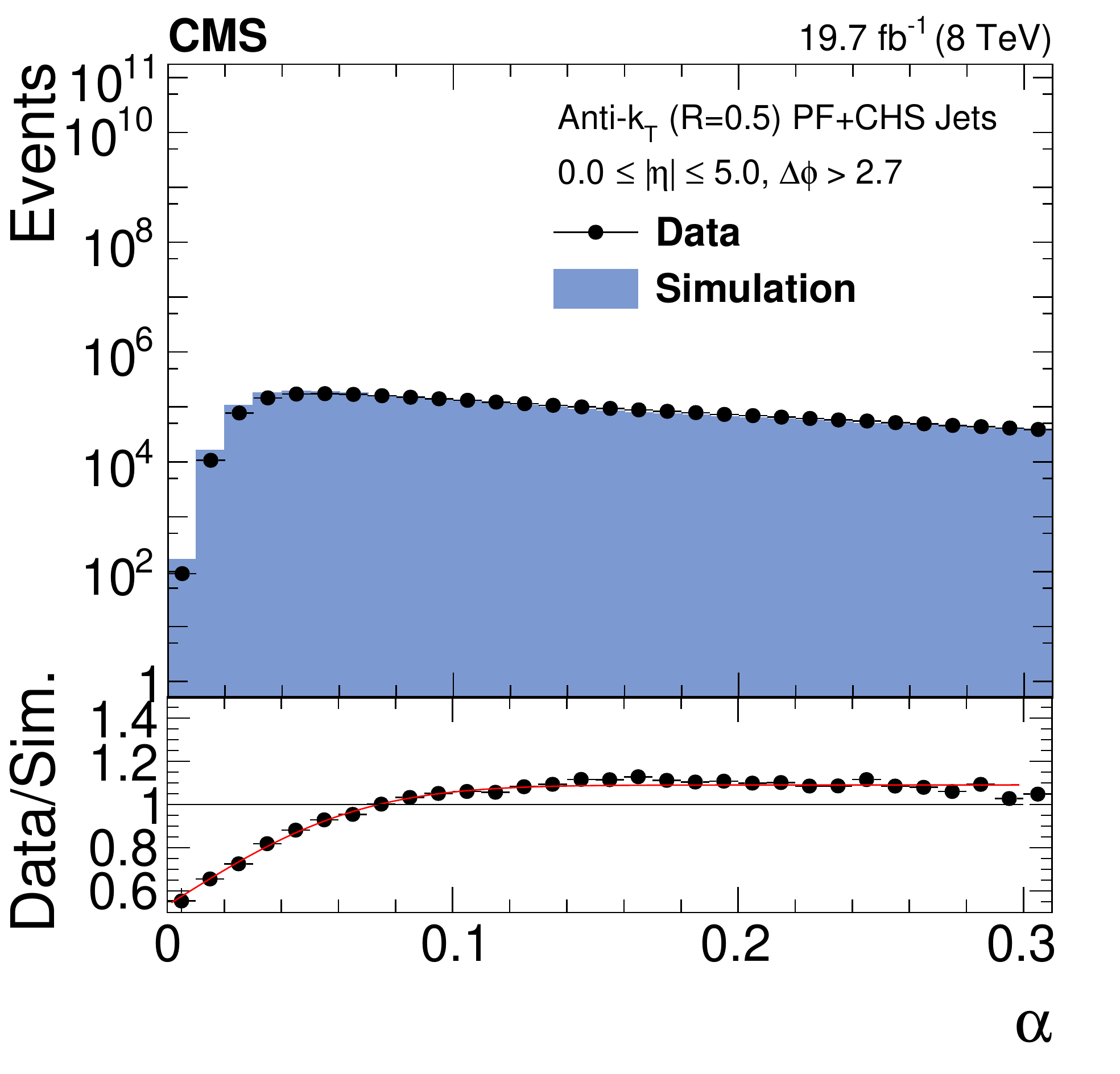}
\includegraphics[width=0.49\textwidth]{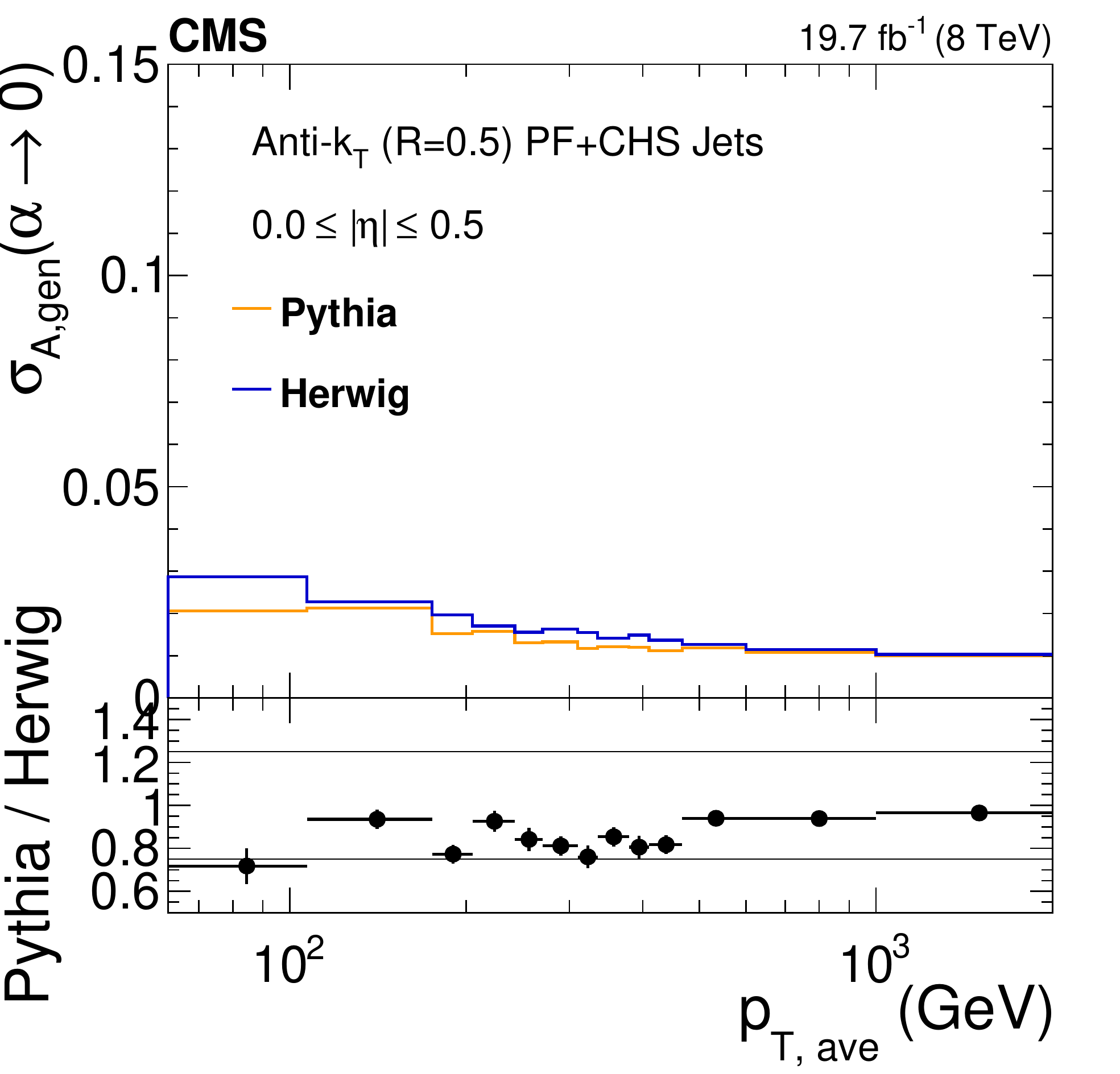}
\caption{\label{fig:alpha_herwig}
Left: The $\alpha$ distribution in data (circles) and simulation (histogram), with the function used for simulation reweighting overlaid on the ratio of data over simulation in the bottom plot.
Right: Comparison of particle-level imbalances $\sigma_{A,\rm gen}(\alpha\to 0)$ in \PYTHIA6.4 tune Z2* and \HERWIGpp~2.3 tune EE3C as a function of jet $p_\mathrm{T,ave}$. The bottom plot shows the ratio of {\PYTHIA} over {\HERWIG}.
}
\end{figure}

The systematic uncertainties in the data-based method come from three main sources: correction for ISR and FSR, particle-level imbalance, and nongaussian tails. The data/MC ratio is also affected by the uncertainties in PU reweighting and JEC.
The parameterization uncertainty comes mainly from the potential difference between the data/MC scale factors for the noise and stochastic terms $k_{NS}$ and the one for the constant terms $k_{C}$.
The systematic uncertainties for the data/MC ratio are evaluated as follows:
\begin{itemize}
\item \textbf{ISR+FSR correction}: The fit shown in Fig.~\ref{fig:asymm} (right) uses a linear extrapolation $\sigma = \sigma_0(1+k\alpha)$, which comes with an associated shape uncertainty beyond the first measured point. As a reminder, for dijet events, $\alpha=p_\text{T, 3rd jet}/p_\mathrm{T, ave}$.
Equation~(\ref{eq:asymm}) suggests that another good functional form would be $\sigma = \sigma_0 \oplus k'\alpha$, which is used as an alternative model to estimate the systematic uncertainty in the ISR+FSR correction.
\item \textbf{FSR+ISR radiation}: The $\alpha$ spectrum is not perfectly modeled in simulation, as shown in Fig.~\ref{fig:alpha_herwig} (left). This affects the extrapolation correction. The systematics are evaluated by reweighting the $\alpha$ spectrum in simulation to match the data, and repeating the analysis.
\item \textbf{Particle-level imbalance}: The PLI uncertainty is estimated by comparing the PLI obtained from two different MC simulations, \PYTHIA6.4 tune Z2* and \HERWIGpp~2.3 tune EE3C, as shown in Fig.~\ref{fig:alpha_herwig} (right). Based on these results we vary the PLI by $\pm 25$\% to estimate the impact of PLI uncertainty on JER.
\item \textbf{Non-Gaussian tails}: The uncertainty from nongaussian tails is estimated by changing the nominal 98.5\% truncation for RMS to 97\%.
\item \textbf{PU reweighting}: The PU reweighting uncertainty is estimated by changing the minimum bias pp cross section from the nominal 69.4 to 73.5\unit{mb} when generating the target PU profile for simulation.
\item \textbf{Jet energy scale}: The uncertainty arising from the knowledge of JES is evaluated by scaling up and down, according to the JEC uncertainty, all jet momenta in the simulation.
\item \textbf{Parameterization uncertainty}: The parameterization uncertainty is estimated by considering different scale factors for the noise term $N$ as well as the stochastic and constant terms $S$ and $C$. Because the available range of data is not sensitive to the noise term, the noise term scale factor is varied by 10\%.
\end{itemize}
The total uncertainty varies between 2--4\% in the tracker covered region $\abs{\eta}<2.3$, and increases up to 6\% in the endcaps and 20\% in the HF, where the uncertainty is driven by the large differences between the results based on \PYTHIA6.4 and on \HERWIGpp.

\subsection{The \texorpdfstring{$\gamma$}{gamma}+jet balance}

Complementary measurements of the jet \pt resolution in data are done with the $\gamma$+jet balancing versus $p_{\mathrm{T},\gamma}$:
\begin{equation}
\mathcal{B} = \frac{p_\mathrm{T,jet}}{p_{\mathrm{T},\gamma}}.
\end{equation}

The width of the balance distribution can be written as a convolution of the jet \pt resolution with additional smearing effects, as in Eq.~(\ref{eq:dijet_reso}) for the case of dijets:
\begin{equation}
\sigma_{\mathcal B} = \sigma \oplus \sigma_\mathrm{ISR+FSR} \oplus \sigma_\mathrm{PLI}.
\end{equation}

The JER is extracted from data and MC simulation that are binned in exclusive bins of $\alpha$.

\begin{figure}
\includegraphics[width=0.49\textwidth]{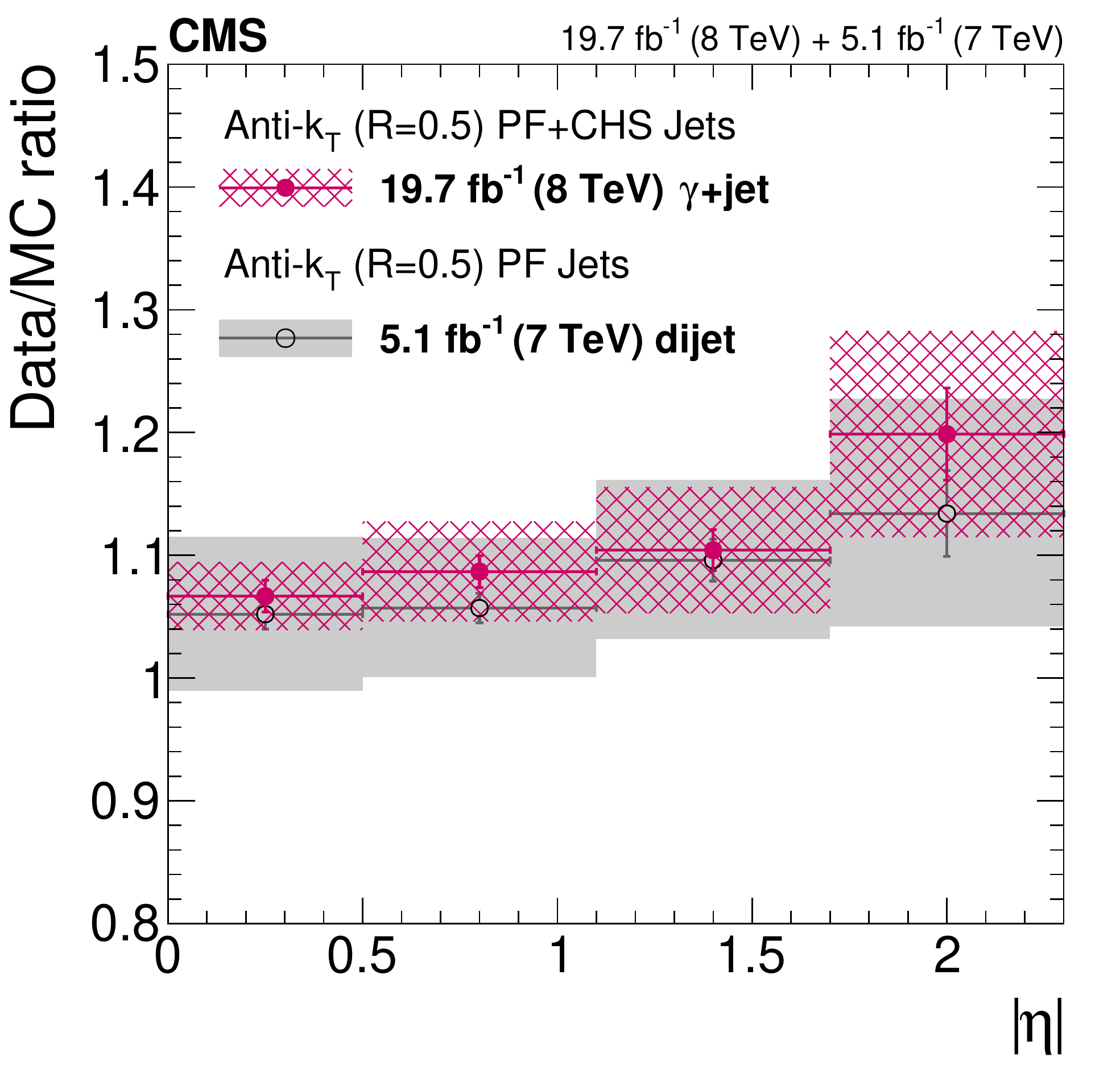}
\includegraphics[width=0.49\textwidth]{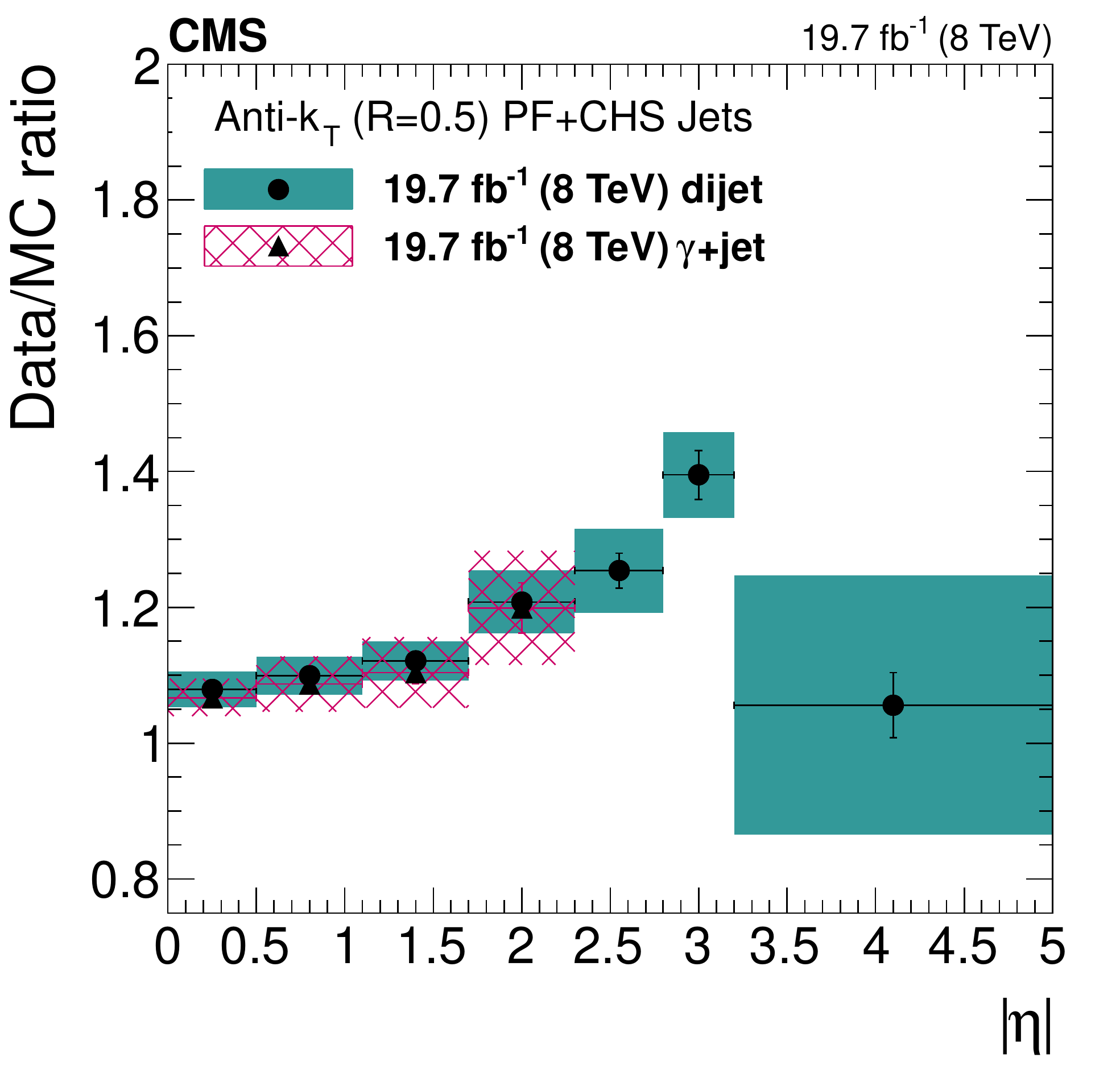}
\caption{\label{fig:gjetjer}
Data/MC scale factors for the jet \pt resolution as a function of $\abs{\eta}$, determined from 8-TeV $\gamma$+jet data (hatched boxes) compared to those obtained from dijet data (solid boxes) at 7\TeV (left) and at 8\TeV (right).
}
\end{figure}

In the procedure, $\sigma_\mathrm{PLI}$ is fixed to the value obtained from a fit to the particle-level imbalance.
The results from the photon+jet analysis are shown in Fig.~\ref{fig:gjetjer}, compared to the 7\TeV dijet data from 2011 (left) and to the 8\TeV dijet results from 2012 (right).

The following systematic uncertainties have been considered for $\gamma$+jet balancing:
\begin{itemize}
\item \textbf{QCD dijet background}: The uncertainty from QCD dijet contamination in the $\gamma$+jet sample is estimated by measuring JER with and without the dijet simulated sample added to the $\gamma$+jet sample.
\item \textbf{Flavor uncertainty}: The poor resolution for $\PQc$ and $\PQb$ jets when including neutrinos at particle level, as shown in Fig.~\ref{fig:jerflavor}, can bias the JER measurement if the flavor fractions in data and simulation differ. To estimate this uncertainty, the quark and gluon fractions are varied by $\pm 10$\%.
\item \textbf{Out-of-cone showering}: Out-of-cone showering is an important contribution to the PLI correction. To evaluate the systematic variation, the analysis was repeated for jet reconstruction with distance parameter $R=0.7$ and the difference to the nominal $R=0.5$ is taken as a systematic uncertainty.
\item \textbf{Jet energy scale}: The uncertainty arising from the knowledge of JES is evaluated by scaling all jet momenta in simulation up and down by the JEC uncertainty.
\item \textbf{PU reweighting}: The PU reweighting uncertainty is estimated by varying the minimum bias pp cross section by $\pm 5$\% from the nominal 69.4\unit{mb} when generating the target PU profile for MC simulation.
\end{itemize}
The total uncertainty varies between 3--8\% in the measured region at $\abs{\eta}<2.3$, increasing toward higher rapidity.

\section{Systematic uncertainties}
\label{Sec:uncert}

\begin{figure}[htbp!]
\centering
\includegraphics[width=0.45\textwidth]{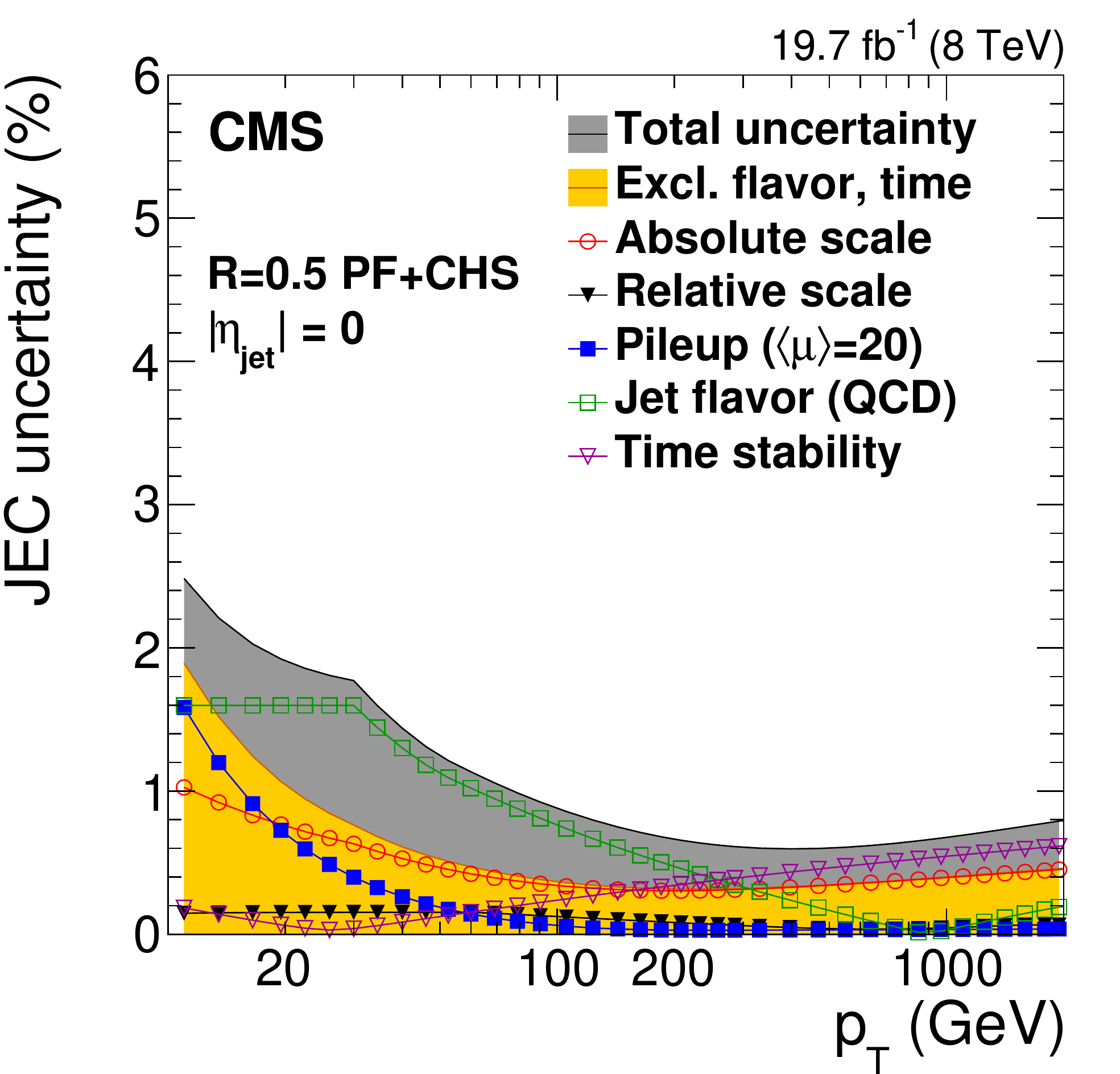}
\includegraphics[width=0.45\textwidth]{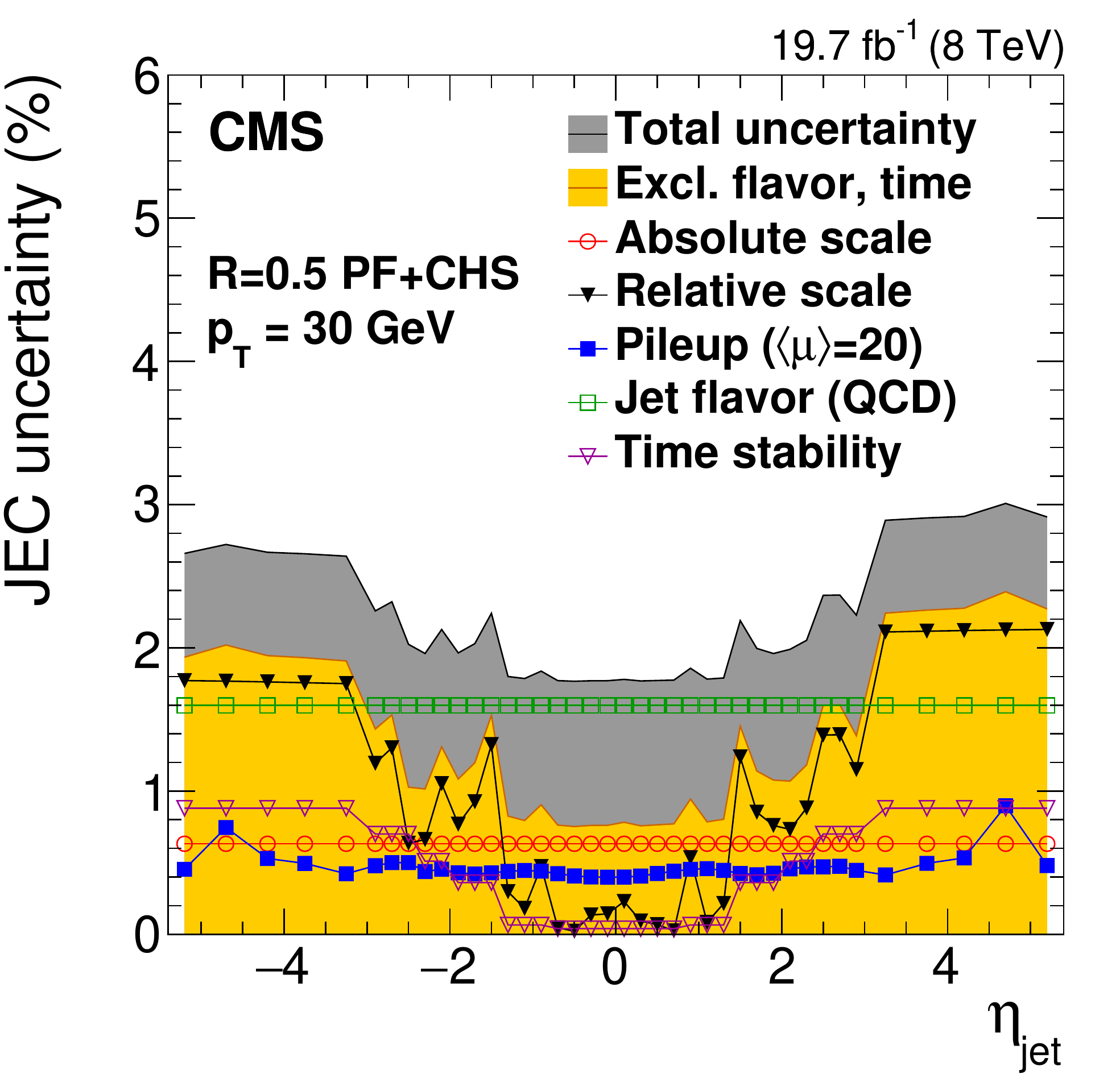}
\includegraphics[width=0.45\textwidth]{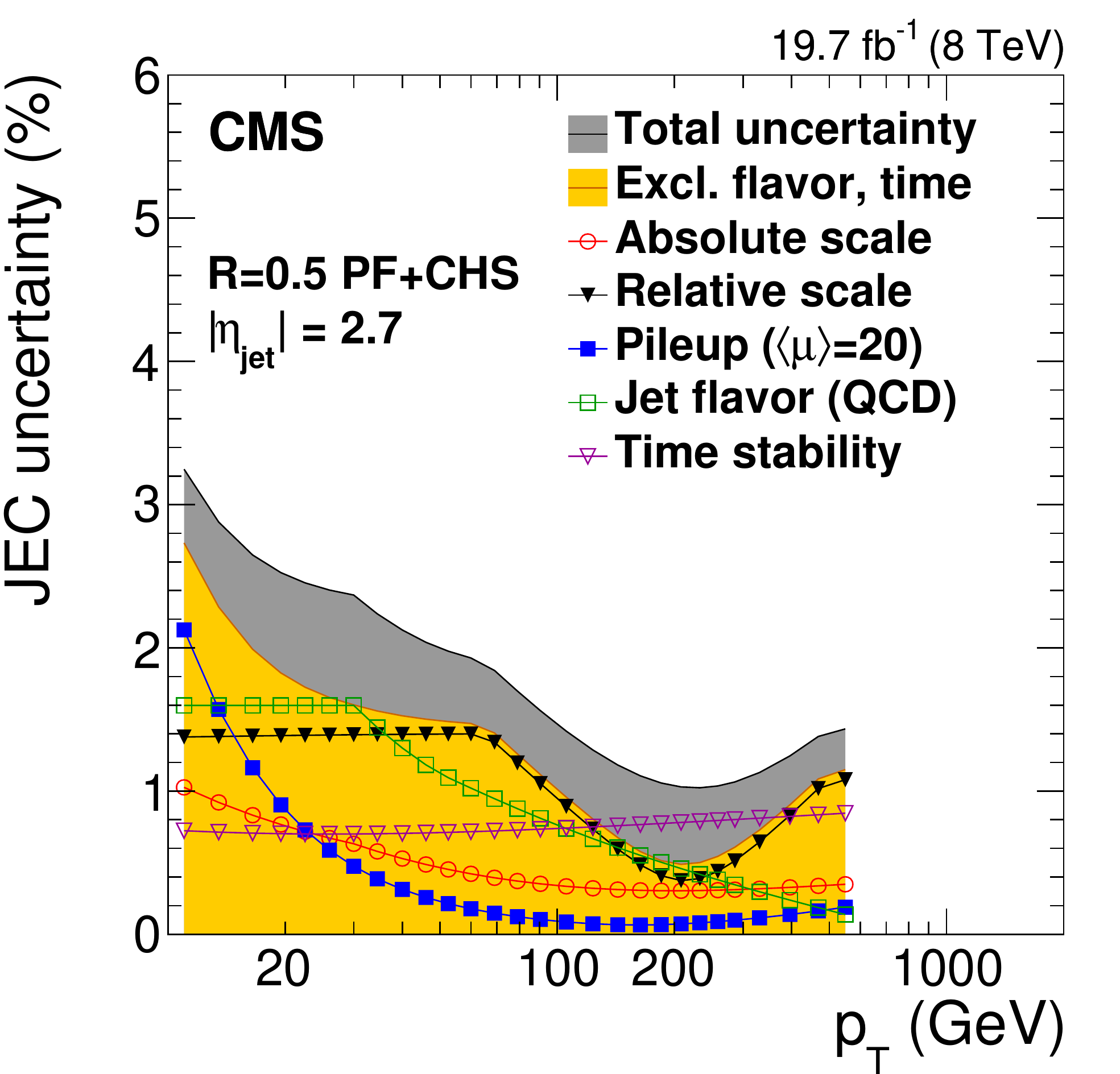}
\includegraphics[width=0.45\textwidth]{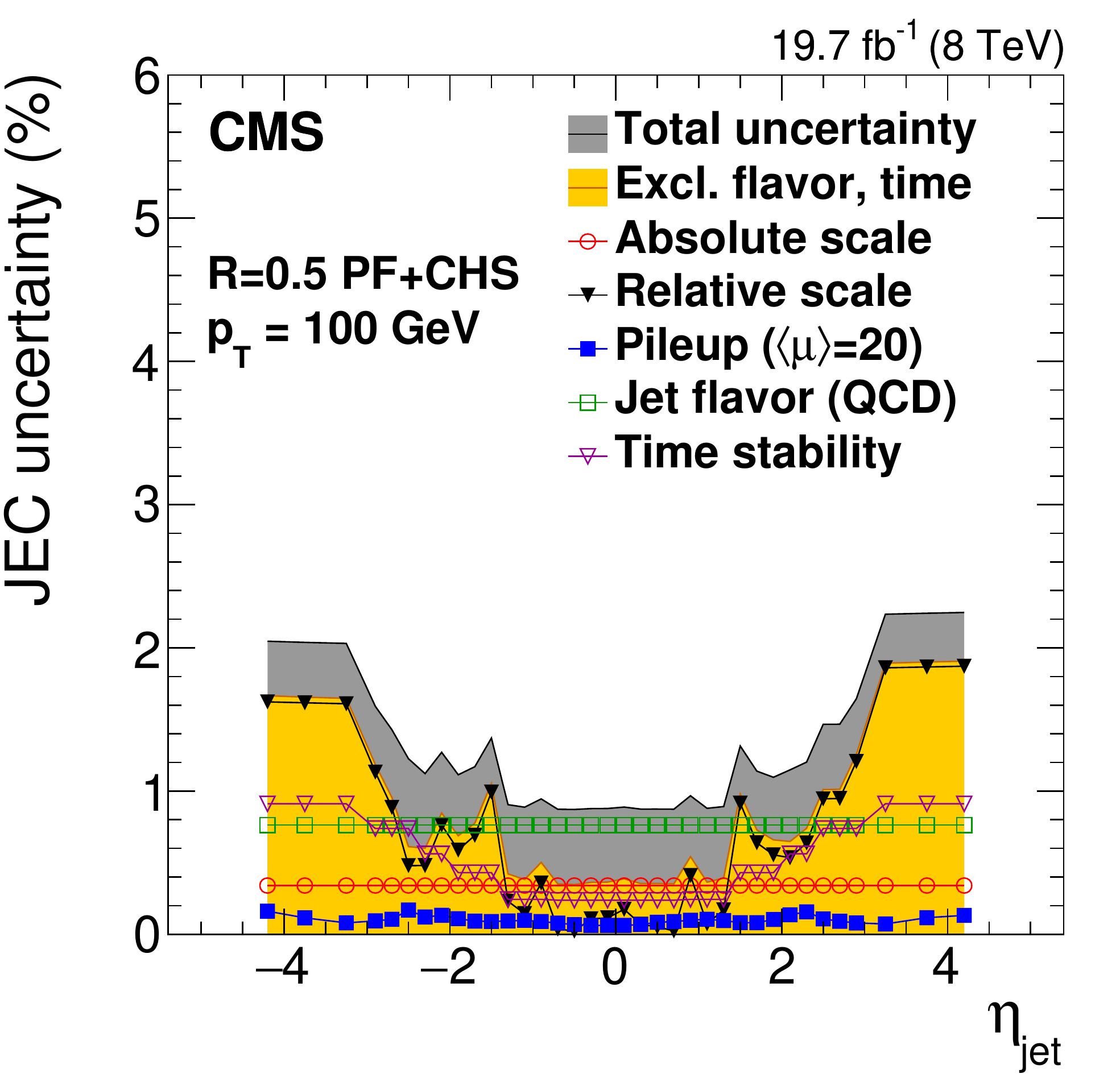}
\includegraphics[width=0.45\textwidth]{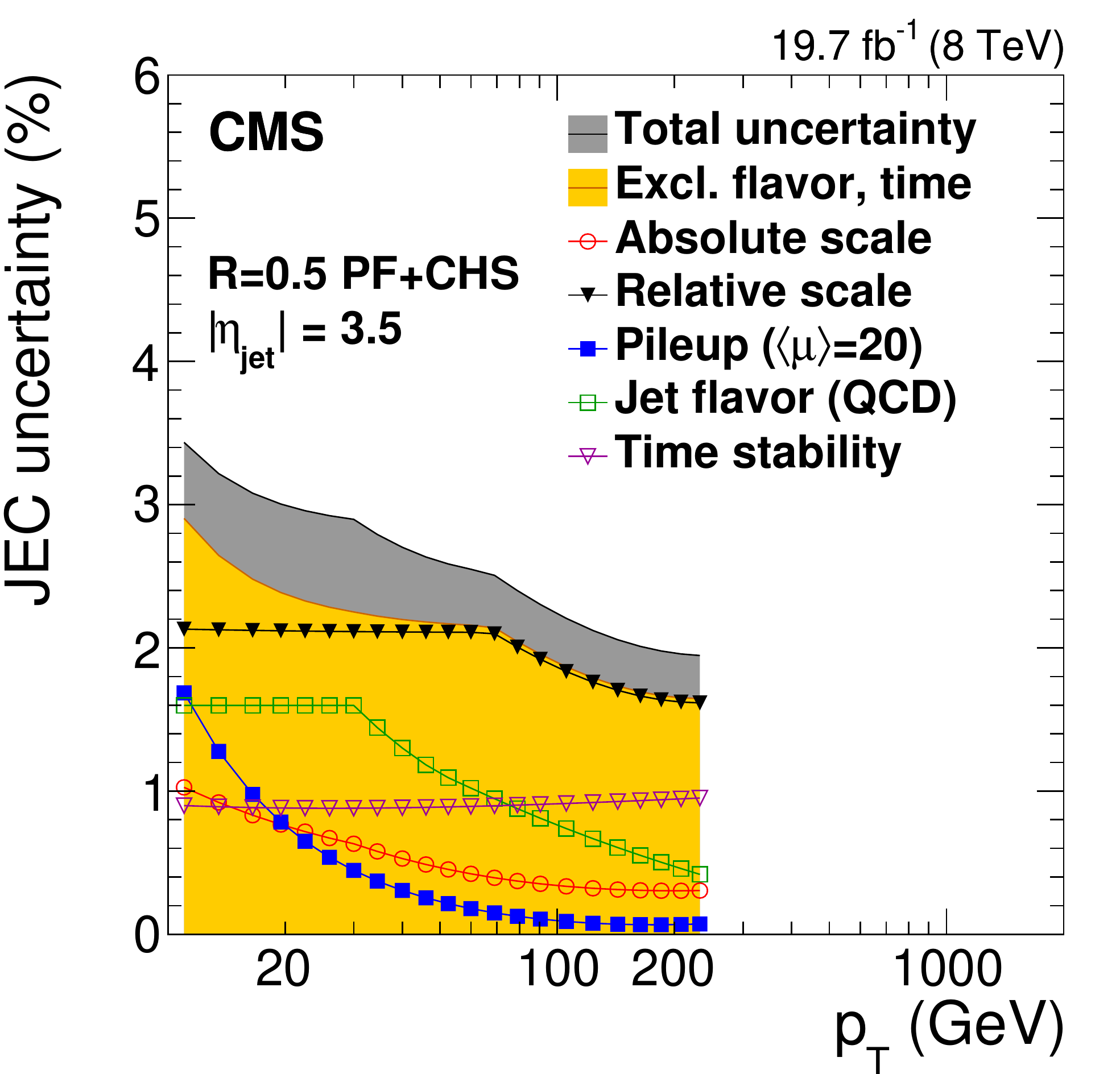}
\includegraphics[width=0.45\textwidth]{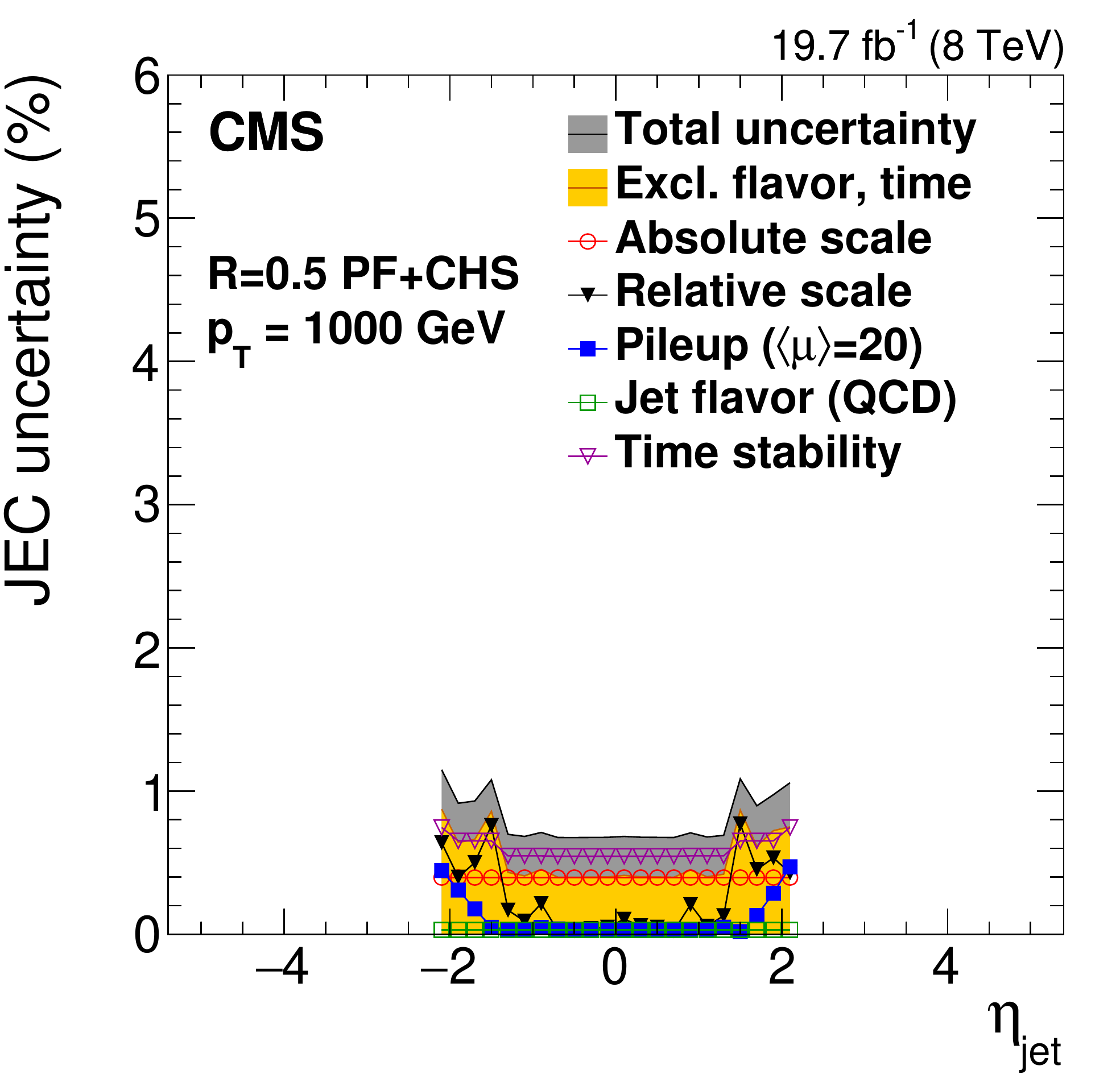}
\caption{\label{fig:total}
Summary of JES systematic uncertainties as a function of jet \pt (for 3 different $|\eta_\text{jet}|$ values, left) and of $\eta_\text{jet}$ (for 3 different \pt values, right).
The markers show the single effect of different sources, the gray dark band the cumulative total uncertainty. The total uncertainty, when excluding the effects of time dependence and flavor, is also shown in yellow light. The plots are limited to a jet energy $E=\pt\cosh\eta=4000\GeV$ so as to show only the correction factors for reasonable \pt in the considered data-taking period.
}
\end{figure}

The JEC uncertainties for each correction have been detailed in their corresponding sections. They are also summarized in Fig.~\ref{fig:total}. For the purposes of physics analyses, the uncertainties are provided as \textit{systematic sources} that include correlations across \pt and $\eta$. Each source represents a 1$\sigma$ systematic shift that is fully correlated in \pt and $\eta$. The decorrelation in $\eta$ (and \pt) is obtained by providing a single uncertainty from multiple sources that span only limited regions of phase space and can also overlap. Their sum in quadrature will always correspond to the original uncertainty, and the quadratic sum of all the sources equals the total JEC uncertainty.
This approach is very similar to the set of eigenvectors provided, \eg, by the CTEQ collaboration~\cite{cteq}.

\begin{figure}[htbp!]
\centering
\includegraphics[width=0.48\textwidth]{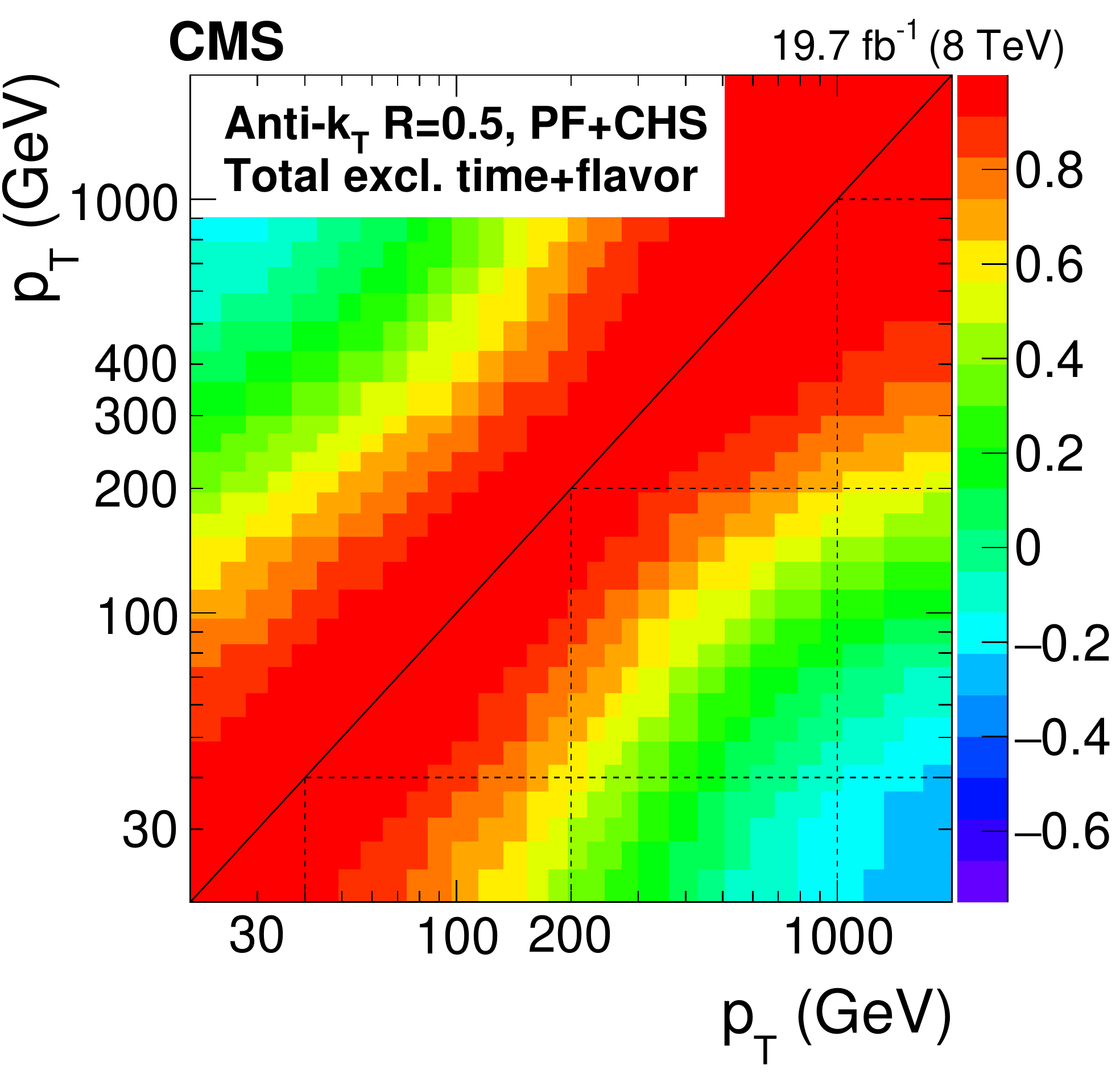}
\includegraphics[width=0.48\textwidth]{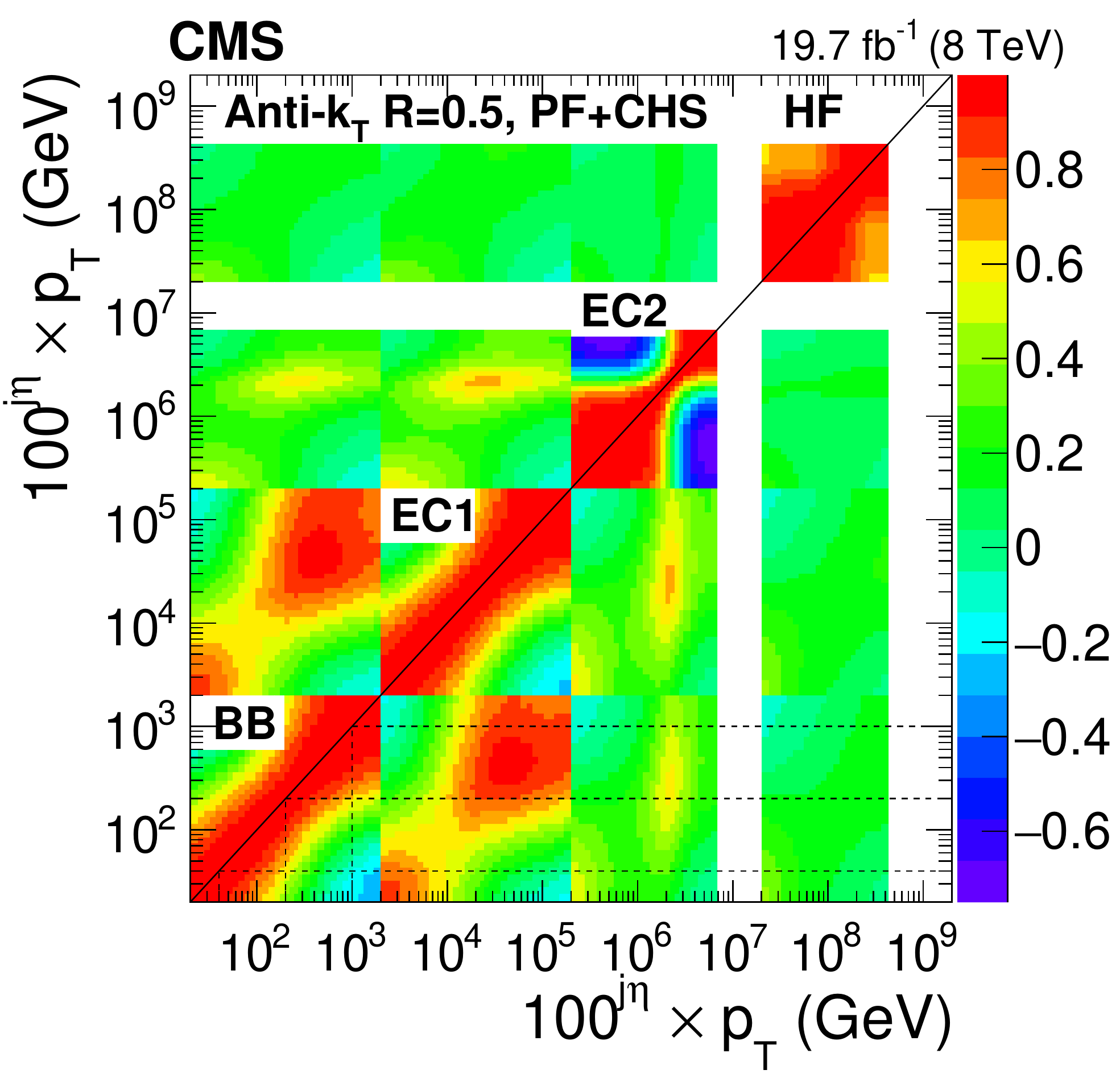}
\caption{\label{fig:mcov}
Correlation of total JES systematic uncertainties excluding time-dependent and flavor uncertainties (\textit{TotalNoTimeNoFlavor}) for PF+CHS versus \pt at $\abs{\eta}<1.3$ (left). The color represents the degree of correlation (between $-1$ and $1$).
Correlation of JES systematic uncertainties (\textit{TotalNoTimeNoFlavor}) for PF+CHS versus \pt (multiplied by $100^{\mathrm{j}\eta}$) and j$\eta$ bin (right). The integer j$\eta$ is introduced for illustration purposes, with j$\eta=0$ for the barrel region (BB), j$\eta=1$ for the endcap inside tracker coverage (EC1), j$\eta=2$ for the endcap outside tracker coverage (EC2), and j$\eta=3$ for the forward region (HF).
}
\end{figure}

The correlation factor $\rho_{ij}$ between any two points $x_i=(p_{\mathrm{T},i},\eta_i)$ and $x_j=(p_{\mathrm{T},j},\eta_j)$ of the phase space can be calculated using the systematic sources $s_k$:
\begin{eqnarray}\label{eq:mcov}\nonumber
S_{i} &=& \sqrt{\sum_k s_{ki}^2},\quad
S_{j} = \sqrt{\sum_k s_{kj}^2},\\
\rho_{ij} &=& \frac{\sum_k s_{ki} s_{kj}}{S_{i}S_{j}},
\end{eqnarray}
where $s_{ki}$ is the relative uncertainty related to the systematic source $k$ and $S_{i}$ the relative total systematic uncertainties, for the point $x_i$.
Figure \ref{fig:mcov} shows the level of correlation between various bins in \pt and $\eta$.

We recommend that in fits to the theoretical predictions ($T_i$) in data ($D_i$), the JEC systematic sources $s_k$ are propagated as multiplicative factors to the theoretical predictions and their nuisance parameters $\epsilon_k$ are added in quadrature to the global $\chi^2$:
\begin{equation}
\chi^2 = \sum_i \frac{ \left[ D_i - T_i(1+\sum_k\epsilon_k s_{ki}) \right]^2}{\hat{\sigma}_{D_i}^2} + \sum_k \epsilon_k^2 ,
\end{equation}
where $\hat{\sigma}_{D_i}$ is the statistical uncertainty ($\hat{\sigma}_{D_i}$ denotes an estimated value for $\sigma_{D_i}$).
The \textit{a priori} expectation for the $\chi^2$ minimum is that the $\epsilon_k$ will be Gaussian distributed with mean 0 and RMS 1.
Alternatively, the global $\chi^2$ can be calculated using the correlations from Eq.~(\ref{eq:mcov}):
\begin{eqnarray}
\chi^2 &=& \sum_{i,j} (D_i-T_i) M^{-1}_{ij} (D_j-T_j),\quad \mathrm{where} \\
M_{ij} &=& \sum_k s_{ki}s_{kj} + \hat{\sigma}_{D_i}\hat{\sigma}_{D_j}\delta_{ij},
\end{eqnarray}
and $\delta_{ij}$ is the Kronecker delta function.
The resulting $\chi^2$ should be identical, but the former method also provides the nuisance parameters and the best fit theoretical predictions for sanity checks.

The implementation of correlations across $\eta$ uses a rough subdivision based on detector structure to limit the final number of sources. The main regions are the barrel (BB: $\abs{\eta}<1.3$), endcap with tracking coverage (EC1: $1.3<\abs{\eta}<2.5$), endcap outside tracking coverage (EC2: $2.5<\abs{\eta}<3.0$), and hadron forward (HF: $3.0<\abs{\eta}<5.2$). This division is applied to the systematic uncertainties sensitive to the detector response (\textit{RelativePt}, \textit{PileUpPt}), resolution (\textit{RelativeJER}) modeling, and coming from statistical uncertainties (\textit{RelativeStat}). Other systematic uncertainties are treated as fully correlated across the $\eta$ bins.

The systematic sources are listed in Table~\ref{tab:sources} together with their applicable $\eta$ ranges. They are also briefly described in the following, summarizing the discussion in the respective sections. The uncertainty sources are also provided in special combinations, specific to different analysis use cases, described in Table~\ref{tab:totals}.

\begin{table}
\topcaption{\label{tab:sources}List of JES uncertainty sources, grouped by categories, with numbering, a short description, and range of validity in $\abs{\eta}$.}
\resizebox{\textwidth}{!}{
\begin{tabular}{c|c|c|c}
\hline
Source & \# & Description & Range \\
\hline
Pileup &&&\\ [0.5ex]
\multirow{2}{*}\textit{PileUpDataMC} &\multirow{2}{*}{01} & Data vs. MC simulation offset & \multirow{2}{*}{$\abs{\eta}<5.2$} \\
&&with random cone (RC) method & \\
\textit{PileUpPtRef} & 02 & True offset vs. RC $\otimes$ absolute \pt & $\abs{\eta}<5.2$ \\
\textit{PileUpPtBB} & 03 & True offset vs. RC $\otimes$ relative $\eta$ & $\abs{\eta}<1.3$ \\
\textit{PileUpPtEC1} & 04 & True offset vs. RC $\otimes$ relative $\eta$ & $1.3<\abs{\eta}<2.5$ \\
\textit{PileUpPtEC2} & 05 & True offset vs. RC $\otimes$ relative $\eta$ & $2.5<\abs{\eta}<3.0$ \\
\textit{PileUpPtHF} & 06 & True offset vs. RC $\otimes$ relative $\eta$ & $3.0<\abs{\eta}<5.2$ \\ [0.5ex]
(alternative source) & & & \\
\multirow{2}{*}{\textit{PileUpMuZero} } & \multirow{2}{*}{02-06b } & True offset vs. RC $\otimes$ residual JES &  \multirow{2}{*}{$\abs{\eta}<5.2$} \\
 && for $\langle\mu\rangle=0$ & \\ [0.5ex]
(benchmark source) & & & \\
\textit{PileUpEnvelope} & 02-06c & True offset vs. RC $\times$ 30\%  & $\abs{\eta}<5.2$ \\ [0.5ex]
\hline
Relative JES (vs. $\eta$) &&&\\ [0.5ex]
\textit{RelativeJEREC1} & 07 & Jet \pt resolution & $1.3<\abs{\eta}<2.5$ \\
\textit{RelativeJEREC2} & 08 & Jet \pt resolution & $2.5<\abs{\eta}<3.0$ \\
\textit{RelativeJERHF} & 09 & Jet \pt resolution & $3.0<\abs{\eta}<5.2$ \\
\textit{RelativeFSR} & 10 & ISR+FSR correction & $\abs{\eta}<5.2$ \\
\textit{RelativeStatFSR} & 11 & ISR+FSR statistical uncertainty &  $\abs{\eta}<5.2$ \\
\textit{RelativeStatEC2} & 12 & Statistical uncertainty & $2.5<\abs{\eta}<3.0$ \\
\textit{RelativeStatHF}  & 13 & Statistical uncertainty & $3.0<\abs{\eta}<5.2$ \\
\textit{RelativePtBB}  & 14 & Log-lin. vs. flat fit $\times$ 50\% & $\abs{\eta}<1.3$ \\
\textit{RelativePtEC1} & 15 & Log-lin. vs. flat fit $\times$ 50\% & $1.3<\abs{\eta}<2.5$\\
\textit{RelativePtEC2} & 16 & Log-lin. vs. flat fit $\times$ 50\% & $2.5<\abs{\eta}<3.0$\\
\textit{RelativePtHF}  & 17 & Log-lin. vs. flat fit $\times$ 50\% & $3.0<\abs{\eta}<5.2$\\
\textit{TimeEta}  & 18 & Relative $\eta$ time dependence & $\abs{\eta}<5.2$ \\ [0.5ex]
\hline
Absolute JES (vs. \pt) &&&\\ [0.5ex]
\textit{AbsoluteScale}  & 19 & Lepton scale, $\pm 0.11$\%  & $\abs{\eta}<5.2$ \\
\multirow{2}{*}{\textit{AbsoluteMPFBias} } & \multirow{2}{*}{20}  & MPF bias, $\pm0.28$\%  & \multirow{2}{*}{$\abs{\eta}<5.2$} \\
&& (from $\nu$'s $\oplus$ ISR acceptance, $0.2{\rm\%}\oplus0.2{\rm\%}$) & \\
\textit{AbsoluteStat} & 21 & Statistical uncertainty vs. \pt  & $\abs{\eta}<5.2$ \\
\textit{SinglePionECAL} & 22 & Single-pion response in ECAL, $\pm4.2$\% & $\abs{\eta}<5.2$\\
\textit{SinglePionHCAL} & 23 & Single-pion response in HCAL, $\pm1.5$\% & $\abs{\eta}<5.2$\\
\textit{Fragmentation} & 24 & Jet fragmentation in \PYTHIA6.4 vs. \HERWIGpp2.3 & $\abs{\eta}<5.2$ \\
\multirow{2}{*}{\textit{TimePt} } & \multirow{2}{*}{25 } & Absolute \pt time dependence & \multirow{2}{*}{$\abs{\eta}<5.2$} \\
&& (indirectly with charged-pion $E_\mathrm{HCAL}/p$)  &\\ [0.5ex]
\hline
Jet flavor &&&\\ [0.5ex]
(only one of these) &&&\\
\textit{FlavorQCD} & 26a & QCD dijet mixture (default) & $\abs{\eta}<5.2$ \\
\textit{FlavorZJet} & 26b & $\PZ$+jet mixture & $\abs{\eta}<5.2$ \\
\textit{FlavorPhoton} & 26c & $\gamma$+jet mixture & $\abs{\eta}<5.2$ \\ [0.5ex]
(or mixture of these) &&&\\
\multirow{2}{*}{\textit{FlavorGluon} } & \multirow{2}{*}{26d1}  & Pure gluon ($\Pg$) & \multirow{2}{*}{$\abs{\eta}<5.2$}  \\
&& (incl.~$ \Pg \to \PQq \PAQq $ and unmatched) & \\
\textit{FlavorQuark} & 26d2 & Pure light quark ($\PQu\PQd\PQs$) &  $\abs{\eta}<5.2$ \\
\textit{FlavorCharm} & 26d3 & Pure charm ($\PQc$) & $\abs{\eta}<5.2$ \\
\multirow{2}{*}{\textit{FlavorBottom}}  &\multirow{2}{*}{ 26d4}  & Pure bottom ($\PQb$) &  \multirow{2}{*}{$\abs{\eta}<5.2$ } \\
&& (Pure flavors refer to the \textit{Physics} definition) & \\ [0.5ex]
\hline
\end{tabular}
}
\end{table}

\begin{table}
\centering
\topcaption{\label{tab:totals}
List of JES uncertainty source combinations with a short description and list of uncertainty components. The numbering of the sources ($3^\mathrm{rd}$ column) corresponds to that used in Table~\ref{tab:sources} (2$^\mathrm{nd}$ column).
}
\resizebox{\textwidth}{!}{
\begin{tabular}{c|c|c}
\hline
Enumerator & Description & List of sources \\
\hline
\textit{Total} & Default uncertainty applicable for most CMS analyses & 01--25, 26a \\
\textit{TotalNoFlavor} & \begin{tabular}[c]{@{}c@{}}Default uncertainty without \textit{FlavorQCD,}\\as basis for other mixtures\end{tabular} & 01--25 \\
\textit{TotalNoTime} & \begin{tabular}[c]{@{}c@{}}Uncertainty for unprescaled analyses\\with full 8\TeV data set\end{tabular} & 01--17, 19--24, 26a \\
\textit{TotalNoFlavorNoTime} & \begin{tabular}[c]{@{}c@{}}Same as above without \textit{FlavorQCD},\\as basis for other mixtures\end{tabular} & 01--17, 19--24 \\
\hline
\textit{SubTotalPileUp} & Combination of pileup offset uncertainties & 01--06 \\
\textit{SubTotalRelative} & Combination of relative $\eta$ correction uncertainties  & 07--17 \\
\textit{SubTotalAbsolute} & Combination of absolute scale uncertainties  & 19--24 \\
\hline
\textit{SubTotalScale} & \begin{tabular}[c]{@{}c@{}}Scale uncertainties independent of \pt,\\plus statistical uncertainty\end{tabular} & 19--21 \\
\textit{SubTotalPt} & \begin{tabular}[c]{@{}c@{}}Combination of absolute scale \pt dependence\\ uncertainties\end{tabular} & 22--24 \\
\textit{SubTotalMC} & Default uncertainty without \textit{PileUpPt} sources & 01,07--25,26a \\
\hline
\end{tabular}
}
\end{table}

The JES uncertainties are classified in four broad categories: pileup offset, relative calibration of JES versus $\eta$, absolute energy scale versus \pt, and jet-flavor response. In addition, the residual JES time-dependence is considered as an extra source of systematic uncertainty for samples other than the full unprescaled 8\TeV data set.

The \textbf{pileup offset} (Section~\ref{sec:pileup}) is mostly important at low \pt. The pileup correction is taken from the true offset in simulation, and then scaled by the ratio of random cone offsets for data and simulation. The main systematic uncertainty is evaluated as $30\%$ of the \pt-dependent difference between the true offset and the random cone offset in simulation.
This difference is propagated through the fit procedure using $\PZ$/$\gamma$+jet (absolute \pt) and dijet (relative $\eta$)  balancing to estimate the remaining residual pileup uncertainty after residual JES (Section~\ref{sec:l3res_globalfit}).
A special alternative systematic (\textit{PileUpMuZero}) is added to estimate the bias introduced by the data-based calibration for no-pileup ($\langle\mu\rangle=0$) conditions.
A residual uncertainty on the scale factor (\textit{PileUpDataMC}) is estimated by  varying $\rho$ within one standard deviation.

The \textbf{relative $\eta$-dependent correction} (Section~\ref{sec:relative}) calibrates forward regions relative to $\abs{\eta}<1.3$ using dijet events. The main systematic uncertainties come from JER (\textit{RelativeJER}) and the ISR+FSR (\textit{RelativeFSR}) bias corrections. The JER uncertainty is evaluated by varying the JER for each detector region independently within the JER uncertainties estimated in Section~\ref{sec:jer}, when applying smearing to the simulation.
The differences obtained from comparisons based on \PYTHIA6.4 and \HERWIGpp~2.3 simulations are used to assign an ISR+FSR uncertainty.
The $\eta$-dependent correction is parameterized with a log-linear function versus \pt at $\abs{\eta}<3.0$, and as a constant in HF ($3\leq\abs{\eta}<5.2$).
Half the difference between a log-linear fit and a constant fit is taken as a \pt-dependent parameterization uncertainty (\textit{RelativePt}).
For $\abs{\eta}>2.5$, the statistical uncertainty (\textit{RelativeStat}) is also a significant contribution.

The \textbf{absolute scale} (Section~\ref{sec:l3res_globalfit_unc}) uncertainties for constant scale (\textit{AbsoluteScale}) and \pt-dependent scale (\textit{AbsoluteStat}) are extracted from a global fit to $\PZ$/$\gamma$+jet and multijet data. The former is essentially related to the lepton momentum scale for muons in \zmmjet, while the latter to the single-pion response in HCAL. Additional constant scale uncertainty is added for the biases of the MPF and \pt-balance methods (\textit{AbsoluteMPFbias}) coming from neutrinos and ISR outside of detector acceptance.
Shape uncertainties for the \pt dependence are considered from the following sources relative to the global fit reference of $\pt=208\GeV$ at $\abs{\eta}<1.3$: single-pion responses in ECAL and HCAL (\textit{SinglePionECAL}, \textit{Single\-Pion\-HCAL}) and jet response differences due to the different fragmentation models implemented in \PYTHIA6.4 and \HERWIGpp~2.3 (\textit{Fragmentation}).

The \textbf{flavor response differences} (Section~\ref{sec:flavor}) are estimated using simulation, and cross-checked with $\PZ$+$\PQb$-jet, and quark- and gluon-tagged photon+jet and $\PZ$+jet events. The flavor uncertainties are assigned based on \PYTHIA6.4 and \HERWIGpp~2.3 differences, which are propagated through the data-based calibration chain with dijet, photon+jet, and $\PZ$+jet events. As a result, the flavor uncertainties are minimized for these flavor mixtures. The flavor differences in \PYTHIA6.4 and \HERWIGpp~2.3 are largest for gluon jets, while the two MC simulations agree well on both light- and heavy-quark jets.

As explained in Section~\ref{sec:time}, the JES in the endcaps shows some residual \textbf{time dependence}, even after correcting for radiation damage to the ECAL and HCAL. Prescaled triggers sample different run periods with different weights, leading to slight scale differences between different data sets. This time-dependence uncertainty is estimated as the RMS variation of the $\eta$-dependent corrections determined with dijet events for different run periods.

\subsection{Uncertainties in \texorpdfstring{7\TeV}{7 TeV} analyses}

For comparisons to published 7\TeV analyses, using the 2011 data set corresponding to an integrated luminosity of 5\fbinv, we briefly summarize the main differences in the JEC and JES uncertainty sources relative to the ones reported in the previous publication \cite{JEC_JINST}:

\begin{itemize}
\item \textbf{Pileup offset:} The 7\TeV uncertainties did not consider the absorption of the pileup offset jet \pt dependence into the residual $\eta$ and \pt-dependent corrections, and the five \textit{PileUpPt} sources (\#02--\#06 in Table~\ref{tab:sources}) were reported as a single large \textit{PileUpPt} uncertainty without $\eta$ decorrelation.
The pileup offset correction was derived from the study of the offset, obtained with the RC method, as a function of the number of primary vertices $N_\mathrm{PV}$ for data only. This lead to a larger \textit{PileUpDataMC} (\#01) uncertainty, necessitating additional \textit{PileUpOOT}, \textit{PileUpJetRate} and \textit{PileUpBias} uncertainties. The \textit{PileUpOOT} covered the variation due to out-of-time pileup, now accounted for by deriving the offset versus $\mu$ before mapping to $N_\mathrm{PV}$. The \textit{ PileUpJetRate} accounted for the offset \pt dependence, now corrected using \pt-dependent true offset from simulation. The 7\TeV offset correction was calibrated for jets in the \pt range of 20--30\GeV using simulation, with \textit{PileUpBias} accounting for the uncertainty in the bias correction for data. In the current scheme, \textit{PileUpBias} is absorbed into the \textit{PileUpPt} sources.

\item \textbf{Relative scale:} The 7\TeV JEC used a constant correction for each $\eta$ bin, and the uncertainties did not provide the four \textit{RelativePt} sources (\#14--\#17) to account for the small residual \pt dependence versus $\eta$. This led to some overestimates in the uncertainty correlation versus $\eta$. The recommended way to address \textit{a posteriori} the underestimated decorrelation in the 7\TeV uncertainties is to clone the 7\TeV \textit{SinglePion} source (roughly corresponding to current uncertainties \#22--\#23) into five separate sources, spanning $\eta$ ranges $\abs{\eta}<0.5$, $0.5<\abs{\eta}<1.0$, $1.0<\abs{\eta}<1.5$, $\abs{\eta}<1.5$ and $\abs{\eta}>1.5$, with weight 1 for the last source and $1/\sqrt{2}$ for the four others. This ensures 50\% decorrelation for the \textit{SinglePion} source within the barrel bins, and 100\% decorrelation between the barrel and endcap, while preserving the total uncertainty. The \textit{RelativeFSR} (\#10) and \textit{RelativeStatFSR} (\#11) sources were reported as a single \textit{RelativeFSR} uncertainty.

\item \textbf{Absolute scale:} The 7\TeV absolute correction was a single constant factor and did not include any \pt dependence except for the uncertainties. The three flat absolute scale factor uncertainties (\#19--\#21) were reported as a single source (\textit{Absolute}), as were the \pt-dependent uncertainties from single-pion response in the ECAL (\#22) and HCAL (\#23). The 7\TeV \textit{SinglePion} uncertainty, which corresponds to the latter two uncertainties, was a factor two larger and is uncorrelated with the 8\TeV \textit{Single\-Pion\-HCAL} (\#23) source, which is now based on the value obtained from the global \pt-dependent fit. The \textit{AbsoluteScale} (\#19) and \textit{AbsoluteStat} (\#21) are also based on the global fit, and therefore uncorrelated with the 7\TeV source \textit{Absolute}.

\item \textbf{Time dependence:} There was neither evidence of residual JEC \pt dependence at 7\TeV, nor of time dependence in the absolute correction. Therefore, the \textit{TimePt} source (\#25) was not provided.
\end{itemize}

In total, the 7\TeV uncertainties included sixteen uncertainty sources, ten fewer than the current recommendations (\#1--\#26): one less for pileup (four new sources and three obsolete), five less for relative scale, three less for absolute scale and one less for time dependence. The current uncertainties provide a more detailed description of correlations versus both $\eta$ and \pt.

\section{The PF jet composition}
\label{sec:jetcomp}

The different detector contributions to the JES can be understood in more detail by looking at the fractions of jet energy reconstructed as the various types of PF candidates. Although this information is not yet used to directly constrain JES, it gives valuable information in monitoring the stability of JES.

The PF jet composition is determined from the dijet sample, selected as described in Section~\ref{sec:event_sel}, using the tag-and-probe method. In this method the tag jet, which is in the barrel and matched to the trigger in data, provides measurement of the jet \pt, while the back-to-back probe jet is used for an unbiased determination for PF candidate energy fractions. This indirect method avoids the selection bias that would enhance energy fractions correlated with upward fluctuations in jet \pt. This is particularly important because data and simulation are known to have different resolutions, and we want to avoid introducing JER biases in the composition measurement.

\begin{figure}[htbp!]
\centering
\includegraphics[width=0.49\textwidth]{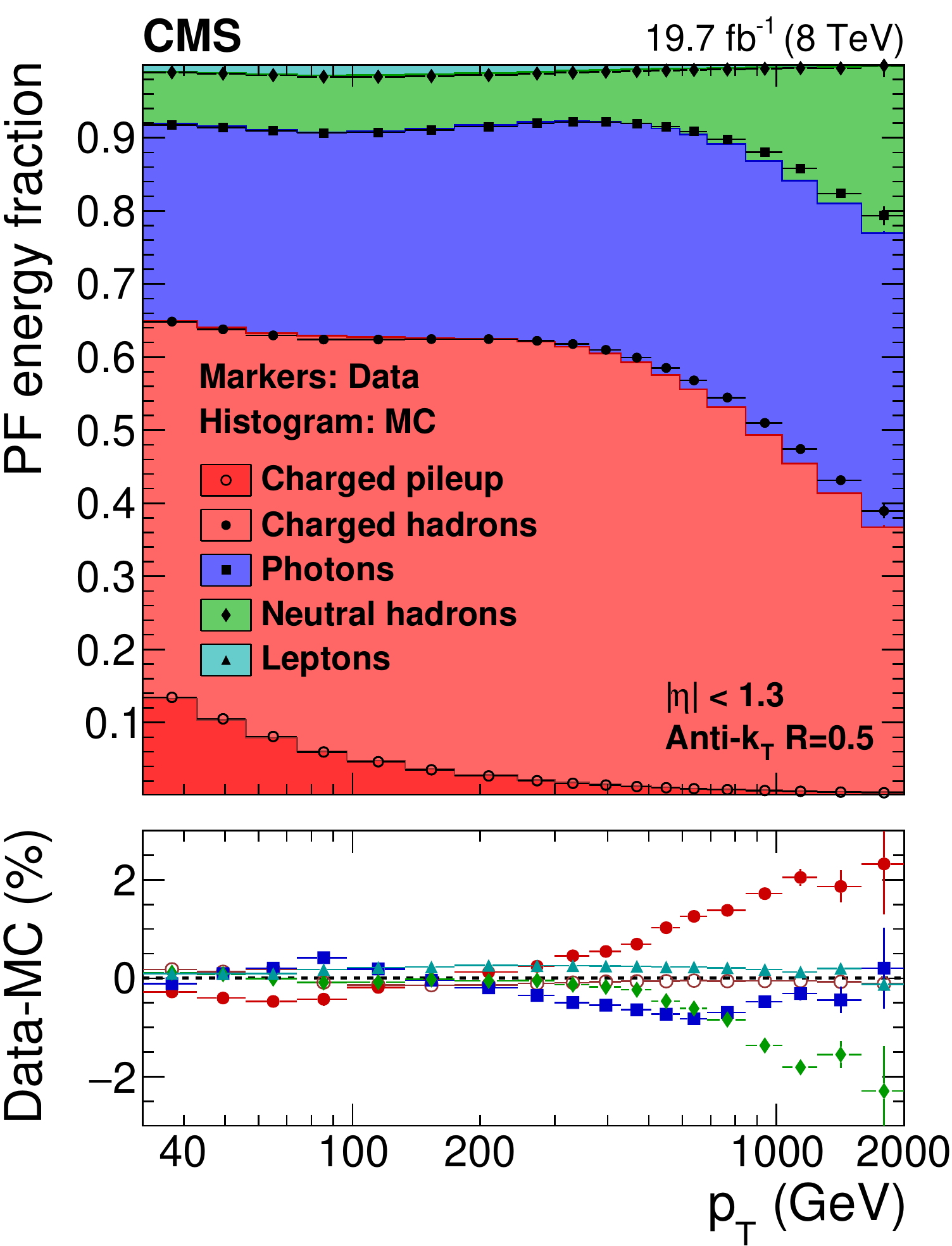}
\includegraphics[width=0.49\textwidth]{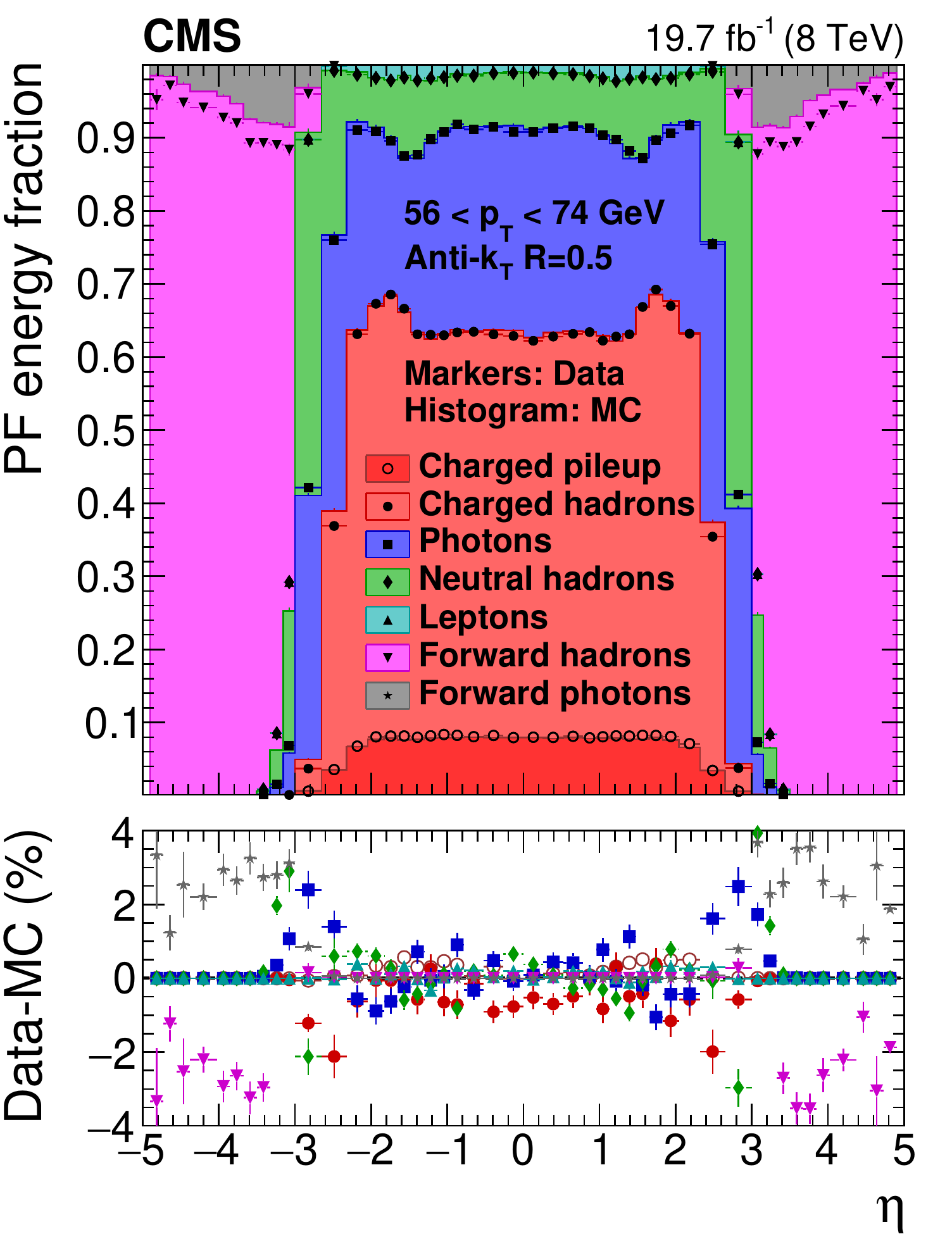}
\caption{\label{fig:l3pf}
PF jet composition in data and simulation versus \pt at $\abs{\eta}<1.3$ (left), and versus $\eta$ at $56<\pt<74\GeV$ (right).
}
\end{figure}

The measured PF energy fractions are shown in Fig.~\ref{fig:l3pf} as a function of \pt and $\eta$.
The agreement between data and simulation is at the level of 1--2\% in the barrel, consistent with the measured residual JEC. In particular, the fraction of charged hadrons associated with pileup agrees well between data and simulation.
The differences increase at high \pt, where the tracking efficiency within the densely populated jet core drops significantly, and PF becomes more sensitive to the calorimeter scale. The differences also increase at higher rapidities, where the observed residual JEC are larger.

\subsection*{Interpretation of the results}

To interpret the implications for JEC of composition differences in data and simulation, we can start from the definition of the jet response, in the absence of pileup, and in bins of particle-level jet $p_\text{T, ptcl}$ and reconstructed $\eta$ ($\pt$ is the reconstructed jet transverse momentum):
\begin{equation}
\label{eq:jes}
R(\langle \pt\rangle,\eta)  \equiv \frac{\langle \pt\rangle}{\langle p_\text{T, ptcl}\rangle} [p_\text{T, ptcl},\eta] \cong \left\langle \frac{\pt}{p_\text{T, ptcl}} \right\rangle [p_\text{T, ptcl},\eta].
\end{equation}

\begin{figure}[htbp!]
\centering
\includegraphics[width=0.58\textwidth]{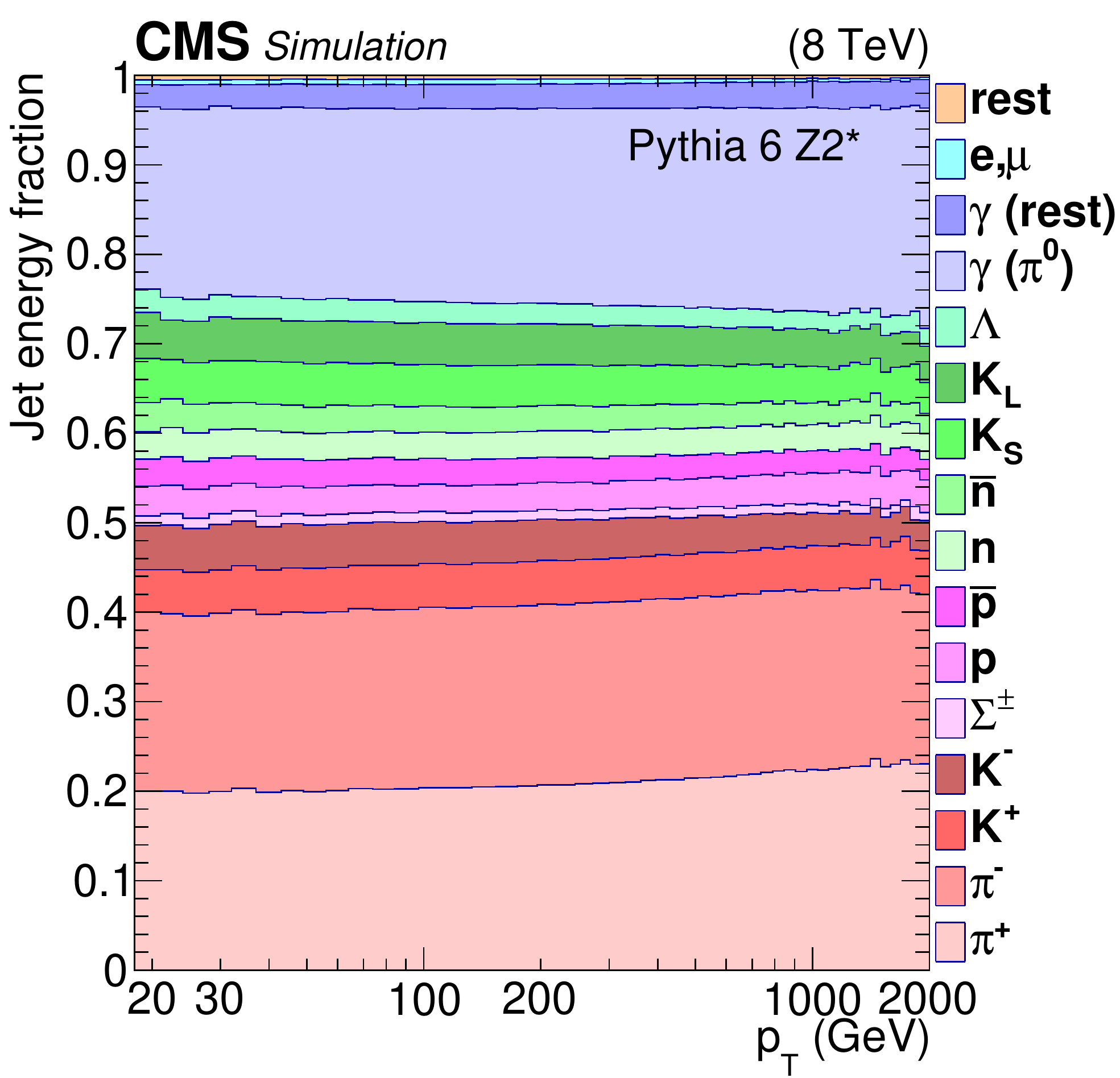}
\caption{\label{fig:genfrac}Jet composition at particle level in the \PYTHIA6.4 tune Z2* for QCD dijet sample, shown versus \pt at $\abs{\eta}<1.3$. The component labeled '$\gamma~(\text{rest})$' denotes all photons not coming from $\pi^0$s, and the component labeled 'rest' refers to all particles not listed specifically.}
\end{figure}

The last equivalence holds if the $p_\text{T, ptcl}$ bins are narrow enough.
Leaving aside corrections for overlapping particles caused, \eg, by calorimeter zero-suppression or neutral hadron shadowing in the PF algorithm, we can write
\begin{equation}\label{eq:jec}
R(\langle \pt\rangle,\eta)
= \frac{\langle\sum_{i} R_f (\langle p_{\mathrm{T}}^{i}\rangle,\eta^i)p_\text{T, ptcl}^{i}\rangle}{p_\text{T, ptcl}},
\quad p_\text{T, ptcl} = \sum_i p_\text{T, ptcl}^{i},
\end{equation}
where the sum runs over all stable ($c\tau>$1\unit{cm}) particles $i$ of different particle species $f$, excluding neutrinos.
Figure~\ref{fig:genfrac} illustrates the average particle jet composition in QCD dijet sample versus \pt at $\abs{\eta}<1.3$, which is about 60\% charged hadrons (red hues), 15\% neutral hadrons (green hues) and 25\% photons (blue hues). About 65\% of the jet energy is carried by pions ($\pi^+$, $\pi^-$, $\pi^0\to\gamma\gamma$).
The nucleons ($\Pp$, $\PAp$, $\Pn$, $\PAn$) and kaons ($\PKp$, $\PKm$, $\PKzL$, $\PKzS$) carry about 15\% each, with the remaining 5\% in fragmentation photons ($\gamma$), lambda ($\Lambda^0$), and sigma ($\Sigma^0$) baryons, leptons ($\Pe$, $\mu$) and other particles. The composition does not significantly depend on $\eta$.

The PF algorithm \cite{CMS-PAS-PFT-09-001, CMS-PAS-PFT-10-001} uses tracking to measure charged particles \pt down to $\pt\approx 0.3\GeV$, which effectively results in response $R_\text{ch}(\pt)\approx1$ for all charged particles (charged hadrons, electron, muons) within the tracking coverage at $\abs{\eta}<2.4$, when neglecting tracking inefficiencies and the low \pt acceptance. The finely segmented ECAL has a linear response to photons down to its acceptance of about $0.1\GeV$, which also results in response $R_\gamma\approx 1$.

Neutral hadrons (predominantly n, $\PKzL$, $\Lambda^0$), hadrons outside the tracking coverage, and hadrons with failed tracking (\eg, $\PKzS$ decaying in the outer layers of tracking, or unreconstructed nuclear interactions within the tracker) are reconstructed with a nonlinear calorimeter response $R_\text{calo}(\pt) \approx 1 - a\pt^{m-1}$ \cite{Groom:2007zz} separately in the ECAL (as PF photons) and the HCAL (as PF neutral hadrons). For  particles with $\abs{\eta}>2.5$, overlapping ECAL and HCAL energy deposits are linked to build neutral hadrons.
The combination of the HCAL reconstruction threshold of $\pt>0.8\GeV$, the low response of $R_\text{calo}\approx 0.3$ for soft hadrons, and many neutral hadrons showering early in ECAL, result in an effective neutral hadron response $R_\mathrm{nh} \ll 1$ at low \pt despite the PF neutral hadron calibration for HCAL deposits.
However, the neutral hadron response approaches unity asymptotically at high \pt.

Rearranging Eq.~(\ref{eq:jec}) in terms of the main particle categories seen by the detector gives
\begin{equation}\label{eq:jec2}
R_{\text{jet}} = f_\text{ch} R_\text{ch} + f_\gamma R_\gamma + f_\mathrm{nh} R_\mathrm{nh},
\end{equation}
where $f_i$ are the average fractions of energy carried by each particle category at particle level, such that $\sum_i f_i = 1$. The effective categories are charged hadrons+electrons+muons (ch), photons ($\gamma$), and neutral hadrons (nh).
These effective categories have some ambiguity in the classification of, \eg, $\PKzS\to \pi^++\pi^-$ and $\PKzS\to \pi^0+\pi^0$ decays within the tracker volume. Typical generated fractions are $f_\text{ch}\approx 60$\%, $f_\gamma\approx 25$\%, and $f_\mathrm{nh}\approx 15$\%. These values differ slightly from typical measured fractions $f_\text{ch, meas}\approx 65$\%, $f_{\gamma, \text{meas}}\approx 30$\%, and $f_\mathrm{nh, meas}\approx 5$\% due to different detector responses ($R_\text{ch}\approx R_\gamma\approx 1$, $R_\mathrm{nh}\approx 0.6$, thus $R_\text{jet}\approx 0.95$) and ambiguities in particle identification, specially between neutral hadrons and photons.

The impact of typical detector mismodeling effects can be estimated with the help of Eq.~(\ref{eq:jec2}). A change of $-1\%$ in the charged hadron fraction through tracking inefficiencies would result in a corresponding but smaller increase of the neutral hadron fraction, for a total relative jet response variation of $-0.4\%$. A variation of $-1\%$ of the ECAL scale in data would change the relative jet response by $-0.3\%$, while a variation of the single-pion response of $-3\%$ would also change it by $-0.3\%$.

Adding these \textit{a priori} uncertainty estimates in quadrature gives $0.6$\%, while summing them up gives $-1.0$\%. Incidentally, these are about the order of magnitude of the minimum energy scale uncertainty and the data/MC correction applied at the moment in the reference region $\abs{\eta}<1.3$. The PF energy fractions between data and simulation in Fig.~\ref{fig:l3pf} are in almost perfect agreement to the level of about $10^{-3}$ at $\pt\approx 200\GeV$, which is consistent with the jet response difference of $-1$\%, assuming the inefficiencies for charged hadrons, photons, and neutral hadrons all go in the same direction.

The corrections and uncertainties increase outside the tracking coverage at $\abs{\eta}>2.5$, where effectively $f_\text{ch}=0$ and $f_\mathrm{nh}=75$\%, resulting in about five times higher sensitivity to single-pion response and threshold effects in $R_\mathrm{nh}$.

\section{Conclusions}
\label{sec:conclusions}

The understanding of jet energy scale and resolution achieved by the CMS experiment exploiting the pp data taken at 8\TeV, corresponding to an integrated luminosity of 19.7\fbinv, has been documented.
The CMS strategy involves the sequential correction of the jets for the pileup offset,
simulated response, and residual $\eta$ and \pt dependencies in data with respect to simulation, with optional corrections depending on the jet flavor.

The pileup offset corrections are determined from QCD dijet simulations processed with and without pileup overlay, with additional
corrections for residual differences between data and detector simulation, extracted using the random-cone method in zero-bias events.
The simulated jet response corrections are determined from a Monte Carlo event sample, as a function of jet \pt and $\eta$, for various jet algorithms
and distance parameter sizes.
The $\eta$-dependent corrections for residual differences between data and MC simulation are determined with dijet events, relative
to a jet in the central region $\abs{\eta}<1.3$.
The \pt-dependent corrections for residual differences within $\abs{\eta}<1.3$ at $30<\pt<800\GeV$
are measured combining photon+jet, \zmmjet, and \zeejet\ events. The \pt dependence at $\pt>800\GeV$ is directly
constrained with multijet events.  The optional jet-flavor corrections are derived from MC simulation and checked for $\PQb$ jets.

The simulated jet \pt resolution is determined, as a function of jet \pt and $\eta$, for different levels of pileup.
The dependence of the resolution on the jet flavor is also studied. The $\eta$-dependent corrections for residual differences between data
and MC simulation in the jet resolution are determined with both dijet and $\gamma$+jet events.

In all methods using MC simulation,
the uncertainties arising from the modeling of jet radiation and hadronization are
constrained from the study of the differences between \PYTHIA6.4 and \HERWIGpp~2.3 generators.
Detailed studies are performed to correct for biases in the data-based methods due to differences with respect to the MC simulation, in initial- and final-state radiation as well as in jet \pt resolution.
The systematic uncertainties in the jet energy corrections, and their correlations, are provided as a function of $\eta$ and \pt.

The final uncertainties on the jet energy scale are below 3\% across the phase space considered by most analyses ($\pt>30\GeV$ and $\abs{\eta}<5.0$).
In the barrel region we reach an uncertainty below 1\% for $\pt>30\GeV$, when excluding the jet-flavor uncertainties, provided separately
for different jet-flavor mixtures. At its lowest, the core uncertainty (excluding optional time-dependent and flavor systematics) is 0.32\% for jets with
$\pt$ between 165 and 330\GeV, and $\abs{\eta}<0.8$. These results set a new benchmark for jet energy scale determination at hadron colliders.

\begin{acknowledgments}

\hyphenation{Bundes-ministerium Forschungs-gemeinschaft Forschungs-zentren} We congratulate our colleagues in the CERN accelerator departments for the excellent performance of the LHC and thank the technical and administrative staffs at CERN and at other CMS institutes for their contributions to the success of the CMS effort. In addition, we gratefully acknowledge the computing centers and personnel of the Worldwide LHC Computing Grid for delivering so effectively the computing infrastructure essential to our analyses. Finally, we acknowledge the enduring support for the construction and operation of the LHC and the CMS detector provided by the following funding agencies: the Austrian Federal Ministry of Science, Research and Economy and the Austrian Science Fund; the Belgian Fonds de la Recherche Scientifique, and Fonds voor Wetenschappelijk Onderzoek; the Brazilian Funding Agencies (CNPq, CAPES, FAPERJ, and FAPESP); the Bulgarian Ministry of Education and Science; CERN; the Chinese Academy of Sciences, Ministry of Science and Technology, and National Natural Science Foundation of China; the Colombian Funding Agency (COLCIENCIAS); the Croatian Ministry of Science, Education and Sport, and the Croatian Science Foundation; the Research Promotion Foundation, Cyprus; the Secretariat for Higher Education, Science, Technology and Innovation, Ecuador; the Ministry of Education and Research, Estonian Research Council via IUT23-4 and IUT23-6 and European Regional Development Fund, Estonia; the Academy of Finland, Finnish Ministry of Education and Culture, and Helsinki Institute of Physics; the Institut National de Physique Nucl\'eaire et de Physique des Particules~/~CNRS, and Commissariat \`a l'\'Energie Atomique et aux \'Energies Alternatives~/~CEA, France; the Bundesministerium f\"ur Bildung und Forschung, Deutsche Forschungsgemeinschaft, and Helmholtz-Gemeinschaft Deutscher Forschungszentren, Germany; the General Secretariat for Research and Technology, Greece; the National Scientific Research Foundation, and National Innovation Office, Hungary; the Department of Atomic Energy and the Department of Science and Technology, India; the Institute for Studies in Theoretical Physics and Mathematics, Iran; the Science Foundation, Ireland; the Istituto Nazionale di Fisica Nucleare, Italy; the Ministry of Science, ICT and Future Planning, and National Research Foundation (NRF), Republic of Korea; the Lithuanian Academy of Sciences; the Ministry of Education, and University of Malaya (Malaysia); the Mexican Funding Agencies (BUAP, CINVESTAV, CONACYT, LNS, SEP, and UASLP-FAI); the Ministry of Business, Innovation and Employment, New Zealand; the Pakistan Atomic Energy Commission; the Ministry of Science and Higher Education and the National Science Centre, Poland; the Funda\c{c}\~ao para a Ci\^encia e a Tecnologia, Portugal; JINR, Dubna; the Ministry of Education and Science of the Russian Federation, the Federal Agency of Atomic Energy of the Russian Federation, Russian Academy of Sciences, and the Russian Foundation for Basic Research; the Ministry of Education, Science and Technological Development of Serbia; the Secretar\'{\i}a de Estado de Investigaci\'on, Desarrollo e Innovaci\'on and Programa Consolider-Ingenio 2010, Spain; the Swiss Funding Agencies (ETH Board, ETH Zurich, PSI, SNF, UniZH, Canton Zurich, and SER); the Ministry of Science and Technology, Taipei; the Thailand Center of Excellence in Physics, the Institute for the Promotion of Teaching Science and Technology of Thailand, Special Task Force for Activating Research and the National Science and Technology Development Agency of Thailand; the Scientific and Technical Research Council of Turkey, and Turkish Atomic Energy Authority; the National Academy of Sciences of Ukraine, and State Fund for Fundamental Researches, Ukraine; the Science and Technology Facilities Council, UK; the US Department of Energy, and the US National Science Foundation.

Individuals have received support from the Marie-Curie program and the European Research Council and EPLANET (European Union); the Leventis Foundation; the A. P. Sloan Foundation; the Alexander von Humboldt Foundation; the Belgian Federal Science Policy Office; the Fonds pour la Formation \`a la Recherche dans l'Industrie et dans l'Agriculture (FRIA-Belgium); the Agentschap voor Innovatie door Wetenschap en Technologie (IWT-Belgium); the Ministry of Education, Youth and Sports (MEYS) of the Czech Republic; the Council of Science and Industrial Research, India; the HOMING PLUS program of the Foundation for Polish Science, cofinanced from European Union, Regional Development Fund, the Mobility Plus program of the Ministry of Science and Higher Education, the OPUS program contract 2014/13/B/ST2/02543 and contract Sonata-bis DEC-2012/07/E/ST2/01406 of the National Science Center (Poland); the Thalis and Aristeia programs cofinanced by EU-ESF and the Greek NSRF; the National Priorities Research Program by Qatar National Research Fund; the Programa Clar\'in-COFUND del Principado de Asturias; the Rachadapisek Sompot Fund for Postdoctoral Fellowship, Chulalongkorn University and the Chulalongkorn Academic into Its 2nd Century Project Advancement Project (Thailand); and the Welch Foundation, contract C-1845.

\end{acknowledgments}

\clearpage

\bibliography{auto_generated}

\cleardoublepage \appendix\section{The CMS Collaboration \label{app:collab}}\begin{sloppypar}\hyphenpenalty=5000\widowpenalty=500\clubpenalty=5000\input{JME-13-004-authorlist.tex}\end{sloppypar}
\end{document}

%% file: JME-13-004-authorlist.tex
\textbf{Yerevan Physics Institute,  Yerevan,  Armenia}\\*[0pt]
V.~Khachatryan, A.M.~Sirunyan, A.~Tumasyan
\vskip\cmsinstskip
\textbf{Institut f\"{u}r Hochenergiephysik der OeAW,  Wien,  Austria}\\*[0pt]
W.~Adam, E.~Asilar, T.~Bergauer, J.~Brandstetter, E.~Brondolin, M.~Dragicevic, J.~Er\"{o}, M.~Flechl, M.~Friedl, R.~Fr\"{u}hwirth\cmsAuthorMark{1}, V.M.~Ghete, C.~Hartl, N.~H\"{o}rmann, J.~Hrubec, M.~Jeitler\cmsAuthorMark{1}, V.~Kn\"{u}nz, A.~K\"{o}nig, M.~Krammer\cmsAuthorMark{1}, I.~Kr\"{a}tschmer, D.~Liko, T.~Matsushita, I.~Mikulec, D.~Rabady\cmsAuthorMark{2}, B.~Rahbaran, H.~Rohringer, J.~Schieck\cmsAuthorMark{1}, R.~Sch\"{o}fbeck, J.~Strauss, W.~Treberer-Treberspurg, W.~Waltenberger, C.-E.~Wulz\cmsAuthorMark{1}
% ...
\vskip\cmsinstskip
% ...
1:~~Also at Vienna University of Technology, Vienna, Austria\\
2:~~Also at CERN, European Organization for Nuclear Research, Geneva, Switzerland\\
% ...


%% file: JME-13-004_temp.bbl
\providecommand{\href}[2]{#2}\begingroup\raggedright\begin{thebibliography}{10}%
\makeatletter
\providecommand{\hrefCMSnoop }[0]{\@secondoftwo}%
\makeatother
\providecommand{\doi}{\texttt{doi:}\begingroup \urlstyle{tt}\Url}

\bibitem{Chatrchyan:2012bja}
\hrefCMSnoop {}{{CMS Collaboration}, ``{Measurements of differential jet cross
  sections in proton-proton collisions at $\sqrt{s}=7\TeV$ with the CMS
  detector}'',} \textit{ Phys. Rev. D} \textbf{ 87} (2013) 112002,
  \href{http://dx.doi.org/10.1103/PhysRevD.87.112002}{\doi{10.1103/PhysRevD.87.112002}},
  \href{http://www.arXiv.org/abs/1212.6660}{\texttt{arXiv:1212.6660}}.
Erratum: \DOI{10.1103/PhysRevD.87.119902}.

\bibitem{CMS:2011ab}
\hrefCMSnoop {}{{CMS Collaboration}, ``{measurement of the inclusive jet cross
  section in pp collisions at $\sqrt{s}=7\TeV$}'',} \textit{ Phys. Rev. Lett.}
  \textbf{ 107} (2011) 132001,
  \href{http://dx.doi.org/10.1103/PhysRevLett.107.132001}{\doi{10.1103/PhysRevLett.107.132001}},
\href{http://www.arXiv.org/abs/1106.0208}{\texttt{arXiv:1106.0208}}.

\bibitem{Chatrchyan:2011qta}
\hrefCMSnoop {}{{CMS Collaboration}, ``{Measurement of the differential dijet
  production cross section in proton-proton collisions at $\sqrt{s}=7\TeV$}'',}
  \textit{ Phys. Lett. B} \textbf{ 700} (2011) 187,
  \href{http://dx.doi.org/10.1016/j.physletb.2011.05.027}{\doi{10.1016/j.physletb.2011.05.027}},
\href{http://www.arXiv.org/abs/1104.1693}{\texttt{arXiv:1104.1693}}.

\bibitem{Chatrchyan:2014gia}
\hrefCMSnoop {}{{CMS Collaboration}, ``{Measurement of the ratio of inclusive
  jet cross sections using the anti-$\kt$ algorithm with radius parameters
  $R=0.5$ and 0.7 in pp collisions at $\sqrt{s}=7\TeV$}'',} \textit{ Phys. Rev.
  D} \textbf{ 90} (2014) 072006,
  \href{http://dx.doi.org/10.1103/PhysRevD.90.072006}{\doi{10.1103/PhysRevD.90.072006}},
\href{http://www.arXiv.org/abs/1406.0324}{\texttt{arXiv:1406.0324}}.

\bibitem{Khachatryan:2015xwa}
\hrefCMSnoop {}{{CMS Collaboration}, ``{Distributions of topological
  observables in inclusive three- and four-jet events in pp collisions at
  $\sqrt{s}=7\TeV$}'',} \textit{ Eur. Phys. J. C} \textbf{ 75} (2015) 302,
  \href{http://dx.doi.org/10.1140/epjc/s10052-015-3491-9}{\doi{10.1140/epjc/s10052-015-3491-9}},
\href{http://www.arXiv.org/abs/1502.04785}{\texttt{arXiv:1502.04785}}.

\bibitem{Chatrchyan:2013xza}
\hrefCMSnoop {}{{CMS Collaboration}, ``{Measurement of the top-quark mass in
  all-jets $\ttbar$ events in pp collisions at $\sqrt{s}=7\TeV$}'',} \textit{
  Eur. Phys. J. C} \textbf{ 74} (2014) 2758,
  \href{http://dx.doi.org/10.1140/epjc/s10052-014-2758-x}{\doi{10.1140/epjc/s10052-014-2758-x}},
\href{http://www.arXiv.org/abs/1307.4617}{\texttt{arXiv:1307.4617}}.

\bibitem{Chatrchyan:2013haa}
\hrefCMSnoop {}{{CMS Collaboration}, ``{Determination of the top-quark pole
  mass and strong coupling constant from the $\ttbar$ production cross section
  in pp collisions at $\sqrt{s}=7\TeV$}'',} \textit{ Phys. Lett. B} \textbf{
  728} (2014) 496,
  \href{http://dx.doi.org/10.1016/j.physletb.2013.12.009}{\doi{10.1016/j.physletb.2013.12.009}},
  \href{http://www.arXiv.org/abs/1307.1907}{\texttt{arXiv:1307.1907}}.
Erratum: \DOI{10.1016/j.physletb.2014.08.040}.

\bibitem{Chatrchyan:2012cz}
\hrefCMSnoop {}{{CMS Collaboration}, ``{Measurement of the top-quark mass in
  $\ttbar$ events with lepton+jets final states in pp collisions at
  $\sqrt{s}=7\TeV$}'',} \textit{ JHEP} \textbf{ 12} (2012) 105,
  \href{http://dx.doi.org/10.1007/JHEP12(2012)105}{\doi{10.1007/JHEP12(2012)105}},
\href{http://www.arXiv.org/abs/1209.2319}{\texttt{arXiv:1209.2319}}.

\bibitem{Khachatryan:2015hba}
\hrefCMSnoop {}{{CMS Collaboration}, ``{Measurement of the top quark mass using
  proton-proton data at $\sqrt{s}=7$ and 8\TeV}'',} \textit{ Phys. Rev. D}
  \textbf{ 93} (2016) 072004,
  \href{http://dx.doi.org/10.1103/PhysRevD.93.072004}{\doi{10.1103/PhysRevD.93.072004}},
\href{http://www.arXiv.org/abs/1509.04044}{\texttt{arXiv:1509.04044}}.

\bibitem{CMS-PAS-PFT-09-001}
\href {http://cdsweb.cern.ch/record/1194487}{{CMS Collaboration},
  ``Particle--flow event reconstruction in {CMS} and performance for jets,
  taus, and {\MET}'',} CMS Physics Analysis Summary CMS-PAS-PFT-09-001, 2009.

\bibitem{CMS-PAS-PFT-10-001}
\href {http://cdsweb.cern.ch/record/1247373}{{CMS Collaboration},
  ``Commissioning of the particle--flow event reconstruction with the first
  {LHC} collisions recorded in the {CMS} detector'',} CMS Physics Analysis
  Summary CMS-PAS-PFT-10-001, 2010.

\bibitem{Cacciari:2008gp}
\hrefCMSnoop {}{M.~Cacciari, G.~P. Salam, and G.~Soyez, ``{The anti-$k_t$ jet
  clustering algorithm}'',} \textit{ JHEP} \textbf{ 04} (2008) 063,
  \href{http://dx.doi.org/10.1088/1126-6708/2008/04/063}{\doi{10.1088/1126-6708/2008/04/063}},
\href{http://www.arXiv.org/abs/0802.1189}{\texttt{arXiv:0802.1189}}.

\bibitem{JEC_JINST}
\hrefCMSnoop {}{{CMS Collaboration}, ``{Determination of jet energy calibration
  and transverse momentum resolution in CMS}'',} \textit{ JINST} \textbf{ 6}
  (2011) P11002,
  \href{http://dx.doi.org/10.1088/1748-0221/6/11/P11002}{\doi{10.1088/1748-0221/6/11/P11002}},
\href{http://www.arXiv.org/abs/1107.4277}{\texttt{arXiv:1107.4277}}.

\bibitem{Aad:2011he}
\hrefCMSnoop {}{{ATLAS Collaboration}, ``{Jet energy measurement with the ATLAS
  detector in proton-proton collisions at $\sqrt{s}=7\TeV$}'',} \textit{ Eur.
  Phys. J. C} \textbf{ 73} (2013) 2304,
  \href{http://dx.doi.org/10.1140/epjc/s10052-013-2304-2}{\doi{10.1140/epjc/s10052-013-2304-2}},
\href{http://www.arXiv.org/abs/1112.6426}{\texttt{arXiv:1112.6426}}.

\bibitem{CDF_JEC}
\hrefCMSnoop {}{{CDF} Collaboration, ``{Determination of the jet energy scale
  at the Collider Detector at Fermilab}'',} \textit{ Nucl. Instrum. Meth. A}
  \textbf{ 566} (2006) 375,
  \href{http://dx.doi.org/10.1016/j.nima.2006.05.269}{\doi{10.1016/j.nima.2006.05.269}},
\href{http://www.arXiv.org/abs/hep-ex/0510047}{\texttt{arXiv:hep-ex/0510047}}.

\bibitem{D0_JEC}
\hrefCMSnoop {}{{D0} Collaboration, ``{Jet energy scale determination in the D0
  experiment}'',} \textit{ Nucl. Instrum. Meth. A} \textbf{ 763} (2014) 442,
  \href{http://dx.doi.org/10.1016/j.nima.2014.05.044}{\doi{10.1016/j.nima.2014.05.044}},
\href{http://www.arXiv.org/abs/1312.6873}{\texttt{arXiv:1312.6873}}.

\bibitem{ATLAS_JEC}
\hrefCMSnoop {}{{ATLAS Collaboration}, ``{Jet energy measurement and its
  systematic uncertainty in proton-proton collisions at $\sqrt{s}=7\TeV$ with
  the ATLAS detector}'',} \textit{ Eur. Phys. J. C} \textbf{ 75} (2015) 17,
  \href{http://dx.doi.org/10.1140/epjc/s10052-014-3190-y}{\doi{10.1140/epjc/s10052-014-3190-y}},
\href{http://www.arXiv.org/abs/1406.0076}{\texttt{arXiv:1406.0076}}.

\bibitem{Agostinelli:2002hh}
\hrefCMSnoop {}{{GEANT4} Collaboration, ``{GEANT4---a simulation toolkit}'',}
  \textit{ Nucl. Instrum. Meth. A} \textbf{ 506} (2003) 250,
\href{http://dx.doi.org/10.1016/S0168-9002(03)01368-8}{\doi{10.1016/S0168-9002(03)01368-8}}.

\bibitem{Sjostrand:2006za}
\hrefCMSnoop {}{T.~Sj{\" o}strand, S.~Mrenna, and P.~Skands, ``{PYTHIA} 6.4
  physics and manual'',} \textit{ JHEP} \textbf{ 05} (2006) 026,
  \href{http://dx.doi.org/10.1088/1126-6708/2006/05/026}{\doi{10.1088/1126-6708/2006/05/026}},
\href{http://www.arXiv.org/abs/hep-ph/0603175}{\texttt{arXiv:hep-ph/0603175}}.

\bibitem{GEN-14-001}
\hrefCMSnoop {}{{CMS Collaboration}, ``{Event generator tunes obtained from
  underlying event and multiparton scattering measurements}'',} \textit{ Eur.
  Phys. J. C} \textbf{ 76} (2016) 155,
  \href{http://dx.doi.org/10.1140/epjc/s10052-016-3988-x}{\doi{10.1140/epjc/s10052-016-3988-x}},
\href{http://www.arXiv.org/abs/1512.00815}{\texttt{arXiv:1512.00815}}.

\bibitem{Bahr:2008pv}
M.~B{\"a}hr\hrefCMSnoop {}{ {et~al.}, ``Herwig++ physics and manual'',}
  \textit{ Eur. Phys. J. C} \textbf{ 58} (2008) 639,
  \href{http://dx.doi.org/10.1140/epjc/s10052-008-0798-9}{\doi{10.1140/epjc/s10052-008-0798-9}},
\href{http://www.arXiv.org/abs/0803.0883}{\texttt{arXiv:0803.0883}}.

\bibitem{Gieseke:2012ft}
\hrefCMSnoop {}{S.~Gieseke, C.~Rohr, and A.~Siodmok, ``{Colour reconnections in
  Herwig++}'',} \textit{ Eur. Phys. J. C} \textbf{ 72} (2012) 2225,
  \href{http://dx.doi.org/10.1140/epjc/s10052-012-2225-5}{\doi{10.1140/epjc/s10052-012-2225-5}},
\href{http://www.arXiv.org/abs/1206.0041}{\texttt{arXiv:1206.0041}}.

\bibitem{Giammanco:2014bza}
\hrefCMSnoop {}{A.~Giammanco, ``{The Fast Simulation of the CMS Experiment}'',}
  \textit{ J. Phys. Conf. Ser.} \textbf{ 513} (2014) 022012,
\href{http://dx.doi.org/10.1088/1742-6596/513/2/022012}{\doi{10.1088/1742-6596/513/2/022012}}.

\bibitem{CMSdetector}
\hrefCMSnoop {}{{CMS Collaboration}, ``The {CMS} experiment at the {CERN}
  {LHC}'',} \textit{ JINST} \textbf{ 3} (2008) S08004,
\href{http://dx.doi.org/10.1088/1748-0221/3/08/S08004}{\doi{10.1088/1748-0221/3/08/S08004}}.

\bibitem{madgraph}
J.~Alwall\hrefCMSnoop {}{ {et~al.}, ``{MadGraph 5: going beyond}'',} \textit{
  JHEP} \textbf{ 06} (2011) 128,
  \href{http://dx.doi.org/10.1007/JHEP06(2011)128}{\doi{10.1007/JHEP06(2011)128}},
\href{http://www.arXiv.org/abs/1106.0522}{\texttt{arXiv:1106.0522}}.

\bibitem{Frixione:2007vw}
\hrefCMSnoop {}{S.~Frixione, P.~Nason, and C.~Oleari, ``{Matching NLO QCD
  computations with Parton Shower simulations: the POWHEG method}'',} \textit{
  JHEP} \textbf{ 11} (2007) 070,
  \href{http://dx.doi.org/10.1088/1126-6708/2007/11/070}{\doi{10.1088/1126-6708/2007/11/070}},
\href{http://www.arXiv.org/abs/0709.2092}{\texttt{arXiv:0709.2092}}.

\bibitem{Chatrchyan:2014fea}
\hrefCMSnoop {}{{CMS Collaboration}, ``{Description and performance of track
  and primary-vertex reconstruction with the CMS tracker}'',} \textit{ JINST}
  \textbf{ 9} (2014) P10009,
  \href{http://dx.doi.org/10.1088/1748-0221/9/10/P10009}{\doi{10.1088/1748-0221/9/10/P10009}},
\href{http://www.arXiv.org/abs/1405.6569}{\texttt{arXiv:1405.6569}}.

\bibitem{CMS-PAS-JME-10-003}
\href {http://cdsweb.cern.ch/record/1279362}{{CMS Collaboration}, ``Jet
  performance in pp collisions at $\sqrt{s}=7\TeV$'',} CMS Physics Analysis
  Summary CMS-PAS-JME-10-003, 2010.

\bibitem{cmsmuons}
\hrefCMSnoop {}{{CMS Collaboration}, ``{Performance of CMS muon reconstruction
  in pp collision events at $\sqrt{s}=7\TeV$}'',} \textit{ JINST} \textbf{ 7}
  (2012) P10002,
  \href{http://dx.doi.org/10.1088/1748-0221/7/10/P10002}{\doi{10.1088/1748-0221/7/10/P10002}},
\href{http://www.arXiv.org/abs/1206.4071}{\texttt{arXiv:1206.4071}}.

\bibitem{Khachatryan:2015hwa}
\hrefCMSnoop {}{{CMS Collaboration}, ``{Performance of electron reconstruction
  and selection with the CMS detector in proton-proton collisions at
  $\sqrt{s}=8\TeV$}'',} \textit{ JINST} \textbf{ 10} (2015) P06005,
  \href{http://dx.doi.org/10.1088/1748-0221/10/06/P06005}{\doi{10.1088/1748-0221/10/06/P06005}},
\href{http://www.arXiv.org/abs/1502.02701}{\texttt{arXiv:1502.02701}}.

\bibitem{btag}
\hrefCMSnoop {}{{CMS Collaboration}, ``{Identification of b-quark jets with the
  CMS experiment}'',} \textit{ JINST} \textbf{ 8} (2013) P04013,
  \href{http://dx.doi.org/10.1088/1748-0221/8/04/P04013}{\doi{10.1088/1748-0221/8/04/P04013}},
\href{http://www.arXiv.org/abs/1211.4462}{\texttt{arXiv:1211.4462}}.

\bibitem{Khachatryan:2015iwa}
\hrefCMSnoop {}{{CMS Collaboration}, ``{Performance of photon reconstruction
  and identification with the CMS detector in proton-proton collisions at
  $\sqrt{s}=8\TeV$}'',} \textit{ JINST} \textbf{ 10} (2015) P08010,
  \href{http://dx.doi.org/10.1088/1748-0221/10/08/P08010}{\doi{10.1088/1748-0221/10/08/P08010}},
\href{http://www.arXiv.org/abs/1502.02702}{\texttt{arXiv:1502.02702}}.

\bibitem{fastjet}
\hrefCMSnoop {}{M.~Cacciari, G.~P. Salam, and G.~Soyez, ``{FastJet} user
  manual'',} \textit{ Eur. Phys. J. C} \textbf{ 72} (2012) 1896,
  \href{http://dx.doi.org/10.1140/epjc/s10052-012-1896-2}{\doi{10.1140/epjc/s10052-012-1896-2}},
\href{http://www.arXiv.org/abs/1111.6097}{\texttt{arXiv:1111.6097}}.

\bibitem{fastjetB}
\hrefCMSnoop {}{M.~Cacciari and G.~P. Salam, ``{Dispelling the $N^{3}$ myth for
  the $k_t$ jet-finder}'',} \textit{ Phys. Lett. B} \textbf{ 641} (2006) 57,
  \href{http://dx.doi.org/10.1016/j.physletb.2006.08.037}{\doi{10.1016/j.physletb.2006.08.037}},
\href{http://www.arXiv.org/abs/hep-ph/0512210}{\texttt{arXiv:hep-ph/0512210}}.

\bibitem{KT1}
\hrefCMSnoop {}{S.~Catani, Y.~L. Dokshitzer, and B.~R. Webber, ``The
  $k_{\perp}$-clustering algorithm for jets in deep inelastic scattering and
  hadron collisions'',} \textit{ Phys. Lett. B} \textbf{ 285} (1992) 291,
  \href{http://dx.doi.org/10.1016/0370-2693(92)91467-N}{\doi{10.1016/0370-2693(92)91467-N}}.

\bibitem{KT2}
\hrefCMSnoop {}{S.~Catani, Y.~L. Dokshitzer, M.~H. Seymour, and B.~R. Webber,
  ``Longitudinally invariant $k_{\perp}$-clustering algorithms for
  hadron-hadron collisions'',} \textit{ Nucl. Phys. B} \textbf{ 406} (1993)
  187,
  \href{http://dx.doi.org/10.1016/0550-3213(93)90166-M}{\doi{10.1016/0550-3213(93)90166-M}}.

\bibitem{KT3}
\hrefCMSnoop {}{S.~D. Ellis and D.~E. Soper, ``Successive combination jet
  algorithm for hadron collisions'',} \textit{ Phys. Rev. D} \textbf{ 48}
  (1993) 3160,
  \href{http://dx.doi.org/10.1103/PhysRevD.48.3160}{\doi{10.1103/PhysRevD.48.3160}}.

\bibitem{Cacciari:2007fd}
\hrefCMSnoop {}{M.~Cacciari and G.~P. Salam, ``Pileup subtraction using jet
  areas'',} \textit{ Phys. Lett. B} \textbf{ 659} (2008) 119,
  \href{http://dx.doi.org/10.1016/j.physletb.2007.09.077}{\doi{10.1016/j.physletb.2007.09.077}},
\href{http://www.arXiv.org/abs/0707.1378}{\texttt{arXiv:0707.1378}}.

\bibitem{Chatrchyan:2012nj}
\hrefCMSnoop {}{{CMS Collaboration}, ``{Measurement of the inelastic
  proton-proton cross section at $\sqrt{s}=7\TeV$}'',} \textit{ Phys. Lett. B}
  \textbf{ 722} (2013) 5,
  \href{http://dx.doi.org/10.1016/j.physletb.2013.03.024}{\doi{10.1016/j.physletb.2013.03.024}},
\href{http://www.arXiv.org/abs/1210.6718}{\texttt{arXiv:1210.6718}}.

\bibitem{CMS-PAS-LUM-13-001}
\href {http://cdsweb.cern.ch/record/1598864}{{CMS Collaboration}, ``CMS
  luminosity based on pixel cluster counting --- Summer 2013 update'',} CMS
  Physics Analysis Summary CMS-PAS-LUM-13-001, 2013.

\bibitem{phystdr1}
\hrefCMSnoop {}{{CMS Collaboration}, ``{CMS} technical design report, volume
  {II}: {Physics} performance'',} \textit{ J. Phys. G} \textbf{ 34} (2007) 995,
\href{http://dx.doi.org/10.1088/0954-3899/34/6/S01}{\doi{10.1088/0954-3899/34/6/S01}}.

\bibitem{dNdeta}
\hrefCMSnoop {}{{CMS Collaboration}, ``{Transverse-momentum and pseudorapidity
  distributions of charged hadrons in pp collisions at $\sqrt{s}=7\TeV$}'',}
  \textit{ Phys. Rev. Lett.} \textbf{ 105} (2010) 022002,
  \href{http://dx.doi.org/10.1103/PhysRevLett.105.022002}{\doi{10.1103/PhysRevLett.105.022002}},
\href{http://www.arXiv.org/abs/1005.3299}{\texttt{arXiv:1005.3299}}.

\bibitem{puppi}
\hrefCMSnoop {}{D.~Bertolini, P.~Harris, M.~Low, and N.~Tran, ``{Pileup per
  particle identification}'',} \textit{ JHEP} \textbf{ 10} (2014) 59,
  \href{http://dx.doi.org/10.1007/JHEP10(2014)059}{\doi{10.1007/JHEP10(2014)059}},
\href{http://www.arXiv.org/abs/1407.6013}{\texttt{arXiv:1407.6013}}.

\bibitem{CMS-PAS-JME-13-005}
\href {http://cdsweb.cern.ch/record/1581583}{{CMS Collaboration}, ``Pileup jet
  identification'',} CMS Physics Analysis Summary CMS-PAS-JME-13-005, 2013.

\bibitem{Chatrchyan:2012tt}
\hrefCMSnoop {}{{CMS Collaboration}, ``{Measurement of the underlying event
  activity in pp collisions at $\sqrt{s} = 0.9$ and 7\TeV with the novel
  jet-area/median approach}'',} \textit{ JHEP} \textbf{ 08} (2012) 130,
  \href{http://dx.doi.org/10.1007/JHEP08(2012)130}{\doi{10.1007/JHEP08(2012)130}},
\href{http://www.arXiv.org/abs/1207.2392}{\texttt{arXiv:1207.2392}}.

\bibitem{CMS-DP-2012-006}
\href {https://cds.cern.ch/record/1454659}{{CMS Collaboration}, ``{Jet energy
  scale performance in 2011}'',} CMS Detector Performance note CMS-DP-2012-006,
  2012.

\bibitem{CMS-PAS-JME-10-008}
\href {https://cds.cern.ch/record/1308178}{{CMS Collaboration},
  ``{Single-particle response in the CMS calorimeters}'',} CMS Physics Analysis
  Summary CMS-PAS-JME-10-008, 2010.

\bibitem{Chatrchyan:2013dga}
\hrefCMSnoop {}{{CMS Collaboration}, ``{Energy calibration and resolution of
  the CMS electromagnetic calorimeter in pp collisions at $\sqrt{s}=7\TeV$}'',}
  \textit{ JINST} \textbf{ 8} (2013) P09009,
  \href{http://dx.doi.org/10.1088/1748-0221/8/09/P09009}{\doi{10.1088/1748-0221/8/09/P09009}},
\href{http://www.arXiv.org/abs/1306.2016}{\texttt{arXiv:1306.2016}}.

\bibitem{CMS-PAS-JME-13-002}
\href {https://cds.cern.ch/record/1599732}{{CMS Collaboration}, ``{Performance
  of quark/gluon discrimination in 8\TeV pp data}'',} CMS Physics Analysis
  Summary CMS-PAS-JME-13-002, 2013.

\bibitem{CMS-PAS-JME-13-001}
\href {https://cds.cern.ch/record/1951028}{{CMS Collaboration}, ``{Calculation
  of Residual Energy Correction for b Jets Using Z+b Events in 8 TeV pp
  Collisions}'',} CMS Physics Analysis Summary CMS-PAS-JME-13-001, 2014.

\bibitem{crystalball}
\href {http://www.slac.stanford.edu/pubs/slacreports/slac-r-236.html}{M.~J.
  Oreglia, ``A study of the reactions $\psi^\prime \to \gamma \gamma \psi$''}.
\newblock PhD thesis, Stanford University, 1980.
\newblock {SLAC} Report {SLAC-R-236}, see Appendix {D}.

\bibitem{cteq}
H.~L. Lai\hrefCMSnoop {}{ {et~al.}, ``{Global QCD analysis and the CTEQ parton
  distributions}'',} \textit{ Phys. Rev. D} \textbf{ 51} (1995) 4763,
  \href{http://dx.doi.org/10.1103/PhysRevD.51.4763}{\doi{10.1103/PhysRevD.51.4763}},
\href{http://www.arXiv.org/abs/hep-ph/9410404}{\texttt{arXiv:hep-ph/9410404}}.

\bibitem{Groom:2007zz}
\hrefCMSnoop {}{D.~E. Groom, ``{A simplistic view of hadron calorimetry}'',}
  \textit{ AIP Conf. Proc.} \textbf{ 896} (2007) 137,
\href{http://dx.doi.org/10.1063/1.2720464}{\doi{10.1063/1.2720464}}.

\end{thebibliography}\endgroup
